%% file: HIG-20-003_temp.tex
\pdfoutput=1
\documentclass[11pt,twoside,a4paper,cmspaper,final,collab]{cms-tdr}
\def\svnVersion{88b1c05}\def\svnDate{2022/05/25}\def\cmsCernNoTag{CERN-EP-2021-273}\def\cmsCernDate{\today}\def\cmsMessage{Published in Physical Review D as \href{http://dx.doi.org/10.1103/PhysRevD.105.092007}{\doi{10.1103/PhysRevD.105.092007}.}}
\begin{document}\cmsNoteHeader{HIG-20-003}

\newlength\cmsFigWidth
\ifthenelse{\boolean{cms@external}}{\setlength\cmsFigWidth{0.95\columnwidth}}{\setlength\cmsFigWidth{0.65\textwidth}}
\ifthenelse{\boolean{cms@external}}{\providecommand{\cmsLeft}{upper\xspace}}{\providecommand{\cmsLeft}{left\xspace}}
\ifthenelse{\boolean{cms@external}}{\providecommand{\cmsRight}{lower\xspace}}{\providecommand{\cmsRight}{right\xspace}}

\ifthenelse{\boolean{cms@external}}{\providecommand{\cmsClearpage}{}}{\providecommand{\cmsClearpage}{\relax}}

\ifthenelse{\boolean{cms@external}}{\providecommand{\cmsTable}[1]{#1}}{\providecommand{\cmsTable}[1]{\resizebox{\textwidth}{!}{#1}}}
\newlength\cmsTabSkip\setlength{\cmsTabSkip}{1ex}

\newcommand{\HAWK}{{\textsc{hawk}}\xspace}
\newcommand{\Vjets}{\ensuremath{\PV\text{+jets}}\xspace}
\newcommand{\Zmm}{\ensuremath{\PZ(\PGm\PGm)}\xspace}
\newcommand{\Zee}{\ensuremath{\PZ(\Pe\Pe)}\xspace}
\newcommand{\Zvv}{\ensuremath{\PZ(\PGn\PAGn)}\xspace}
\newcommand{\Wlv}{\ensuremath{\PW(\Pell\PGn)}\xspace}
\newcommand{\Wmn}{\ensuremath{\PW(\PGm\PGn)}\xspace}
\newcommand{\Wen}{\ensuremath{\PW(\Pe\PGn)}\xspace}
\newcommand{\Zmmjets}{\ensuremath{\PZ(\PGm\PGm)\text{+jets}}\xspace}
\newcommand{\Zeejets}{\ensuremath{\PZ(\Pe\Pe)\text{+jets}}\xspace}
\newcommand{\Zlljets}{\ensuremath{\PZ(\Pell\Pell)\text{+jets}}\xspace}
\newcommand{\Zlljet}{\ensuremath{\PZ(\Pell\Pell)\text{+jet}}\xspace}
\newcommand{\Zjet}{\ensuremath{\PZ\text{+jet}}\xspace}
\newcommand{\Zjets}{\ensuremath{\PZ\text{+jets}}\xspace}
\newcommand{\Wjets}{\ensuremath{\PW\text{+jets}}\xspace}
\newcommand{\Zvvjets}{\ensuremath{\PZ(\PGn\PAGn)\text{+jets}}\xspace}
\newcommand{\Wlvjets}{\ensuremath{\PW(\Pell\PGn)\text{+jets}}\xspace}
\newcommand{\Wmvjets}{\ensuremath{\PW(\PGm\PGn)\text{+jets}}\xspace}
\newcommand{\Wevjets}{\ensuremath{\PW(\Pe\PGn)\text{+jets}}\xspace}
\newcommand{\phojets}{\ensuremath{\Pgg\text{+jets}}\xspace}
\newcommand{\phojet}{\ensuremath{\Pgg\text{+jet}}\xspace}
\newcommand{\kappaz}{\ensuremath{\kappa\smash[b]{^{\PGn\PAGn}}}\xspace}
\newcommand{\kappazi}{\ensuremath{\kappa_{i}\smash[b]{^{\PGn\PAGn}}}\xspace}
\newcommand{\brhinv}{\ensuremath{{\mathcal{B}(\PH \to \text{inv})}}\xspace}
\newcommand{\brinv}{\ensuremath{{\mathcal{B}(\PH \to \text{inv})}}\xspace}
\newcommand{\hinv}{\ensuremath{\PH \to \text{inv}}\xspace}
\newcommand{\mjj}{\ensuremath{m_{\mathrm{jj}}}\xspace}
\newcommand{\detajj}{\ensuremath{\Delta\eta_{\mathrm{jj}}}\xspace}
\newcommand{\dphijj}{\ensuremath{\Delta\phi_{\mathrm{jj}}}\xspace}
\newcommand{\dphijmet}{\ensuremath{\Delta\phi(\ptvecmiss,\vec{p}_{\mathrm{T}}^{\kern1pt\text{jet}})}\xspace}
\newcommand{\sigmabr}{\ensuremath{(\sigma_{\PH}/\sigma_{\PH}^{\mathrm{SM}}) \brhinv}\xspace}
\newcommand{\NNPDF}{\textsc{nnpdf}\xspace}
\newcommand{\sieie}{\ensuremath{\sigma_{i\eta i\eta}}\xspace}
\newcommand{\hltptmiss}{\ensuremath{p_{\mathrm{T, HLT}}^\text{miss}}\xspace}
\newcommand{\tkptmiss}{\ensuremath{p_{\mathrm{T, trk}}^\text{miss}}\xspace}
\newcommand{\loneptmiss}{\ensuremath{p_{\mathrm{T, L1}}^\text{miss}}\xspace}
\newcommand{\lonept}{\ensuremath{p_{\mathrm{T}}^{\mathrm{L1}}}\xspace}
\newcommand{\lonemjj}{\ensuremath{m_{\mathrm{jj}}^{\mathrm{L1}}}\xspace}

\newcommand{\brexp}{\ensuremath{12}\xspace}
\newcommand{\brexpRun}{\ensuremath{10}\xspace}
\newcommand{\brobs}{\ensuremath{18}\xspace}
\newcommand{\brobsRun}{\ensuremath{18}\xspace}

\newcommand{\ggH}{\ensuremath{\Pg\Pg\PH}\xspace}
\newcommand{\vh}{\ensuremath{\PV\PH}\xspace}
\newcommand{\tth}{\ensuremath{\PQt{}\PQt{}\PH}\xspace}
\newcommand{\vbf}{VBF\xspace}
\newcommand{\pp}{\ensuremath{\Pp\Pp}\xspace}

\newcommand{\lumiRunOne}{19.7\fbinv}
\newcommand{\lumiRunTwo}{140\fbinv}
\newcommand{\lumiSevEi}{101\fbinv}

\ifthenelse{\boolean{cms@external}}
{\providecommand{\suppMaterial}{the supplemental material, additional figures and tables~\cite{suppMatBib}}}
{\providecommand{\suppMaterial}{Appendix~\ref{app:suppMat}\nocite{suppMatBib}}}

\ifthenelse{\boolean{cms@external}}
{\providecommand{\suppMaterialbis}{the supplemental material~\cite{suppMatBib}}}
{\providecommand{\suppMaterialbis}{Appendix~\ref{app:suppMat}}}

\cmsNoteHeader{HIG-20-003}
\title{Search for invisible decays of the Higgs boson produced via vector boson fusion in proton-proton collisions at \texorpdfstring{$\sqrt{s}=13\TeV$}{sqrt(s) = 13 TeV}}

\date{\today}

\abstract{
A search for invisible decays of the Higgs boson produced via vector boson fusion (VBF) has been performed with \lumiSevEi of proton-proton collisions delivered by the LHC at $\sqrt{s}=13\TeV$ and collected by the CMS detector in 2017 and 2018. The sensitivity to the VBF production mechanism is enhanced by constructing two analysis categories, one based on missing transverse momentum, and a second based on the properties of jets. In addition to control regions with \PZ and \PW boson candidate events, a highly populated control region, based on the production of a photon in association with jets, is used to constrain the dominant irreducible background from the invisible decay of a \PZ boson produced in association with jets. The results of this search are combined with all previous measurements in the VBF topology, based on data collected in 2012 (at $\sqrt{s}=8\TeV$), 2015, and 2016, corresponding to integrated luminosities of 19.7, 2.3, and 36.3\fbinv, respectively. The observed (expected) upper limit on the invisible branching fraction of the Higgs boson is found to be 0.\brobsRun (0.\brexpRun) at the 95\% confidence level, assuming the standard model production cross section. The results are also interpreted in the context of Higgs-portal models.
}

\hypersetup{pdfauthor={CMS Collaboration},pdftitle={Search for invisible decays of the Higgs boson produced via vector boson fusion in proton-proton collisions at sqrt(s)=13 TeV.},pdfsubject={CMS},pdfkeywords={CMS,  Higgs boson, dark matter}}

\maketitle

\section{Introduction}

A particle compatible with the standard model (SM) Higgs boson
(\PH)~\cite{PhysRevLett.13.321, Higgs:1964ia, PhysRevLett.13.508,
PhysRevLett.13.585, Higgs:1966ev, Kibble:1967sv} was discovered at the
CERN LHC in
2012~\cite{Aad:2012tfa,Chatrchyan:2012xdj,Chatrchyan:2013lba}. Since
then, extensive studies of this particle have been performed with data taken at
$\sqrt{s}=7$, 8, and 13\TeV, in particular
to understand how it couples to other SM particles.

In the SM, the branching fraction to invisible final states, \brhinv,
is only about 0.1\%~\cite{Dittmaier:2011ti}, from the decay of the
Higgs boson via $\PZ\PZ^* \to 4\PGn$. Several theories beyond the SM
(BSM), however, predict much higher values of \brhinv
(\cite{Belanger:2001am,Datta:2004jg,Dominici:2009pq,SHROCK1982250}, as
well as Ref.~\cite{Argyropoulos:2021sav} and references therein). In
particular, in Higgs portal models, the Higgs boson acts as the
mediator between SM particles and dark matter
(DM)~\cite{Djouadi:2011aa,Baek:2012se,Djouadi:2012zc,Beniwal:2015sdl},
strongly enhancing \brhinv.

Direct searches for \hinv decays have already been performed by the
ATLAS~\cite{Aaboud:2019rtt,ATLAS:2021gcn} and
CMS~\cite{Sirunyan:2018owy,CMS:2021far,CMS:2020ulv} Collaborations
using data collected at $\sqrt{s}=7$, 8, and 13\TeV, and combining the
three main Higgs boson production modes, namely gluon-gluon fusion
(\ggH), production of a Higgs boson in association with vector bosons
(\vh, with $\PV=\PWpm$ or \PZ), and vector boson fusion
(\vbf). Assuming SM production of the Higgs boson, the best observed
(expected) 95\% confidence level (\CL) upper limits on \brhinv are set
at 0.19 (0.15) by CMS, using data collected at $\sqrt{s}=7$, 8, and
13\TeV, and at 0.26 (0.17) by ATLAS using data collected at
13\TeV. In both cases, the data at 13\TeV were collected in
2016. Combining the latest CMS constraints on both visible and
invisible decays within the $\kappa$
framework~\cite{Sirunyan:2018koj}, the upper bound on \brhinv is 0.22
at the 95\% \CL, using only the data set collected at 13\TeV in 2016.

Thanks to its large production cross section~\cite{deFlorian:2016spz}
and distinctive event topology, the \vbf production mechanism drives
the overall sensitivity in the direct search for invisible decays of
the Higgs boson. This paper focuses exclusively on the search
for \hinv in the VBF production mode using the LHC proton-proton (\pp)
collision data set collected during 2017--2018, corresponding to an
integrated luminosity of up to \lumiSevEi, and on the combination of
this search with analyses performed on previous data
sets~\cite{Khachatryan:2016whc,Sirunyan:2018owy}.

\begin{figure*}[htb!]
\centering
\includegraphics[width=0.28\textwidth]{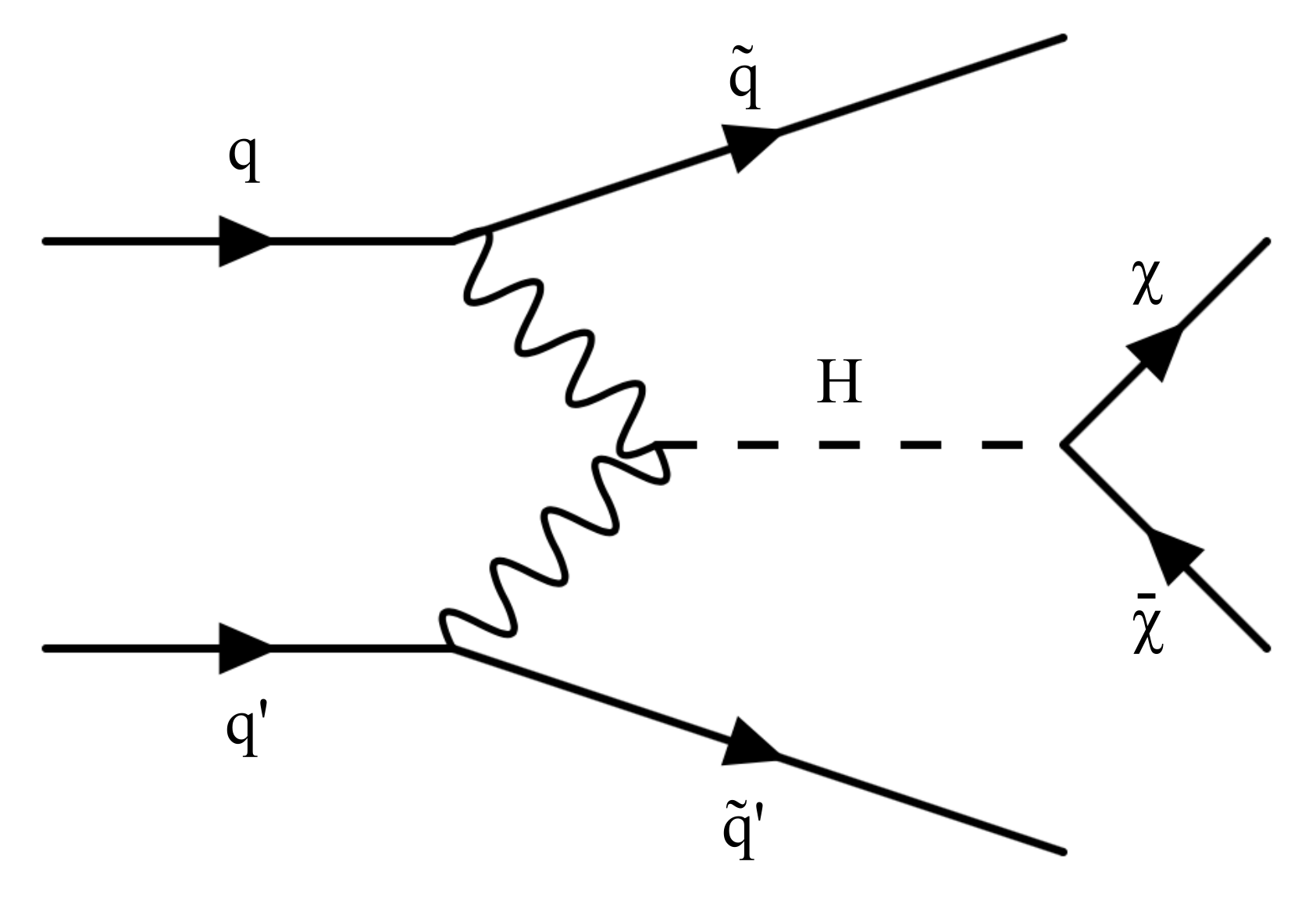}\hfill
\includegraphics[width=0.3\textwidth]{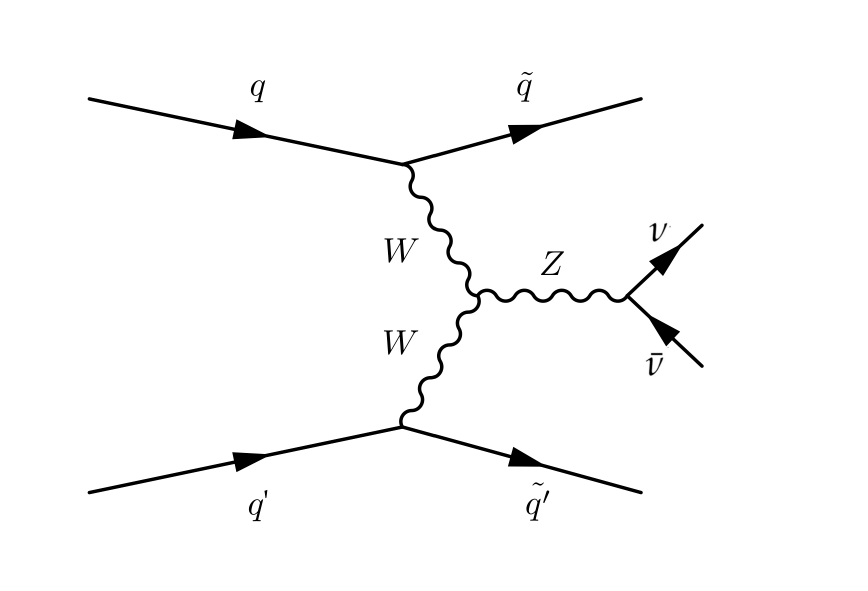}
\includegraphics[width=0.3\textwidth]{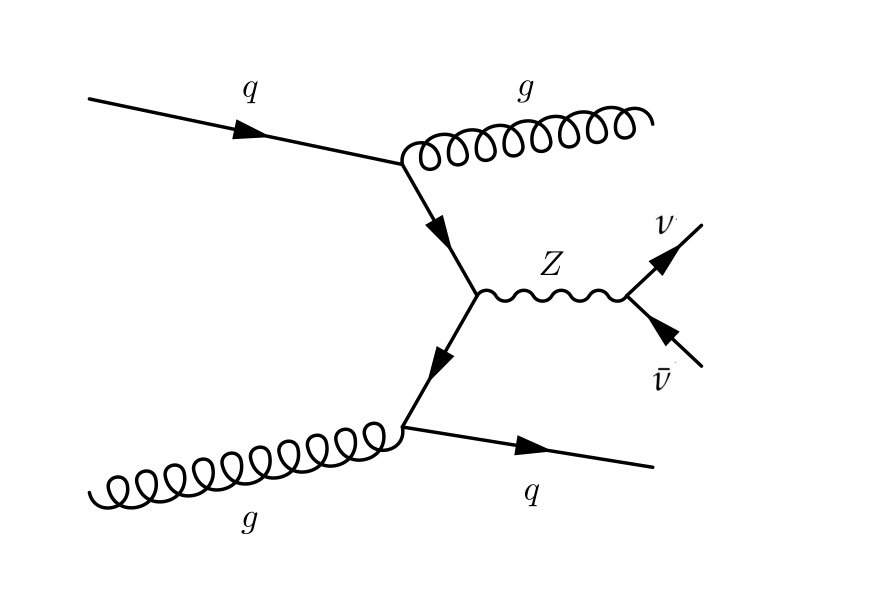}
\caption{Leading-order Feynman diagrams for the production of the Higgs boson in 
association with two jets from \vbf (left), and representative leading-order Feynman 
diagrams for the production of a \PZ boson in association with two jets either through 
VBF production (middle) or strong production (right). Diagrams for the production of a 
\PW boson in association with two jets are similar.}
\label{fig:Feynman}
\end{figure*}

Employing a strategy similar to the one used in the previously published
analysis~\cite{Sirunyan:2018owy}, the invariant mass of the jet pair
produced by \vbf, \mjj, is used as a discriminating variable to
separate the signal and the dominant backgrounds arising from vector
boson production in association with two jets (\Vjets). Representative
Feynman diagrams for the signal and main background processes are
shown in Fig.~\ref{fig:Feynman}.

Control regions enriched in \Vjets processes are used to constrain
the associated background contributions in the signal region. Additional
sensitivity is obtained by using \phojets events to further constrain
the \Zvv background. In the previous CMS publication, the trigger
strategy was based exclusively on the invisible Higgs boson decay
products, requiring a high threshold on the missing transverse
momentum. With the availability of a trigger based on the jet
properties from \vbf production, in this analysis, additional
sensitivity is achieved by including events with lower missing
transverse momentum.

This article is organized as follows. Section~\ref{sec:CMSdet}
introduces the CMS detector. Section~\ref{sec:samples} summarizes the
data and simulated samples. The event reconstruction is detailed in
Section~\ref{sec:obj}, followed by the analysis strategy in
Section~\ref{sec:strategy}. Section~\ref{sec:systs} describes the
systematic uncertainties. Finally, the results are presented in
Section~\ref{sec:res}, with tabulated versions provided in
HEPData~\cite{hepdata}, followed by a summary in
Section~\ref{sec:concl}.

\section{The CMS detector}
\label{sec:CMSdet}

The central feature of the CMS apparatus is a superconducting solenoid
of 6\unit{m} internal diameter, providing a magnetic field of
3.8\unit{T}. Within the solenoid volume are a silicon pixel and strip
tracker, a lead tungstate crystal electromagnetic calorimeter (ECAL),
and a brass and scintillator hadron calorimeter, each composed of a
barrel and two endcap sections. Hadron forward (HF) steel and quartz
fiber calorimeters extend the pseudorapidity $\eta$ coverage provided
by the barrel and endcap detectors. Muons are measured in
gas-ionization detectors embedded in the steel flux-return yoke
outside the solenoid. A more detailed description of the CMS detector,
together with a definition of the coordinate system used and the
relevant kinematic variables, can be found in
Ref.~\cite{Chatrchyan:2008zzk}.

Events of interest are selected using a two-tiered trigger
system~\cite{Khachatryan:2016bia}. The first level (L1) is composed of
custom hardware processors, which use information from the
calorimeters and muon detectors to select events at a rate of about
100\unit{kHz}~\cite{Sirunyan:2020zal}. The second level, known as the
high-level trigger (HLT), is a software-based system that runs a
version of the full event reconstruction optimized for fast
processing, reducing the event rate to about 1\unit{kHz}.

At the end of 2016, the first part of the CMS detector upgrade
program (Phase 1) was undertaken, with the replacement of the inner
tracking pixel detector and the L1 trigger system. During the 2016 and
2017 data-taking periods, partial mistiming of signals in the forward
region of the ECAL endcaps (${2.5 < \abs{\eta} < 3.0}$) led to a large
reduction in the L1 trigger efficiency~\cite{Sirunyan:2020zal}. A
separate correction was determined using an unbiased data sample and
applied to simulated events to reproduce the loss of efficiency. This
problem was resolved before the 2018 data-taking period.

\section{Data and simulated samples}
\label{sec:samples}

Data were recorded by several triggers, as detailed in
Section~\ref{sec:sel}, during 2017 and 2018, for maximum integrated
luminosities corresponding to 41.5 and 59.8\fbinv, respectively.

The signal and background processes are simulated using similar Monte
Carlo (MC) generator configurations as described in detail in
Ref.~\cite{Sirunyan:2018owy}, and summarized below. Separate
independent samples were produced for each data-taking year. The
same generator settings were used for the 2017 and 2018 samples.

The Higgs boson signal events, produced through \ggH, \vbf, \vh, and
in association with top quarks (\tth), are generated with \POWHEG
v2.0~\cite{Nason:2004rx,Frixione:2007vw,Alioli:2010xd,Bagnaschi:2011tu,Nason:2009ai}
at next-to-leading order (NLO) approximation in perturbative quantum
chromodynamics (pQCD). The signal yields are normalized to the
inclusive Higgs boson production cross sections, calculated in Ref.~\cite{deFlorian:2016spz} at 
approximate next-to-NLO (NNLO) in pQCD, with NLO electroweak (EW)
corrections. For the \vbf production process, an additional event weight is
applied to the simulated events to account for EW NLO effects,
dependent on the boson transverse momentum (\pt). The correction
factor is determined using \HAWK~\cite{Denner:2014cla} and
parameterized as $(1-0.000372\pt/\GeVns-0.0304)/0.95$, with the
numerator accounting for the full leading order (LO)-to-NLO correction, and the
denominator representing the overall normalization effect of the EW
correction. The latter is already included in the inclusive cross section,
and therefore has to be removed here to avoid double counting.

The $\PZ/\Pgg^{*}(\Pell^{+}\Pell^{-})$+jets,~\Zvvjets, and \Wlvjets
(with $\Pell = \Pe, \PGm, \PGt$) processes are simulated at NLO
in pQCD using \MGvATNLO v2.6.5~\cite{Alwall:2014hca}, with the
5-flavor scheme and the FxFx~\cite{Frederix:2012ps} merging scheme,
in several bins of boson \pt. Up to two additional partons are
included in the final state in the matrix element calculations. These
processes are referred to as \Vjets (strong) in what follows.

The \phojets background is simulated at LO in pQCD
using \MGvATNLO v2.4.2~\cite{Alwall:2014hca}, with up to four partons
in the final state included in the matrix element
calculations~\cite{Alwall:2007fs}. We refer to this process
as \phojets (strong) in the rest of this paper.

The \MGvATNLO generator is also used for the production of a vector
boson, or a photon, in association with two jets exclusively through
EW interactions at LO. These are referred to as \Vjets (VBF) and \phojets (VBF),
respectively, in the following.

The LO simulation for the \phojets (strong) process is corrected using
boson \pt- and \mjj-depen\-dent NLO pQCD $K$-factors derived
with \MGvATNLO v2.4.2~\cite{Alwall:2014hca}. The simulations for
\phojets (strong) and \Vjets (strong) processes are also corrected as a function of
boson \pt with NLO EW $K$-factors derived in
Ref.~\cite{Lindert:2017olm}. Similarly, \Vjets (VBF) processes are
corrected with NLO pQCD $K$-factors derived using the \textsc{vbfnlo}
v2.7.0 event generator~\cite{Arnold:2008rz,Baglio:2014uba}, as
functions of boson \pt and \mjj. For the \phojets (VBF) process, the
NLO pQCD corrections, evaluated using \MGvATNLO, are found to be
negligible.

Samples of QCD multijet events are generated at LO using \MGvATNLO. The \ttbar
and single top quark background samples are produced at NLO in pQCD
using \POWHEG v2.0 and v1.0,
respectively~\cite{Campbell:2014kua,Alioli:2009je,Re:2010bp}. Samples
of $\PW\PZ$ and $\PZ\PZ$ events are simulated at LO
with \PYTHIA v8.205~\cite{Sjostrand:2014zea}, while the $\PV\PGg$ and
$\PW\PW$ processes are simulated at NLO in pQCD using \MGvATNLO
and \POWHEG~\cite{Melia:2011tj}, respectively.

The \NNPDF v3.1~\cite{Ball:2014uwa} NNLO parton distribution functions
(PDFs) are used for all the matrix element calculations. All
generators are interfaced with \PYTHIA v8.205 for the parton shower
simulation, hadronization, and fragmentation processes. The underlying
event description uses the \textsc{CP5} parameter
tune~\cite{Sirunyan:2019dfx}.

Interactions of the final-state particles with the CMS detector are
simulated using \GEANTfour~\cite{Agostinelli:2002hh}. Additional
\pp interactions (pileup) are included in the simulation,
and simulated events are weighted to reproduce the pileup distribution
observed in data, separately for each data-taking year. The average
number of pileup vertices is 32 in the 2017 and 2018 data.

\section{Event reconstruction}
\label{sec:obj}

The event reconstruction and object definitions closely follow those of
the previous publication~\cite{Sirunyan:2018owy}. The main
aspects are summarized below.

A global event description is available using the particle-flow (PF)
algorithm~\cite{Sirunyan:2017ulk}. Using a combination of the
information provided by the tracker, calorimeters, and muon systems,
the PF algorithm aims to reconstruct individual particles (PF
candidates), classifying them as electrons, photons, muons, or charged
and neutral hadrons. The final state for this analysis is composed
solely of jets from gluons or light-flavored quarks, and missing
transverse momentum. We employ explicit vetoes on events containing
all other identified types of objects (electrons, photons, muons,
hadronically-decaying \PGt leptons, heavy-flavored jets), which help
to reject background processes with leptonic decays (\PW and \PZ
bosons), and those containing top quarks.

Electron and photon candidates~\cite{CMS:2020uim} are selected in the
range $\abs{\eta}< 2.5$, while muon~\cite{CMS:2018rym} candidates are
selected with $\abs{\eta}< 2.4$. When considered for event vetoes,
candidates are required to satisfy loose identification and isolation
criteria. These requirements ensure genuine leptons and photons are
discarded with high efficiency.  For electrons, the loose working
point is referred to as ``loose'' (``veto'' in
Ref.~\cite{CMS:2020uim}), and has $\simeq$95\% efficiency. For photons
and muons the loose working point corresponds to efficiencies of
$\simeq$90\% and $>$99\%, respectively. The \pt threshold on loose
objects is set to 10\GeV for electrons and muons, and 15\GeV for
photons. When leptons (photons) are explicitly selected to enhance the
contributions from \Vjets (\phojets) processes, which is done to
populate control regions in data, ``tight'' identification and
isolation criteria are required. These enhance the purity at the price
of lower efficiency ($\simeq$70\% for electrons and photons,
$\simeq$96\% for muons). The \pt thresholds are then set to 20\GeV for
the leading muon, and higher values for the leading electron (40\GeV)
and photon (230\GeV) because of trigger requirements. The subleading
electron or muon is required to have $\pt>10\GeV$. In 2018, a section
of the hadron calorimeter endcap (HE) was not functional for part of
the year, leading to the inability to properly identify electrons and
photons in the region $\eta<-1.39$ and azimuthal angle $-1.6 < \phi <
-0.9$. For data collected during this time, specific
electron and photon selection criteria are applied. These are
described in more detail in Section~\ref{sec:Vbkg}.

Jets are reconstructed by clustering all PF candidates associated with
the primary interaction vertex using the anti-\kt clustering
algorithm~\cite{Cacciari:2008gp}, with a distance parameter of 0.4, as
implemented in the \FASTJET
package~\cite{Cacciari:2011ma}. The candidate vertex with the largest
value of summed physics-object $\pt^2$ is taken to be the primary \pp
interaction vertex. The physics objects used for this determination
are the jets, clustered using the jet finding
algorithm~\cite{Cacciari:2008gp,Cacciari:2011ma} with the tracks
assigned to candidate vertices as inputs, and the associated missing
transverse momentum, taken as the negative vector sum of the \pt of
those jets. Pileup mitigation techniques~\cite{Cacciari:2007fd} are
used to correct the objects for energy deposits belonging to pileup
vertices, as well as to remove objects not associated with the primary
interaction vertex. Loose identification criteria are applied on the
jet composition to remove contributions from calorimeter noise. To
correct the average measured energy of the jets to that of
particle-level jets, jet energy corrections are derived using
simulated events, as a function of the reconstructed jet \pt and
$\eta$. In situ measurements of the momentum balance in
dijet, \phojet, \Zjet, and multijet events are used to determine any
residual differences between the jet energy scale in data and in
simulation, and appropriate corrections are
made~\cite{Khachatryan:2016kdb}. In simulated events, the jet energy
is also smeared to reproduce the jet energy resolution measured in the
data~\cite{Khachatryan:2016kdb}. For jets with $\pt<50\GeV$, a
multivariate discriminant against pileup jets is applied, using a
loose working point~\cite{Sirunyan:2020foa}. Jets are selected in the
range $\abs{\eta}< 4.7$ and with $\pt>30\GeV$.  Jets with an
identified electron, muon, or photon within $\Delta R < 0.4$ are
rejected, where $\Delta R = \sqrt{\smash[b]{(\Delta\eta)^2+(\Delta\phi)^2}}$.

The missing transverse momentum vector (\ptvecmiss) is computed as the
negative vector \pt sum of all the PF candidates in an event, and its
magnitude is denoted as \ptmiss. Any correction applied to individual
objects is propagated correspondingly to
the \ptmiss~\cite{CMS:2019ctu}. Specific event filters have
been designed to reduce the contamination arising from large
misreconstructed \ptmiss from noncollision
backgrounds~\cite{CMS:2019ctu}. For the analysis of the 2018 data, the
missing HE section affects the PF reconstruction. The inability
to distinguish electrons and photons from jets leads to
spurious \ptmiss in the corresponding $\phi$ region as a result of the
suboptimal reconstruction of charged and neutral
hadrons. Consequently, events with $-1.8 < \phi(\ptmiss) <
-0.6$ are rejected in data for the part of this data set,
around 65\% of the total, that is affected.
Simulated events in this region are reweighted accordingly to model the
efficiency loss.

It has been observed during data taking that the HF detector can,
on rare occasions, give rise to unphysical high-energy signals. This
occurs in particular when a muon or a charged particle coming from a
late showering hadron directly hits one of the photomultiplier tubes
that are used to read out the quartz fibers. The photomultiplier tubes are
located behind the HF detector, in readout boxes gathering quartz fibers of a
given $\phi$ region. The resulting energy is therefore typically
spread across several channels of constant $\phi$. Spurious jets can 
also arise when a high energy muon from machine-induced
backgrounds~\cite{Bruce:2013tea} undergoes bremsstrahlung in the HF
detector. The associated energy deposit is then narrow in both $\eta$
and $\phi$. Although these two effects are uncommon, they lead to
large \ptmiss, and a dedicated mitigation technique is therefore
applied to reject events with such calorimeter noise. For jets
reconstructed in the HF detector, with $\abs{\eta}>2.99$, shower shape
variables are constructed based on the associated PF candidates found
within $\Delta R < 0.4$ of the jet. A central $\eta$ strip size is
defined by counting the number of associated PF candidates,
$N_{\text{PFCand}}^{\text{cent}}$, with transverse momentum
${\pt>10\GeV}$ within $\Delta \phi < 0.05$ of the jet direction.
This corresponds to the number of candidates within the same $\phi$ HF
tower. The shower widths in both directions are defined as
$\sigma_{\eta\eta}$ and $\sigma_{\phi\phi}$, using the
pileup-corrected energy-weighted sums of the separations in $\eta$ and
$\phi$ between the associated PF candidates with $\pt>3\GeV$ and the
jet axis directions.

As stated above, jets stemming from calorimeter noise, called HF noise
jets in the following, tend to be either more spread in $\eta$ than in
$\phi$, or narrow in both directions. They lead to 
spurious \ptmiss in the opposite direction in $\phi$. Events are hence
rejected if they contain any jet with $\abs{\eta}>2.99$, $\pt > 80\GeV$,
and $\dphijmet>2.5$ that does not satisfy the criteria summarized in
Table~\ref{tab:hfnoise}. The requirements of this selection are chosen to have
a mistagging rate smaller than 10\% for signal-like jets, while being
more than 60--90\% efficient at rejecting noise-like jets, depending
on their \pt and $\eta$. To correct for mismodelling of
these selections in simulation, the selection efficiency on signal-like jets is
measured in both data and simulated \Zlljet and \phojet events,
and scale factors are applied to correct the simulation. The scale
factors are measured as functions of jet \pt and $\eta$, and are
consistent with unity to within 10\%.

\begin{table}[htb!]
  \centering \topcaption{Selection applied in the 2017 and 2018 data
    sets to remove HF jets stemming from calorimeter noise.}
    \renewcommand{\arraystretch}{1.1}
    \begin{scotch}{l c c}
       Observable                       & $2.99 < \abs{\eta_{\text{jet}}} < 4.00$ &  $4.00 < \abs{\eta_{\text{jet}}} < 5.00$  \\
       \hline
       $\sigma_{\eta\eta} - \sigma_{\phi\phi}$ &  $<$0.02 & \NA \\
       $\sigma_{\eta\eta}$ &  $>$0.02 & $<$0.10 \\
       $\sigma_{\phi\phi}$ & \multicolumn{2}{c}{$>$0.02} \\
       $N_{\text{PFCand}}^{\text{cent}}$ & \multicolumn{2}{c}{$<$3} \\
\end{scotch}
    \label{tab:hfnoise}
\end{table}

Hadronically-decaying \PGt leptons (\tauh) are identified from
reconstructed jets through the multivariate \textsc{DeepTau}
algorithm~\cite{CMS-DP-2019-033}, using a working point that has an
average efficiency above 96\% for a jet misidentification rate of less
than 10\%. The \tauh candidates are reconstructed with $\pt>20\GeV$
and $\abs{\eta}<2.3$. Jets with an identified loose electron or muon
within $\Delta R < 0.4$ are rejected before applying
the \textsc{DeepTau} algorithm.

The specific features of heavy-flavored jets, in particular the
presence of displaced vertices, are used in a multivariate jet tagging
method. The ``medium'' working point of the \textsc{DeepCSV} algorithm
from Ref.~\cite{Sirunyan:2017ezt} is used to tag \PQb quark jets with
${\pt > 20\GeV}$ and ${\abs{\eta} < 2.4}$ with 68\% efficiency, and
1.1\,(12)\% probability of misidentifying a light-flavor or gluon
(\PQc~quark) jet as a bottom quark jet.

\section{Analysis strategy}
\label{sec:strategy}

The distinctive feature of \vbf production is a pair of jets
originating from light-flavor quarks, with a large separation in
$\eta$ (\detajj) and therefore a large \mjj. The signal region (SR) in
this analysis uses selection requirements on the jet pair together
with the presence of a significant amount of \ptmiss.

\begin{figure*}[htb!]
\centering
\includegraphics[width=0.48\textwidth]{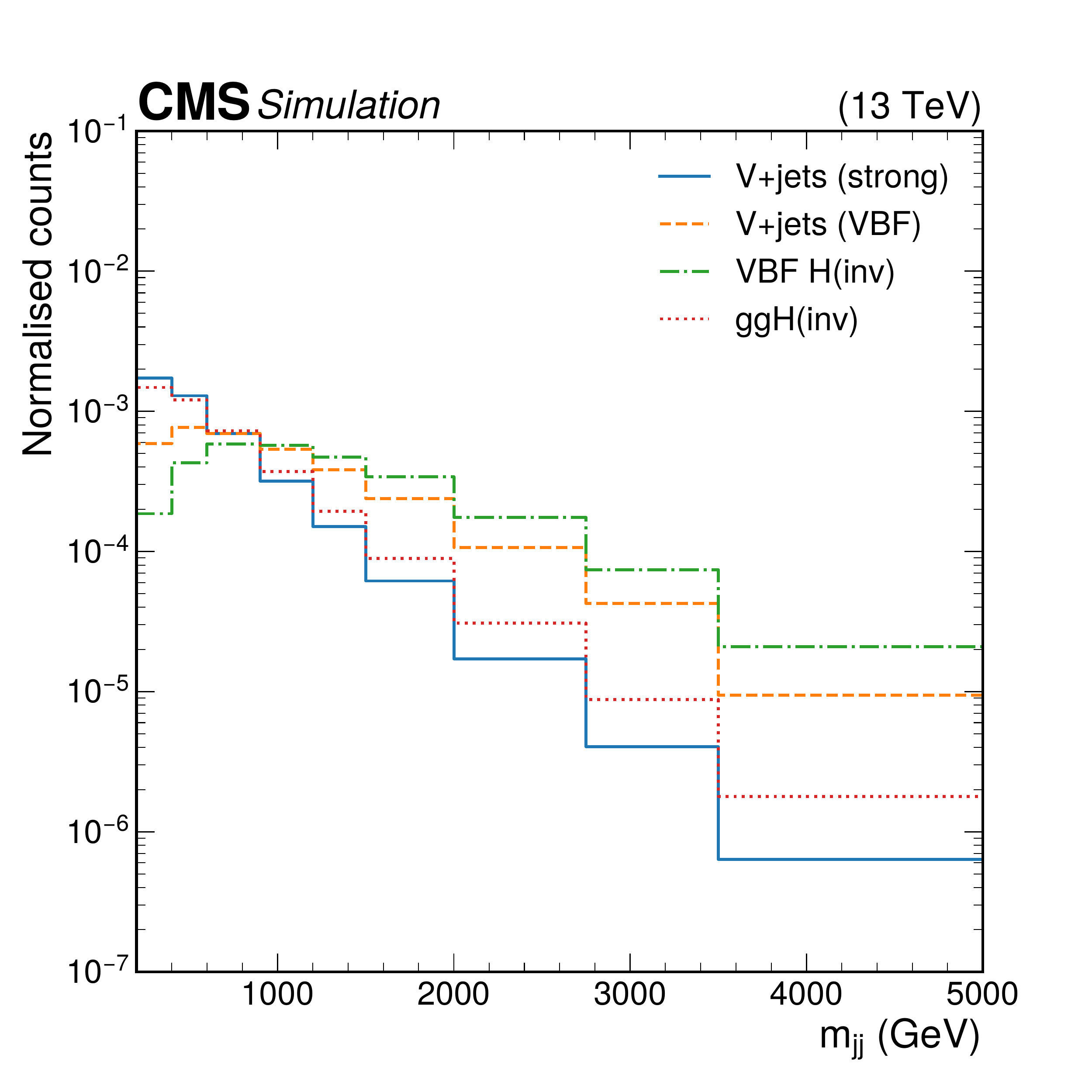}\\
\includegraphics[width=0.48\textwidth]{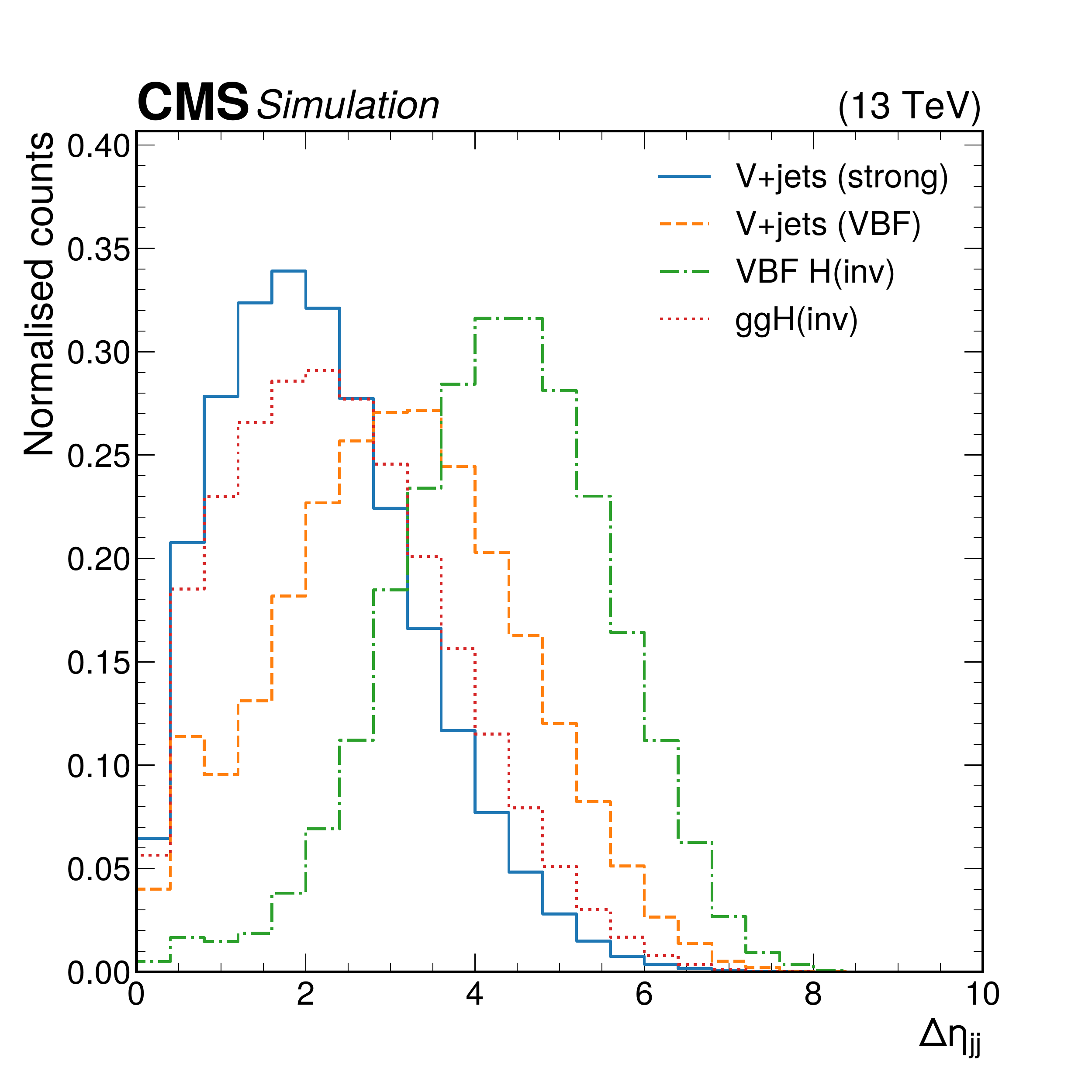}\hfill
\includegraphics[width=0.48\textwidth]{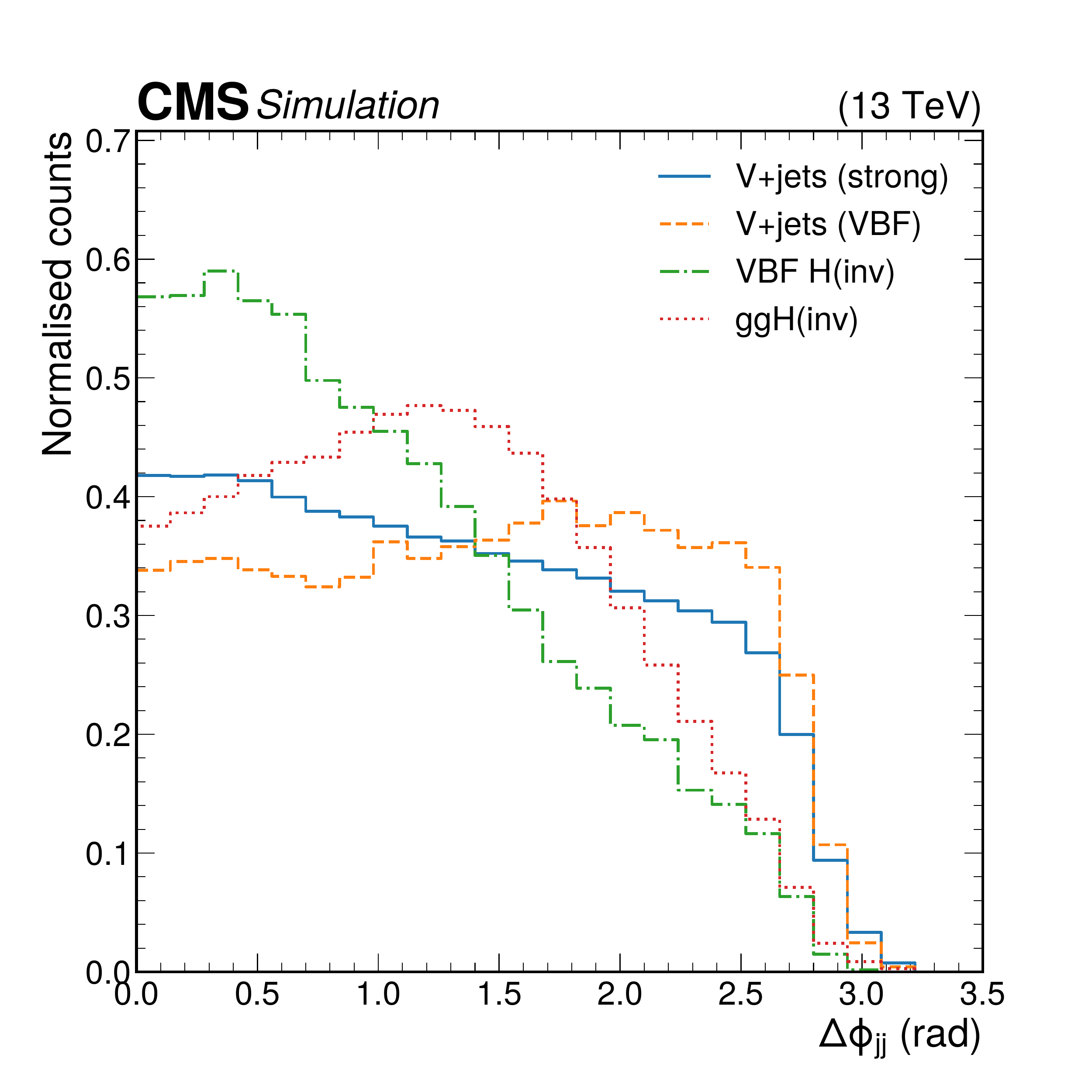}
\caption{Comparison of shapes of quantities related to the dijet pair. Shown are \mjj (upper), \detajj (lower left), and \dphijj (lower right), 
as predicted by simulation, separating strong and VBF production for \Vjets and signal processes.}
\label{fig:simDijet}
\end{figure*}

The shape of the \mjj distribution is used to disentangle jet pairs
produced in VBF production from other SM processes. When fitting the
shape of this distribution, the strong production of the \Vjets
processes together with the \ggH signal dominate at low \mjj, whereas
the VBF-produced \Vjets processes populate the high-\mjj tail,
together with the \vbf \PH signal. The shapes of \mjj,
$\abs{\detajj}$, and the dijet separation in azimuthal angle
($\abs{\dphijj}$) predicted by the simulation are compared between
strong and VBF production of both \Vjets and signal processes in
Fig.~\ref{fig:simDijet}.
Whereas similarities are seen between the VBF
production of vector bosons and Higgs bosons, there are some
differences. First, the respective bosons have different coupling
structures, identified as the main reason for the different behaviour
in the \dphijj distribution~\cite{Hankele:2006ma}. Second, the \Vjets
(VBF) samples include additional diagrams compared with the VBF \PH
production, in which the vector boson is produced through a coupling
to quarks, and jets are produced from additional EW vertices. For the
VBF \PH production, such diagrams are strongly suppressed through the
Yukawa mechanism, also leading to differences in the kinematic
behaviour.

The dominant \Vjets backgrounds are measured using control regions (\PW,
\PZ, and \PGg CRs), in which one or more charged leptons (electrons
and muons), or a photon, are required, but the selection on the jets
and \ptmiss is kept identical to that in the SR. The MC simulation is
used to define transfer factors. These make it possible to predict 
the \Vjets yields in the SR from both the yields measured and
predicted in the CRs. The full procedure is described in
Section~\ref{sec:res}.

The \PZ CR suffers from a lack of statistical precision, particularly in
the high-\mjj region, due to the low branching fraction of the \PZ boson to a
pair of leptons. Because of their similarities, the ratios of the
yields of \PWpm or \PGg to \PZ production in the SR are
constrained, within theoretical uncertainties, to those predicted by 
the simulation.

In the following subsections, the SR selection is first
presented. Then, the data-driven methods used to estimate the \Vjets and
QCD multijet backgrounds are described. The remaining small
contributions expected from diboson and top quark processes are
estimated using simulation.

\subsection{Signal region selection}
\label{sec:sel}

Two complementary trigger strategies were used to select events
online. The first category only uses the \ptmiss information at L1 and
in the HLT, and is referred to as the ``missing momentum triggered
region'' (MTR) category in the following. During 2017 and 2018, the HF
region was included in the definition of \loneptmiss, leading to an
increase in trigger acceptance when the VBF jets are reconstructed
with $3 \leq \abs{\eta} \leq 5$, compared with 2016. The \loneptmiss
thresholds varied between 65 and 90\GeV, with a \hltptmiss threshold
of 120\GeV. After correcting for the L1 mistiming inefficiency
described in Section~\ref{sec:CMSdet}, this trigger is more than 90\%
efficient for $\ptmiss > 250\GeV$. Loose muon candidates are ignored
at L1 and in the HLT when calculating the \loneptmiss and \hltptmiss
variables, ensuring that the same trigger can be used in the muon
CRs. The data collected by these triggers in 2017
and 2018 correspond to integrated luminosities of 41.5 and 59.8\fbinv, respectively.

The second category uses a combination of \ptmiss and jet information,
and is referred to as the ``VBF jets triggered region'' (VTR) category
in the following. After the upgrade of the L1 trigger
system~\cite{Tapper:2013yva} it was possible to develop an algorithm
targeting jets originating from VBF production. This trigger was added
after the start of data taking in 2017, and collected a data set
corresponding to an integrated luminosity of 36.7\fbinv during that
year. The L1 VBF algorithm requires the presence of at least two jets
passing \lonept thresholds of 115 (110)\GeV for the leading jet and 40
(35)\GeV for the subleading jet in 2017 (2018).  Pairs are formed from
the selected jets, and the pair with the highest invariant
mass \lonemjj must satisfy $\lonemjj > 620\GeV$. The pair is allowed
to be formed by two jets with $\lonept>40\,(35)\GeV$ if there is a
third jet passing the leading-\lonept threshold requirement, in 2017
(2018). A corresponding HLT algorithm was designed specifically for
the \hinv analysis, adding a requirement on
\hltptmiss with a minimum threshold of 110\GeV. At the analysis level,
after correcting for the L1 mistiming inefficiency, this trigger is
more than 85\% efficient for $\ptmiss > 160\GeV$ and the leading-\mjj
jet pair passing the following requirements: $\mjj>900\GeV$, with \pt
thresholds on the two jets forming the leading-\mjj pair of
$p_{\mathrm{T}}^{\mathrm{1,2}}>140,70\GeV$. The trigger efficiencies for  the two  
trigger algorithms are available in \suppMaterial.  
Again, loose muon
candidates are ignored when calculating the \ptmiss variables at all
stages.

We ensure the MTR and VTR categories are orthogonal by requiring $160 < \ptmiss \leq
250\GeV$ in the VTR category, and \ptmiss$ > 250\GeV$ in the MTR
category.

To enhance the selection of jets with VBF properties at the analysis
level, and to reduce the contamination arising from jet pairs in QCD
multijet events, the two leading-\pt jets (or the jets forming the
highest-\mjj pair in the VTR category) are required to be in opposite
hemispheres of the detector ($\eta_{\mathrm{j1}}\eta_{\mathrm{j}2} <
0$). The two selected jets are also required to have $\abs{\detajj}>1$
and $\abs{\dphijj}<1.5$ ($\abs{\dphijj}<1.8$ in
the VTR category). In the MTR category, the \mjj threshold is set to
200\GeV to use the full shape of the spectrum to better separate the
signal from the background formed by strong \Vjets production.

In QCD multijet events, large \ptmiss may arise from mismeasurements
of the jet momenta, in which case some jets in the event could be
aligned in $\phi$ with the \ptvecmiss. To reduce the contamination
from such events, the minimum value of the azimuthal angle between
the \ptvecmiss vector and any of the first four leading jets
($\pt>30\GeV$), $\min(\dphijmet)$, is required to be above 0.5
(1.8) in the MTR (VTR) category. Events with
possible mismeasurements due to calorimeter noise, which would lead
to jets with anomalously large (small) energy fractions coming from 
neutral (charged) particles, are rejected. This is done by 
rejecting the event if either of the
selected VBF jets has $\abs{\eta} < 2.5$ and a neutral (charged) hadron
energy fraction $\text{NHEF} > 0.8$ ($\text{CHEF} < 0.1$). This
selection rejects at most 2\% of events, independent of the process
and uniformly in \mjj. A criterion is also applied on the difference
between the \ptmiss measured using the PF algorithm and that using
only the calorimeters. This difference is required to be less than 50\% of
the \ptmiss. This selection rejects at most 2 (1)\% of all events,
mostly at low \mjj, for the 2017 (2018) data-taking conditions.

A second source of QCD multijet background is due to the remaining
impact of jets originating from HF noise, where the \ptvecmiss is
balanced in $\phi$ with such a jet, and the jet still passes the
selection criteria from Table~\ref{tab:hfnoise}. Combined with genuine
jets from QCD multijet production, such events can pass the SR
selection. These large energy deposits are generally close to the
outer HF boundary ($\abs{\eta}<3.25$), where the readout boxes are
located, though can extend up to $\abs{\eta}=5$.

Finally, a veto on all other types of loosely identified objects
(electrons, muons, photons, \tauh candidates, and \PQb-tagged jets),
as described in Section~\ref{sec:obj}, is applied.

The criteria for the SR selections are summarized in
Table~\ref{tab:sel}, for the MTR and VTR categories.
\begin{table*}[htb!]
    \centering
    \topcaption{Summary of the kinematic selections used to define the SR for both the MTR and the VTR categories.}
    \begin{scotch}{l c c }
       Observable                       & MTR & VTR \\
       \hline
       Choice of pair & leading-\pt jets & leading-\mjj jets \\
       Leading (subleading) jet         & $\pt>80\,(40)\GeV$, $\abs{\eta}<4.7$ & $\pt>140\,(70)\GeV$, $\abs{\eta}<4.7$ \\
       \ptmiss                          & $>$250\GeV & $160 < \ptmiss < 250\GeV$         \\
       $\min(\dphijmet)$                & $>$0.5 & $>$1.8   \\
       $\abs{\dphijj}$                  & $<$1.5   & $<$1.8   \\
       \mjj                             & $>$200\GeV & $>$900\GeV \\
       $\abs{\ptmiss-\text{calo }\ptmiss}/\ptmiss$ & \multicolumn{2}{c}{$<$0.5} \\
       Leading/subleading jets $\abs{\eta}<2.5$         & \multicolumn{2}{c}{$\text{NHEF} < 0.8$, $\text{CHEF} > 0.1$} \\
       HF noise jet candidates          & \multicolumn{2}{c}{0 (using the requirements from Table~\ref{tab:hfnoise})} \\
       \tauh candidates                 & \multicolumn{2}{c}{$\text{N}_{\tauh}=0$ with $\pt>20\GeV$, $\abs{\eta}<2.3$}         \\
       \cPqb~quark jet                  & \multicolumn{2}{c}{$\text{N}_{\text{jet}}=0$ with $\pt>20\GeV$, DeepCSV Medium}\\
       $\eta_{\mathrm{j1}}\eta_{\mathrm{j2}}$      & \multicolumn{2}{c}{$<$0} \\
       $\abs{\detajj}$                  & \multicolumn{2}{c}{$>$1}             \\ [\cmsTabSkip]

       Electrons (muons)                & \multicolumn{2}{c}{$\text{N}_{\Pe,\Pgm}=0$ with $\pt>10\GeV$, $\abs{\eta}<2.5\,(2.4)$}\\
       Photons                          & \multicolumn{2}{c}{$\text{N}_{\Pgg}=0$ with $\pt>15\GeV$, $\abs{\eta}<2.5$}        \\
\end{scotch}
    \label{tab:sel}
\end{table*}
After these selections, the dominant backgrounds come from the \Vjets
processes. Due to the large branching fraction of the \PZ boson
decay to neutrinos, the \Zvvjets process accounts for about two-thirds
of the total background. After the lepton vetoes and the \ptmiss
requirement, the contributions from other decay modes are
negligible. The next largest background arises from \Wlvjets
production in which the charged lepton from the \PWpm boson decay is
outside of the acceptance of the tracking detector, leading to
additional \ptmiss. In the case of muons, which deposit very little
energy in the calorimeters, the \ptmiss is significant. The hadronic
decay modes of the \PWpm boson are rejected by the large \ptmiss
requirement. The VBF production of \Vjets contributes about 2\% of
the total \Vjets background for \mjj around 200\GeV. This increases to
about 11\% at $\mjj \approx 1.5\TeV$, and to more than 48\% for $\mjj >
3.5\TeV$.

\subsection{Lepton-based control regions}
\label{sec:Vbkg}

As the boson recoil properties are driven by the production mode and
are independent of the boson decay mode, the dominant \Zvvjets background
is modelled using CRs with leptonic decays of the \PZ boson (\Zee
and \Zmm). To reduce the contribution from Drell--Yan $\Pgg^*$ decays
to leptons, the invariant mass of the selected leptons is required to
lie in the range 60--120\GeV. The lepton selection is chosen to maximize
the event yield while still ensuring leptonic \PZ boson decays are 
selected with high purity.

To stay as close as possible to the SR selection, the same trigger as in the SR is
used for the \Zmm CR. As a result, systematic uncertainties in 
the trigger efficiencies largely cancel when estimating the
corresponding transfer factors. Instead of the muon veto, a pair of
oppositely charged muons, consisting of a tight muon with $\pt>20\GeV$
and a loose muon with $\pt>10\GeV$, is required. All other criteria
from Table~\ref{tab:sel} are applied. The \ptmiss variable is
recalculated ignoring the muons, to mimic the boson recoil.

For the \Zee CR the triggers used in the SR are inefficient, as
electrons deposit their energy in the calorimeter.  Single-electron
triggers are therefore used. The lowest-threshold trigger requires a
minimum \pt of 35\GeV and imposes isolation requirements. It is
supplemented by an electron trigger with a \pt threshold of 115\GeV,
but no isolation requirements, as well as by a photon trigger
requiring $\pt > 200\GeV$. For the last, no isolation criteria are
applied, and it does not rely on track reconstruction. Taken together,
this set of triggers optimizes the efficiency over the full \pt
range. Instead of the electron veto, a pair of oppositely charged
electrons, consisting of a tight electron with $\pt>40\GeV$ and a
loose electron with $\pt>10\GeV$, is required. The \ptmiss variable is
recalculated ignoring the electrons, to mimic the boson recoil.

For the \Wlvjets background, single-lepton CRs are
used. It should be noted that in this case, the \Wen and \Wmn CRs
favour \PWpm boson decays with a high-\pt central lepton ($\abs{\eta}<2.5$
(electrons) or 2.4 (muons)), whereas the background expected in the SR
consists of \PWpm boson decays in which the leptons
(including \PGt leptons) are outside of the acceptance. This has an
impact on the \ptmiss distribution. As explained in greater detail in
Section~\ref{sec:res}, the impact of the lepton acceptance is
accounted for in the definition of the transfer factors between the CRs
and the SR, using the simulation. For the \Wen and \Wmn CRs, the
lepton veto is replaced by the selection of a tight electron (muon) with $\pt>40$ (20)\GeV, and a veto on any other
identified loose electron or loose muon. To reduce the contribution
from misidentified electrons stemming from QCD multijet production,
the \ptmiss associated with the \Wen decay (\ie not ignoring the
electron contribution) is required to be above 80\GeV. In the 2018
analysis, due to the missing HE sector in the data, jets in the
corresponding $\eta$-$\phi$ area are often identified as electrons or
photons, and hence events with an electron in this region are
rejected.

\subsection{\texorpdfstring{The \phojets control region}{The photon plus jets control region}}
\label{sec:phobkg}

To further constrain the \Zvv background in the SR, a photon CR is
used. At large \pt, the kinematic properties of photon
production become similar to those of the \Zvv
process~\cite{Lindert:2017olm}, and can therefore be used to estimate
the latter. Events are selected using a trigger requiring an online
photon \pt of at least 200\GeV. In the offline analysis, photons are
required to be located in the central part of the detector
($\abs{\eta}<1.4442$), have $\pt>230\GeV$ to
ensure full trigger efficiency, and pass additional identification
criteria based on the properties of the associated energy deposit
(supercluster) in the ECAL, as well as the isolation of the photon
relative to nearby objects. Exactly one such photon is required, and
all other criteria from Table~\ref{tab:sel} are applied. The \ptmiss
variable is recalculated ignoring the photon, to mimic the
boson recoil.

In addition to the desired \phojets events with a genuine
well-identified and isolated (``prompt'') photon, small contributions
from QCD multijet events with hadronic jets misidentified as photons
are present in this region (``nonprompt''). To estimate this
background contribution, a purity measurement is performed. The
measurement is based on the distribution of the lateral
width, \sieie~\cite{CMS:2020uim}, of the ECAL supercluster
associated with the photon. For prompt photons, the distribution
of \sieie peaks sharply around values of 0.01 and below, while
nonprompt photons show a much smaller peak and a shoulder towards
values larger than 0.01. To extract the contamination, a template fit
to the \sieie distribution is performed in data collected with a
looser version of the CR selection. In this looser region, 
instead of the usual selection that is applied to the VBF jets, we require
the presence of a single jet with $\pt>100\GeV$ to enhance the available number
of events. Additionally, the photon identification criteria are
modified by removing a requirement on \sieie that is otherwise
included. A template for prompt photons is obtained from
simulated \phojets events, while a nonprompt template is derived from
a data sample that is enriched in nonprompt events by inverting the
isolation requirements that are part of the photon identification criteria. The nonprompt
fraction is defined as the fraction of nonprompt photons present below
the \sieie threshold set by the identification criteria. The template
fit is performed separately in bins of the photon \pt and yields a
nonprompt fraction between around 4\% at $\pt=200\GeV$ and 2--3\%
at $\pt=800\GeV$, depending on the data-taking period. The final QCD
multijet contribution is then determined by weighting the events
observed in the data by the nonprompt fraction. A 25\% uncertainty is
assigned to the normalization of the QCD multijet background to
account for any mismodelling of \sieie in the simulation. The uncertainty is
estimated by repeating the measurement while varying the binning of
the \sieie distribution used for fitting, capturing the effect of the
mismodelling. The statistical uncertainty in the determination of the
differential \mjj shape is negligible.

\subsection{Multijet background}
\label{sec:qcdBkg}

The QCD multijet background is estimated using events
in data in which the \ptmiss arises from mismeasured jets. Depending
on the source of the mismeasurement, the jet that was mismeasured is
either balanced in the case of additional HF noise, or aligned with
the \ptmiss in $\phi$.

\begin{figure}[htb!]
\centering
\includegraphics[width=\cmsFigWidth]{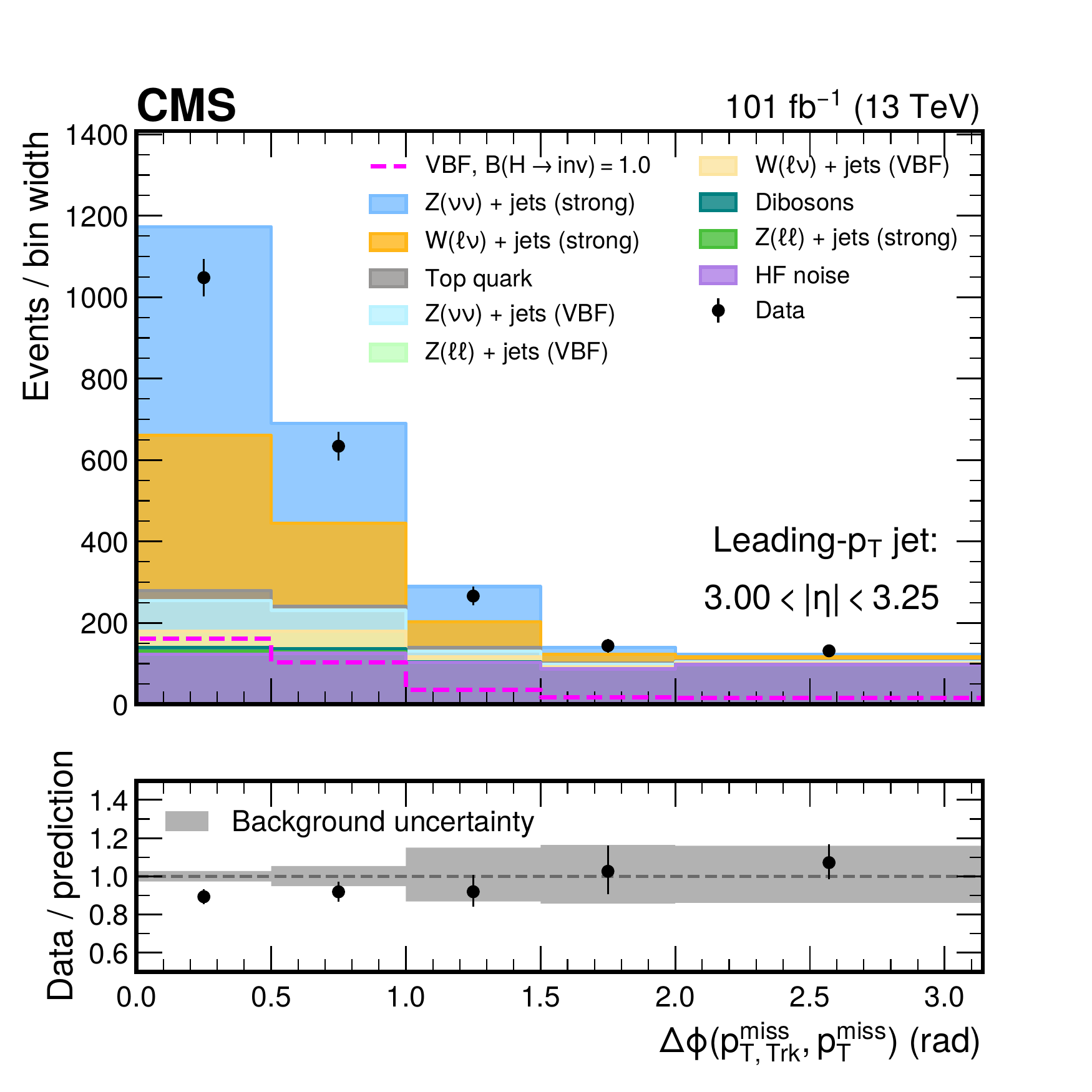}
\caption{The $\Delta\phi$(\tkptmiss, \ptmiss) distribution in the SR with the additional requirement that the leading-\pt jet passes $3 < \abs{\eta} < 3.25$. The 2017 and 2018 data are compared with the sum of the HF noise template and other backgrounds from simulation. The uncertainty band includes only statistical uncertainties in the simulation.}
\label{fig:hfclosure}
\end{figure}

For the multijet background stemming from HF noise, an \mjj template
is extracted by inverting the requirements on the HF jet shape
variables, hence requiring at least one jet in the event to fail the
selection criteria given in Table~\ref{tab:hfnoise}. The probability
for an HF noise jet candidate to pass or fail the criteria is
parameterized as a function of the jet \pt and $\eta$, using events
selected to have large \ptmiss and to contain only one HF jet balanced
in $\phi$ with the \ptvecmiss. An event-by-event weight is applied to
estimate the contribution in the SR from the ``failing'' events. The
estimated contamination from other SM processes is then removed
bin-by-bin for the distribution under study. A closure test is
performed by selecting events with the leading-\pt jet within $3
< \abs{\eta} < 3.25$. With this selection, the signal contamination is
$<$2\% assuming $\brinv=0.19$, as previously excluded in
Ref.~\cite{Sirunyan:2018owy}.  For events with spurious \ptmiss from
noise, one expects a full decorrelation between the \ptmiss measured
from the tracker acceptance only, \tkptmiss, and from the full event,
whereas for events with true \ptmiss a correlation exists. Noise
events are therefore expected to dominate in the region with large
$\Delta\phi$(\tkptmiss,\ptmiss).  Data are compared with the estimated
template for the $\Delta\phi$(\tkptmiss,\ptmiss) distribution in
Fig.~\ref{fig:hfclosure}.  From the agreement observed in the closure
test in both years, a 20\% systematic uncertainty is assigned to the
template shape in the SR.

For the second category of multijet events, the requirement on
$\min(\dphijmet)$ is inverted to define a control region enriched
in QCD multijet events (QCD CR). The \mjj shape for the contribution of this
background in the SR is derived 
from the yields in data in each \mjj bin in the QCD CR, after
subtracting the contributions from \Vjets, diboson, and top quark
processes estimated from simulation, as well as HF noise contributions estimated in the data. The template is normalized as follows. The
distribution of $\min(\dphijmet)$ in data is fit with the sum of
templates derived from the simulated \Vjets, diboson, and top quark 
events, an HF noise template derived in data; and a
functional form $f_{\text{QCD}}$ representing the QCD multijet
contribution. The functional form is
\begin{linenomath}
\begin{equation}
    f_{\text{QCD}}(x) =  p_{0}\re^{-p_{1}x}
\end{equation}
\end{linenomath}
for the MTR region, and 
\begin{linenomath}
\begin{equation}
    f_{\text{QCD}}(x) =  p_{0}\re^{-\frac{(x-p_{1})^{2}}{2p_{2}^{2}}}
\end{equation}
\end{linenomath}
for the VTR region. The parameters $p_{i}$ are allowed to float,
and $x=\min(\dphijmet)$. The choices of these functions are
validated by fitting this model to simulated QCD multijet events, and they are found
to describe the distributions in the MTR and VTR categories well.

The normalizations of the \Wlvjets, \Zvvjets, and HF noise
contributions are allowed to vary independently in the fit. They are
constrained within 20\% of the prediction from simulation to account for systematic
uncertainties related to jet energy calibrations, missing higher
orders in the \Vjets cross section calculations, and the closure of the HF 
noise contribution between data and simulation. The fitted values of
the normalizations are used when subtracting the \Wlvjets, \Zvvjets,
and HF noise contributions from the data to obtain the \mjj template.
Their fitted uncertainties are included in the final
systematic uncertainty in the QCD multijet estimate.

\begin{figure}[htb!]
    \centering
    \includegraphics[width=0.48\textwidth]{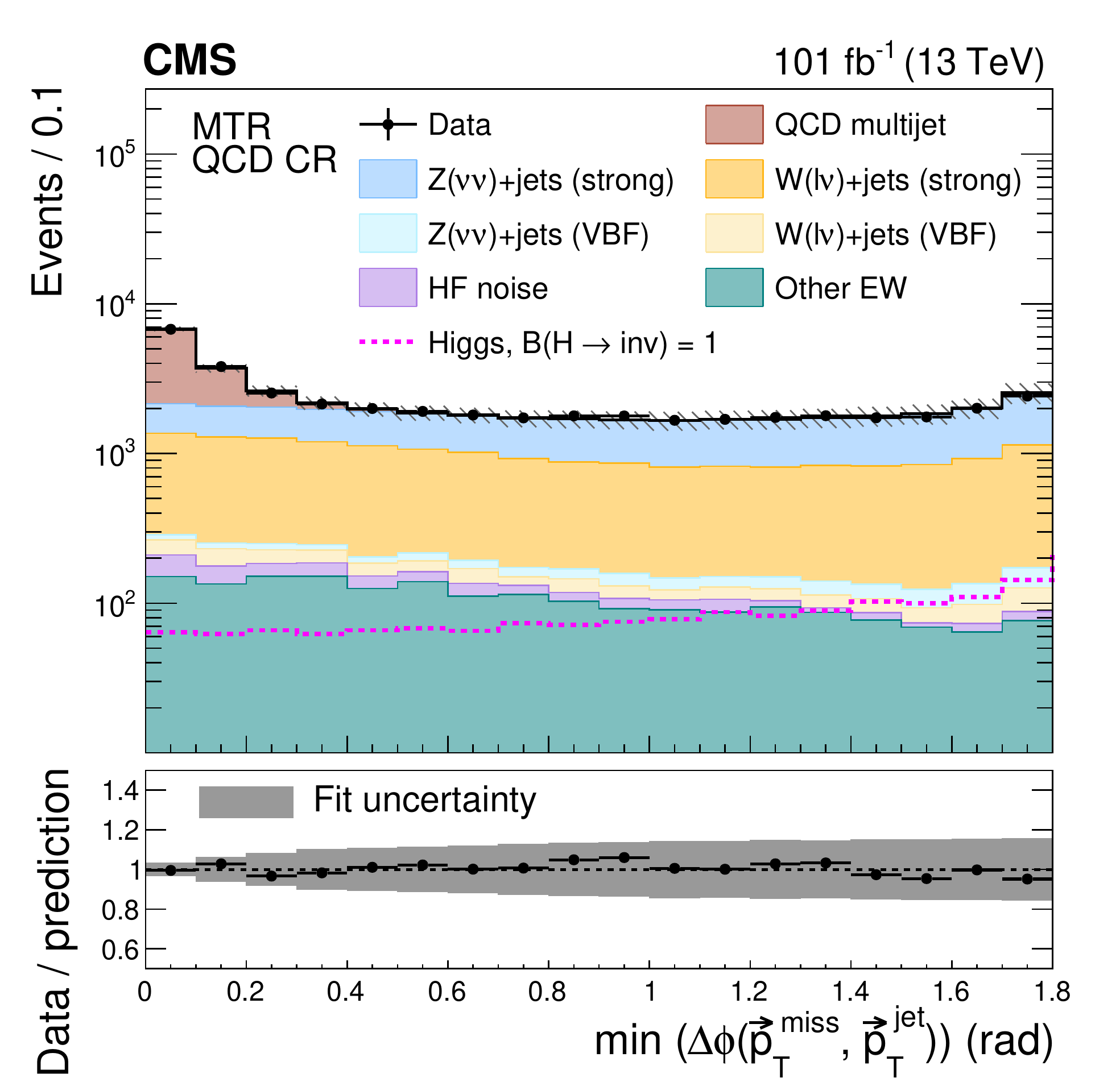}\hfill
    \includegraphics[width=0.48\textwidth]{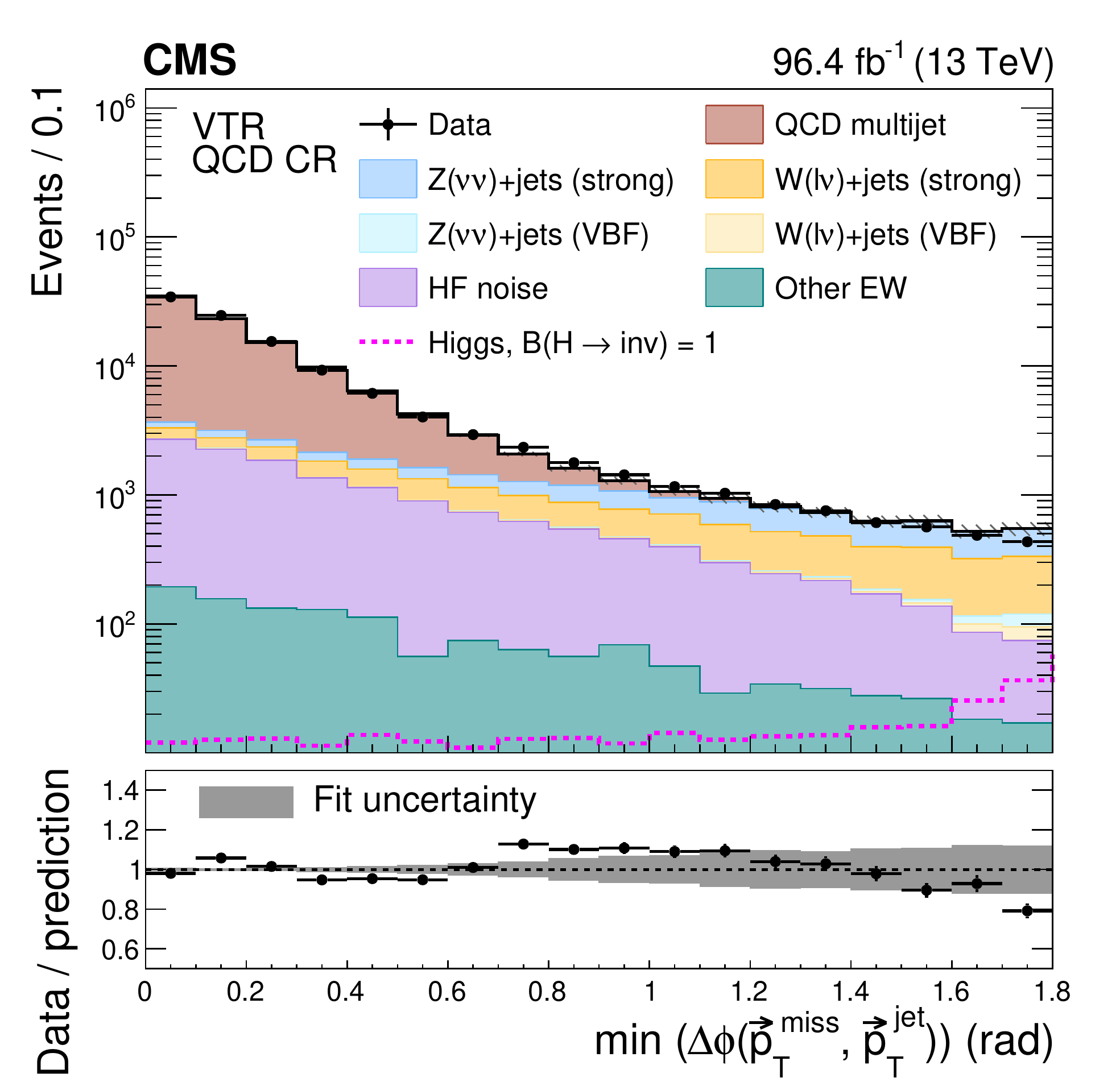}
    \caption{The
    $\min(\dphijmet)$ distribution in data from 2017 and 2018, and
    contributions from \Vjets, diboson plus top quark processes, HF noise,
    and QCD multijet events for the MTR (\cmsLeft) and VTR (\cmsRight)
    categories. The uncertainty band shows the uncertainty from the
    fit used to determine the normalization of the QCD multijet
    template in the corresponding SR. The yields from the 2017 and 2018 samples are summed and the correlations between their uncertainties are neglected.}
    \label{fig:qcdmmultijetfitted}
\end{figure}

In both the MTR and VTR categories, the fit range is $0
< \min(\dphijmet) < 1.8$, chosen to minimize the overlap of events
in data with the SR. The fits are performed separately for the 2017 and
2018 data sets. Figure~\ref{fig:qcdmmultijetfitted} shows the
$\min(\dphijmet)$ distribution in data used in the fit, and the
contributions from \Vjets, diboson plus top quark processes,
HF noise, and QCD multijet events resulting from the fit. The sums of
the 2017 and 2018 data sets are shown for the MTR and VTR categories.

The function $f_{\text{QCD}}(x)$ provides an estimate of the QCD multijet
template normalization, $N$, in the SR via
\begin{linenomath}
\begin{equation}
N = \int_{X}^{\pi} f_{\text{QCD}}(x)\, \rd{x}, 
\end{equation}
\end{linenomath}
where $X=0.5$ for the MTR categories and $X=1.8$ for the VTR
categories. An uncertainty in $N$ is derived by generating, and then
refitting, pseudodata to extract the standard deviation of
$\log N$. This uncertainty includes the statistical uncertainty in
the fit for the QCD multijet normalization and the uncertainty in the
templates from simulation used to subtract the backgrounds due to the limited
number of simulated events.

\section{Sources of uncertainty}
\label{sec:systs}

Several sources of systematic uncertainty affect the predictions of
the signal and background components. They are separated into two
categories, experimental and theoretical sources.
Their effect is propagated either directly to the yields
expected in the SR (for signal and backgrounds estimated directly from
simulation), or to the transfer factors (for the \Vjets processes
estimated from the CRs).

\subsection{Experimental uncertainties}

The reconstruction and identification efficiencies of electrons,
photons, muons, \tauh candidates, and \PQb-tagged jets have been
measured in both data and
simulation~\cite{CMS:2020uim,CMS:2018rym,CMS:2018jrd,Sirunyan:2017ezt}.
The simulated events are corrected by scale factors, which are usually
dependent on the \pt and $\eta$ of the object, and have associated
systematic uncertainties. The scale factors, and their uncertainties,
are also propagated when vetoing events with identified objects. For
the electrons and muons, when different working points are chosen
between leptons selected in the CRs and vetoed in the SR, the
corresponding uncertainties are kept uncorrelated. All these
uncertainties are correlated between the MTR and VTR
categories. Except for the muons, where the dominant source of
systematic uncertainty comes from the experimental method employed,
these uncertainties are considered uncorrelated between the 2017 and
2018 data sets.

The effects of the jet energy scale (JES) and jet energy resolution
(JER) uncertainties are studied explicitly by varying them by one
standard deviation~\cite{Khachatryan:2016kdb}, propagating the changes
to the \ptmiss accordingly, and checking the impact on the transfer
factors for the \Vjets processes as functions of \mjj. For the signal
and the minor backgrounds, the impact is studied on the simulated
yields in the SR as a function of \mjj, with a fit procedure to remove
the statistical contribution in the less-populated high-\mjj bins. The
JER uncertainties impact the signal yields by 4--9 (2--15)\%,
increasing with \mjj, for \vbf (\ggH) production. The impacts of the
JES uncertainties are 5--25 (7--35)\%, increasing with \mjj, for
the \vbf (\ggH) process. Eleven independent JES sources are
considered, with partial correlations between the 2017 and 2018 data
sets. The dominant source is the $\eta$ dependence of the
corrections. The corrections become particularly large for forward
jets, which explains the large increase at high \mjj. The JER
uncertainties are uncorrelated between the two data-taking years. Both
the JER and JES uncertainties are correlated between the MTR and VTR
categories.

Simulated events are weighted to match the distribution of reconstructed
vertices to the distribution observed in the data. An uncertainty associated
with this procedure is obtained by varying the total inelastic
\pp cross section by $\pm 5\%$~\cite{CMS:2018mlc}, and
repeating the background estimation procedure. This uncertainty is
correlated across categories and data sets.

The trigger efficiencies are measured in both data and simulation, and
a scale factor is extracted. For the signal triggers, the scale factor
is parameterized as a function of the \ptmiss. The associated
systematic uncertainty partially cancels in the transfer factors
between the muon CRs and the SR. The associated uncertainty is
uncorrelated across categories as different triggers are used, but partially
correlated between years. For the electron CRs, the scale factor is
parameterized as a function of the lepton \pt and $\eta$, and the
impact of the uncertainties is propagated to the corresponding
expectations in the SR. The uncertainty in the electron trigger scale
factor is uncorrelated between years but correlated across categories.

The uncertainty in the integrated luminosity of the 2017 (2018) data
set is 2.3\%~\cite{CMS:2018elu} (2.5\%~\cite{CMS:2019jhq}). When
combined together with the 2016 data set~\cite{CMS:2021xjt}, the
uncertainty is reduced to 1.6\%. The improvement in the precision
reflects the (uncorrelated) time evolution of some systematic
effects. Eight independent sources are identified to take into account
the correlations across data sets. These uncertainties are considered correlated
between categories.

\subsection{Theoretical uncertainties}

The uncertainties in the \ggH and \vbf predictions due to
PDFs and renormalization and factorization scale variations are taken from
Ref.~\cite{deFlorian:2016spz}. For the \ggH process, an
additional uncertainty of 40\% is assigned to take into account the
limited knowledge of the \ggH production cross section in association with
two or more jets, as well as the uncertainty in the prediction of the
\ggH differential cross section for large Higgs boson
\pt, $\pt^{\PH} > 250\GeV$, following the recipe
described in Ref.~\cite{Sirunyan:2018owy}. The uncertainties in the
signal acceptance due to the choice of the PDF set, and the
renormalization and factorization scales, are evaluated independently
for the different signal processes~\cite{Sirunyan:2018owy}, and are
treated as independent nuisance parameters in the fit.

Some of the theoretical uncertainties in the \Vjets and \phojets
processes are expected to mostly cancel in the ratio of the \PWpm
(\PGg) to \PZ processes. The uncertainties are estimated using
the strong production, and applied to both the VBF and strong
production processes. As a conservative choice, for the pQCD NLO corrections,
the renormalization and factorization scales are varied independently
by a factor of two. They are also varied independently for the \Wlvjets and \Zvvjets
processes. The maximum variation in the ratio of the \Wlvjets
to \Zvvjets yields, per \mjj bin, is taken as the uncertainty. 
The maximum variation is generally given by that of the \PW
process. Uncertainties in the PDFs are directly propagated to the
ratio of \Wlvjets to \Zvvjets in a correlated way, and are also
applied per \mjj bin. The full EW correction is taken as an additional
uncertainty in the ratio of \Wlvjets (\phojets) to \Zvvjets for the
strong production of those processes, and is assumed to be uncorrelated
between \mjj bins. The theoretical uncertainties are assumed uncorrelated
between the VBF- and strong-produced \Vjets processes, as
well as between the MTR and VTR categories.

All theoretical uncertainty sources are fully correlated between years.

\section{Results}
\label{sec:res}

A binned maximum likelihood fit is performed simultaneously across the SR and all CRs
in both categories and for both data sets.
In the fit, one parameter per bin $i$ of the \mjj
distribution, for each category and year, is left freely floating.
This parameter represents the expected rate of the background from the strong production of \Zvvjets events, 
and it is labelled \kappazi. The \mjj
distribution is binned in the same way as in
Ref.~\cite{Sirunyan:2018owy}. This binning has been found to be optimal
for the 2017 and 2018 data sets as well. Similar to the method described in
Ref.~\cite{Sirunyan:2018owy}, the likelihood function is defined as:
\begin{linenomath}
\begin{widetext}
\begin{equation}
\label{eq:mlf}
\begin{aligned}
\mathcal{L} (\mu,\kappaz, \boldsymbol{\theta}) = & \prod_{i} \mathrm{P}\left(d_{i} \Big{|} B_{i}(\boldsymbol{\theta}) + \PZ_{i}(\kappazi)+\PW_{i}(\kappazi,\boldsymbol{\theta}) + \mu S_{i}(\boldsymbol{\theta})\right) \\
& \prod_{\mathrm{CR}} \left( \prod_{i} \mathrm{P} \left(d^{\mathrm{CR}}_{i} \Big{|} B^{\mathrm{CR}}_{i}(\boldsymbol{\theta}) + \PV_{i}^{\mathrm{CR,strong}}(\kappazi,\boldsymbol{\theta}) + \PV_{i}^{\mathrm{CR,VBF}}(\kappazi,\boldsymbol{\theta}) \right)\right) \\
 & \prod_{j} \mathrm{P}(\theta_{j}), \\
\PZ_{i}(\kappazi) = & (1+Z^{\frac{\mathrm{VBF}}{\mathrm{strong}}}_{i}) \kappazi, \\
\PW_{i}(\kappazi,\boldsymbol{\theta}) = & (f^{\PW/\PZ\mathrm{,strong}}_{i}(\boldsymbol{\theta}) + Z^{\frac{\mathrm{VBF}}{\mathrm{strong}}}_{i} f^{\PW/\PZ\mathrm{,VBF}}_{i}(\boldsymbol{\theta})) \kappazi, \\
\PV_{i}^{\mathrm{CR,strong}}(\kappazi,\boldsymbol{\theta}) = & C^{\mathrm{CR,strong}}_{i}(\boldsymbol{\theta}) R^{\mathrm{CR,strong}}_{i}(\boldsymbol{\theta}) \kappazi, \\
\PV_{i}^{\mathrm{CR,VBF}}(\kappazi,\boldsymbol{\theta}) = &C^{\mathrm{CR,VBF}}_{i}(\boldsymbol{\theta}) Z^{\frac{\mathrm{VBF}}{\mathrm{strong}}}_{i} R^{\mathrm{CR,VBF}}_{i}(\boldsymbol{\theta}) \kappazi, \\
\end{aligned}
\end{equation}
\end{widetext} 
\end{linenomath}
where ${\mathrm{P}(x|y) = y^{x}\re^{-y}/x!}$, and
$d^{\mathrm{CR}}_{i}$ and $d_{i}$ are the observed number of events in
each bin $i$ of the \mjj distribution in each of the CRs and in the
SR, respectively. The index $i$ runs over the \mjj bins in the two
years and all categories. The symbol $\boldsymbol{\theta}$ refers to
constrained nuisance parameters used for the modelling of the
systematic uncertainties. The signal term $S_i$ represents the
expected signal prediction from the sum of the main Higgs boson
production mechanisms (\ggH, \vbf, \vh, \tth) assuming the cross
sections predicted in the SM. The parameter $\mu = \sigmabr$ denotes
the signal strength parameter, and is also left freely floating in the
fit.

In a given bin, the \Vjets background yields expected in the SR are
obtained from transfer factors relating the yields in the different
CRs to the yields in the SR, separately for the VBF and strong
production processes. These transfer factors are denoted
$R^{\mathrm{CR,proc}}_{i}(\boldsymbol{\theta})$, where ``proc'' can be
strong or VBF, and are obtained from the
simulation. For the single-lepton (dilepton) CRs, the factors
$R^{\mathrm{CR,proc}}_{i}(\boldsymbol{\theta})$ refer to the ratio
of \Wjets (\Zjets) yields from the corresponding CR to the SR. In the
photon CR, which is only available in the MTR category,
$R^{\mathrm{CR,proc}}_{i}(\boldsymbol{\theta}) = 1$.

In addition, transfer factors are defined between the \PW (\PGg) and
the \PZ processes, separately for the VBF and strong production processes.
These transfer factors are denoted as $f^{\PW/\PZ\mathrm{,proc}}_{i}(\boldsymbol{\theta})$
($f^{\Pgg/\PZ\mathrm{,proc}}_{i}(\boldsymbol{\theta})$), where
``proc'' can be strong or VBF. Finally, a transfer factor, 
denoted as $Z^{\frac{\mathrm{VBF}}{\mathrm{strong}}}_{i}$, relates the VBF production
to the strong production of \Zvvjets. The factors
$C^{\mathrm{CR,strong}}_{i}(\boldsymbol{\theta})$ and
$C^{\mathrm{CR,VBF}}_{i}(\boldsymbol{\theta})$ are dependent on the
nature of the CR, with $C^{(\Pe\Pe,\PGm\PGm)\mathrm{,proc}}_{i} =
1$, $C^{(\Pe,\PGm)\mathrm{,proc}}_{i} =
f^{\PW/\PZ\mathrm{,proc}}_{i}(\boldsymbol{\theta})$, and
$C^{\Pgg\mathrm{,proc}}_{i} =
f^{\Pgg/\PZ\mathrm{,proc}}_{i}(\boldsymbol{\theta})$.

The contributions from subleading backgrounds in each region are
estimated directly from simulation and they are denoted by
$B^{\mathrm{CR}}_{i}(\boldsymbol{\theta})$ in the CRs, and
$B_{i}(\boldsymbol{\theta})$ in the SR.

Systematic uncertainties are modelled as constrained nuisance
parameters ($\boldsymbol{\theta}$), with a log-normal distribution for
those that affect the overall normalization of a given process, and
Gaussian priors for those that directly affect the transfer factors,
indicated by $\mathrm{P}(\theta_{j})$ in Eq.~(\ref{eq:mlf}). The
impact on the transfer factors of each of the uncertainties described
in Section~\ref{sec:systs} is summarized in Table~\ref{tab:systs}.

\begin{table*}[!htbp]
    \centering
    \renewcommand{\arraystretch}{1.1}
    \topcaption{Experimental and theoretical sources of systematic uncertainty in the \Vjets transfer factors. The second column indicates which ratio a given source of uncertainty is applied to. The impact on \mjj is given in the 3rd column, as a single value if no dependence on \mjj is observed, or as a range. When a range is shown, the uncertainty increases with the value of \mjj. The quoted uncertainty values represent the absolute value of the relative change in the transfer factor corresponding to a variation of 
${\pm}1$ standard deviation in the systematic uncertainty, or equivalently to a variation 
of ${\pm}1$ in the corresponding nuisance parameter $\theta$ in Eq.~(\ref{eq:mlf}).
    }
    \cmsTable{
    \begin{scotch}{l l c}
       Source of uncertainty & Ratios & Uncertainty vs. \mjj\\
       \hline
       \multicolumn{3}{c}{\textit{Theoretical uncertainties}} \\ [\cmsTabSkip]
       Ren. scale \Vjets (VBF)   & $f^{\PW/\PZ\mathrm{,VBF}}_{i}$   & 7.5\%  \\
       Ren. scale \Vjets (strong)  & $f^{\PW/\PZ\mathrm{,strong}}_{i}$  & 8.2\%  \\
       Fac. scale \Vjets (VBF)   & $f^{\PW/\PZ\mathrm{,VBF}}_{i}$   & 1.5\%   \\
       Fac. scale \Vjets (strong)  & $f^{\PW/\PZ\mathrm{,strong}}_{i}$  & 1.3\%   \\
       PDF \Vjets (VBF)          & $f^{\PW/\PZ\mathrm{,VBF}}_{i}$   & 0\% \\
       PDF \Vjets (strong)         & $f^{\PW/\PZ\mathrm{,strong}}_{i}$  & 0\% \\
       NLO EW corr. \Vjets (strong) & $f^{\PW/\PZ\mathrm{,strong}}_{i}$ & 0.5\%   \\
       Ren. scale \phojets (VBF)   & $f^{\Pgg/\PZ\mathrm{,VBF}}_{i}$   & 6--10\%  \\
       Ren. scale \phojets (strong)  & $f^{\Pgg/\PZ\mathrm{,strong}}_{i}$  & 6--10\%  \\
       Fac. scale \phojets (VBF)   & $f^{\Pgg/\PZ\mathrm{,VBF}}_{i}$   & 2.5\%   \\
       Fac. scale \phojets (strong)  & $f^{\Pgg/\PZ\mathrm{,strong}}_{i}$  & 2.5\%   \\
       PDF \phojets (VBF)          & $f^{\Pgg/\PZ\mathrm{,VBF}}_{i}$   & 2.5\% \\
       PDF \phojets (strong)         & $f^{\Pgg/\PZ\mathrm{,strong}}_{i}$  & 2.5\% \\
       NLO EW corr. \phojets & $f^{\Pgg/\PZ\mathrm{,strong}}_{i}$ & 3\%   \\ [\cmsTabSkip]

       \multicolumn{3}{c}{\textit{Experimental uncertainties}}  \\ [\cmsTabSkip]
       Electron reco. eff.       & $R^{\mathrm{CR,proc}}_{i}$, CR$=$\Zee or \Wen & $\approx$0.5\% (per lepton)  \\
       Electron id. eff.         & $R^{\mathrm{CR,proc}}_{i}$, CR$=$\Zee or \Wen & $\approx$1\% (per lepton) \\
       Muon id. eff.             & $R^{\mathrm{CR,proc}}_{i}$, CR$=$\Zmm or \Wmn & $\approx$0.5\% (per lepton)  \\
       Muon iso. eff.            & $R^{\mathrm{CR,proc}}_{i}$, CR$=$\Zmm or \Wmn & $\approx$0.1\% (per lepton)  \\
       Photon id. eff.           & $f^{\Pgg/\PZ\mathrm{,proc}}_{i}$       & 5\% \\
       
       Electron veto (reco)      & $f^{\PW/\PZ\mathrm{,proc}}_{i}$, $R^{\mathrm{CR,proc}}_{i}$, CR$=$\Wlv & $\approx$1.5 (1)\% for VBF (strong) \\
       Electron veto (id)        & $f^{\PW/\PZ\mathrm{,proc}}_{i}$, $R^{\mathrm{CR,proc}}_{i}$, CR$=$\Wlv & $\approx$2.5 (2)\% for VBF (strong) \\
       Muon veto                 & $f^{\PW/\PZ\mathrm{,proc}}_{i}$, $R^{\mathrm{CR,proc}}_{i}$, CR$=$\Wlv & $\approx$0.5\% \\
       \tauh veto                & $f^{\PW/\PZ\mathrm{,proc}}_{i}$, $R^{\mathrm{CR,proc}}_{i}$, CR$=$\Wlv & $\approx$1\% \\

       Electron trigger          & $R^{\mathrm{CR,proc}}_{i}$, CR$=$\Zee or \Wen  & $\approx1\%$  \\
       \ptmiss trigger           & $R^{\mathrm{CR,proc}}_{i}$, CR$=$\Zmm or \Wmn & $\approx2\%$  \\
       Photon trigger            & $f^{\Pgg/\PZ\mathrm{,proc}}_{i}$        & 1\% \\ [\cmsTabSkip]

       \multirow{4}{*}{JES}     & $f^{\PW/\PZ\mathrm{,proc}}_{i}$         & 1--2\% \\
                                &  $R^{\mathrm{CR,proc}}_{i}$, CR$=$\Wen or \Wmn     & 1.0--1.5\%  \\
                               	& $R^{\mathrm{CR,proc}}_{i}$, CR$=$\Zee or \Zmm	& 1\% \\
                               	& $f^{\Pgg/\PZ\mathrm{,proc}}_{i}$	    	& 3\% \\ [\cmsTabSkip]

       \multirow{4}{*}{JER}  	& $f^{\PW/\PZ\mathrm{,proc}}_{i}$         & 1.0--2.5\%  \\
                                & $R^{\mathrm{CR,proc}}_{i}$, CR$=$\Wen or \Wmn               & 1.0--1.5\% \\
                                & $R^{\mathrm{CR,proc}}_{i}$, CR$=$\Zee or \Zmm		& 1\% \\
                                & $f^{\Pgg/\PZ\mathrm{,proc}}_{i}$	    	& 1--4\%  \\
    \end{scotch}
}
    \label{tab:systs}
\end{table*}

\cmsClearpage

In the following, the expected background yields used as input to the
fit procedure are denoted as ``prefit'', while the yields after a fit
to the CRs, or the CRs and SR, are denoted as ``CR-postfit'', or
``postfit'', respectively.

The results are presented for the MTR and VTR categories separately. They are shown
separately for the 2017 and 2018 data sets when presenting the transfer factors,
and for the data sets added together otherwise. The corresponding figures and tables 
split into individual control regions and data-taking periods are available in \suppMaterialbis.

All results are obtained
from a combined fit across all categories and data sets.

For the maximum likelihood fit, the different data sets and categories
are treated separately. The final results are obtained from a fit
using the combined likelihood function including all categories and
data sets, which takes into account nuisance correlations.

\subsection{Control regions for the MTR category}
\label{sec:MTR}

\begin{figure*}[htb!]
\centering
\includegraphics[width=0.48\textwidth]{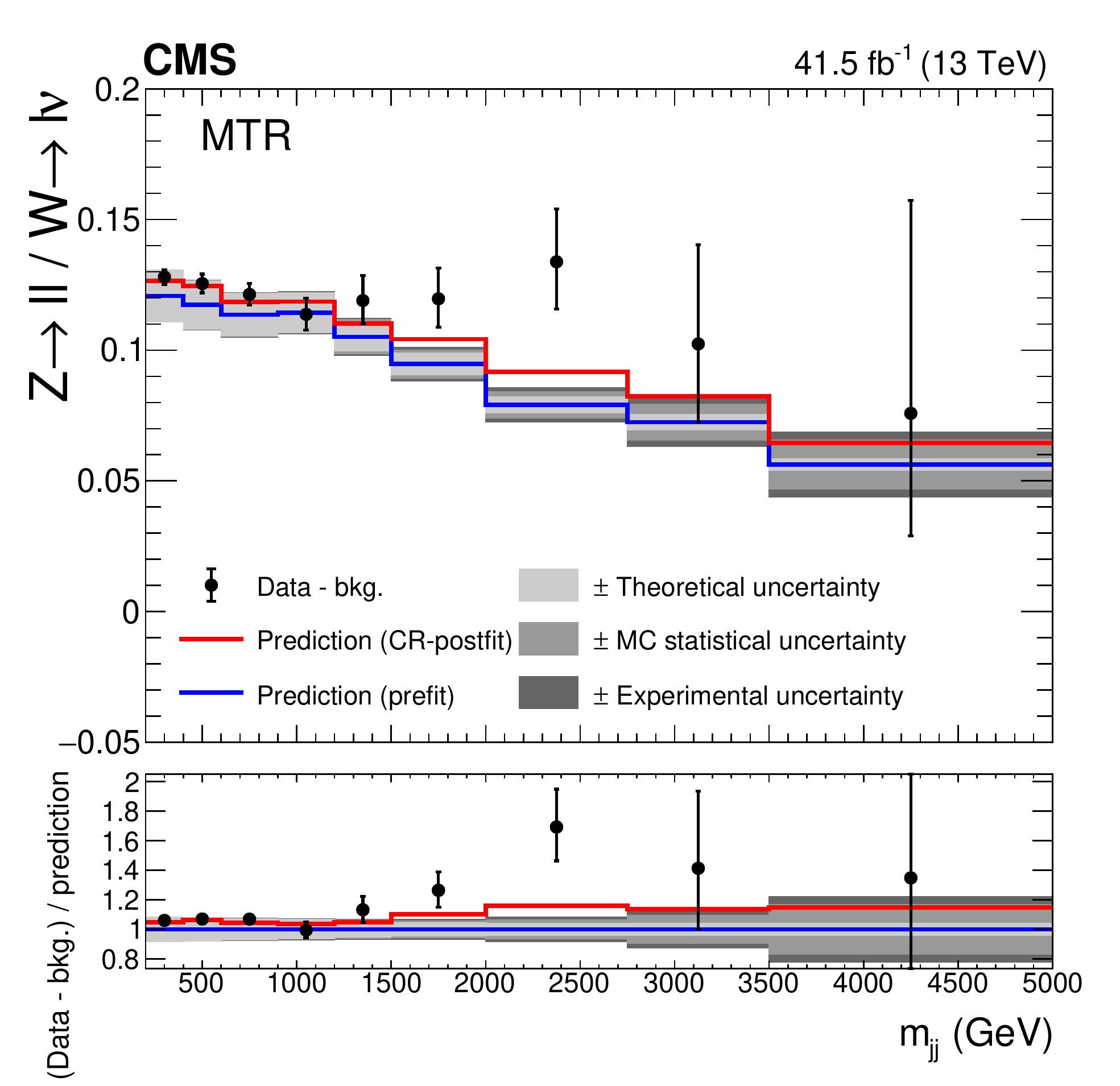}
\includegraphics[width=0.48\textwidth]{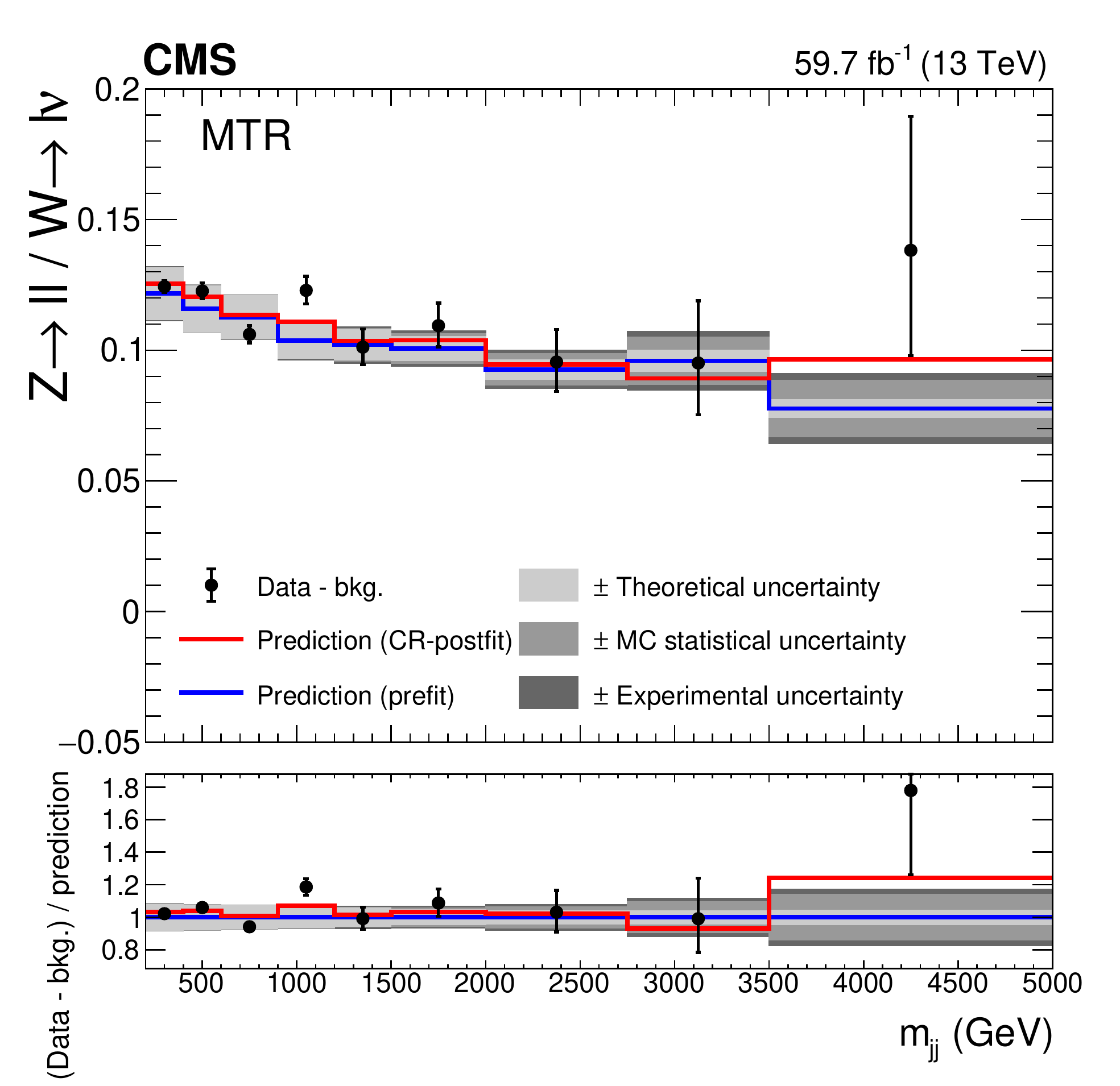}\\
\centering
\includegraphics[width=0.48\textwidth]{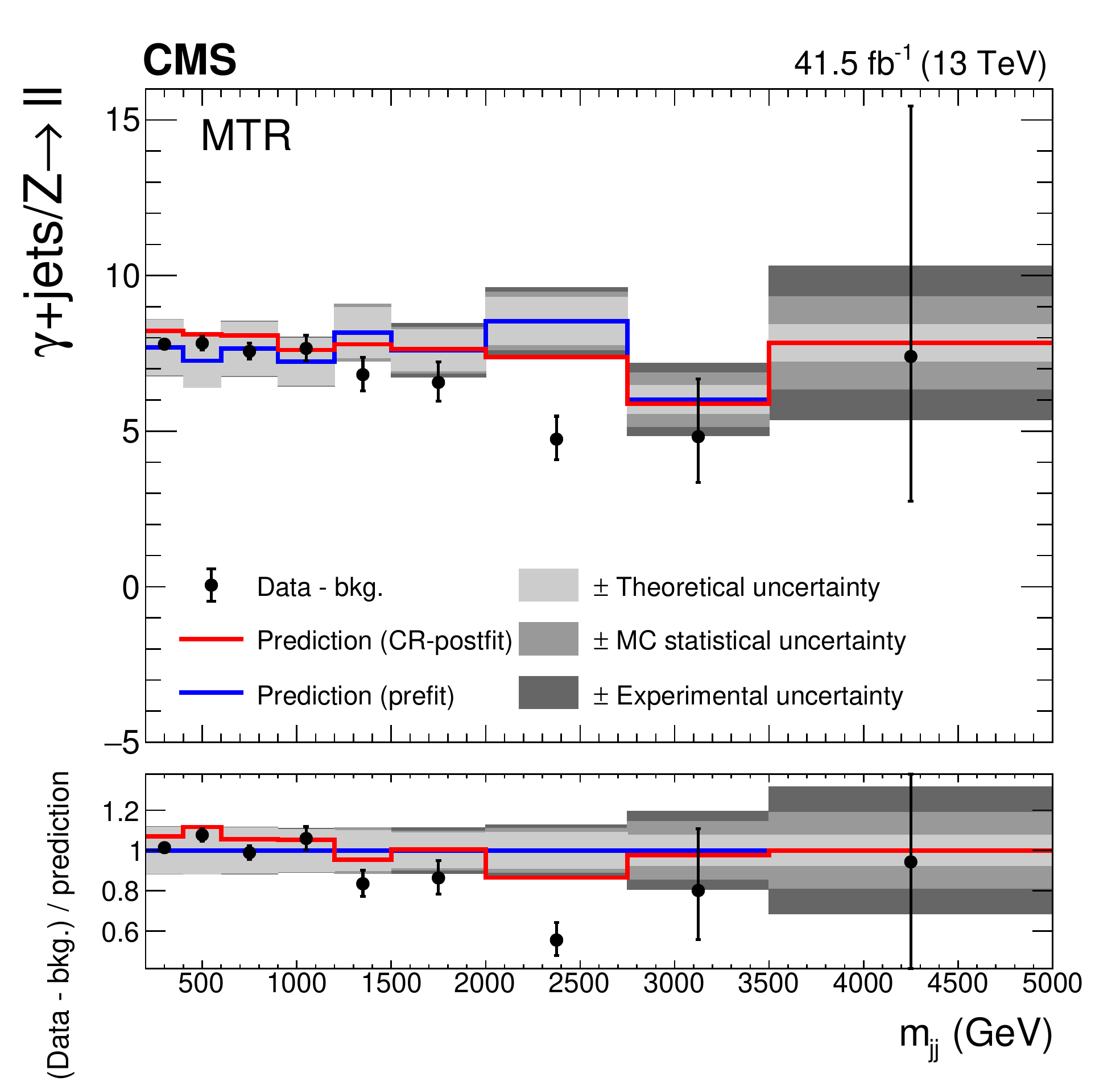}
\includegraphics[width=0.48\textwidth]{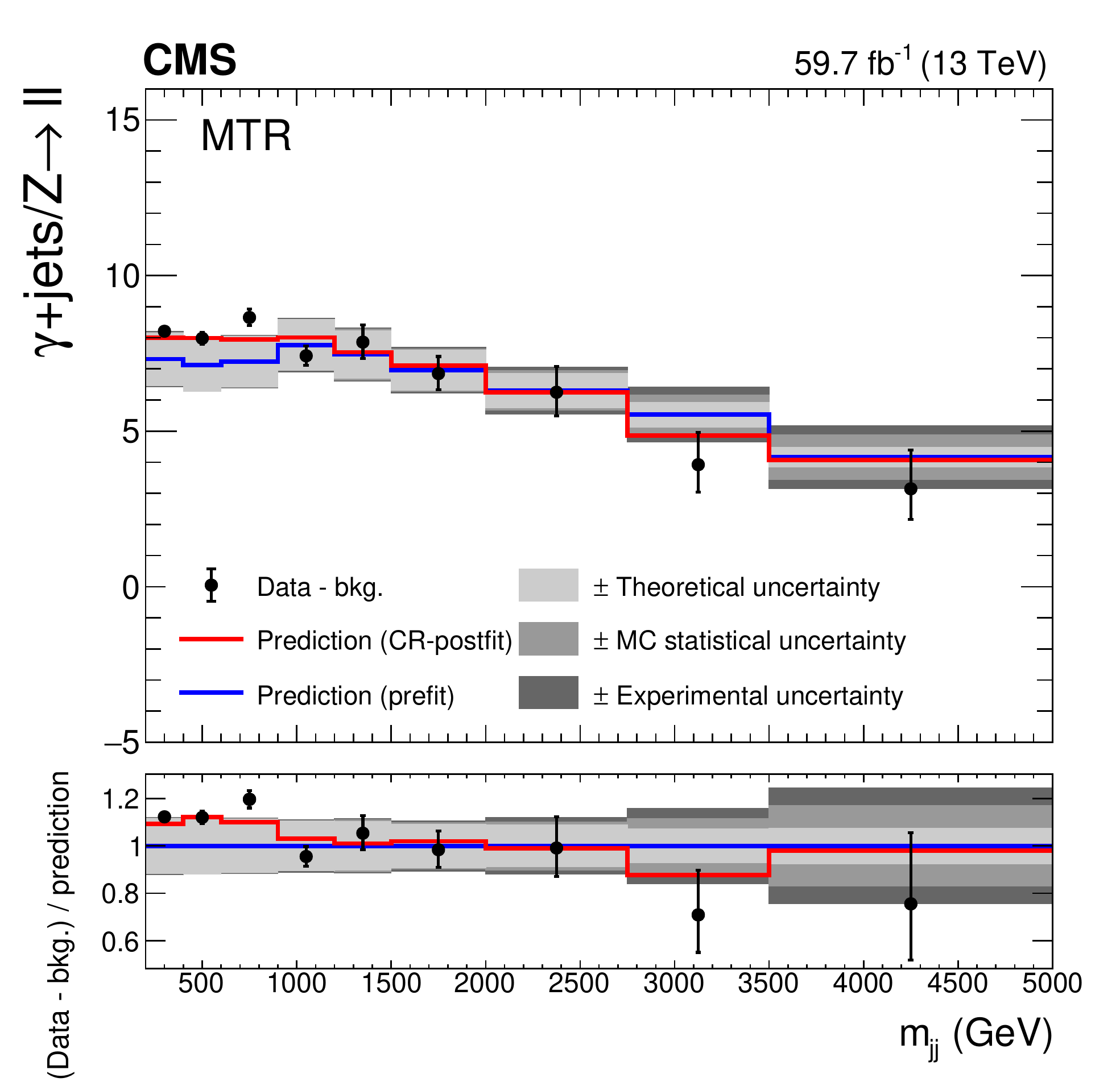}
\caption{Comparison between data and simulation for the \Zlljets/\Wlvjets (upper row) and \phojets/\Zlljets (lower row) prefit and CR-postfit ratios, as functions of \mjj, for the MTR category using the 2017 (left) and 2018 (right) event samples. The minor backgrounds (bkg.) in each CR are subtracted from the data using estimates from simulation. The grey bands include the theoretical and experimental systematic uncertainties listed in Table~\ref{tab:systs}, as well as the statistical uncertainty in the simulation.}
\label{fig:ZWG_MTR}
\end{figure*}

The ratio of dilepton to single-lepton events is studied to validate
the predictions from simulation and uncertainties used to model the ratio of \Zjets
to \Wjets events in the SR (parameters
$f^{\PW/\PZ\mathrm{,proc}}_{i}(\boldsymbol{\theta})$ in
Eq.~(\ref{eq:mlf})). The prefit ratio between the number of \Zjets
and \Wjets events in the CRs in bins of \mjj is shown in
Fig.~\ref{fig:ZWG_MTR} (upper row) for the 2017 (left) and 2018
(right) data sets. The prediction is compared to the ratio of observed
data yields, from which the estimates of minor background processes
have been subtracted. The CR-postfit results are shown together with
the prefit ratio. A reasonable agreement is observed between data and
simulation, with differences in most bins covered by the systematic
uncertainties listed in Table~\ref{tab:systs}. In 2017, the simulation
predicts a lower ratio than observed in data. The compatibility of the
data with the prefit prediction is measured, in this particular
category and year only, using a $\chi^2$ test accounting for
correlations between the 2017 MTR CRs and each \mjj bin. This test
indicates that there is a local discrepancy of approximately two
standard deviations. The disagreement is attributed to the \Zjets
regions with low event yields, and is partially compensated in the fit
through the movement of nuisance parameters representing uncertainties
such as the renormalization and factorization scales. The significance
of the discrepancy is low and none of the nuisance parameters move by
more than one standard deviation from their prefit value in the
combined fit across all categories and years. The
$p$-value~\cite{pvalue} for the 2017 data set in the MTR category
after the combined fit is 38.4\%, and it is 37.0\% for all categories
when combining the 2017 and 2018 data sets. Closure tests have also
been performed comparing the decays to electrons or muons separately
for the \PW and \PZ CRs, again showing reasonable agreement between
data and simulation.

The ratio of photon to dilepton events is also studied to validate the
predictions from simulation and uncertainties used to model the ratio of \phojets
to \Zjets events in the SR (parameters
$f^{\Pgg/\PZ\mathrm{,proc}}_{i}(\boldsymbol{\theta})$ in
Eq.~(\ref{eq:mlf})). The prefit and CR-postfit ratios between the
number of events in the \phojets over \Zjets CRs in bins of \mjj are
shown in Fig.~\ref{fig:ZWG_MTR} (lower row) for the 2017 (left) and
2018 (right) data sets.
The prediction is compared to the ratio of observed data yields, from
which the estimates of minor background processes have been
subtracted. A similar observation can be made for the 2017 MTR
category as for the \Zjets over \Wjets ratio. This effect is again
attributed to the \Zjets CRs with low event yields.

\begin{figure*}[htb!]
\centering
\includegraphics[width=0.48\textwidth]{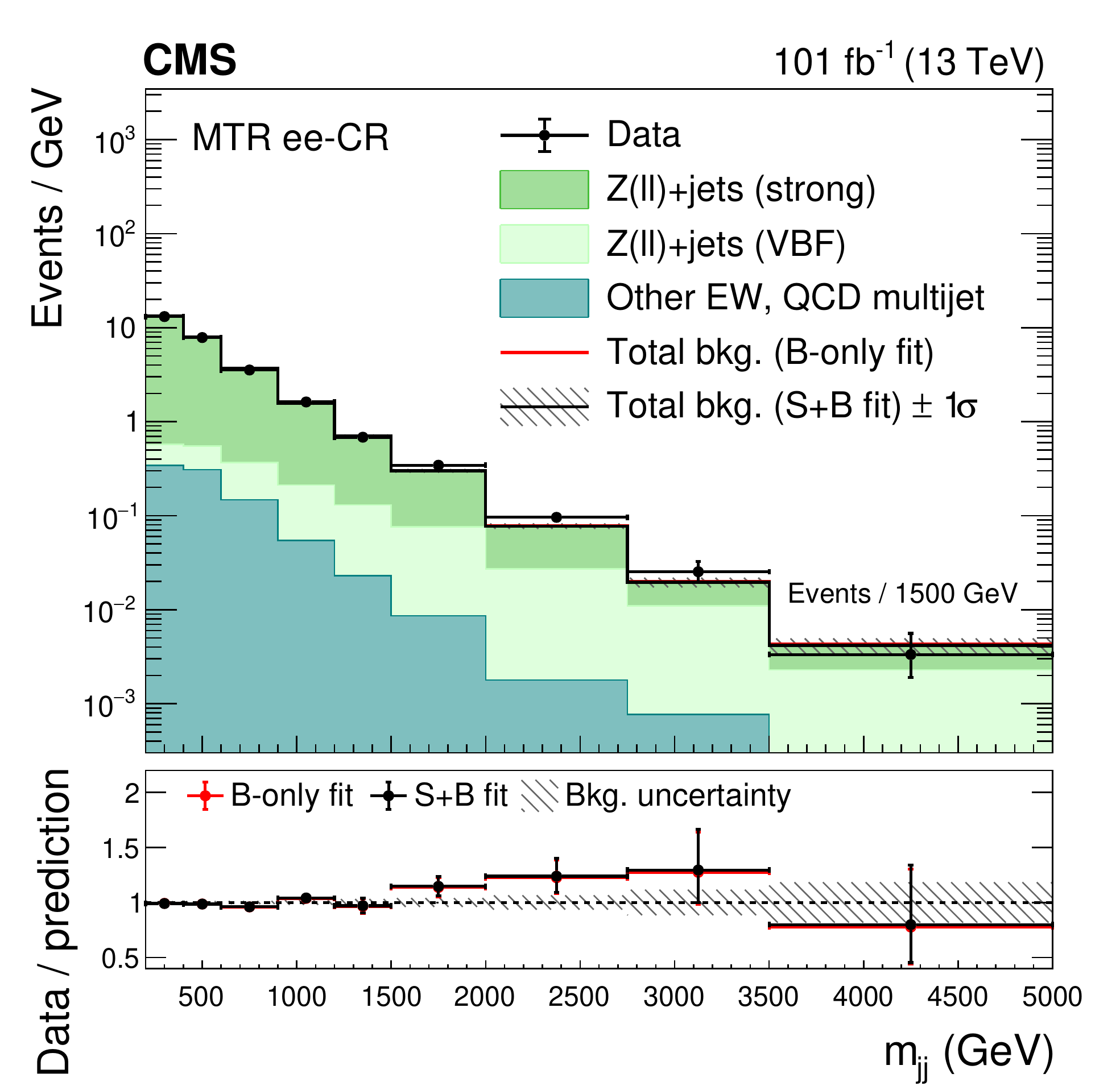}
\includegraphics[width=0.48\textwidth]{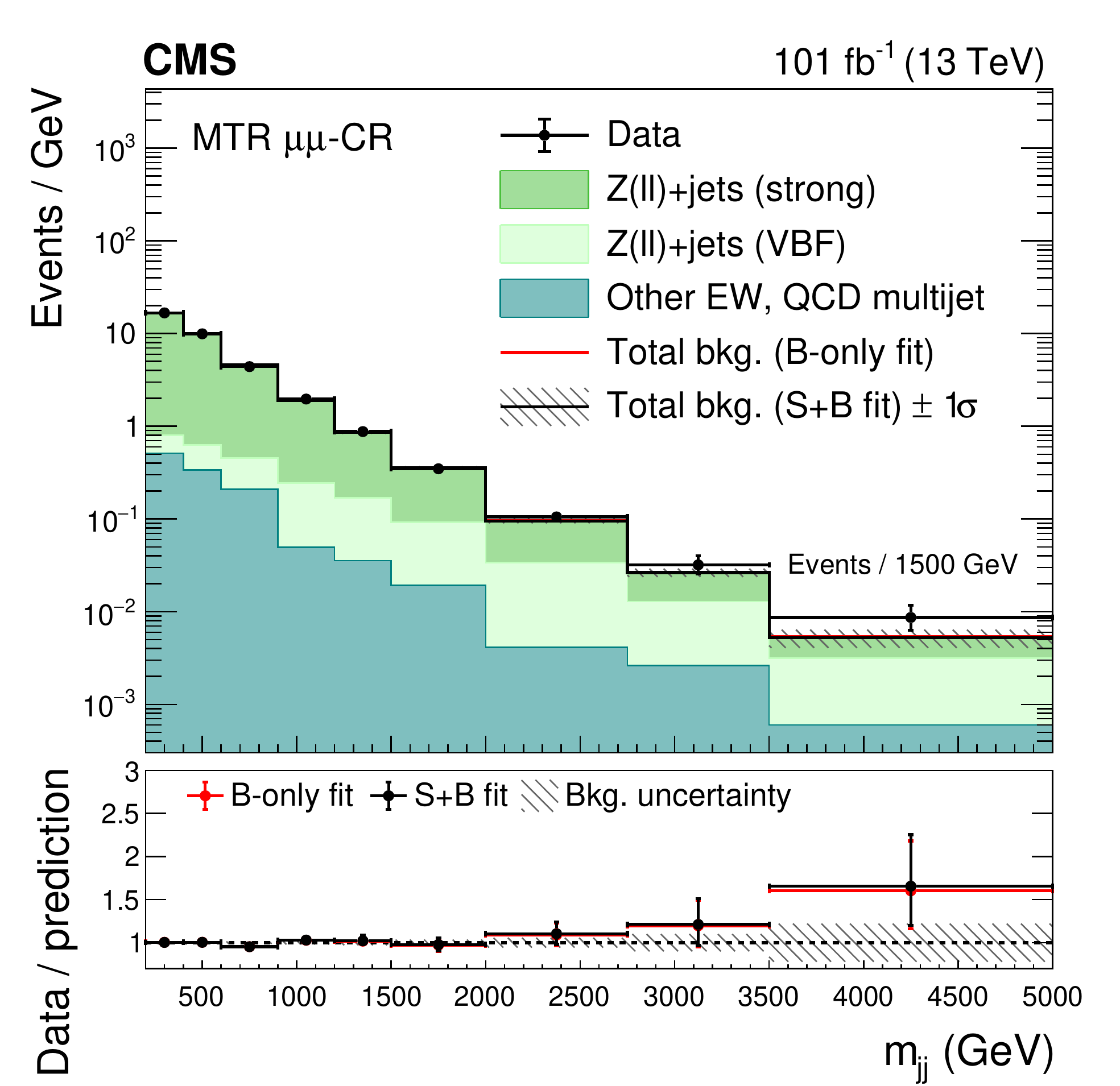}\\
\includegraphics[width=0.48\textwidth]{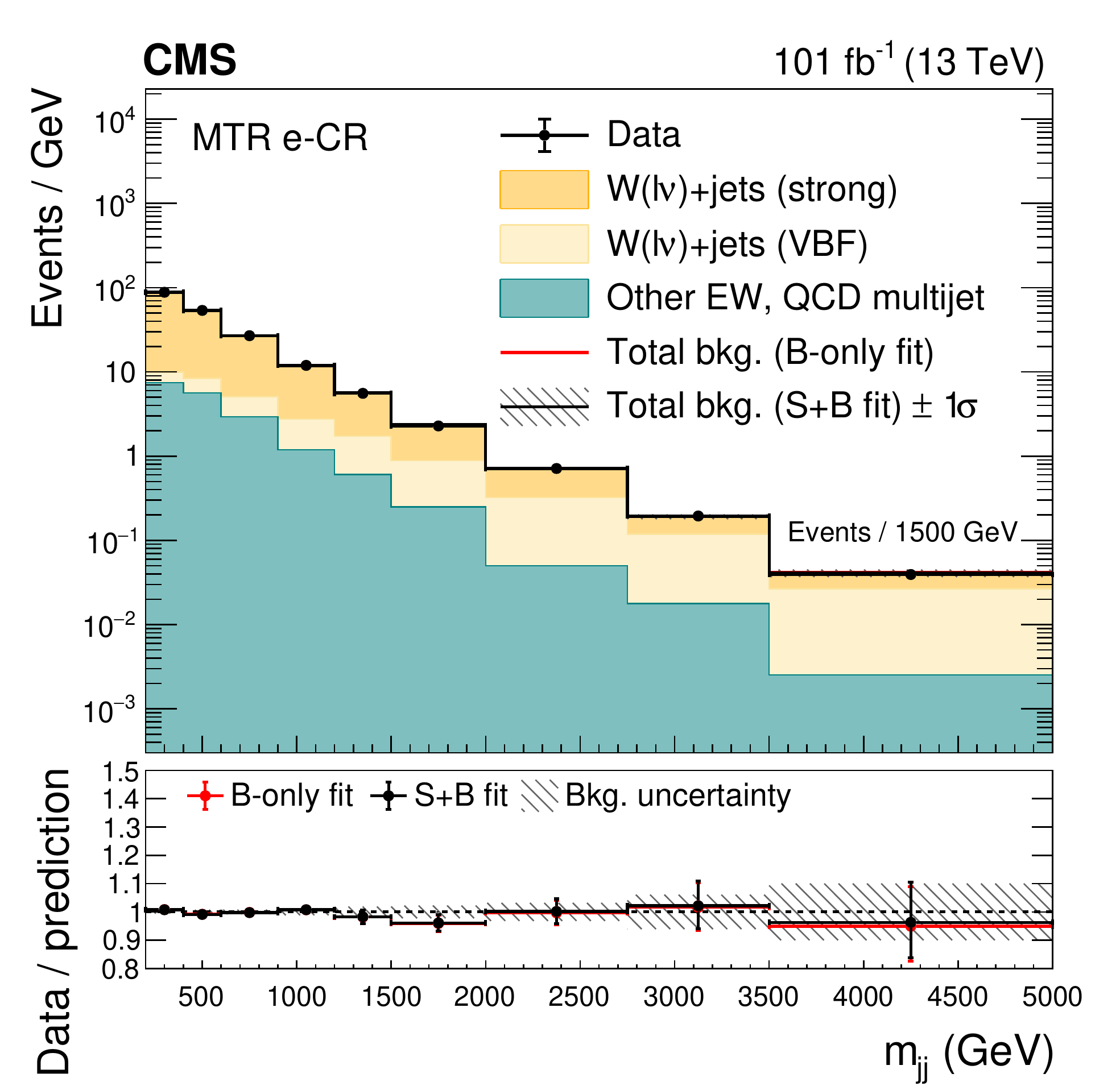}
\includegraphics[width=0.48\textwidth]{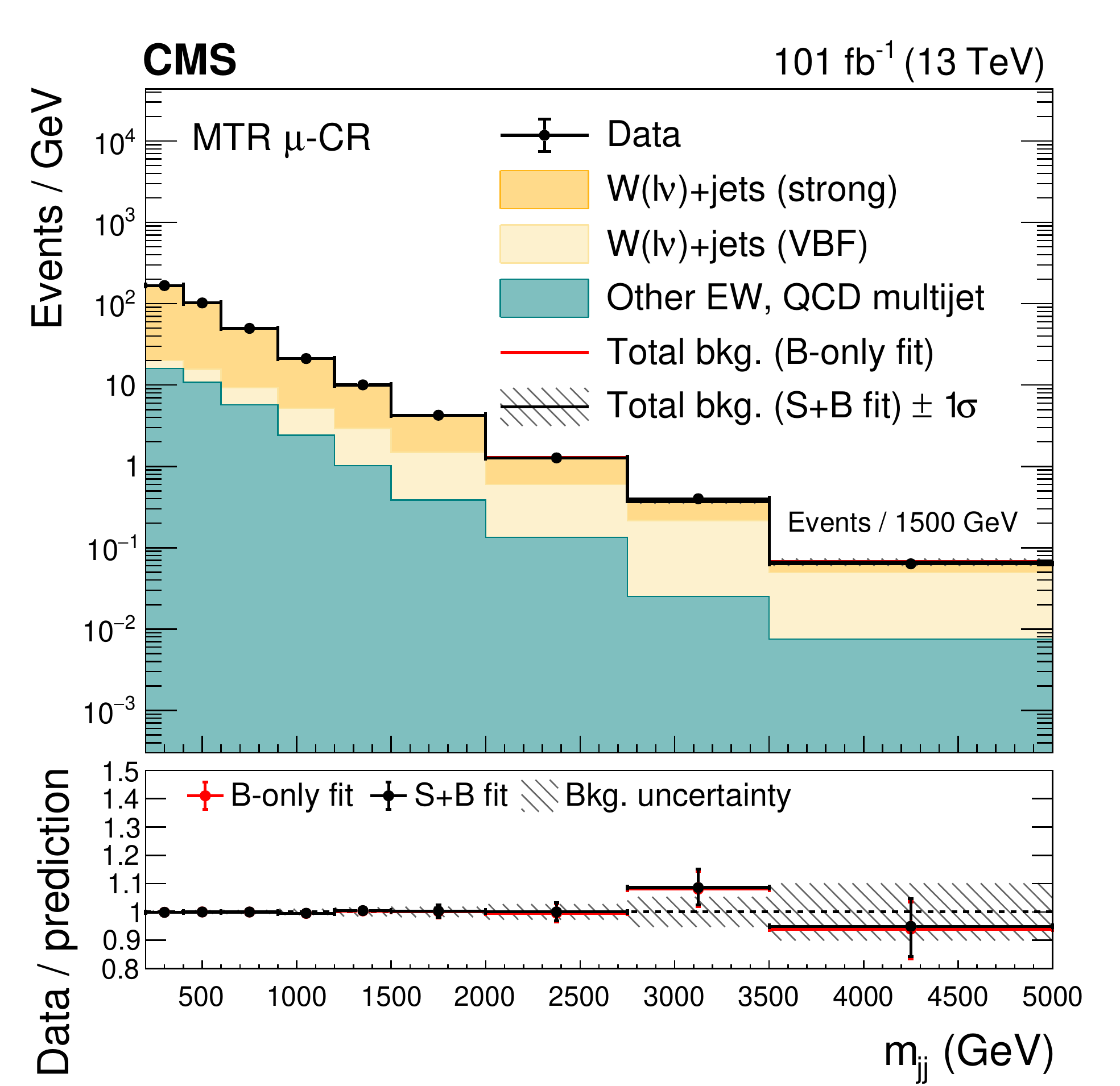}
\caption{
The postfit \mjj distributions in the dielectron (upper
left), dimuon (upper right), single-electron (lower left), and
single-muon (lower right) CR for the MTR category, showing the summed 2017
and 2018 data samples and the background processes. The background contributions are estimated from 
the fit to the data described in the text (S+B fit). The total background (bkg.) estimated from a fit 
assuming $\brinv=0$ (B-only fit) is also shown.
The yields from the 2017 and 2018 samples are summed and the correlations between their uncertainties are neglected.
The last bin of each distribution integrates events above the bin threshold divided by the bin width.}
\label{fig:CR_MTR}
\end{figure*}

\begin{figure}[htb!]
\centering
\includegraphics[width=\cmsFigWidth]{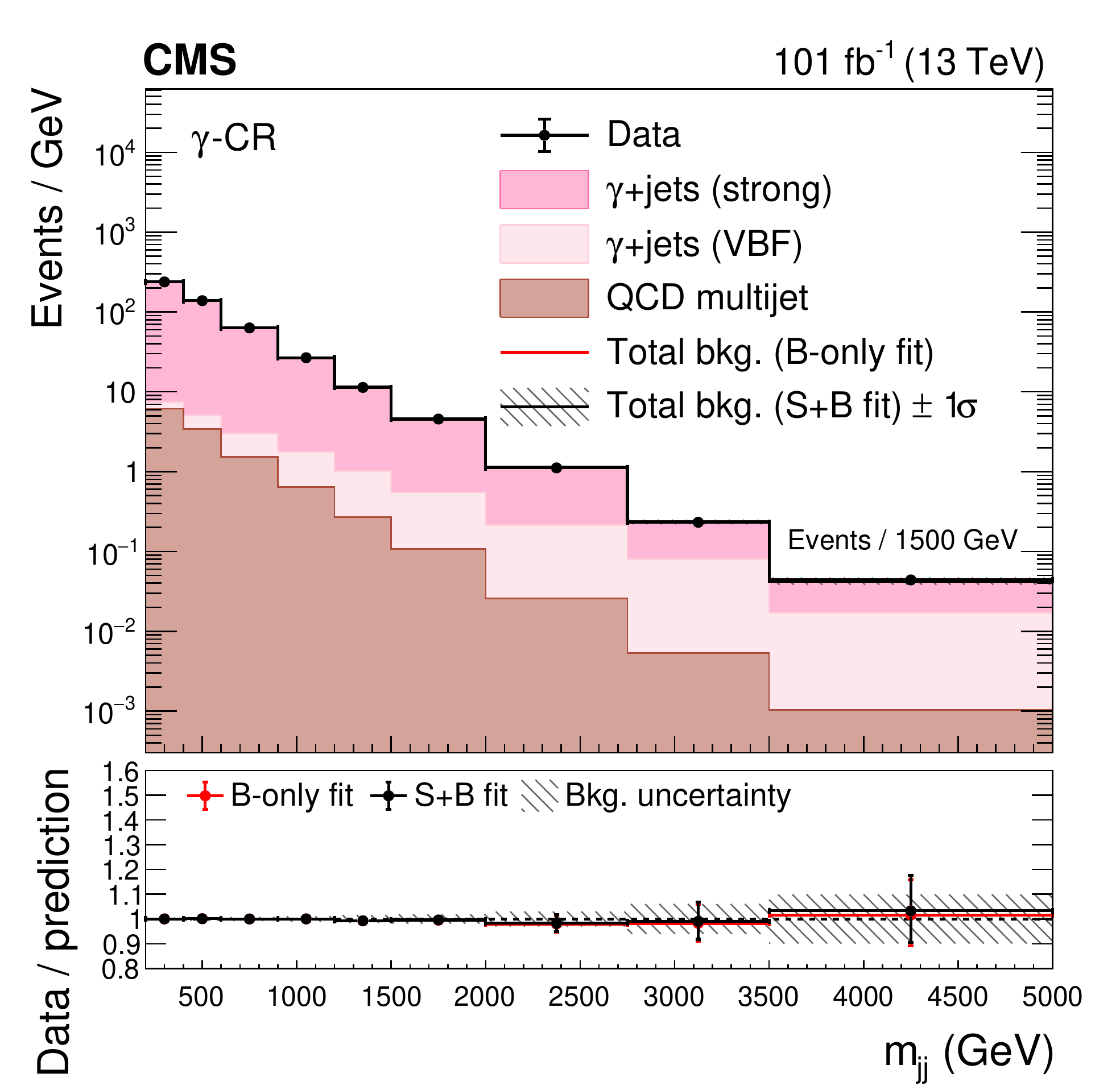}
\caption{The postfit \mjj distribution in the photon CR for 
the MTR category, showing the summed 2017 and 2018 data samples and the background processes. The background contributions are estimated from 
the fit to the data described in the text (S+B fit). The total background (bkg.) estimated from a fit 
assuming $\brinv=0$ (B-only fit) is also shown.
The yields from the 2017 and 2018 samples are summed and the correlations between their uncertainties are neglected.
The last bin of each distribution integrates events above the bin threshold divided by the bin width.}
\label{fig:Zgamma_MTR}
\end{figure}

The \mjj distributions in data in the dilepton and single-lepton CRs,
along with the postfit estimates of the background contributions, are
shown in Fig.~\ref{fig:CR_MTR}.
Similar distributions in the photon CR
are shown in Fig.~\ref{fig:Zgamma_MTR}.
The total background estimated from a fit assuming $\brinv=0$ is also
shown.  The distributions show the sum of the 2017 and 2018 data sets
in each region. The postfit predictions are in good agreement with the
data within one standard deviation for most of the bins, with
discrepancies of just over one standard deviation in the dielectron
MTR CR for two bins at $\mjj\approx2\TeV$. As discussed previously in
the context of the validation of the \Zjets over \Wjets and
the \phojets over \Zjets ratios in the CRs, the prefit disagreement
observed in the 2017 MTR is partially compensated by the nuisance
parameters in the fit, resulting in an overall $p$-value of 37.0\%.

\subsection{Control regions for the VTR category}
\label{sec:VTR}

\begin{figure}[htb!p]
\centering
\includegraphics[width=0.48\textwidth]{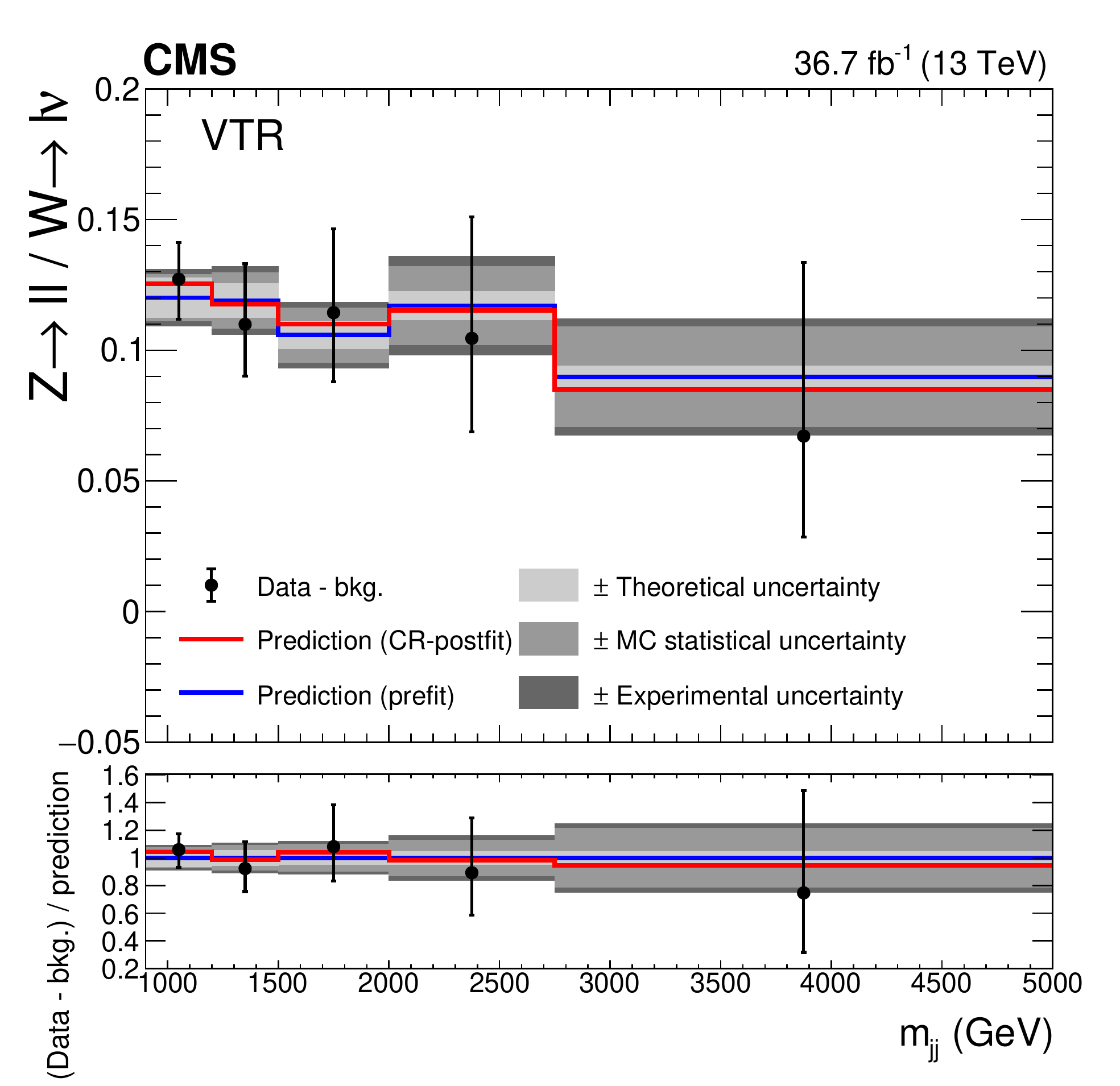}\hfill
\includegraphics[width=0.48\textwidth]{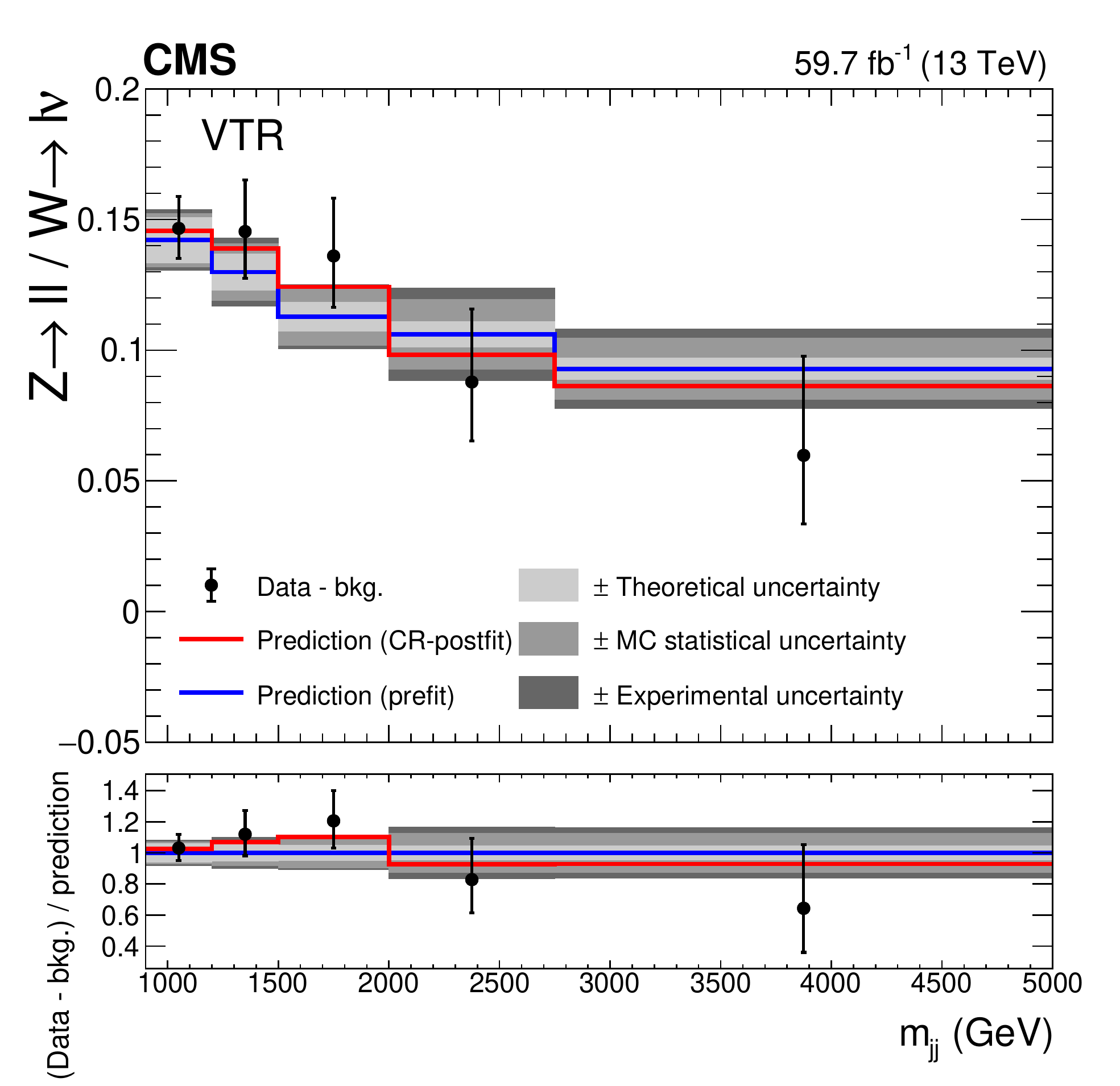}
\caption{Comparison between data and simulation for the \Zlljets/\Wlvjets prefit and CR-postfit ratios, as functions of \mjj, for the VTR category in the 2017
(\cmsLeft) and 2018 (\cmsRight) data samples. The minor backgrounds (bkg.) in
each CR are subtracted from the data using estimates from simulation. The grey
bands include the theoretical and experimental systematic
uncertainties listed in Table~\ref{tab:systs}, as well as the
statistical uncertainty in the simulation.}
\label{fig:ZW_VTR}
\end{figure}

The prefit and CR-postfit ratios between the number of \Zjets
and \Wjets events in the CRs in bins of \mjj are shown in
Fig.~\ref{fig:ZW_VTR}. The predictions from simulated events are found to model the data
in all bins, within the quoted uncertainties.

\begin{figure*}[htb!p]
\centering
\includegraphics[width=0.48\textwidth]{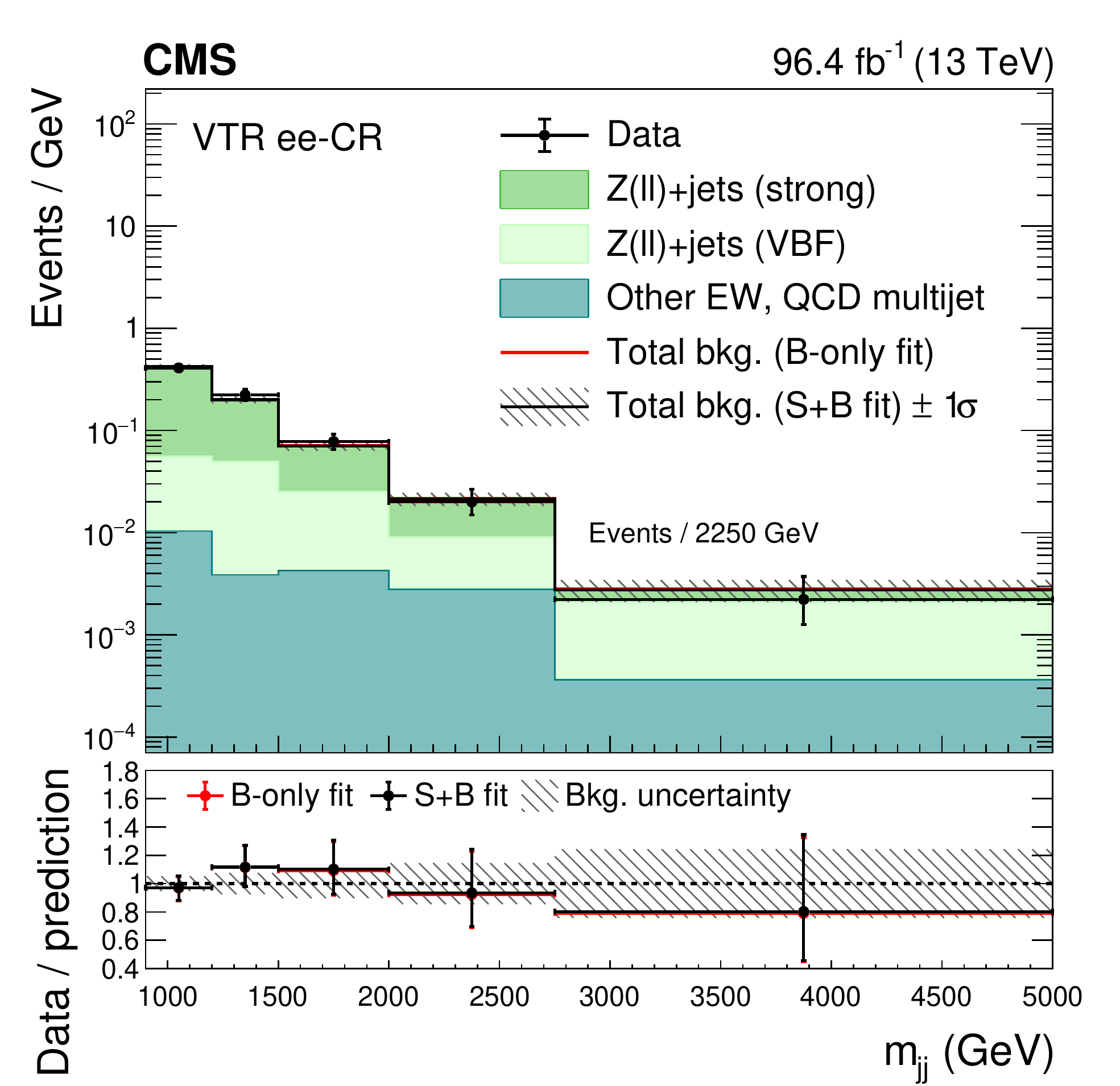}
\includegraphics[width=0.48\textwidth]{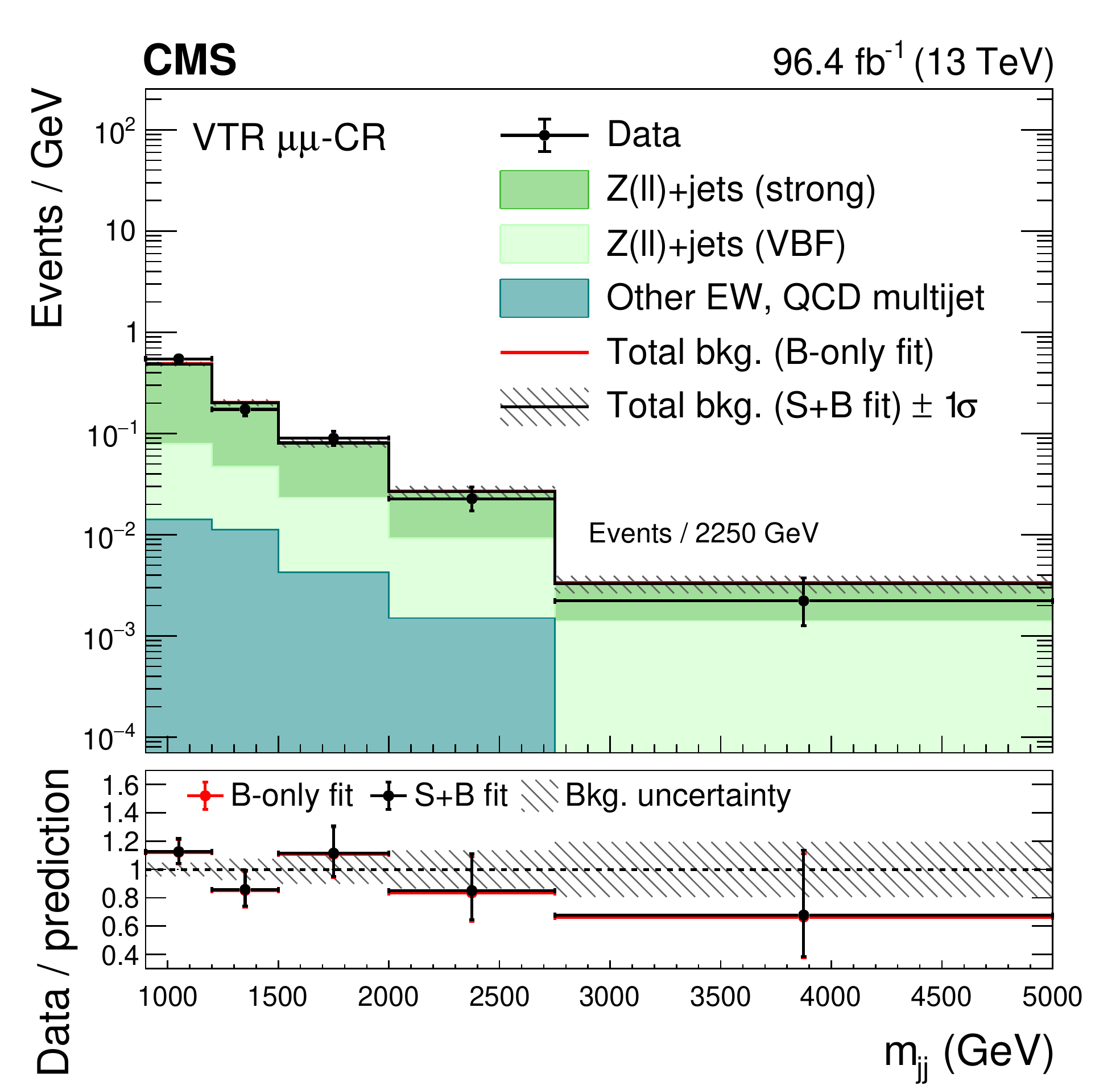}\\
\includegraphics[width=0.48\textwidth]{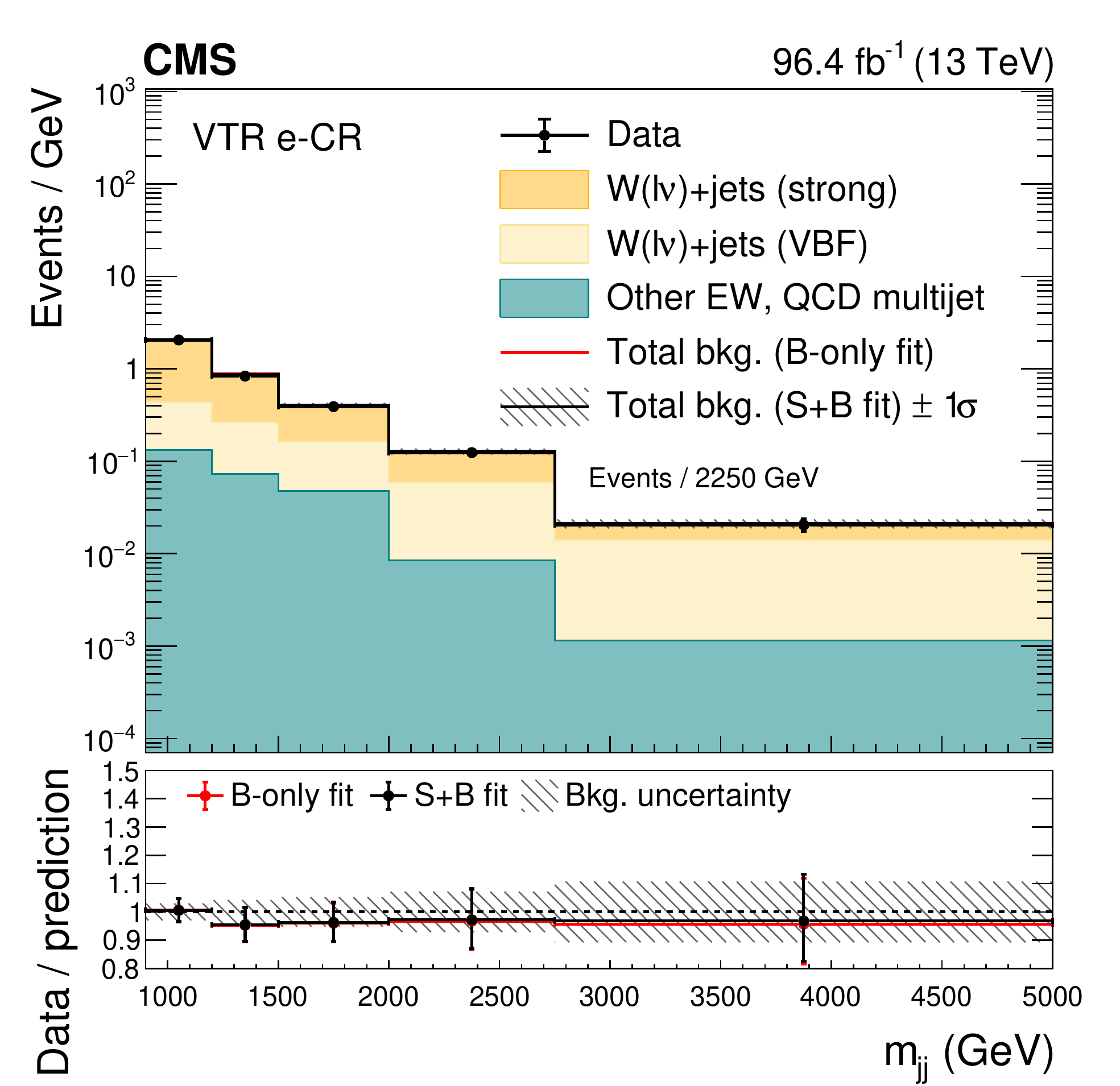}
\includegraphics[width=0.48\textwidth]{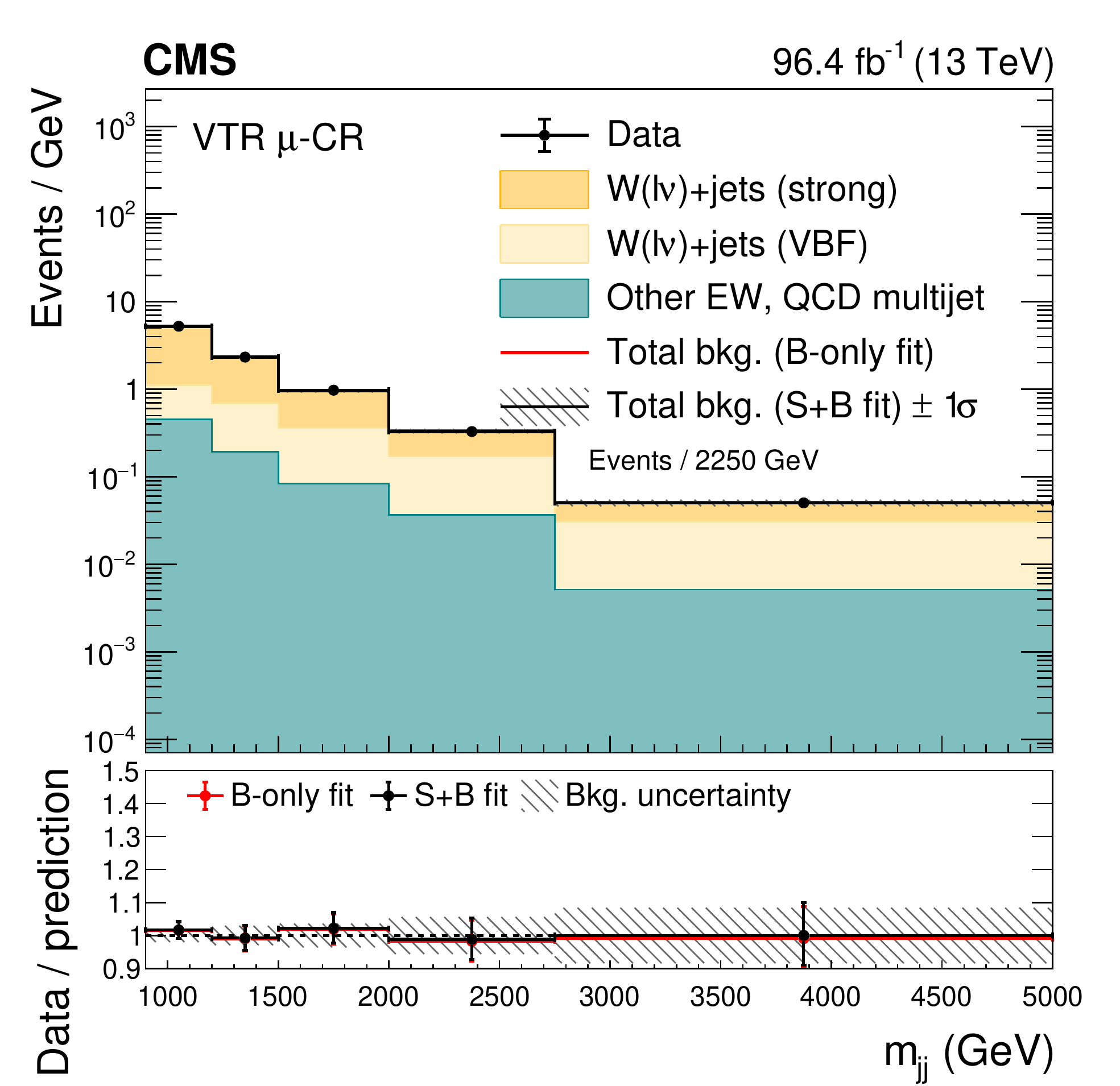}
\caption{
The postfit \mjj distributions in the dielectron (upper left), dimuon
(upper right), single-electron (lower left), and single-muon (lower
right) CRs for the VTR category, showing the summed 2017 and 2018 data
samples and the SM background processes.  The background contributions
are estimated from the fit to the data described in the text (S+B
fit). The total background (bkg.) estimated from a fit assuming $\brinv=0$
(B-only fit) is also shown.  The yields from the 2017 and 2018 samples
are summed and the correlations between their uncertainties are
neglected.  The last bin of each distribution integrates events above
the bin threshold divided by the bin width.}
\label{fig:CR_VTR}
\end{figure*}

The \mjj distributions in the dilepton and single-lepton CRs are shown
in Fig.~\ref{fig:CR_VTR}, along with the CR-postfit and postfit estimates.
Again, a good agreement between the data and the predictions is shown,
within the uncertainties.

\subsection{Signal region fits}

\begin{figure}[htb!]
\centering
\includegraphics[width=0.48\textwidth]{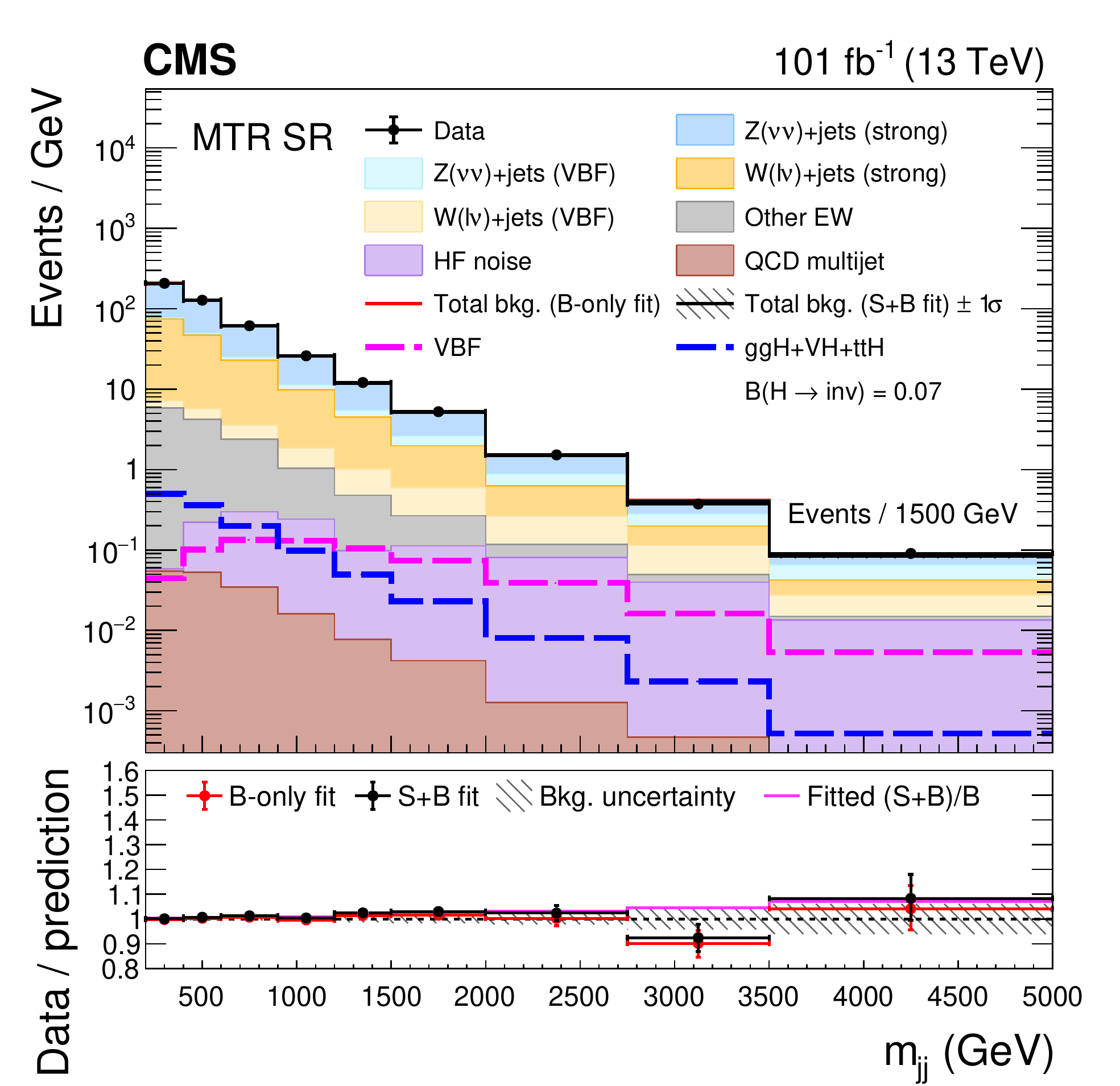}\hfill
\includegraphics[width=0.48\textwidth]{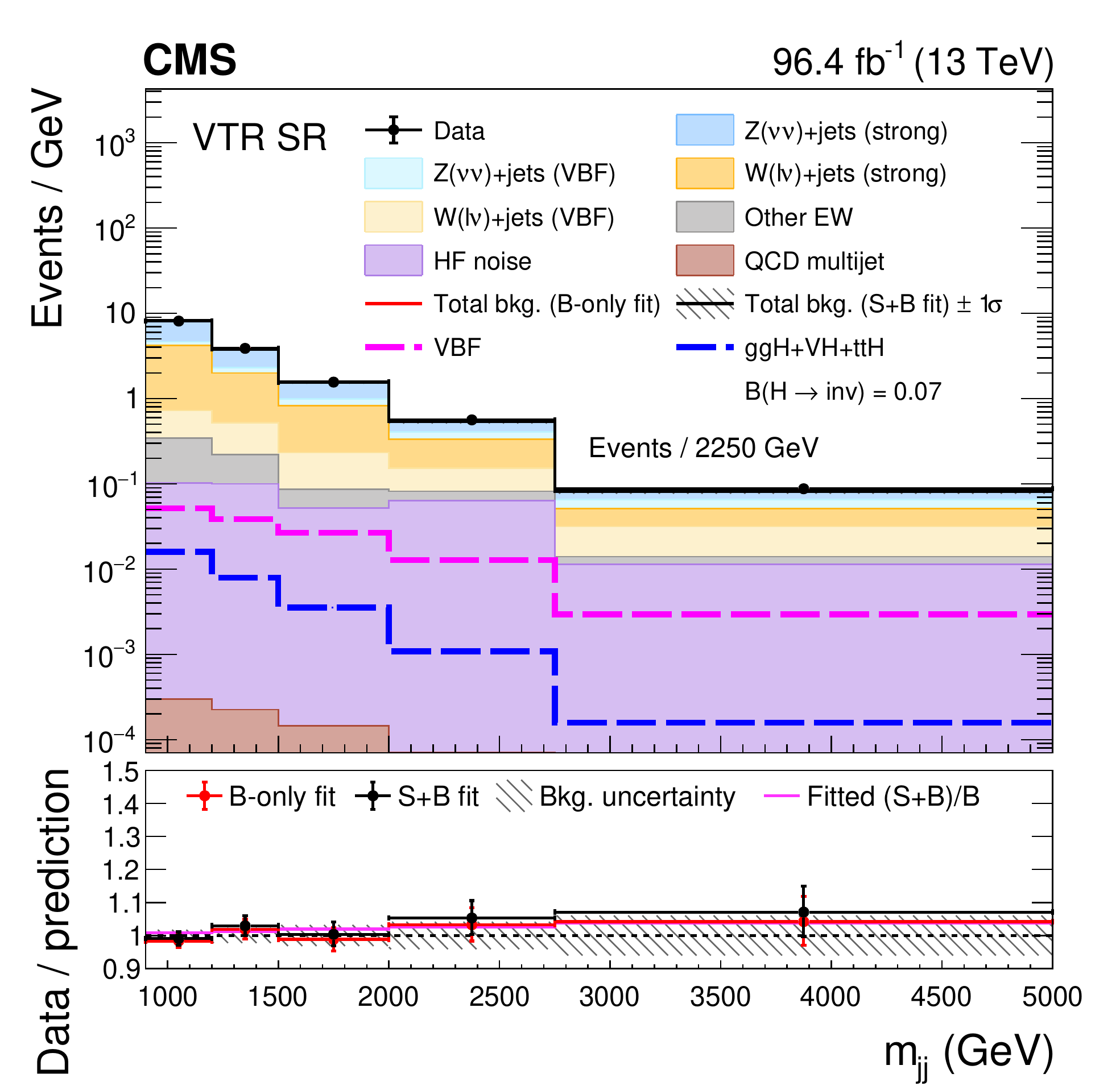}
\caption{The observed \mjj distribution in the MTR (\cmsLeft) and VTR (\cmsRight) SR compared with 
the postfit backgrounds, showing the summed 2017 and 2018 samples. The
signal processes are scaled by the fitted value of \brinv, shown in
the legend. The background contributions are estimated from the fit to
the data described in the text (S+B fit). The total background (bkg.)
estimated from a fit assuming $\brinv=0$ (B-only fit) is also shown.
The yields from the 2017 and 2018 samples are summed and the correlations between their uncertainties are neglected.
The last bin of each distribution integrates events above the bin
threshold divided by the bin width.}
\label{fig:SR_MTRVTR}
\end{figure}

The background estimates in the SR are reported for each \mjj bin
of the MTR category in Table~\ref{tab:yields_MTR_1718}, and for
each \mjj bin of the VTR category in
Table~\ref{tab:yields_VTR_1718}.
The observed and expected \mjj distributions in the
SR are shown in Fig.~\ref{fig:SR_MTRVTR} for the MTR (\cmsLeft) and VTR
(\cmsRight) categories.

\begin{table*}[htb!p]
\centering
\topcaption[]{
Expected event yields in each \mjj bin for the different background
processes in the SR of the MTR category, summing the 2017 and 2018 samples. The
background yields and the corresponding uncertainties are obtained
after performing a combined fit across all of the CRs and the SR. The
expected signal contributions for the Higgs boson, produced in the non-\vbf
and \vbf modes, decaying to invisible particles with a branching
fraction of $\brinv = 1$, and the observed event yields are also
reported. The yields from the 2017 and 2018 samples are summed and the correlations between their uncertainties are neglected. }
\label{tab:yields_MTR_1718}
\cmsTable{
\begin{scotch}{lccccc}
\mjj bin range (\GeVns{}) & 200--400 & 400--600 & 600--900 & 900--1200 & 1200--1500 \\
\hline
$\PZ(\PGn\PGn)+\text{jets}$ (strong)  & $26107.6\pm82.7$ & $15521.0\pm62.1$ & $10747.3\pm48.8$ & $4404.4\pm25.3$ & $1923.4\pm16.7$ \\
$\PZ(\PGn\PGn)+\text{jets}$ (VBF)  & $431.3\pm6.1$ & $498.2\pm6.6$ & $620.6\pm7.0$ & $452.0\pm6.2$ & $294.9\pm5.3$  \\
$\PW(\Pell\PGn)+\text{jets}$ (strong)  & $13571.4\pm76.8$ & $8293.4\pm53.0$ & $5868.4\pm43.2$ & $2409.6\pm23.4$ & $1053.6\pm16.1$  \\
$\PW(\Pell\PGn)+\text{jets}$ (VBF)  & $268.0\pm10.5$ & $301.5\pm11.3$ & $353.5\pm12.7$ & $242.8\pm8.5$ & $163.0\pm5.9$  \\
$\ttbar$ + single \PQt quark  & $498.8\pm21.2$ & $370.6\pm15.6$ & $275.5\pm11.8$ & $115.3\pm5.1$ & $59.6\pm2.8$  \\
Diboson  & $464.9\pm40.0$ & $305.1\pm26.2$ & $246.3\pm21.3$ & $85.4\pm7.5$ & $39.4\pm3.5$  \\
$\PZ/\Pgg^{*}(\Pell^{+}\Pell^{-})+\text{jets}$  & $192.3\pm4.4$ & $126.3\pm2.9$ & $102.0\pm2.5$ & $38.2\pm1.0$   & $16.1\pm0.5$\\
Multijet  & $10.9\pm2.0$ & $10.6\pm1.9$ & $10.4\pm1.8$ & $4.8\pm0.9$ & $2.3\pm0.4$  \\
HF noise  & $0.8\pm0.1$ & $35.1\pm3.0$ & $82.7\pm7.3$ & $70.3\pm6.2$ & $28.1\pm2.5$ \\ [\cmsTabSkip]

$\PQq\PQq\PH(\to \text{inv})$  & 130.5 & 297.0 & 586.1 & 571.7 & 460.5 \\
Other $\PH(\to \text{inv})$ signals  & 1430.9 & 1027.1 & 848.7 & 414.3 & 209.5  \\ [\cmsTabSkip]

Total bkg.  & $41546.0\pm122.3$ & $25461.7\pm88.2$ & $18306.6\pm71.5$ & $7822.8\pm37.6$ & $3580.2\pm25.1$ \\ [\cmsTabSkip]

Observed & 41450 & 25536 & 18438 & 7793 & 3629 \\
\end{scotch}
}
\ifthenelse{\boolean{cms@external}}{\medskip}{\medskip\hrulefill\medskip}
\begin{scotch}{lcccc}
\mjj bin range (\GeVns{})  & 1500--2000 & 2000--2750 & 2750--3500 & $>$3500  \\
\hline
$\PZ(\PGn\PGn)+\text{jets}$ (strong)  & $1261.7\pm12.7$ & $462.4\pm7.4$ & $95.6\pm4.7$ & $28.8\pm1.4$\\
$\PZ(\PGn\PGn)+\text{jets}$ (VBF)  & $317.8\pm6.0$ & $197.3\pm5.3$ & $62.1\pm3.6$ & $35.8\pm2.3$\\
$\PW(\Pell\PGn)+\text{jets}$ (strong) & $704.3\pm10.9$ & $276.7\pm7.6$ & $65.4\pm4.1$ & $23.5\pm2.4$\\
$\PW(\Pell\PGn)+\text{jets}$ (VBF)  & $163.9\pm6.1$ & $111.9\pm4.6$ & $49.4\pm3.2$ & $19.2\pm1.6$\\
$\ttbar$ + single \PQt quark   & $38.7\pm2.2$ & $14.9\pm1.2$ & $5.3\pm0.5$ & $1.8\pm0.2$\\
Diboson  & $27.4\pm2.7$ & $7.9\pm0.8$& $0.6\pm0.1$ & $0.0\pm0.1$\\
$\PZ/\Pgg^{*}(\Pell^{+}\Pell^{-})+\text{jets}$ & $11.9\pm0.6$ & $4.9\pm0.3$ &  $1.5\pm0.1$ & $0.3\pm0.1$\\
Multijet  & $2.1\pm0.4$ & $1.0\pm0.2$& $0.4\pm0.1$ & $0.2\pm0.1$\\
HF noise   & $56.4\pm5.0$ & $62.2\pm5.6$& $30.5\pm2.7$ & $20.9\pm1.8$\\ [\cmsTabSkip]

$\PQq\PQq\PH(\to \text{inv})$   & 539.6 & 427.2 & 177.9 & 118.0 \\
Other $\PH(\to \text{inv})$ signals    & 161.8 & 84.7 & 24.3 & 11.0 \\ [\cmsTabSkip]

Total bkg.  & $2584.4\pm19.8$ & $1139.2\pm14.0$  & $310.7\pm8.3$ & $130.6\pm4.4$\\ [\cmsTabSkip]

Observed  & 2623 & 1142 & 279 & 136\\
\end{scotch}
\end{table*}

\begin{table*}[htb!]
\centering
\topcaption[]{
Expected event yields in each \mjj bin for the different background
processes in the SR of the VTR category, summing the 2017 and 2018
samples. The background yields and the corresponding uncertainties are
obtained after performing a combined fit across all of the CRs and
the SR. The expected signal contributions for the Higgs boson, produced in
the non-\vbf and \vbf modes, decaying to invisible particles with a
branching fraction of $\brinv = 1$, and the observed event yields are
also reported. The yields from the 2017 and 2018 samples are summed and the correlations between their uncertainties are neglected.}
\label{tab:yields_VTR_1718}
\cmsTable{
\begin{scotch}{lccccc}
\mjj bin range (\GeVns{}) & 900--1200 & 1200--1500 & 1500--2000 & 2000--2750 & $>$2750  \\
\hline
$\PZ(\PGn\PGn)+\text{jets}$ (strong)  & $1075.2\pm14.9$ & $444.3\pm9.4$ & $286.5\pm6.6$ & $97.2\pm3.5$ & $38.0\pm1.9$\\
$\PZ(\PGn\PGn)+\text{jets}$ (VBF)  & $132.2\pm4.1$ & $95.2\pm4.2$ & $85.1\pm4.4$ & $56.8\pm3.7$ & $33.6\pm2.5$\\
$\PW(\Pell\PGn)+\text{jets}$ (strong)  & $1048.7\pm20.8$ & $446.4\pm13.5$ & $299.8\pm11.0$ & $139.0\pm9.1$ & $45.1\pm4.6$\\
$\PW(\Pell\PGn)+\text{jets}$ (VBF)  & $114.4\pm6.7$ & $89.9\pm6.0$ & $74.0\pm5.3$ & $53.7\pm4.4$ & $40.5\pm3.5$\\
$\ttbar$ + single \PQt quark  & $25.6\pm1.4$ & $14.5\pm0.9$ & $6.5\pm0.5$ & $9.3\pm0.6$ & $4.2\pm0.8$\\
Diboson  & $15.7\pm1.4$ & $7.6\pm0.8$ & $4.2\pm0.4$ & $0.4\pm0.1$ & $0.0\pm0.1$\\
$\PZ/\Pgg^{*}(\Pell^{+}\Pell^{-})+\text{jets}$  & $31.2\pm1.0$ & $14.5\pm0.5$ & $6.8\pm0.3$ & $4.1\pm0.2$ & $1.6\pm0.2$\\
Multijet  & $0.1\pm0.1$ & $0.1\pm0.1$ & $0.1\pm0.1$ & $0.1\pm0.1$ & $0.0\pm0.1$\\
HF noise  & $30.9\pm3.0$ & $30.4\pm3.2$ & $26.4\pm2.5$ & $48.2\pm4.8$ & $26.0\pm2.6$\\ [\cmsTabSkip]

$\PQq\PQq\PH(\to \text{inv})$  & 226.7 & 169.9 & 195.0 & 140.9 & 97.4\\
Other $\PH(\to \text{inv})$ signals  & 67.1 & 33.2 & 24.9 & 11.4 & 5.0\\ [\cmsTabSkip]

Total bkg.  & $2474.1\pm27.0$ & $1142.8\pm18.3$ & $789.2\pm14.8$ & $408.7\pm12.3$ & $189.1\pm7.1$\\ [\cmsTabSkip]

Observed & 2433 & 1164 & 780 & 422 & 197\\
\end{scotch}
}\end{table*}

\cmsClearpage

\subsection{Combination of results}

No significant deviations from the SM expectations are observed. The
results of this search are interpreted in terms of an upper
limit on the product of the Higgs boson production cross section and
its branching fraction to invisible particles, $\sigma_{\PH}\brhinv$,
relative to the predicted cross section assuming SM interactions,
$\sigma_{\PH}^{\text{SM}}$. Observed and expected 95\% \CL upper
limits are computed using an asymptotic approximation of the \CLs
method detailed in Refs.~\cite{Junk:1999kv,Read:2002av}, with a
profile likelihood ratio test statistic~\cite{Cowan:2010js} in which
systematic uncertainties are modelled as nuisance parameters following
a frequentist approach~\cite{CMS-NOTE-2011-005}.

Both \vbf and non-\vbf signal production modes are included, with
their relative contributions fixed to the SM prediction within their
uncertainties.

Between the 2017 and 2018 data sets, and the two analysis categories
(MTR and VTR), the uncertainties are correlated according to the
description given in Section~\ref{sec:systs}. To combine with the data
taken in 2016, the same correlation scheme as between 2017 and 2018 is
used, except for the jet energy calibration uncertainties (JES and JER), which are
kept fully uncorrelated. The integrated luminosity of the 2016 data set was
updated to 36.3\fbinv to reflect the latest
improvements in the luminosity measurement~\cite{CMS:2021xjt}. To be consistent with the treatment
of the \vbf signal in the 2017 and 2018 analyses, the Higgs
boson \pt-dependent EW NLO corrections are also applied to the 2016
signal shape. The VBF results obtained with the earlier data sets,
namely the data set from 2012 (2015), at $\sqrt{s}=8\,(13)\TeV$, with
19.7 (2.3)\fbinv, from Ref.~\cite{Khachatryan:2016whc}, are combined
taking into account uncertainty correlations where appropriate.
Theoretical uncertainties related to signal modelling are correlated for
data-taking periods with the same center-of-mass energy.  Partial
correlations between data sets exist for the uncertainty in the
luminosity measurements. All other experimental uncertainties are
decorrelated between the run periods before and after 2015. The results of the fit to the data across 
all data-taking periods are available in \suppMaterialbis.

\subsubsection{Constraints on an SM-like Higgs boson}

\begin{figure}[htb!]
\centering
\includegraphics[width=\cmsFigWidth]{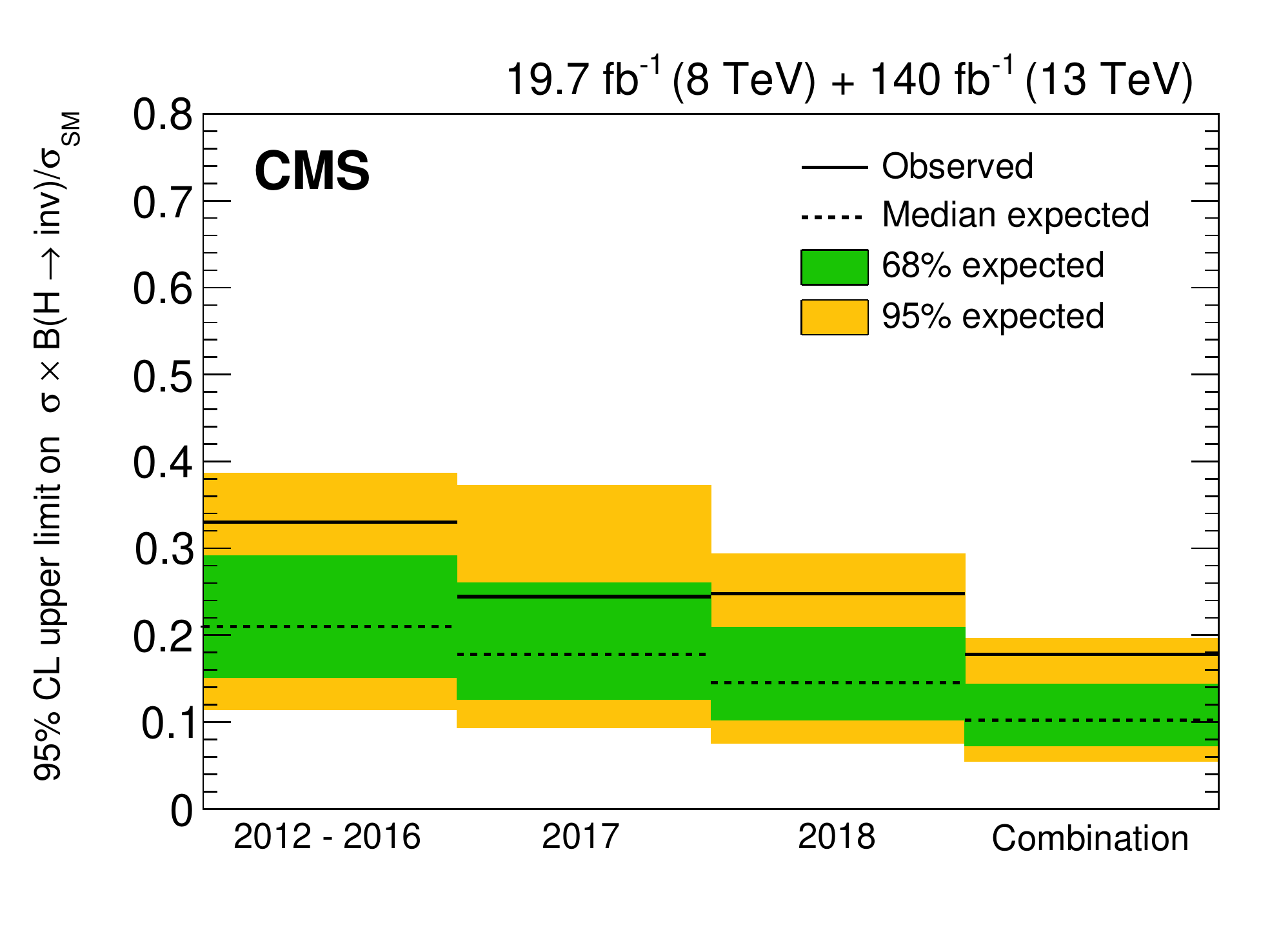}
\caption{Observed and expected 95\% \CL upper limits on \sigmabr for all data-taking years considered, as well as their combination, assuming an SM Higgs boson with a mass of 125.38\GeV.}
\label{fig:limits}
\end{figure}

Observed and expected upper limits on \sigmabr at 95\% \CL are
presented in Fig.~\ref{fig:limits} and Table~\ref{tab:limits}. The
limits are computed for the combination of all data sets, as well as
for individual categories and data-taking periods. By itself, the
addition of the \phojets CR (the addition of the VTR category)
improves the expected limits by about 11 (8)\% compared with the
2016-like analysis selections. Considered together, the improvements reach
about 17\% in both years. The upper limits for the individual categories entering the combination 
are available in \suppMaterialbis.  

{\tolerance=800 The combination of the 2017 and 2018 results yields an
observed (expected) upper limit of ${\brhinv < 0.\brobs\,(0.\brexp)}$
at the 95\% \CL, assuming an SM Higgs boson with a mass of
125.38\GeV~\cite{CMS:2020xrn}. Figure~\ref{fig:limits} additionally
shows a combination with data collected in 2012 and
2015~\cite{Khachatryan:2016whc}, and in
2016~\cite{Sirunyan:2018owy}. This combination yields an observed
(expected) upper limit of ${\brhinv < 0.\brobsRun\,(0.\brexpRun)}$ at
the 95\% \CL, which is currently the most stringent limit
on \brinv.\par}

\begin{table*}[htb!]
    \centering
    \topcaption{The 95\% \CL upper limits on \sigmabr, assuming an SM Higgs boson with a mass of 125.38\GeV. 
    The observed and median expected results are shown, along with the 68\% and 95\%  interquartile ranges for each 
    category and for the combinations.}
    \label{tab:limits}
    \begin{scotch}{l c c c c}
       Category & Observed & Median expected  & 65\% expected & 95\% expected \\
       \hline
          2012--2016  & 0.33  & 0.21  & [0.15, 0.29]  & [0.11, 0.39] \\ [\cmsTabSkip]

          VTR 2017  & 0.57  & 0.45  & [0.32, 0.66]  & [0.24, 0.94] \\ 
          VTR 2018  & 0.44  & 0.34  & [0.24, 0.49]  & [0.18, 0.69] \\ 
     VTR 2017+2018  & 0.40  & 0.28  & [0.20, 0.40]  & [0.15, 0.56] \\ [\cmsTabSkip]

          MTR 2017  & 0.25  & 0.19  & [0.14, 0.28]  & [0.10, 0.40] \\ 
          MTR 2018  & 0.24  & 0.15  & [0.11, 0.22]  & [0.08, 0.31] \\ 
     MTR 2017+2018  & 0.17  & 0.13  & [0.09, 0.18]  & [0.07, 0.25] \\ [\cmsTabSkip] 

          all 2017  & 0.24  & 0.18  & [0.13, 0.26]  & [0.09, 0.37] \\ 
          all 2018  & 0.25  & 0.15  & [0.10, 0.21]  & [0.08, 0.29] \\ 
     all 2017+2018  & 0.18  & 0.12  & [0.08, 0.17]  & [0.06, 0.23] \\  [\cmsTabSkip]

          2012--2018  & 0.18  & 0.10  & [0.07, 0.14]  & [0.05, 0.20] \\ 
    \end{scotch}
\end{table*}

\begin{figure}[htb!]
\centering
\includegraphics[width=\cmsFigWidth]{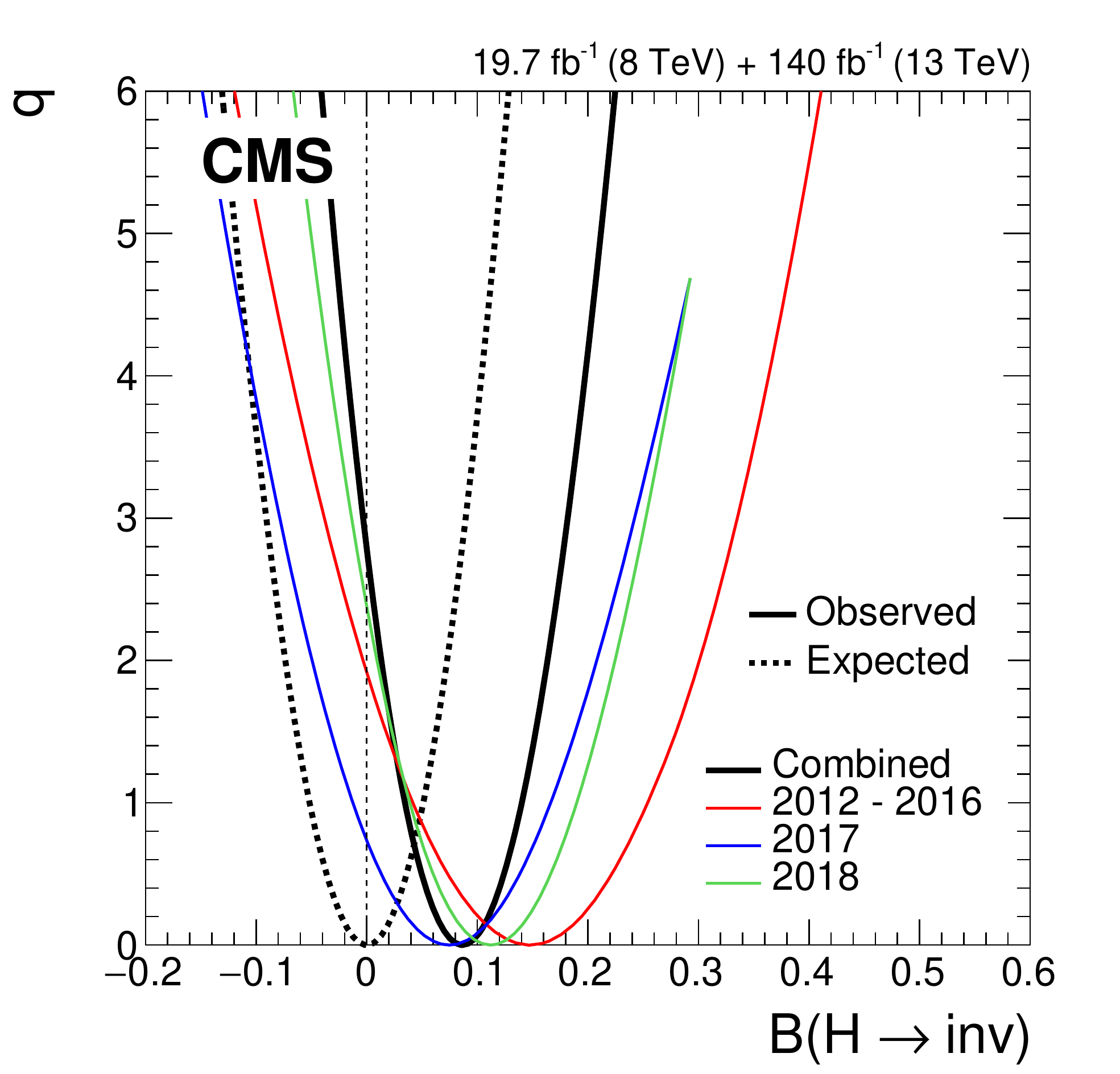}
\caption{Profile likelihood ratios, as functions of \brhinv. The observed likelihood scans are reported for the full combination of 2012--2018 data, as well as for the individual years. The expected results for the combination are obtained using an Asimov data set~\cite{Cowan:2010js} with $\brhinv=0$.}
\label{fig:scanComb}
\end{figure}

Figure~\ref{fig:scanComb} shows the profile likelihood ratio ($q$) as
a function of \brinv, for the individual data-taking periods and for
their combination. The observed (expected) combined 2012--2018 best
fit signal strength is found to be $0.086^{+0.054}_{-0.052}$
($0.00^{+0.051}_{-0.052}$). Table~\ref{tab:scanGroups} summarizes the
uncertainties in the measured \brinv, separating the contributions
from different groups of uncertainties. The systematic uncertainties
with the largest impacts in the \brhinv measurement are the
theoretical uncertainties affecting the
$f^{\PW/\PZ\mathrm{,proc}}_{i}$ ratio, followed by the statistical
uncertainties in the simulated samples, the trigger uncertainties, jet
calibration effects, and the uncertainties in the QCD multijet
modelling.

\begin{table*}[htb!]
    \centering
    \topcaption{Uncertainty breakdown in \brinv. The sources of uncertainty are separated into different groups. Observed and expected results are quoted for the full combination of 2012--2018 data. The expected results are obtained using an Asimov data set~\cite{Cowan:2010js} with $\brhinv=0$.}
    \label{tab:scanGroups}
    \renewcommand{\arraystretch}{1.3}
    \begin{scotch}{l c c}
       \multirow{2}{*}{Group of systematic uncertainties} & \multicolumn{2}{c}{Impact on \brinv} \\ [-3pt]
       & Observed & Expected \\
      \hline
      Theory              & $^{+ 0.026}_{-0.025}$ & $\pm 0.024$ \\
      Simulated event count & $\pm 0.022$ & $^{+ 0.021}_{-0.022}$ \\
      Triggers            & $^{+ 0.018}_{-0.019}$ & $\pm 0.018$ \\
      Jet calibration     & $^{+ 0.014}_{-0.012}$ & $\pm 0.011$ \\
      QCD multijet mismodelling & $\pm 0.012$ & $\pm 0.013$ \\
      Leptons/photons/\PQb-tagged jets   & $^{+ 0.011}_{-0.010}$ & $^{+ 0.009}_{-0.010}$ \\
      Integrated luminosity/pileup & $\pm 0.004$ & $\pm 0.004$ \\
      Other systematic uncertainties & $^{+ 0.013}_{-0.009}$ & $\pm 0.009$ \\ [\cmsTabSkip]
      Statistical uncertainty  & $\pm 0.028 $ & $\pm 0.028$ \\
    \end{scotch}
\end{table*}

\begin{figure}[htb]
\centering
\includegraphics[width=\cmsFigWidth]{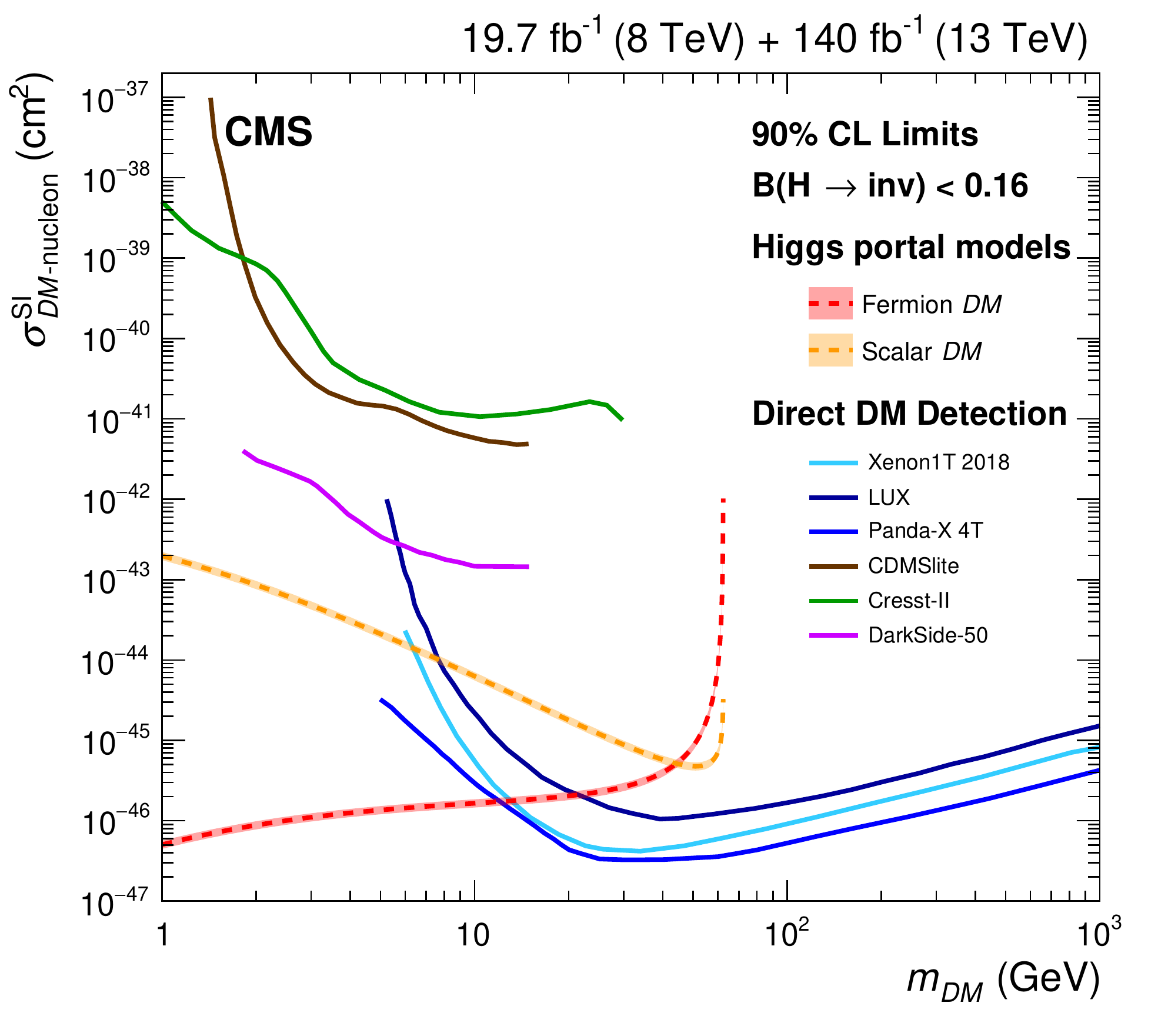}
\caption{The 90\% \CL upper limits on the spin-independent DM-nucleon scattering cross 
section in Higgs-portal models, assuming a scalar (dashed orange) or
fermion (dashed red) DM candidate. Limits are computed as functions
of $m_{\mathrm{DM}}$ and are compared to those from the
XENON1T~\cite{Aprile:2018dbl}, CRESST-II~\cite{Angloher:2015ewa},
CDMSlite~\cite{Agnese:2015nto}, LUX~\cite{Akerib:2016vxi}, Panda-X
4T~\cite{PandaX-4T:2021bab}, and DarkSide-50~\cite{Agnes:2018ves}
experiments, which are shown as solid lines.}
\label{fig:DMlim}
\end{figure}

The upper limit on \brinv, obtained from the combination of 2012--2018
data, is interpreted in the context of Higgs-portal models of DM
interactions, in which a stable DM particle couples to the SM Higgs
boson. The interaction between a DM particle and an atomic nucleus may
be mediated by the exchange of a Higgs boson, producing nuclear recoil
signatures, such as those investigated by direct detection
experiments. The sensitivity of these experiments depends mainly on
the DM particle mass ($m_{\mathrm{DM}}$). If $m_{\mathrm{DM}}$ is
smaller than half of the Higgs boson mass, the partial width of the
invisible Higgs boson decay ($\Gamma_{\text{inv}}$) can be translated, within an effective
field theory approach, into a spin-independent DM-nucleon elastic
scattering cross section, as outlined in
Ref.~\cite{Djouadi:2011aa}. This translation is performed assuming
that the DM candidate is either a scalar or a Majorana fermion, and
both the central value and the uncertainty in the dimensionless
nuclear form factor $f_{N}$ are taken from the recommendations of
Ref.~\cite{Hoferichter:2017olk}. The conversion from \brinv to
$\Gamma_{\text{inv}}$ uses the relation ${\brinv = \Gamma_{\text{inv}}
/ (\Gamma_{\mathrm{SM}}+\Gamma_{\text{inv}})}$, where
$\Gamma_{\mathrm{SM}}$ is set to
4.07\MeV~\cite{Heinemeyer:2013tqa}. We do not perform the
translation under the assumption of a vector DM
candidate in this paper, since it
requires an extended dark Higgs sector, which may lead to
modifications of kinematic distributions assumed for the invisibly
decaying Higgs boson signal. Figure~\ref{fig:DMlim} shows the 90\% \CL upper
limits on the spin-independent DM-nucleon scattering cross section as
a function of $m_{\mathrm{DM}}$, for both the scalar and the fermion
DM scenarios. The corresponding 90\% \CL upper limit on \brinv is
0.16. These limits are computed at the 90\% \CL so that they can be
compared with those from direct detection experiments such as
XENON1T~\cite{Aprile:2018dbl}, CRESST-II~\cite{Angloher:2015ewa},
CDMSlite~\cite{Agnese:2015nto}, LUX~\cite{Akerib:2016vxi}, Panda-X
4T~\cite{PandaX-4T:2021bab}, and DarkSide-50~\cite{Agnes:2018ves},
which provide the strongest constraints in the $m_{\mathrm{DM}}$ range
probed by this search. The collider-based results complement the
direct-detection experiments in the range $m_{\mathrm{DM}}$ smaller
than 12\,(6)\GeV, assuming a fermion (scalar) DM candidate.

\clearpage

\section{Summary}
\label{sec:concl}

A search for the Higgs boson (\PH) decaying invisibly, produced in the
vector boson fusion mode, is performed with \lumiSevEi of
proton-proton collisions delivered by the LHC at $\sqrt{s}=13\TeV$ and
collected by the CMS detector during 2017--2018. Building upon the
previously published results, an additional category targeting events
at lower Higgs boson transverse momentum is added. An additional
highly populated control region, based on production of a photon
associated with jets, is used to constrain the dominant irreducible
background from invisible decays of a \PZ boson produced in
association with jets. Compared with the strategy of the previously
published analysis, these additions improve the expected limits by
approximately 17\%. The observed (expected) upper limit on the
invisible branching fraction of the Higgs boson, \brinv, is found to
be $0.\brobs$ ($0.\brexp$) at the 95\% confidence level (\CL),
assuming the standard model production cross section. The results are
combined with previous measurements in the vector boson fusion
topology, for total integrated luminosities of \lumiRunOne at
$\sqrt{s}=8\TeV$ and \lumiRunTwo at $\sqrt{s}=13\TeV$, yielding an
observed (expected) upper limit of $0.\brobsRun$ ($0.\brexpRun$) at
the 95\% \CL. This is currently the most stringent limit
on \brinv. Finally, the results are interpreted in the context of
Higgs-portal models. The 90\% \CL upper limits on the spin-independent
dark-matter-nucleon scattering cross section obtained from the
observed LHC data collected during 2012--2018 complement the direct
detection experiments in the range of dark matter particle masses
smaller than 12\,(6)\GeV, assuming a fermion (scalar) dark matter
candidate.

\begin{acknowledgments}

    We congratulate our colleagues in the CERN accelerator departments for the excellent performance of the LHC and thank the technical and administrative staffs at CERN and at other CMS institutes for their contributions to the success of the CMS effort. In addition, we gratefully acknowledge the computing centers and personnel of the Worldwide LHC Computing Grid and other centers for delivering so effectively the computing infrastructure essential to our analyses. Finally, we acknowledge the enduring support for the construction and operation of the LHC, the CMS detector, and the supporting computing infrastructure provided by the following funding agencies: BMBWF and FWF (Austria); FNRS and FWO (Belgium); CNPq, CAPES, FAPERJ, FAPERGS, and FAPESP (Brazil); MES and BNSF (Bulgaria); CERN; CAS, MoST, and NSFC (China); MINCIENCIAS (Colombia); MSES and CSF (Croatia); RIF (Cyprus); SENESCYT (Ecuador); MoER, ERC PUT and ERDF (Estonia); Academy of Finland, MEC, and HIP (Finland); CEA and CNRS/IN2P3 (France); BMBF, DFG, and HGF (Germany); GSRI (Greece); NKFIA (Hungary); DAE and DST (India); IPM (Iran); SFI (Ireland); INFN (Italy); MSIP and NRF (Republic of Korea); MES (Latvia); LAS (Lithuania); MOE and UM (Malaysia); BUAP, CINVESTAV, CONACYT, LNS, SEP, and UASLP-FAI (Mexico); MOS (Montenegro); MBIE (New Zealand); PAEC (Pakistan); MSHE and NSC (Poland); FCT (Portugal); JINR (Dubna); MON, RosAtom, RAS, RFBR, and NRC KI (Russia); MESTD (Serbia); MCIN/AEI and PCTI (Spain); MOSTR (Sri Lanka); Swiss Funding Agencies (Switzerland); MST (Taipei); ThEPCenter, IPST, STAR, and NSTDA (Thailand); TUBITAK and TAEK (Turkey); NASU (Ukraine); STFC (United Kingdom); DOE and NSF (USA).

    \hyphenation{Rachada-pisek} Individuals have received support from the Marie-Curie program and the European Research Council and Horizon 2020 Grant, contract Nos.\ 675440, 724704, 752730, 758316, 765710, 824093, 884104, and COST Action CA16108 (European Union); the Leventis Foundation; the Alfred P.\ Sloan Foundation; the Alexander von Humboldt Foundation; the Belgian Federal Science Policy Office; the Fonds pour la Formation \`a la Recherche dans l'Industrie et dans l'Agriculture (FRIA-Belgium); the Agentschap voor Innovatie door Wetenschap en Technologie (IWT-Belgium); the F.R.S.-FNRS and FWO (Belgium) under the ``Excellence of Science -- EOS" -- be.h project n.\ 30820817; the Beijing Municipal Science \& Technology Commission, No. Z191100007219010; the Ministry of Education, Youth and Sports (MEYS) of the Czech Republic; the Deutsche Forschungsgemeinschaft (DFG), under Germany's Excellence Strategy -- EXC 2121 ``Quantum Universe" -- 390833306, and under project number 400140256 - GRK2497; the Lend\"ulet (``Momentum") Program and the J\'anos Bolyai Research Scholarship of the Hungarian Academy of Sciences, the New National Excellence Program \'UNKP, the NKFIA research grants 123842, 123959, 124845, 124850, 125105, 128713, 128786, and 129058 (Hungary); the Council of Science and Industrial Research, India; the Latvian Council of Science; the Ministry of Science and Higher Education and the National Science Center, contracts Opus 2014/15/B/ST2/03998 and 2015/19/B/ST2/02861 (Poland); the Funda\c{c}\~ao para a Ci\^encia e a Tecnologia, grant CEECIND/01334/2018 (Portugal); the National Priorities Research Program by Qatar National Research Fund; the Ministry of Science and Higher Education, projects no. 0723-2020-0041 and no. FSWW-2020-0008, and the Russian Foundation for Basic Research, project No.19-42-703014 (Russia); MCIN/AEI/10.13039/501100011033, ERDF ``a way of making Europe", and the Programa Estatal de Fomento de la Investigaci{\'o}n Cient{\'i}fica y T{\'e}cnica de Excelencia Mar\'{\i}a de Maeztu, grant MDM-2017-0765 and Programa Severo Ochoa del Principado de Asturias (Spain); the Stavros Niarchos Foundation (Greece); the Rachadapisek Sompot Fund for Postdoctoral Fellowship, Chulalongkorn University and the Chulalongkorn Academic into Its 2nd Century Project Advancement Project (Thailand); the Kavli Foundation; the Nvidia Corporation; the SuperMicro Corporation; the Welch Foundation, contract C-1845; and the Weston Havens Foundation (USA).
\end{acknowledgments}
\bibliography{auto_generated}

\ifthenelse{\boolean{cms@external}}{}{
\clearpage
\renewcommand{\cmsClearpage}{\clearpage}
\newcommand{\tabsysts}{\ref{tab:systs}}
\numberwithin{table}{section}
\numberwithin{figure}{section}
\appendix
\section{Additional figures and tables}
\label{app:suppMat}

\input{supplemental_material}

}
\cleardoublepage \section{The CMS Collaboration \label{app:collab}}\begin{sloppypar}\hyphenpenalty=5000\widowpenalty=500\clubpenalty=5000\input{HIG-20-003-authorlist.tex}\end{sloppypar}
\end{document}

%% file: supplemental_material.tex
\subsection*{Trigger efficiencies}

\begin{figure*}[!htb]
    \centering
    \includegraphics[width=0.48\textwidth]{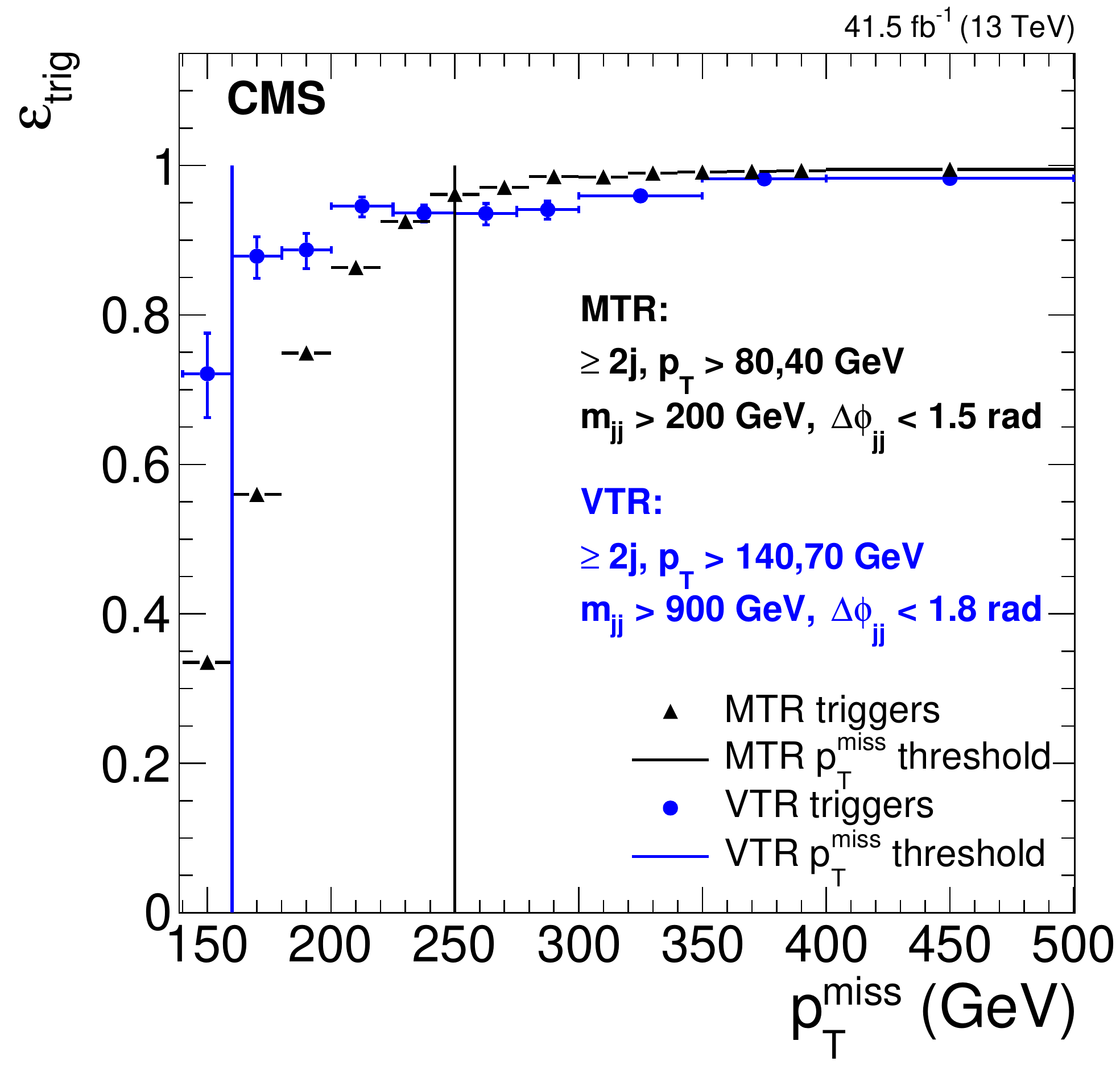}
    \includegraphics[width=0.48\textwidth]{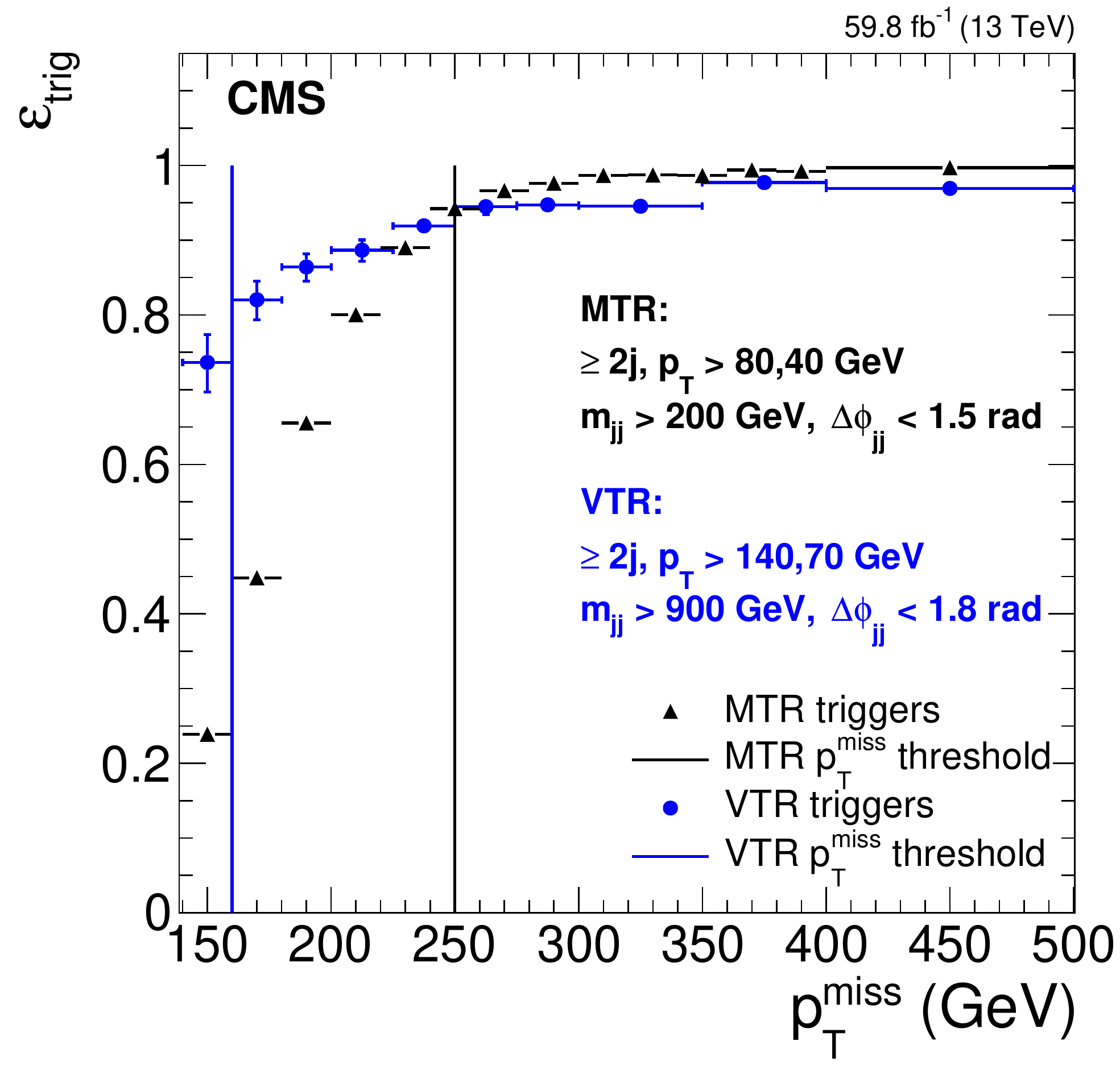}
    \caption{Trigger efficiency measured in data for the 2017 and 2018 samples,
        comparing the two sets of triggers used in the analysis (\ptmiss-based in MTR, VBF jet-based in VTR).
        The lower thresholds are indicated by the vertical lines. The lower threshold of the MTR category also corresponds to the upper threshold applied in the VTR category.}
    \label{fig:trigeff}
\end{figure*}

Trigger efficiencies are shown in Fig.~\ref{fig:trigeff}.

\cmsClearpage
\subsection*{Detailed limits}

\begin{figure*}[h!]
    \centering
    \includegraphics[width=\cmsFigWidth]{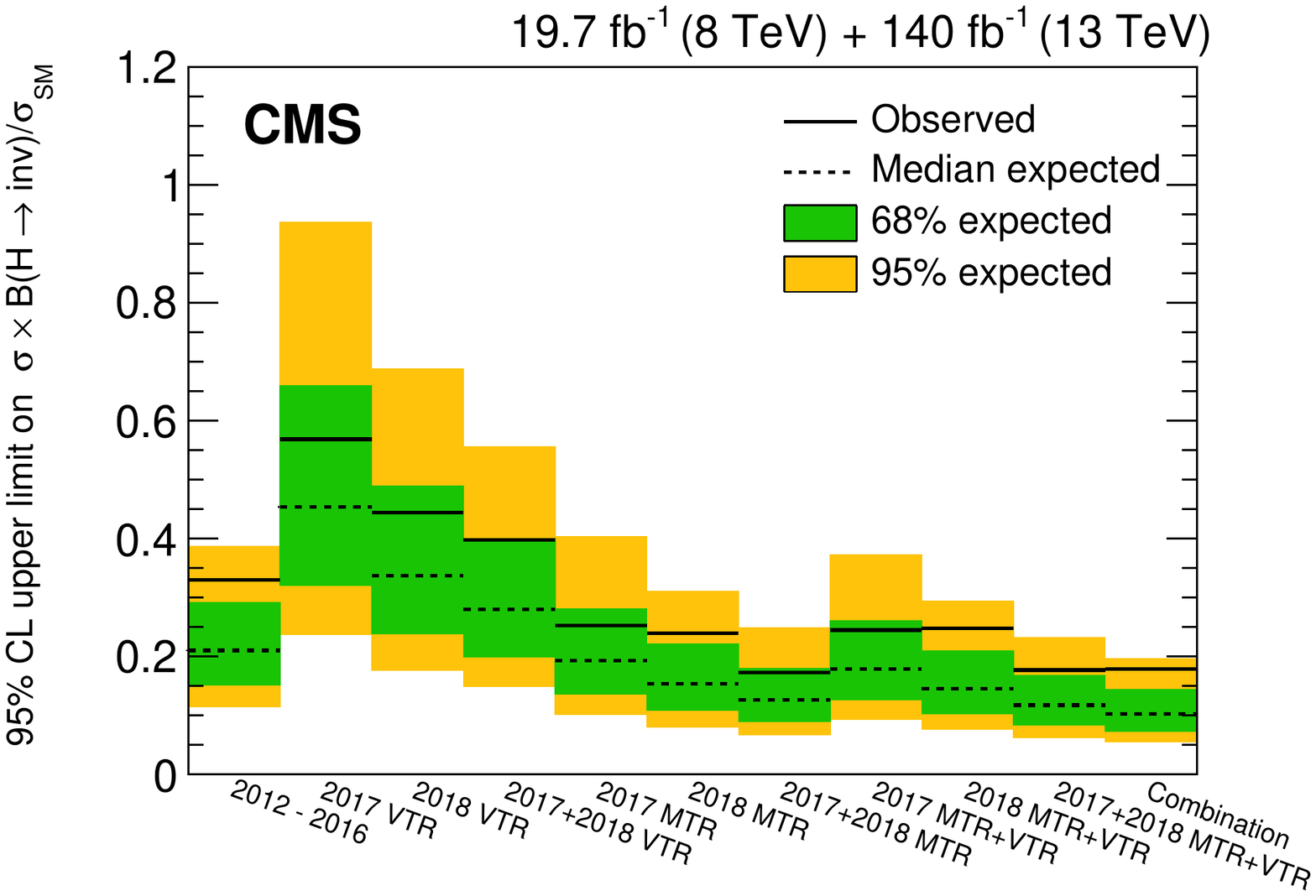}
    \caption{Observed and expected 95\% \CL upper limits on \sigmabr for both individual categories and all data-taking years, as well as their combination, assuming an SM Higgs boson with a mass of 125.38\GeV.}
    \label{fig:limitsSplit}
\end{figure*}

Limits separated by category and year as well as their combinations are shown in Fig.~\ref{fig:limitsSplit}.
 
\cmsClearpage
\subsection*{Combination for MTR adding 2016 data}

\begin{figure*}[htb]
    \centering
    \includegraphics[width=\cmsFigWidth]{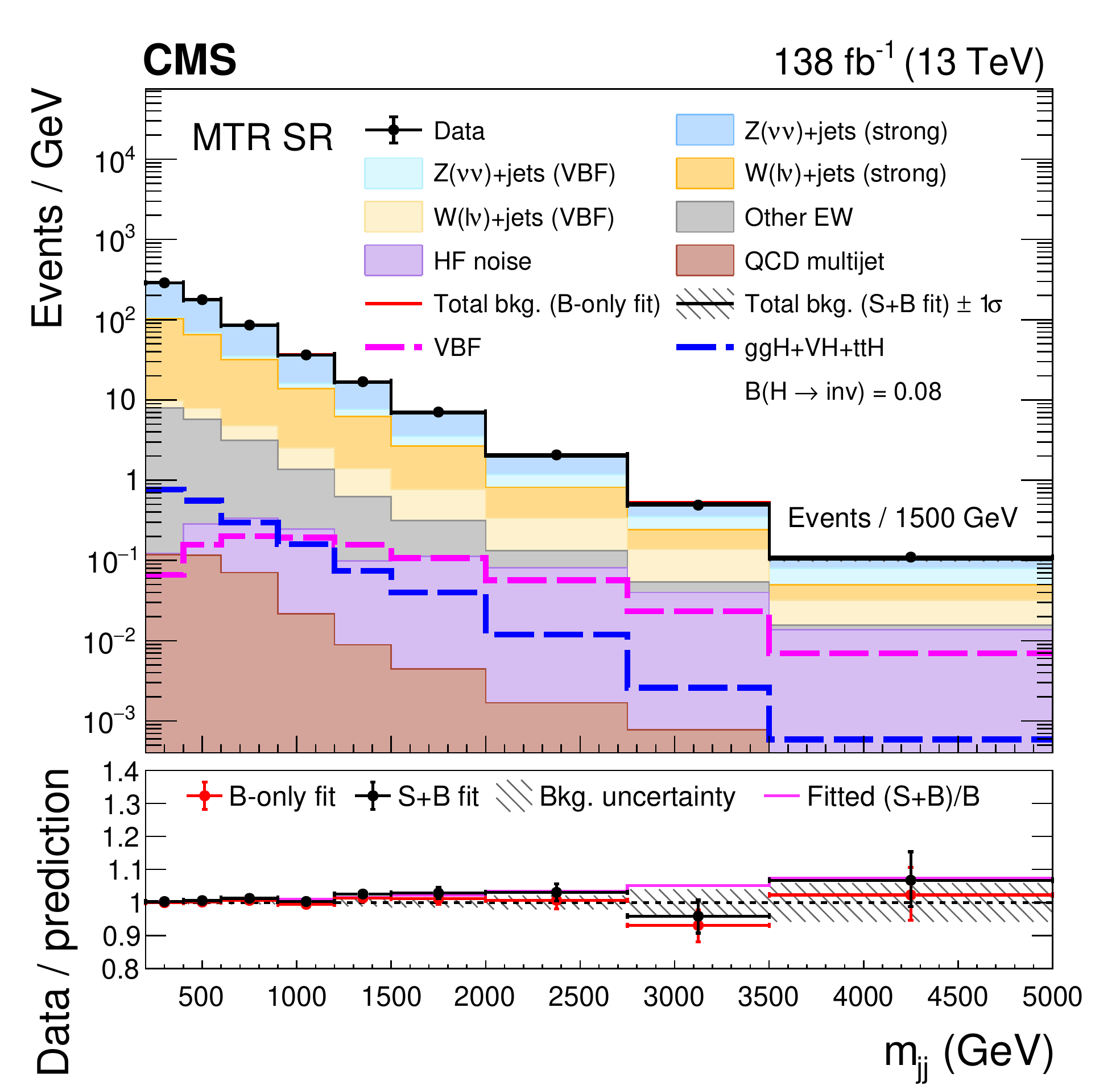}
    \caption{The observed \mjj distribution in the MTR SR compared to the postfit backgrounds, with the 2016, 2017, and 2018 samples. The signal processes are scaled by the fitted value of \brinv, shown in the legend.
        The background contributions are estimated from
        the fit to the data described in the text (S+B fit) and the total background estimated from a fit
        assuming $\brinv=0$ (B-only fit) is also shown. The yields from the 2016, 2017, and 2018 samples are summed and the correlations between their uncertainties are neglected.
        The last bin of each distribution integrates events above the bin threshold divided by the bin width.}
    \label{fig:MjjSRfullRun}
\end{figure*}

 The observed \mjj distribution in the MTR SR for the sum of the 2016--2018 samples is shown in Fig.~\ref{fig:MjjSRfullRun}.
 
\cmsClearpage
\subsection*{CR-only fits}

\begin{figure*}[!htb]
    \centering
    \includegraphics[width=0.48\textwidth]{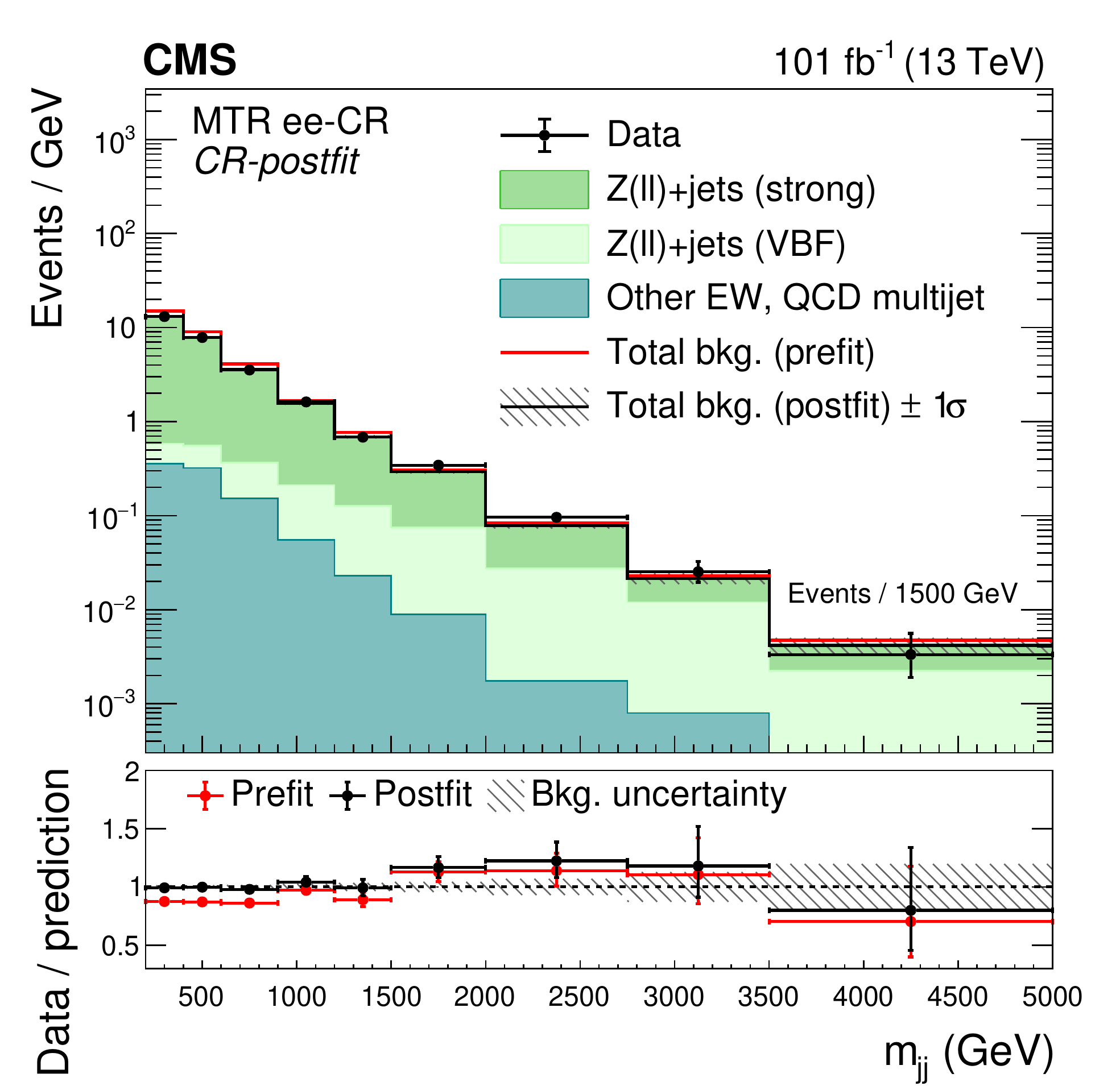}
    \includegraphics[width=0.48\textwidth]{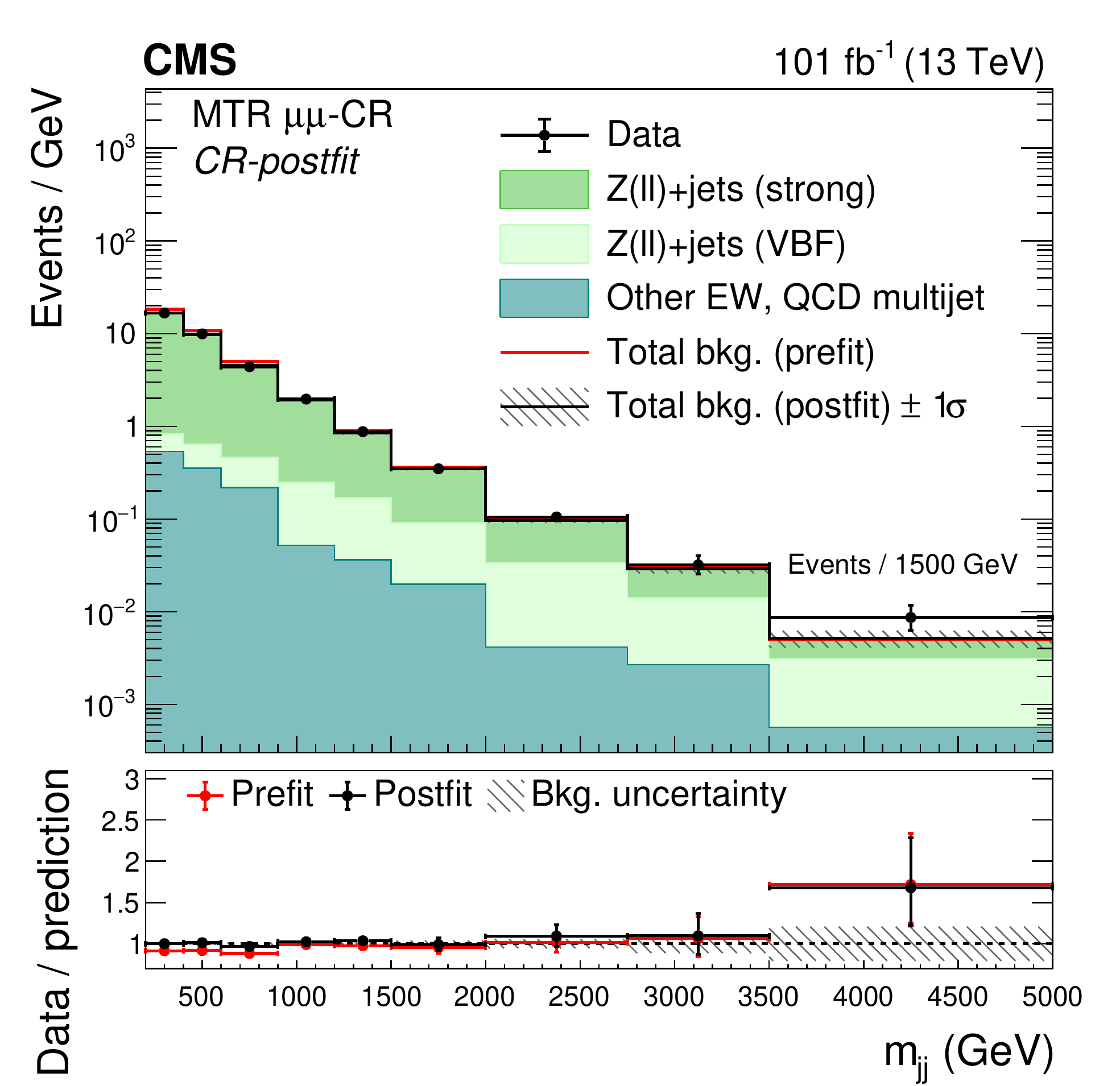}\\
    \includegraphics[width=0.48\textwidth]{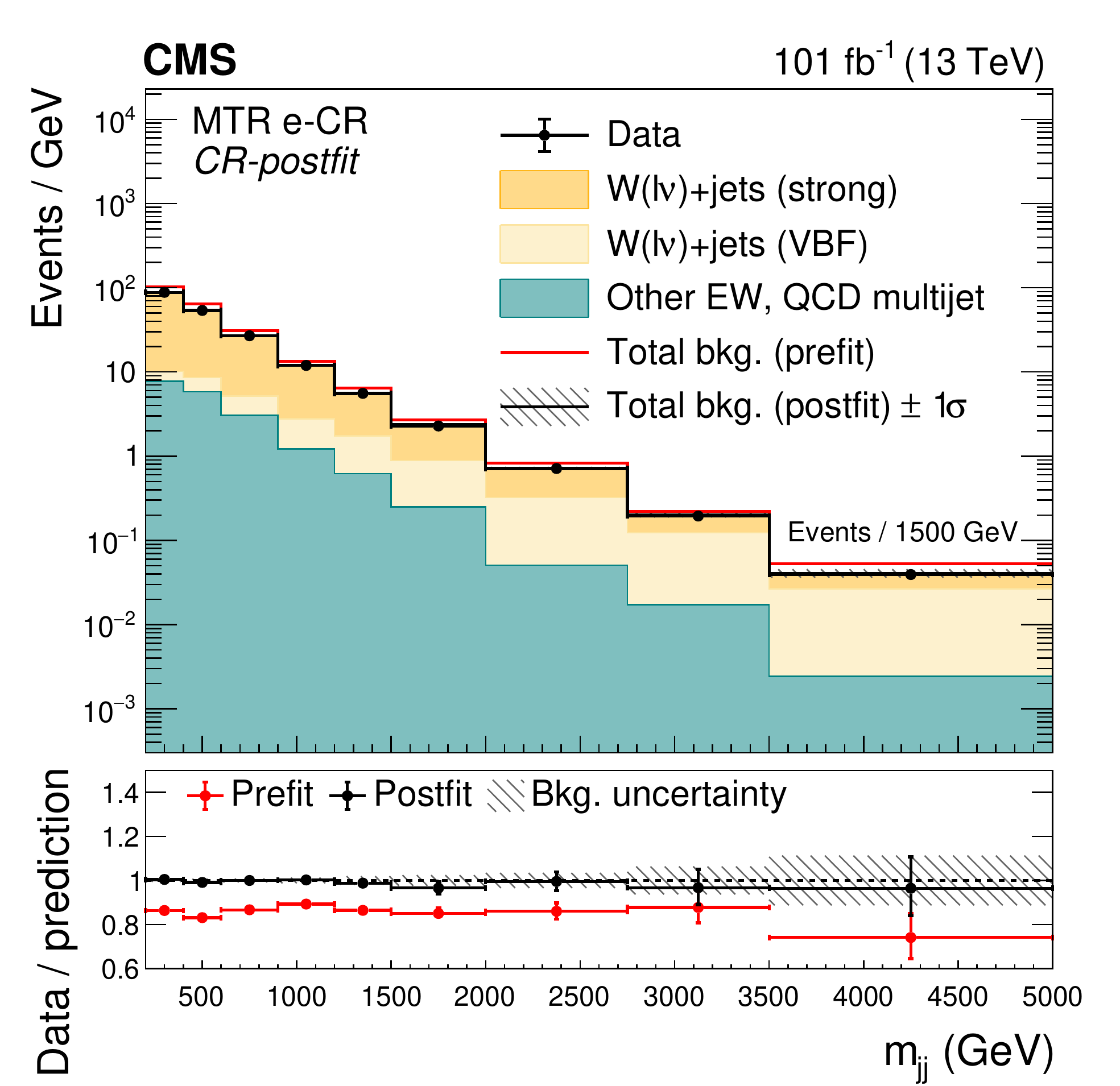}
    \includegraphics[width=0.48\textwidth]{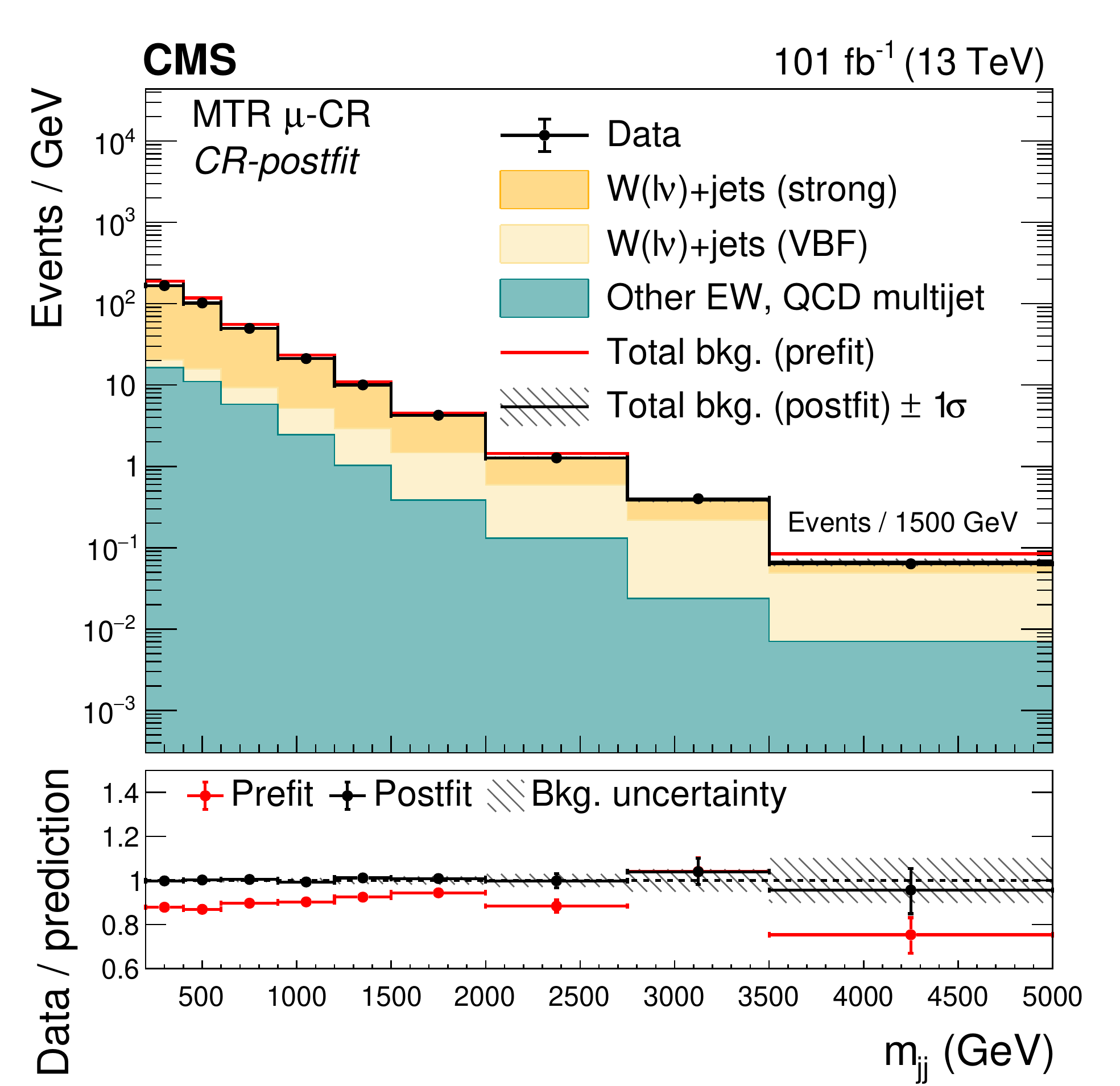}
    \caption{
        The \mjj distributions (prefit and CR-postfit) in the dielectron (upper
        left), dimuon (upper right), single-electron (lower left),
        and single-muon (lower right) CR for the MTR category, with the 2017
        and 2018 samples. The yields from the 2017 and 2018 samples are summed and the correlations between their uncertainties are neglected.}
    \label{fig:CR_MTR_CRonly}
\end{figure*}

\begin{figure*}[htb]
    \centering
    \includegraphics[width=\cmsFigWidth]{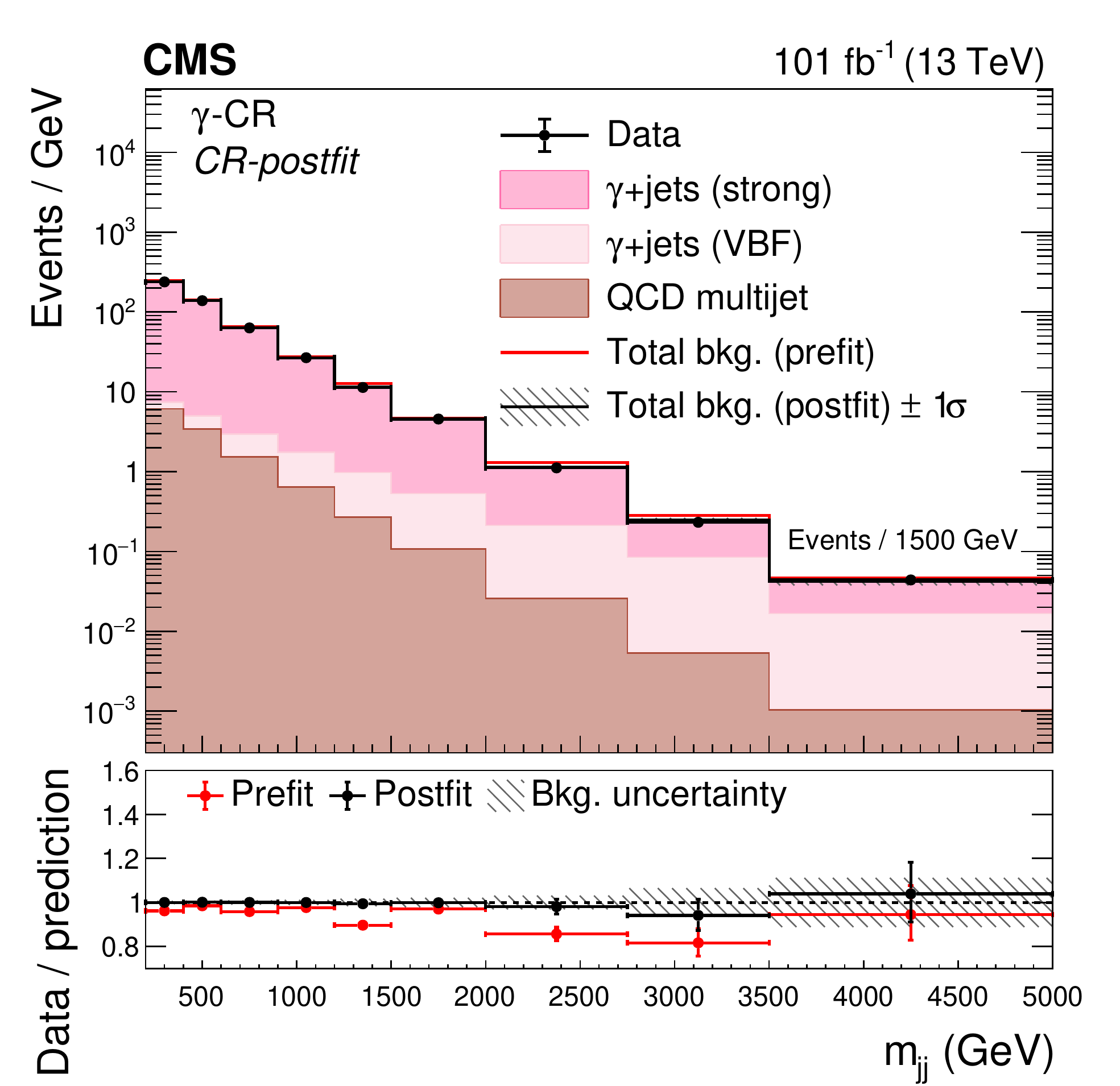}
    \caption{The \mjj distributions (prefit and CR-postfit) in the photon CR for the MTR category, with the 2017 and 2018 samples. The yields from the 2017 and 2018 samples are summed and the correlations between their uncertainties are neglected. }
    \label{fig:Zgamma_MTR_CRonly}
\end{figure*}

\begin{figure*}[!htb]
    \centering
    \includegraphics[width=0.48\textwidth]{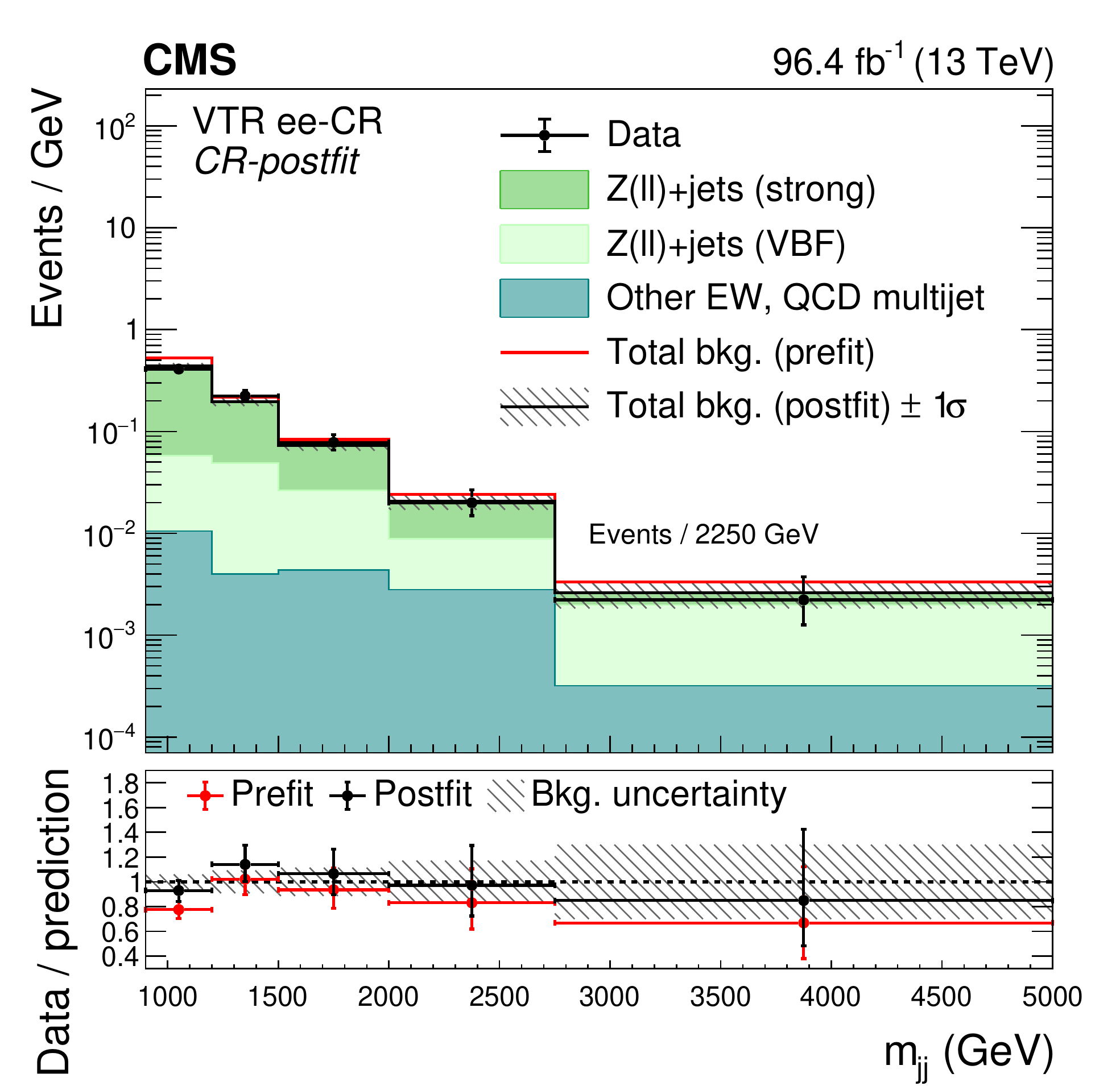}
    \includegraphics[width=0.48\textwidth]{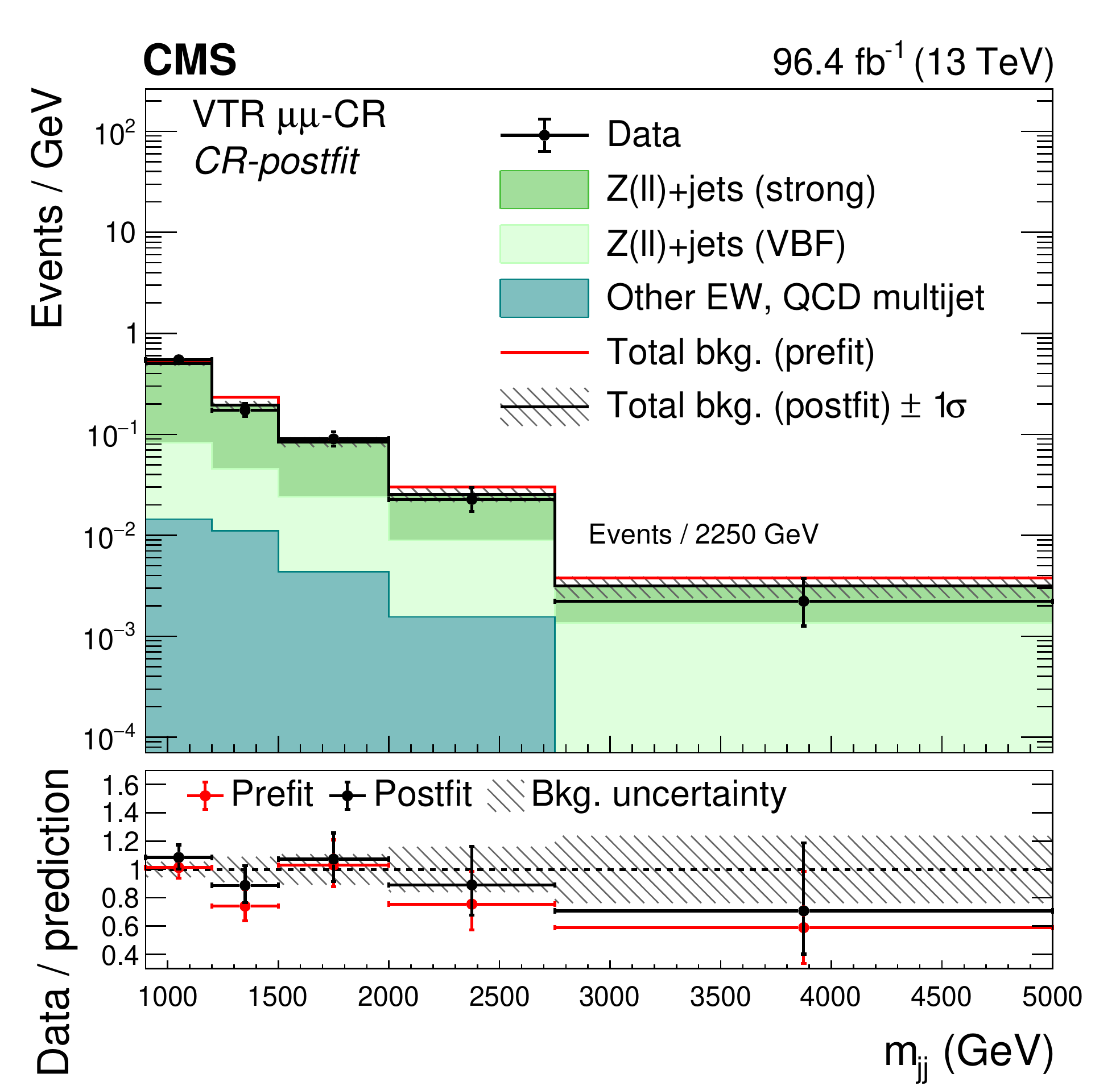}\\
    \includegraphics[width=0.48\textwidth]{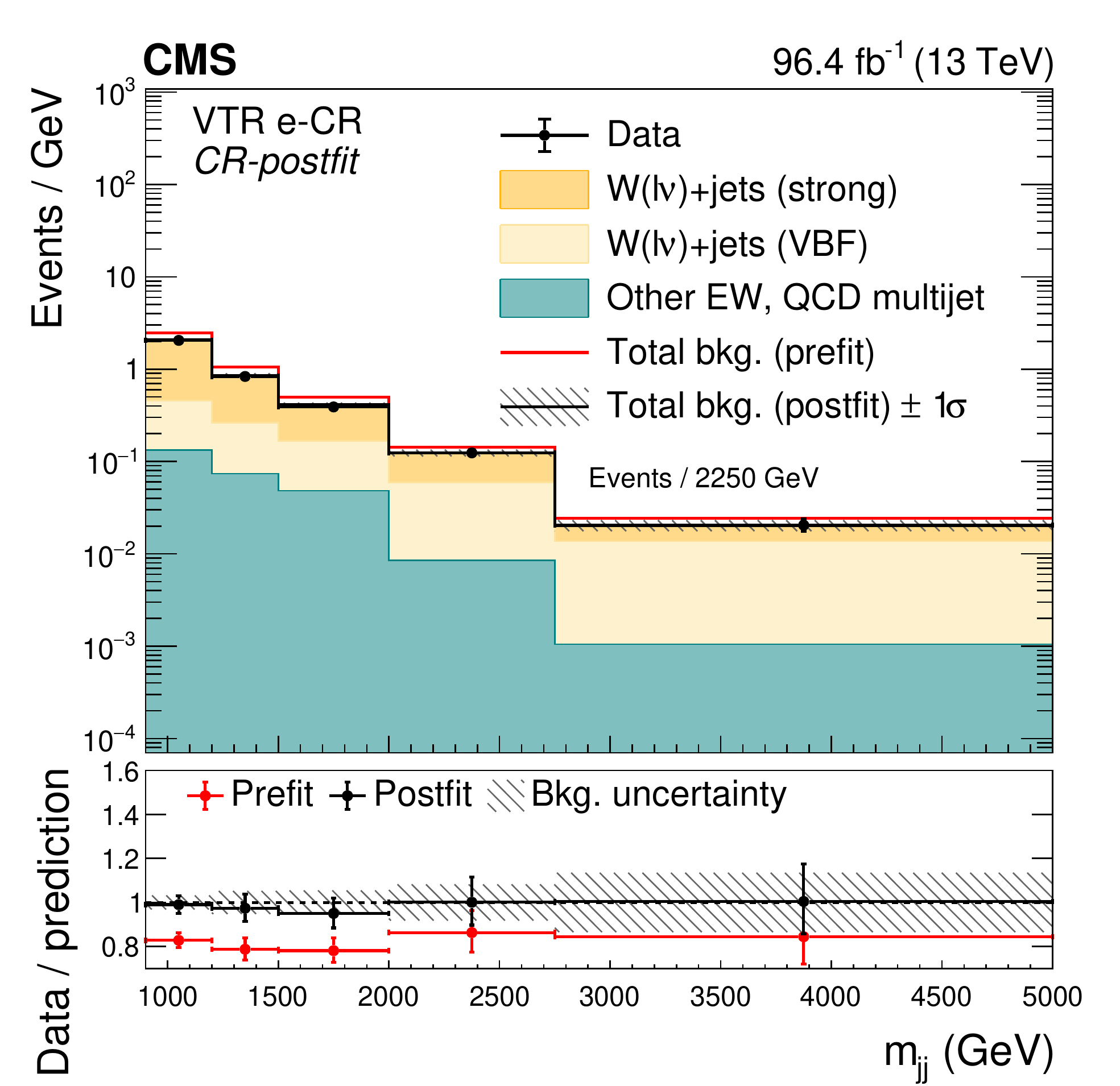}
    \includegraphics[width=0.48\textwidth]{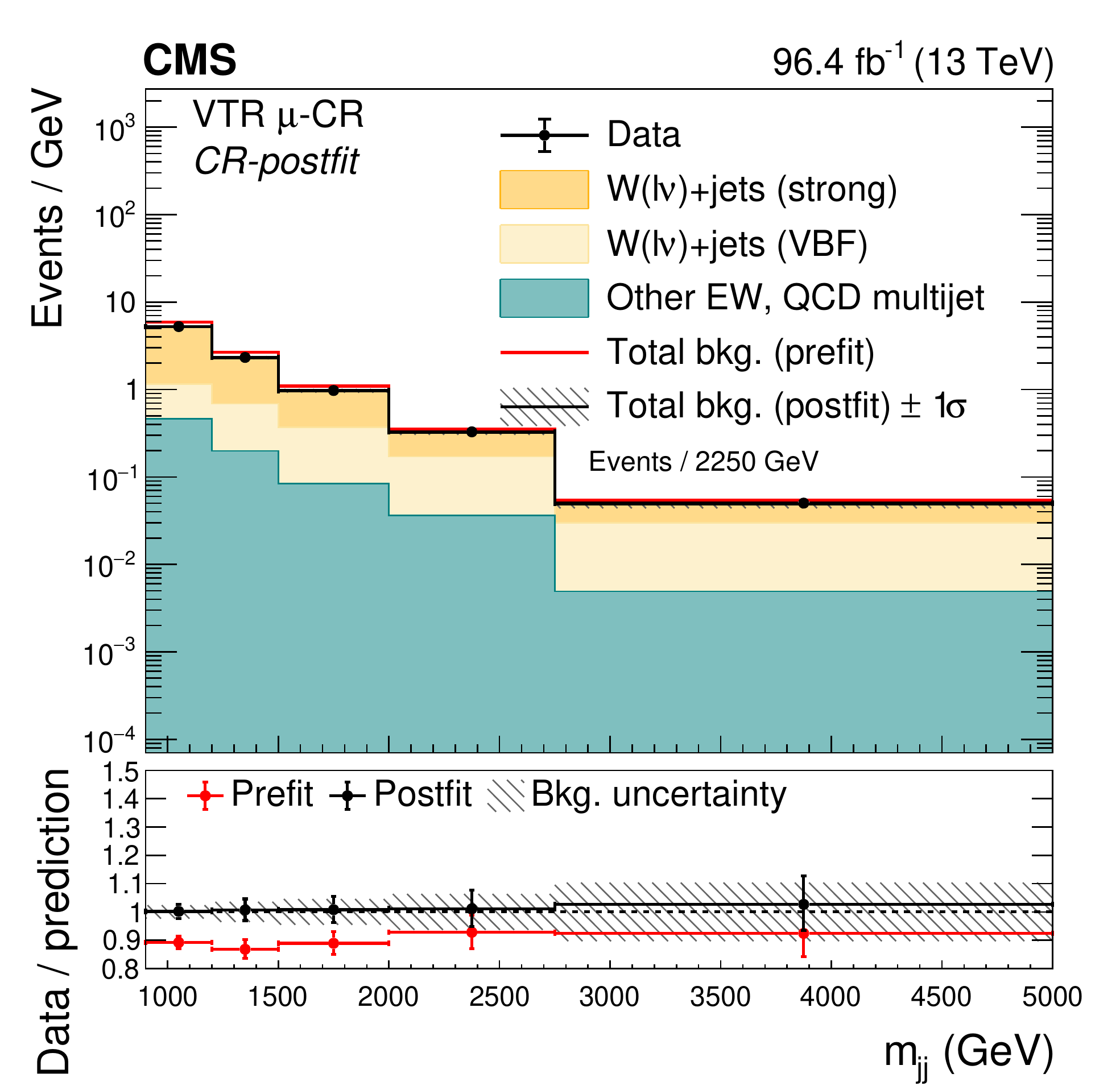}
    \caption{
        The \mjj distributions (prefit and CR-postfit) in the dielectron (upper
        left), dimuon (upper right), single-electron (lower left),
        and single-muon (lower right) CR for the VTR category, with the 2017
        and 2018 samples. The yields from the 2017 and 2018 samples are summed and the correlations between their uncertainties are neglected.}
    \label{fig:CR_VTR_CRonly}
\end{figure*}

\begin{figure*}[!htb]
    \centering
    \includegraphics[width=0.48\textwidth]{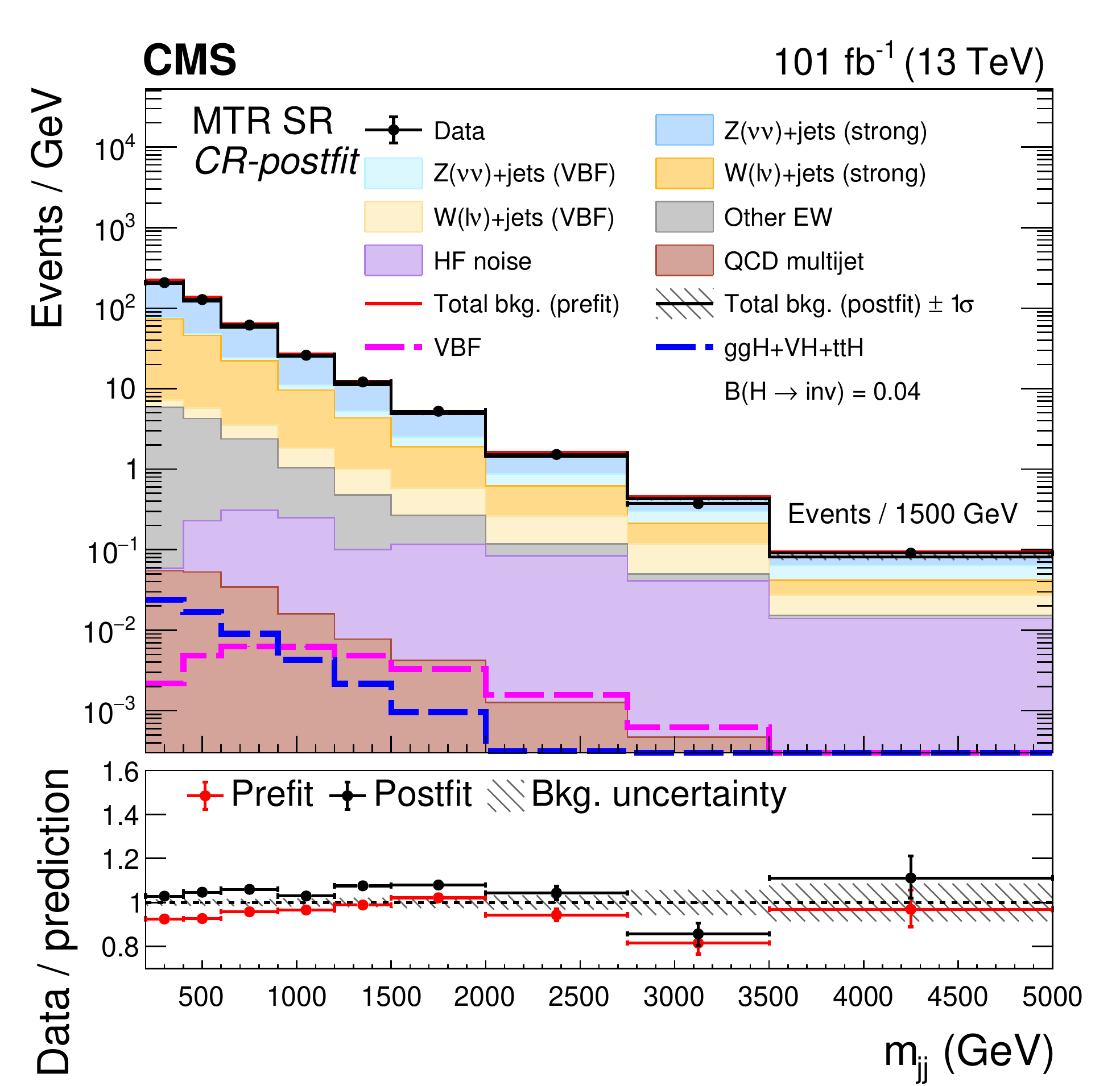}\hfill
    \includegraphics[width=0.48\textwidth]{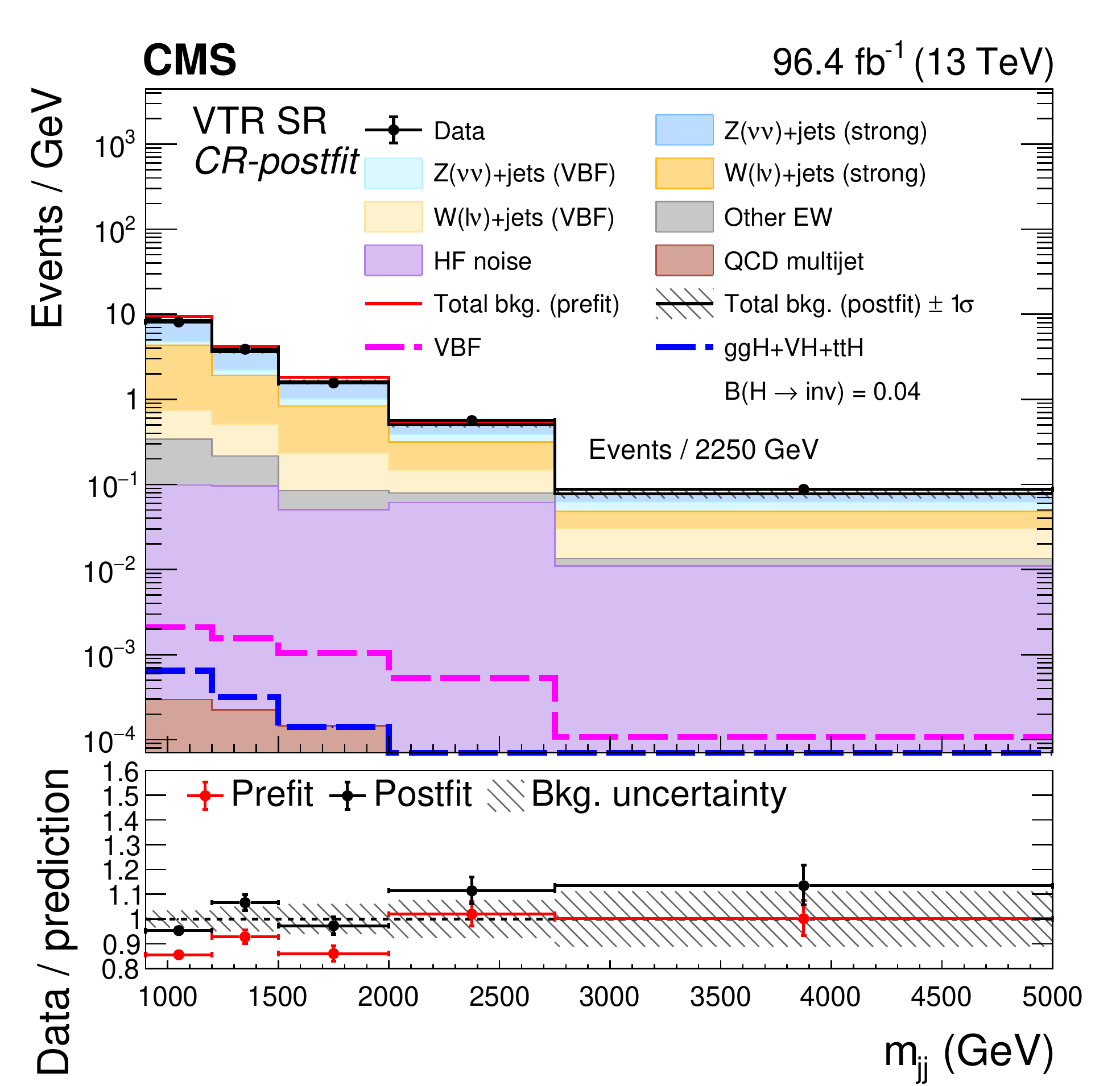}
    \caption{The observed \mjj distribution in the MTR (left) and VTR (right) SRs compared to
        the CR-postfit backgrounds, with the 2017 and 2018 samples. The yields from the 2017 and 2018 samples are summed and the correlations between their uncertainties are neglected.
        The signal processes are scaled by the fitted value of \brinv, shown in the legend.}
    \label{fig:SR_MTRVTR_CRonly}
\end{figure*}

CR-only fits are shown in Figs.~\ref{fig:CR_MTR_CRonly} to~\ref{fig:SR_MTRVTR_CRonly}, and corresponding tables of yields in Tables~\ref{tab:yields_MTR_2017_CRonly} to~\ref{tab:yields_VTR_1718_CRonly}.

\begin{table*}[htb!]
    \centering
    \topcaption[]{
        Expected event yields in each \mjj bin for the different background
        processes in the SR of the MTR category, in the 2017 samples. The
        background yields and the corresponding uncertainties are obtained
        after performing a combined fit across all of the CRs and SR. The
        expected signal contributions for the Higgs boson, produced in the non-\vbf
        and \vbf modes, decaying to invisible particles with a branching
        fraction of $\brinv = 1$, and the observed event yields are also
        reported.  }
    \label{tab:yields_MTR_2017_CRonly}
    \cmsTable{
        \begin{tabular}{lccccccccc}
            \hline
            \mjj bin range (\GeVns{})                        & 200--400          & 400--600          & 600--900         & 900--1200       & 1200--1500      & 1500--2000     & 2000--2750     & 2750--3500    & $>$3500      \\
            \hline
            $\PZ(\PGn\PGn)+\text{jets}$ (strong)             & $11686.7\pm281.9$ & $6720.5\pm185.2$  & $4639.8\pm129.8$ & $1843.5\pm54.5$ & $783.6\pm29.4$  & $499.2\pm22.6$ & $176.5\pm11.6$ & $44.8\pm4.9$  & $7.7\pm1.6$  \\
            $\PZ(\PGn\PGn)+\text{jets}$ (VBF)                & $197.8\pm7.3$     & $212.6\pm8.2$     & $260.3\pm8.9$    & $190.2\pm7.4$   & $120.6\pm5.7$   & $118.7\pm6.6$  & $71.1\pm5.4$   & $28.6\pm3.4$  & $10.5\pm2.3$ \\
            $\PW(\Pell\PGn)+\text{jets}$ (strong)            & $6132.9\pm145.5$  & $3616.2\pm89.3$   & $2549.9\pm66.6$  & $1023.4\pm34.3$ & $433.2\pm17.7$  & $272.2\pm12.3$ & $113.2\pm8.1$  & $30.6\pm4.5$  & $4.7\pm1.4$  \\
            $\PW(\Pell\PGn)+\text{jets}$ (VBF)               & $122.4\pm13.1$    & $135.4\pm13.6$    & $158.6\pm15.4$   & $107.4\pm9.9$   & $70.3\pm6.0$    & $63.3\pm5.4$   & $44.6\pm4.3$   & $19.2\pm2.4$  & $5.5\pm1.2$  \\
            $\ttbar$ + single \PQt quark                     & $246.0\pm35.0$    & $140.1\pm20.3$    & $128.2\pm18.9$   & $62.1\pm9.5$    & $31.6\pm5.3$    & $9.7\pm2.0$    & $2.7\pm0.8$    & $0.9\pm0.4$   & $0.4\pm0.2$  \\
            Diboson                                          & $200.9\pm43.9$    & $133.1\pm28.8$    & $101.5\pm22.3$   & $34.2\pm7.8$    & $15.6\pm3.7$    & $9.1\pm2.3$    & $3.3\pm1.0$    & $0.2\pm0.1$   & $0.0\pm0.1$  \\
            $\PZ/\Pgg^{*}(\Pell^{+}\Pell^{-})+\mathrm{jets}$ & $84.9\pm6.5$      & $53.4\pm4.7$      & $43.7\pm4.1$     & $15.3\pm1.7$    & $5.7\pm0.9$     & $4.2\pm0.6$    & $2.5\pm0.5$    & $0.5\pm0.2$   & $0.1\pm0.1$  \\
            Multijet                                         & $6.6\pm2.1$       & $6.1\pm1.9$       & $6.6\pm2.1$      & $2.7\pm0.9$     & $1.3\pm0.4$     & $1.1\pm0.3$    & $0.4\pm0.1$    & $0.2\pm0.1$   & $0.1\pm0.1$  \\
            HF noise                                         & $0.8\pm0.2$       & $17.0\pm3.2$      & $28.9\pm5.5$     & $25.7\pm4.9$    & $9.5\pm1.8$     & $18.8\pm3.6$   & $18.6\pm3.5$   & $11.0\pm2.1$  & $7.6\pm1.4$  \\ [\cmsTabSkip]

            $\Pg\Pg\PH(\to \mathrm{inv})$                    & 110.3             & 79.3              & 65.0             & 31.2            & 15.8            & 11.7           & 5.7            & 1.5           & 0.7          \\
            $\PQq\PQq\PH(\to \mathrm{inv})$                  & 10.9              & 24.2              & 47.2             & 46.7            & 36.5            & 41.4           & 29.6           & 11.7          & 8.6          \\
            $\PW\PH(\to \mathrm{inv})$                       & 5.7               & 3.3               & 2.2              & 0.7             & 0.4             & 0.2            & 0.1            & 0.0           & 0.0          \\
            $\PQq\PQq\PZ\PH(\to \mathrm{inv})$               & 2.7               & 1.3               & 0.8              & 0.2             & 0.1             & 0.1            & 0.1            & 0.0           & 0.0          \\
            $\Pg\Pg\PZ\PH(\to \mathrm{inv})$                 & 2.7               & 1.6               & 1.0              & 0.4             & 0.2             & 0.1            & 0.0            & 0.0           & 0.0          \\
            $\PQt\PQt\PH(\to \mathrm{inv})$                  & 0.7               & 0.5               & 0.3              & 0.1             & 0.1             & 0.0            & 0.0            & 0.0           & 0.0          \\ [\cmsTabSkip]

            Total bkg.                                       & $18679.0\pm322.5$ & $11034.3\pm209.3$ & $7917.5\pm150.0$ & $3304.5\pm66.9$ & $1471.5\pm36.0$ & $996.3\pm27.5$ & $433.0\pm16.2$ & $136.1\pm8.1$ & $36.7\pm3.7$ \\ [\cmsTabSkip]

            Observed                                         & 18945             & 11500             & 8218             & 3419            & 1549            & 1068           & 447            & 104           & 41           \\
            \hline
        \end{tabular}
    }\end{table*}

\begin{table*}[htb!]
    \centering
    \topcaption[]{
        Expected event yields in each \mjj bin for the different background
        processes in the SR of the MTR category, in the 2018 samples. The
        background yields and the corresponding uncertainties are obtained
        after performing a combined fit across all of the CRs and SR. The
        expected signal contributions for the Higgs boson, produced in the non-\vbf
        and \vbf modes, decaying to invisible particles with a branching
        fraction of $\brinv = 1$, and the observed event yields are also
        reported.  }
    \label{tab:yields_MTR_2018_CRonly}
    \cmsTable{
        \begin{tabular}{lccccccccc}
            \hline
            \mjj bin range (\GeVns{})                        & 200--400          & 400--600          & 600--900         & 900--1200       & 1200--1500      & 1500--2000      & 2000--2750     & 2750--3500     & $>$3500      \\
            \hline
            $\PZ(\PGn\PGn)+\text{jets}$ (strong)             & $13580.6\pm334.2$ & $8064.7\pm225.5$  & $5496.4\pm148.6$ & $2401.7\pm71.9$ & $1014.3\pm38.9$ & $678.9\pm29.4$  & $265.3\pm15.7$ & $56.9\pm5.5$   & $18.9\pm2.7$ \\
            $\PZ(\PGn\PGn)+\text{jets}$ (VBF)                & $219.6\pm8.1$     & $262.1\pm10.1$    & $325.2\pm11.0$   & $245.3\pm9.7$   & $155.3\pm7.6$   & $178.1\pm9.3$   & $117.4\pm8.2$  & $37.5\pm4.0$   & $22.4\pm3.5$ \\
            $\PW(\Pell\PGn)+\text{jets}$ (strong)            & $7038.7\pm153.2$  & $4376.3\pm101.3$  & $3073.8\pm74.7$  & $1316.2\pm37.7$ & $566.1\pm22.4$  & $393.7\pm16.6$  & $153.8\pm10.6$ & $39.2\pm5.4$   & $17.1\pm3.7$ \\
            $\PW(\Pell\PGn)+\text{jets}$ (VBF)               & $142.6\pm15.1$    & $159.3\pm15.5$    & $184.8\pm18.6$   & $130.3\pm12.2$  & $86.9\pm7.9$    & $93.5\pm8.6$    & $64.0\pm6.0$   & $32.1\pm4.2$   & $12.6\pm2.0$ \\
            $\ttbar$ + single \PQt quark                     & $268.1\pm36.4$    & $241.7\pm34.5$    & $152.5\pm23.2$   & $56.0\pm9.7$    & $28.4\pm5.4$    & $28.0\pm6.0$    & $11.0\pm3.2$   & $4.1\pm1.4$    & $1.3\pm0.5$  \\
            Diboson                                          & $259.9\pm53.4$    & $168.1\pm35.3$    & $139.8\pm30.2$   & $49.2\pm11.1$   & $22.3\pm5.4$    & $16.9\pm4.5$    & $4.3\pm1.3$    & $0.3\pm0.1$    & $0.0\pm0.1$  \\
            $\PZ/\Pgg^{*}(\Pell^{+}\Pell^{-})+\mathrm{jets}$ & $102.2\pm7.0$     & $68.7\pm5.4$      & $54.1\pm5.2$     & $21.2\pm2.3$    & $9.4\pm1.2$     & $6.6\pm1.3$     & $2.2\pm0.5$    & $0.8\pm0.2$    & $0.2\pm0.1$  \\
            Multijet                                         & $4.3\pm1.7$       & $4.5\pm1.8$       & $3.8\pm1.5$      & $2.1\pm0.8$     & $1.0\pm0.4$     & $1.0\pm0.4$     & $0.5\pm0.2$    & $0.2\pm0.1$    & $0.1\pm0.1$  \\
            HF noise                                         & $0.0\pm0.1$       & $18.2\pm3.3$      & $53.3\pm9.6$     & $44.3\pm7.9$    & $18.4\pm3.3$    & $37.3\pm6.7$    & $43.1\pm7.7$   & $19.4\pm3.5$   & $13.2\pm2.4$ \\ [\cmsTabSkip]

            $\Pg\Pg\PH(\to \mathrm{inv})$                    & 140.4             & 104.0             & 89.3             & 44.7            & 22.8            & 18.1            & 9.8            & 2.9            & 1.3          \\
            $\PQq\PQq\PH(\to \mathrm{inv})$                  & 14.6              & 33.5              & 66.4             & 63.8            & 52.2            & 62.1            & 52.0           & 21.9           & 13.7         \\
            $\PW\PH(\to \mathrm{inv})$                       & 7.7               & 4.5               & 2.4              & 1.0             & 0.4             & 0.2             & 0.1            & 0.0            & 0.0          \\
            $\PQq\PQq\PZ\PH(\to \mathrm{inv})$               & 3.6               & 1.7               & 0.8              & 0.4             & 0.1             & 0.1             & 0.0            & 0.0            & 0.0          \\
            $\Pg\Pg\PZ\PH(\to \mathrm{inv})$                 & 3.4               & 2.2               & 1.5              & 0.6             & 0.2             & 0.2             & 0.1            & 0.0            & 0.0          \\
            $\PQt\PQt\PH(\to \mathrm{inv})$                  & 0.9               & 0.7               & 0.5              & 0.2             & 0.1             & 0.1             & 0.0            & 0.0            & 0.0          \\ [\cmsTabSkip]

            Total bkg.                                       & $21616.0\pm373.8$ & $13363.5\pm252.9$ & $9483.8\pm172.4$ & $4266.4\pm84.3$ & $1902.2\pm47.0$ & $1434.0\pm37.5$ & $661.6\pm23.1$ & $190.5\pm10.3$ & $85.7\pm6.6$ \\ [\cmsTabSkip]

            Observed                                         & 22505             & 14036             & 10220            & 4374            & 2080            & 1555            & 695            & 176            & 95           \\
            \hline
        \end{tabular}
    }\end{table*}

\begin{table*}[htb!]
    \centering
    \topcaption[]{
        Expected event yields in each \mjj bin for the different background
        processes in the SR of the MTR category, in the 2017 and 2018
        samples. The background yields and the corresponding uncertainties are
        obtained after performing a combined fit across all of the CRs and
        SR. The expected signal contributions for the Higgs boson, produced in
        the non-\vbf and \vbf modes, decaying to invisible particles with a
        branching fraction of $\brinv = 1$, and the observed event yields are
        also reported. The yields from the 2017 and 2018 samples are summed and the correlations between their uncertainties are neglected.  }
    \label{tab:yields_MTR_1718_CRonly}
    \cmsTable{
        \begin{tabular}{lccccccccc}
            \hline
            \mjj bin range (\GeVns{})                        & 200--400          & 400--600          & 600--900          & 900--1200        & 1200--1500      & 1500--2000      & 2000--2750      & 2750--3500     & $>$3500       \\
            \hline
            $\PZ(\PGn\PGn)+\text{jets}$ (strong)             & $25267.3\pm437.2$ & $14785.2\pm291.8$ & $10136.3\pm197.3$ & $4245.3\pm90.2$  & $1797.8\pm48.8$ & $1178.1\pm37.1$ & $441.8\pm19.5$  & $101.8\pm7.4$  & $26.5\pm3.2$  \\
            $\PZ(\PGn\PGn)+\text{jets}$ (VBF)                & $417.4\pm10.9$    & $474.7\pm13.1$    & $585.5\pm14.1$    & $435.5\pm12.2$   & $276.0\pm9.5$   & $296.8\pm11.4$  & $188.5\pm9.9$   & $66.1\pm5.3$   & $32.9\pm4.2$  \\
            $\PW(\Pell\PGn)+\text{jets}$ (strong)            & $13171.6\pm211.3$ & $7992.4\pm135.0$  & $5623.6\pm100.1$  & $2339.7\pm51.0$  & $999.3\pm28.6$  & $665.9\pm20.7$  & $267.0\pm13.3$  & $69.8\pm7.0$   & $21.8\pm3.9$  \\
            $\PW(\Pell\PGn)+\text{jets}$ (VBF)               & $265.0\pm19.9$    & $294.7\pm20.7$    & $343.5\pm24.2$    & $237.7\pm15.7$   & $157.2\pm9.9$   & $156.8\pm10.1$  & $108.6\pm7.4$   & $51.3\pm4.9$   & $18.1\pm2.4$  \\
            $\ttbar$ + single \PQt quark                     & $514.1\pm50.5$    & $381.8\pm40.0$    & $280.7\pm29.9$    & $118.1\pm13.6$   & $60.0\pm7.6$    & $37.7\pm6.3$    & $13.7\pm3.3$    & $5.0\pm1.4$    & $1.7\pm0.6$   \\
            Diboson                                          & $460.8\pm69.1$    & $301.1\pm45.6$    & $241.3\pm37.6$    & $83.4\pm13.6$    & $38.0\pm6.5$    & $26.0\pm5.1$    & $7.5\pm1.6$     & $0.6\pm0.2$    & $0.0\pm0.1$   \\
            $\PZ/\Pgg^{*}(\Pell^{+}\Pell^{-})+\mathrm{jets}$ & $187.1\pm9.6$     & $122.2\pm7.2$     & $97.8\pm6.7$      & $36.5\pm2.9$     & $15.1\pm1.5$    & $10.8\pm1.5$    & $4.7\pm0.7$     & $1.3\pm0.2$    & $0.3\pm0.1$   \\
            Multijet                                         & $10.9\pm2.7$      & $10.6\pm2.6$      & $10.3\pm2.6$      & $4.8\pm1.2$      & $2.3\pm0.6$     & $2.1\pm0.5$     & $0.9\pm0.2$     & $0.4\pm0.1$    & $0.2\pm0.1$   \\
            HF noise                                         & $0.8\pm0.2$       & $35.2\pm4.6$      & $82.2\pm11.0$     & $70.0\pm9.3$     & $27.9\pm3.8$    & $56.1\pm7.6$    & $61.8\pm8.5$    & $30.4\pm4.0$   & $20.8\pm2.8$  \\ [\cmsTabSkip]

            $\PQq\PQq\PH(\to \mathrm{inv})$                  & 25.5              & 57.7              & 113.6             & 110.4            & 88.6            & 103.5           & 81.5            & 33.7           & 22.2          \\
            Other $\PH(\to \mathrm{inv})$ signals            & 278.1             & 199.0             & 163.7             & 79.6             & 40.1            & 30.8            & 16.0            & 4.5            & 2.0           \\ [\cmsTabSkip]

            Total bkg.                                       & $40295.0\pm493.7$ & $24397.8\pm328.3$ & $17401.2\pm228.5$ & $7570.9\pm107.7$ & $3373.7\pm59.2$ & $2430.3\pm46.5$ & $1094.6\pm28.2$ & $326.6\pm13.1$ & $122.4\pm7.5$ \\ [\cmsTabSkip]

            Observed                                         & 41450             & 25536             & 18438             & 7793             & 3629            & 2623            & 1142            & 279            & 136           \\
            \hline
        \end{tabular}
    }\end{table*}

\begin{table*}[htb!]
    \centering
    \topcaption[]{
        Expected event yields in each \mjj bin for the different background
        processes in the SR of the VTR category, in the 2017 samples. The
        background yields and the corresponding uncertainties are obtained
        after performing a combined fit across all of the CRs and SR. The
        expected signal contributions for the Higgs boson, produced in the non-\vbf
        and \vbf modes, decaying to invisible particles with a branching
        fraction of $\brinv = 1$, and the observed event yields are also
        reported.  }
    \label{tab:yields_VTR_2017_CRonly}
    \cmsTable{
        \begin{tabular}{lccccc}
            \hline
            \mjj bin range (\GeVns{})                        & 900--1200       & 1200--1500     & 1500--2000     & 2000--2750     & $>$2750      \\
            \hline
            $\PZ(\PGn\PGn)+\text{jets}$ (strong)             & $458.2\pm31.7$  & $175.5\pm17.5$ & $109.8\pm13.2$ & $32.6\pm6.1$   & $11.8\pm4.2$ \\
            $\PZ(\PGn\PGn)+\text{jets}$ (VBF)                & $60.2\pm5.3$    & $38.0\pm4.4$   & $27.7\pm3.8$   & $20.9\pm3.9$   & $10.1\pm3.4$ \\
            $\PW(\Pell\PGn)+\text{jets}$ (strong)            & $450.5\pm26.8$  & $173.7\pm16.0$ & $108.4\pm13.2$ & $45.0\pm9.6$   & $11.4\pm6.3$ \\
            $\PW(\Pell\PGn)+\text{jets}$ (VBF)               & $49.6\pm6.3$    & $34.7\pm5.1$   & $24.2\pm4.2$   & $23.4\pm4.8$   & $14.4\pm4.4$ \\
            $\ttbar$ + single \PQt quark                     & $6.5\pm1.1$     & $1.5\pm0.4$    & $2.0\pm0.5$    & $3.7\pm1.0$    & $2.8\pm1.5$  \\
            Diboson                                          & $6.4\pm1.5$     & $1.8\pm0.4$    & $1.3\pm0.3$    & $0.2\pm0.1$    & $0.0\pm0.1$  \\
            $\PZ/\Pgg^{*}(\Pell^{+}\Pell^{-})+\mathrm{jets}$ & $8.5\pm1.0$     & $5.6\pm0.9$    & $1.4\pm0.3$    & $0.8\pm0.2$    & $0.6\pm0.2$  \\
            Multijet                                         & $0.0\pm0.1$     & $0.0\pm0.1$    & $0.0\pm0.1$    & $0.0\pm0.1$    & $0.0\pm0.1$  \\
            HF noise                                         & $12.1\pm2.4$    & $7.8\pm1.6$    & $12.8\pm2.5$   & $17.0\pm3.4$   & $8.5\pm1.7$  \\ [\cmsTabSkip]

            $\Pg\Pg\PH(\to \mathrm{inv})$                    & 4.8             & 2.4            & 1.8            & 0.8            & 0.3          \\
            $\PQq\PQq\PH(\to \mathrm{inv})$                  & 15.7            & 11.7           & 13.0           & 9.9            & 6.0          \\
            $\PW\PH(\to \mathrm{inv})$                       & 0.0             & 0.0            & 0.0            & 0.0            & 0.0          \\
            $\PQq\PQq\PZ\PH(\to \mathrm{inv})$               & 0.0             & 0.0            & 0.0            & 0.0            & 0.0          \\
            $\Pg\Pg\PZ\PH(\to \mathrm{inv})$                 & 0.0             & 0.0            & 0.0            & 0.0            & 0.0          \\
            $\PQt\PQt\PH(\to \mathrm{inv})$                  & 0.0             & 0.0            & 0.0            & 0.0            & 0.0          \\ [\cmsTabSkip]

            Total bkg.                                       & $1052.0\pm42.4$ & $438.6\pm24.7$ & $287.6\pm19.7$ & $143.6\pm13.4$ & $59.6\pm9.7$ \\ [\cmsTabSkip]

            Observed                                         & 1020            & 432            & 298            & 154            & 68           \\
            \hline
        \end{tabular}
    }\end{table*}

\begin{table*}[htb!]
    \centering
    \topcaption[]{
        Expected event yields in each \mjj bin for the different background
        processes in the SR of the VTR category, in the 2018 samples. The
        background yields and the corresponding uncertainties are obtained
        after performing a combined fit across all of the CRs and SR. The
        expected signal contributions for the Higgs boson, produced in the non-\vbf
        and \vbf modes, decaying to invisible particles with a branching
        fraction of $\brinv = 1$, and the observed event yields are also
        reported.  }
    \label{tab:yields_VTR_2018_CRonly}
    \cmsTable{
        \begin{tabular}{lccccc}
            \hline
            \mjj bin range (\GeVns{})                        & 900--1200       & 1200--1500     & 1500--2000     & 2000--2750     & $>$2750        \\
            \hline
            $\PZ(\PGn\PGn)+\text{jets}$ (strong)             & $670.8\pm41.5$  & $249.0\pm22.9$ & $186.2\pm19.0$ & $56.2\pm9.4$   & $22.3\pm4.2$   \\
            $\PZ(\PGn\PGn)+\text{jets}$ (VBF)                & $78.2\pm6.1$    & $53.4\pm6.3$   & $60.0\pm7.9$   & $32.5\pm5.7$   & $20.6\pm3.8$   \\
            $\PW(\Pell\PGn)+\text{jets}$ (strong)            & $613.9\pm32.9$  & $250.3\pm19.8$ & $191.8\pm18.0$ & $80.7\pm16.1$  & $29.2\pm6.2$   \\
            $\PW(\Pell\PGn)+\text{jets}$ (VBF)               & $69.5\pm9.1$    & $52.6\pm8.0$   & $51.6\pm7.7$   & $28.1\pm5.9$   & $23.4\pm4.2$   \\
            $\ttbar$ + single \PQt quark                     & $20.3\pm3.8$    & $13.1\pm2.5$   & $4.3\pm1.1$    & $5.6\pm1.5$    & $1.3\pm0.6$    \\
            Diboson                                          & $9.1\pm2.1$     & $5.4\pm1.5$    & $2.8\pm0.8$    & $0.1\pm0.1$    & $0.0\pm0.1$    \\
            $\PZ/\Pgg^{*}(\Pell^{+}\Pell^{-})+\mathrm{jets}$ & $21.6\pm2.7$    & $8.6\pm1.1$    & $5.0\pm0.9$    & $3.1\pm0.7$    & $1.0\pm0.3$    \\
            Multijet                                         & $0.1\pm0.1$     & $0.0\pm0.1$    & $0.1\pm0.1$    & $0.0\pm0.1$    & $0.0\pm0.1$    \\
            HF noise                                         & $17.4\pm3.6$    & $21.0\pm4.3$   & $12.5\pm2.6$   & $28.8\pm5.9$   & $16.2\pm3.3$   \\ [\cmsTabSkip]

            $\Pg\Pg\PH(\to \mathrm{inv})$                    & 7.9             & 3.9            & 2.9            & 1.3            & 0.6            \\
            $\PQq\PQq\PH(\to \mathrm{inv})$                  & 28.2            & 21.1           & 24.5           & 17.0           & 12.4           \\
            $\PW\PH(\to \mathrm{inv})$                       & 0.1             & 0.0            & 0.1            & 0.0            & 0.0            \\
            $\PQq\PQq\PZ\PH(\to \mathrm{inv})$               & 0.0             & 0.0            & 0.0            & 0.0            & 0.0            \\
            $\Pg\Pg\PZ\PH(\to \mathrm{inv})$                 & 0.1             & 0.0            & 0.0            & 0.0            & 0.0            \\
            $\PQt\PQt\PH(\to \mathrm{inv})$                  & 0.0             & 0.0            & 0.0            & 0.0            & 0.0            \\ [\cmsTabSkip]

            Total bkg.                                       & $1500.8\pm54.4$ & $653.4\pm32.3$ & $514.3\pm28.6$ & $235.3\pm21.3$ & $114.0\pm10.0$ \\ [\cmsTabSkip]

            Observed                                         & 1413            & 732            & 482            & 268            & 129            \\
            \hline
        \end{tabular}
    }\end{table*}

\begin{table*}[htb!]
    \centering
    \topcaption[]{
        Expected event yields in each \mjj bin for the different background
        processes in the SR of the VTR category, in the 2017 and 2018
        samples. The background yields and the corresponding uncertainties are
        obtained after performing a combined fit across all of the CRs and
        SR. The expected signal contributions for the Higgs boson, produced in
        the non-\vbf and \vbf modes, decaying to invisible particles with a
        branching fraction of $\brinv = 1$, and the observed event yields are
        also reported. The yields from the 2017 and 2018 samples are summed and the correlations between their uncertainties are neglected.  }
    \label{tab:yields_VTR_1718_CRonly}
    \cmsTable{
        \begin{tabular}{lccccc}
            \hline
            \mjj bin range (\GeVns{})                        & 900--1200       & 1200--1500      & 1500--2000     & 2000--2750     & $>$2750        \\
            \hline
            $\PZ(\PGn\PGn)+\text{jets}$ (strong)             & $1129.0\pm52.2$ & $424.4\pm28.8$  & $295.9\pm23.2$ & $88.9\pm11.2$  & $34.1\pm6.0$   \\
            $\PZ(\PGn\PGn)+\text{jets}$ (VBF)                & $138.4\pm8.1$   & $91.4\pm7.7$    & $87.8\pm8.7$   & $53.3\pm7.0$   & $30.8\pm5.0$   \\
            $\PW(\Pell\PGn)+\text{jets}$ (strong)            & $1064.3\pm42.4$ & $424.0\pm25.4$  & $300.2\pm22.4$ & $125.6\pm18.7$ & $40.5\pm8.9$   \\
            $\PW(\Pell\PGn)+\text{jets}$ (VBF)               & $119.1\pm11.1$  & $87.3\pm9.5$    & $75.8\pm8.8$   & $51.5\pm7.5$   & $37.8\pm6.1$   \\
            $\ttbar$ + single \PQt quark                     & $26.8\pm4.0$    & $14.6\pm2.5$    & $6.3\pm1.2$    & $9.3\pm1.8$    & $4.1\pm1.6$    \\
            Diboson                                          & $15.6\pm2.6$    & $7.3\pm1.5$     & $4.1\pm0.8$    & $0.3\pm0.1$    & $0.0\pm0.1$    \\
            $\PZ/\Pgg^{*}(\Pell^{+}\Pell^{-})+\mathrm{jets}$ & $30.1\pm2.9$    & $14.2\pm1.4$    & $6.4\pm1.0$    & $3.9\pm0.7$    & $1.6\pm0.4$    \\
            Multijet                                         & $0.1\pm0.1$     & $0.1\pm0.1$     & $0.1\pm0.1$    & $0.1\pm0.1$    & $0.0\pm0.1$    \\
            HF noise                                         & $29.5\pm4.3$    & $28.8\pm4.6$    & $25.2\pm3.6$   & $45.9\pm6.8$   & $24.7\pm3.7$   \\ [\cmsTabSkip]

            $\PQq\PQq\PH(\to \mathrm{inv})$                  & 43.9            & 32.8            & 37.5           & 27.0           & 18.4           \\
            Other $\PH(\to \mathrm{inv})$ signals            & 12.9            & 6.4             & 4.8            & 2.2            & 0.9            \\ [\cmsTabSkip]

            Total bkg.                                       & $2552.9\pm69.0$ & $1092.0\pm40.7$ & $801.8\pm34.7$ & $378.9\pm25.2$ & $173.7\pm13.9$ \\ [\cmsTabSkip]

            Observed                                         & 2433            & 1164            & 780            & 422            & 197            \\
            \hline
        \end{tabular}
    }\end{table*}

\cmsClearpage
\subsection*{MTR 2017}
\begin{figure*}[!htb]
    \centering
    \includegraphics[width=0.48\textwidth]{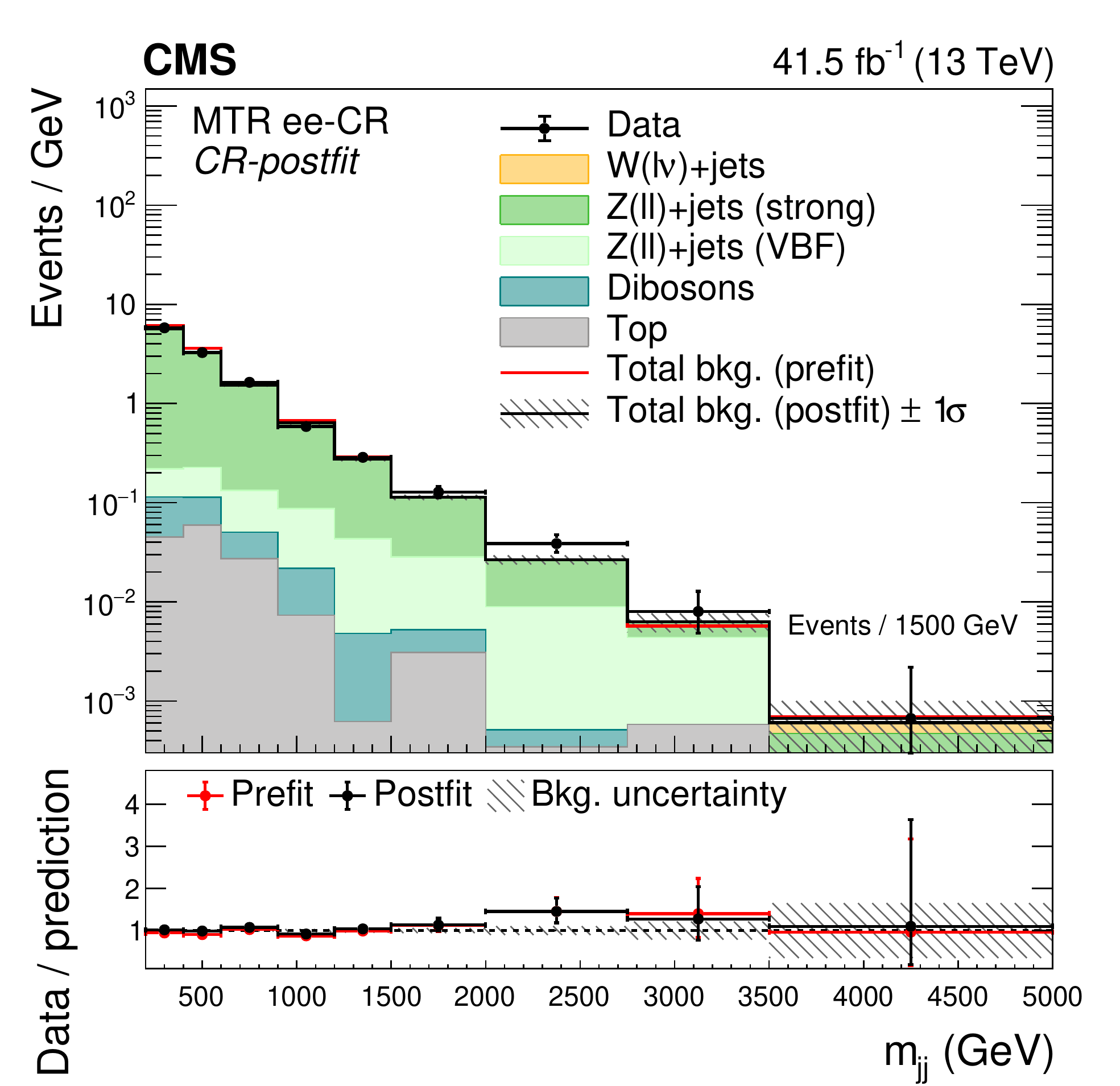}
    \includegraphics[width=0.48\textwidth]{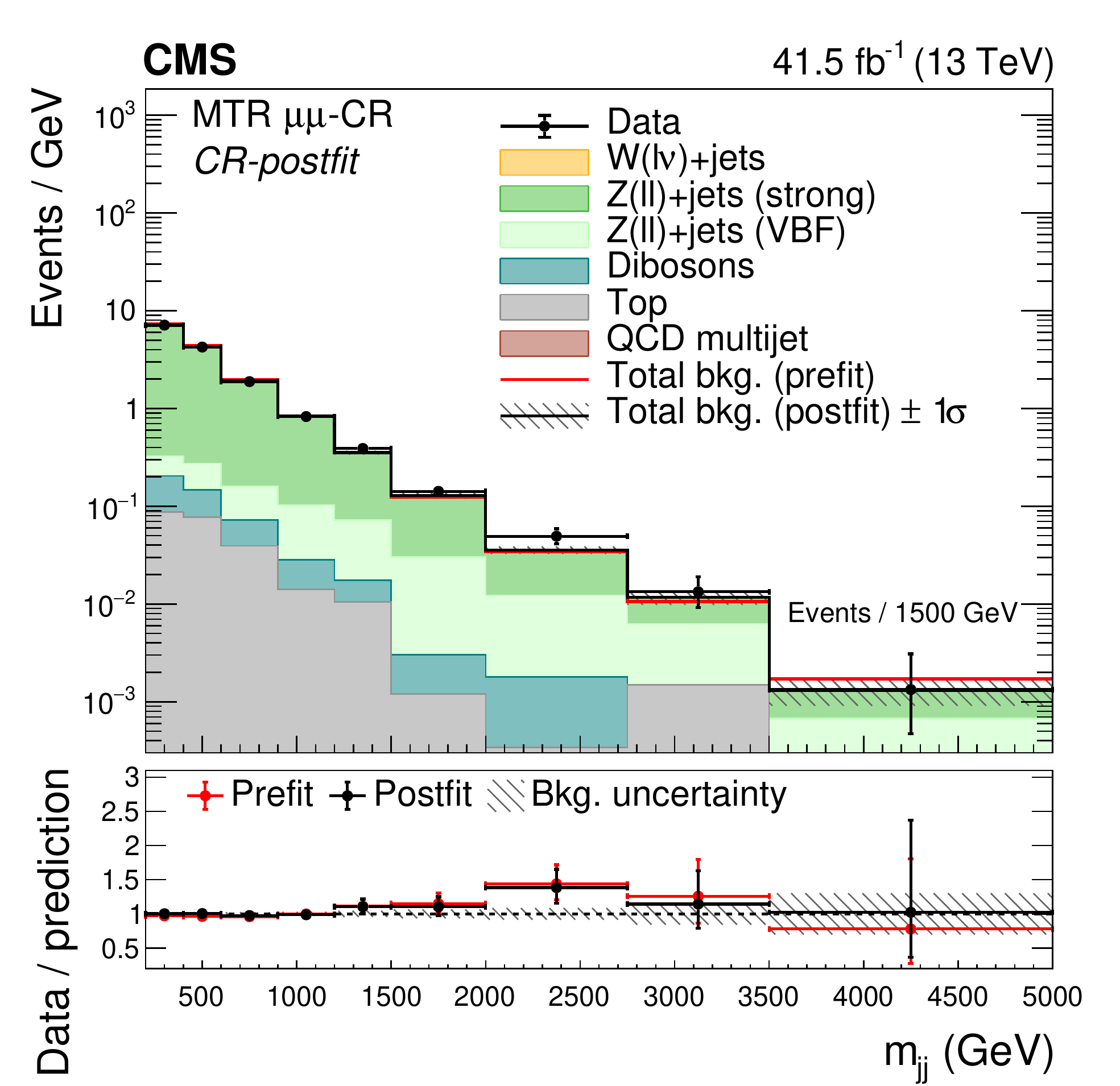}\\
    \includegraphics[width=0.48\textwidth]{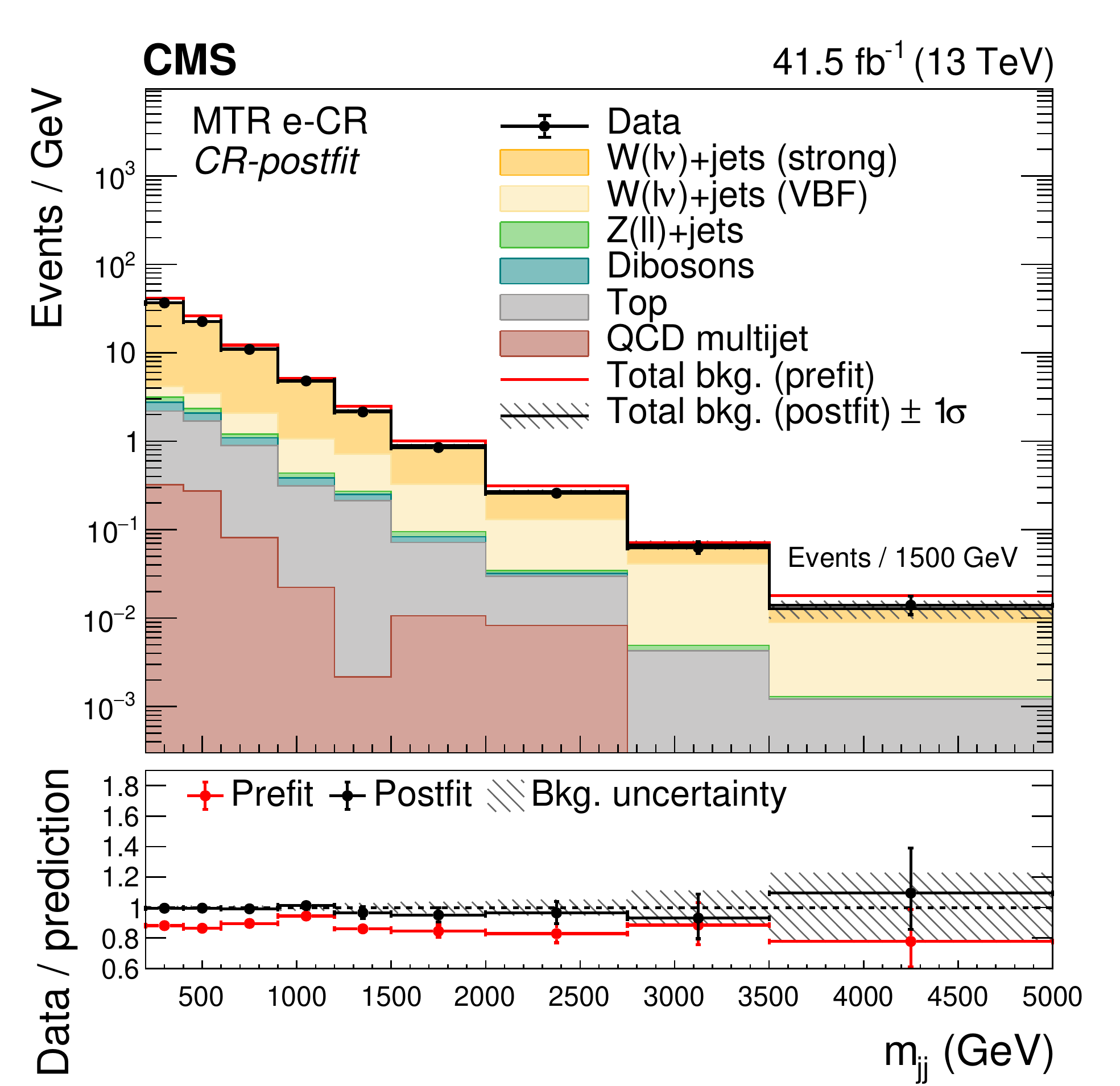}
    \includegraphics[width=0.48\textwidth]{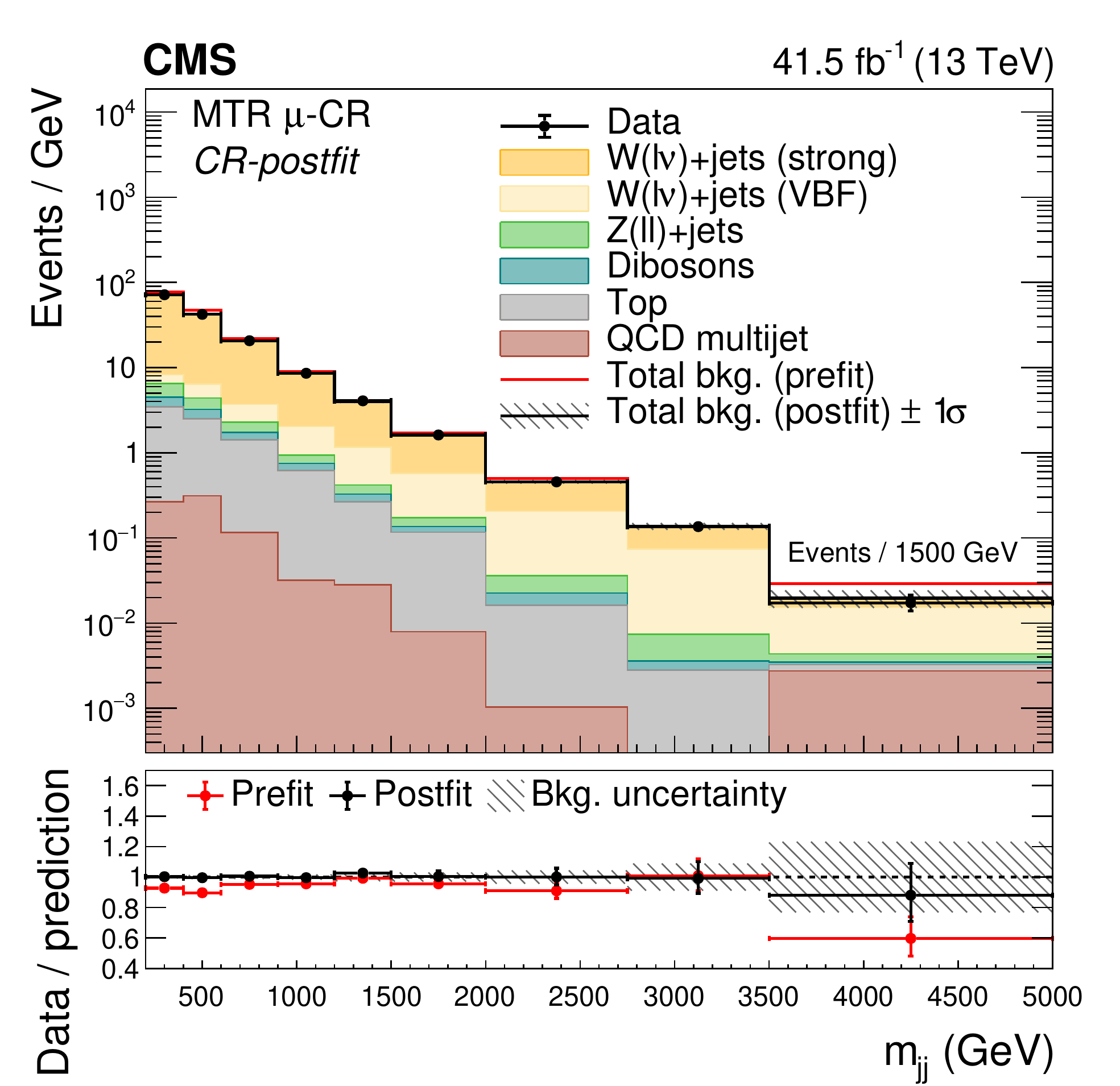}
    \caption{
        The \mjj distributions (prefit and CR-postfit) in the dielectron (upper
        left), dimuon (upper right), single-electron (lower left),
        and single-muon (lower right) CR for the MTR category, with the 2017
        sample.}
    \label{fig:CR_MTR_2017_CRonly}
\end{figure*}

\begin{figure*}[!htb]
    \centering
    \includegraphics[width=0.48\textwidth]{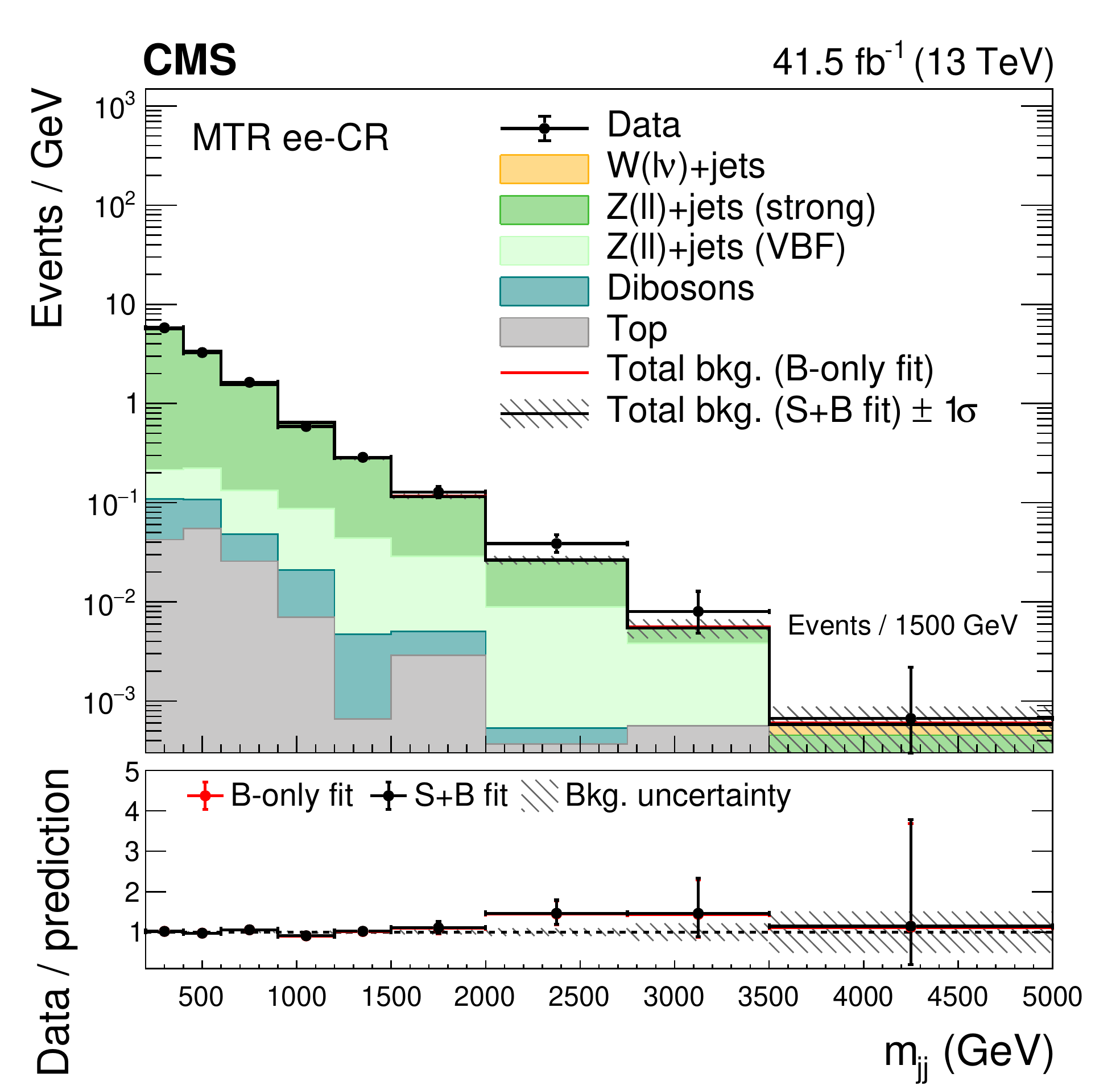}
    \includegraphics[width=0.48\textwidth]{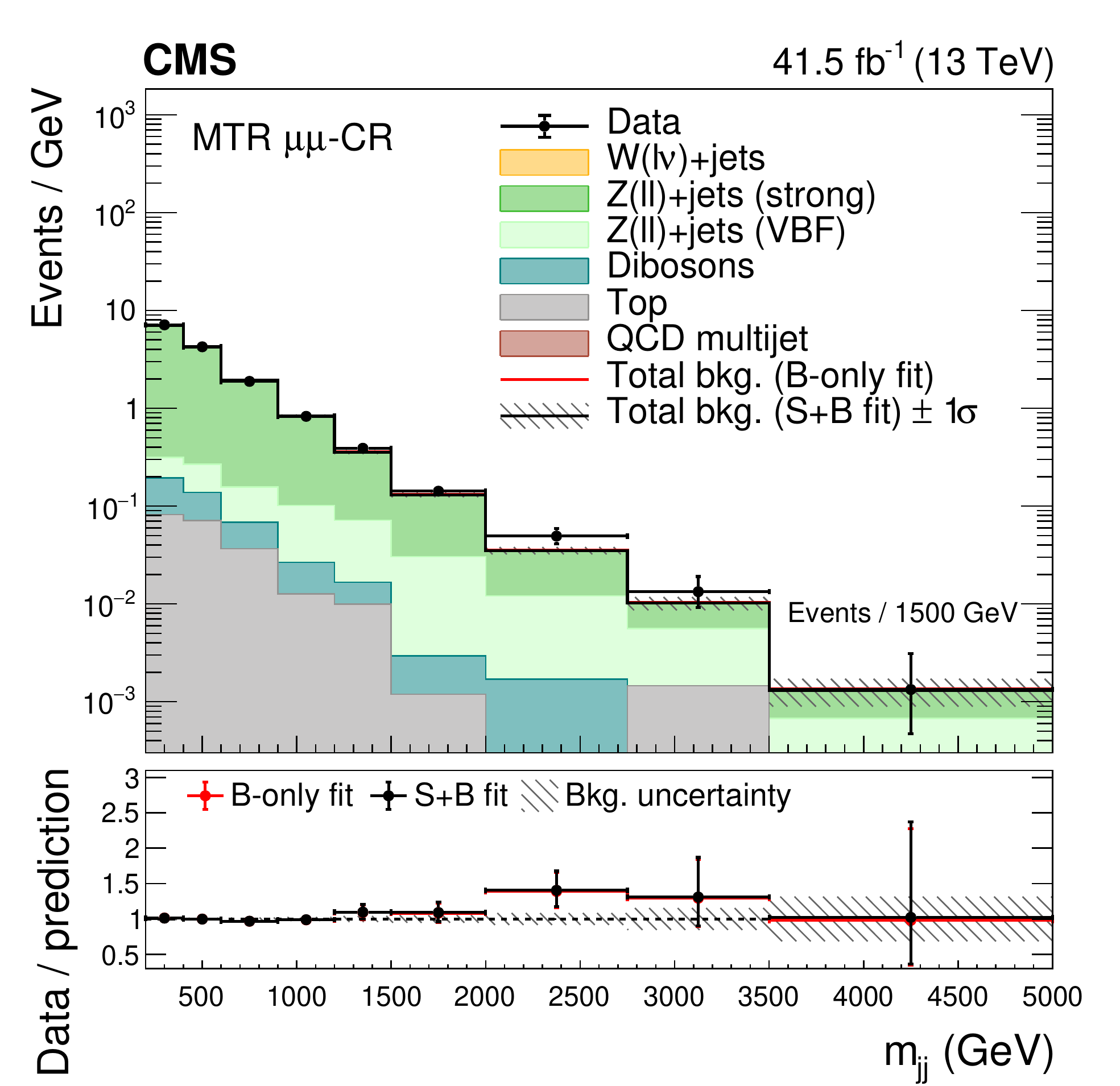}\\
    \includegraphics[width=0.48\textwidth]{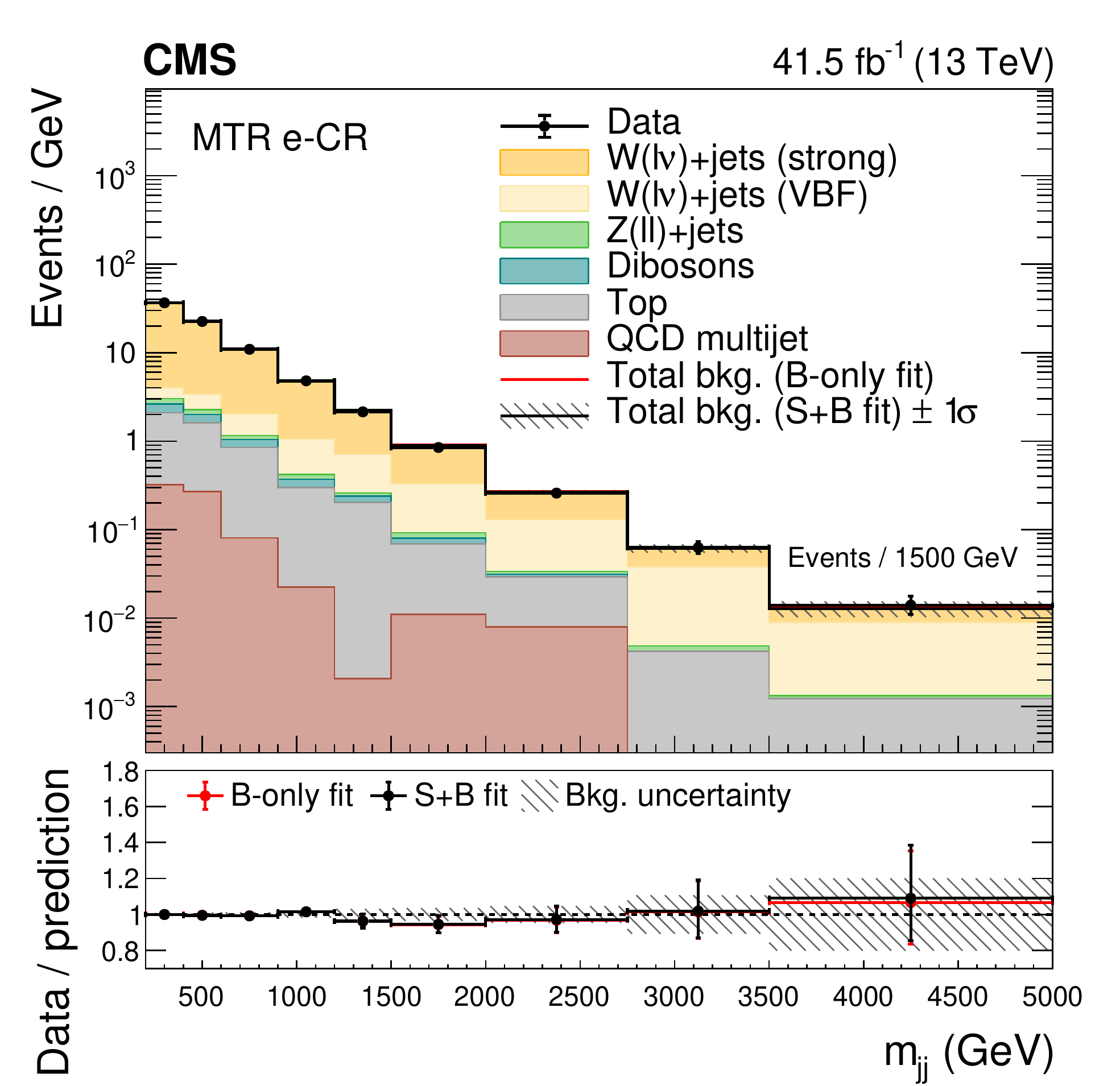}
    \includegraphics[width=0.48\textwidth]{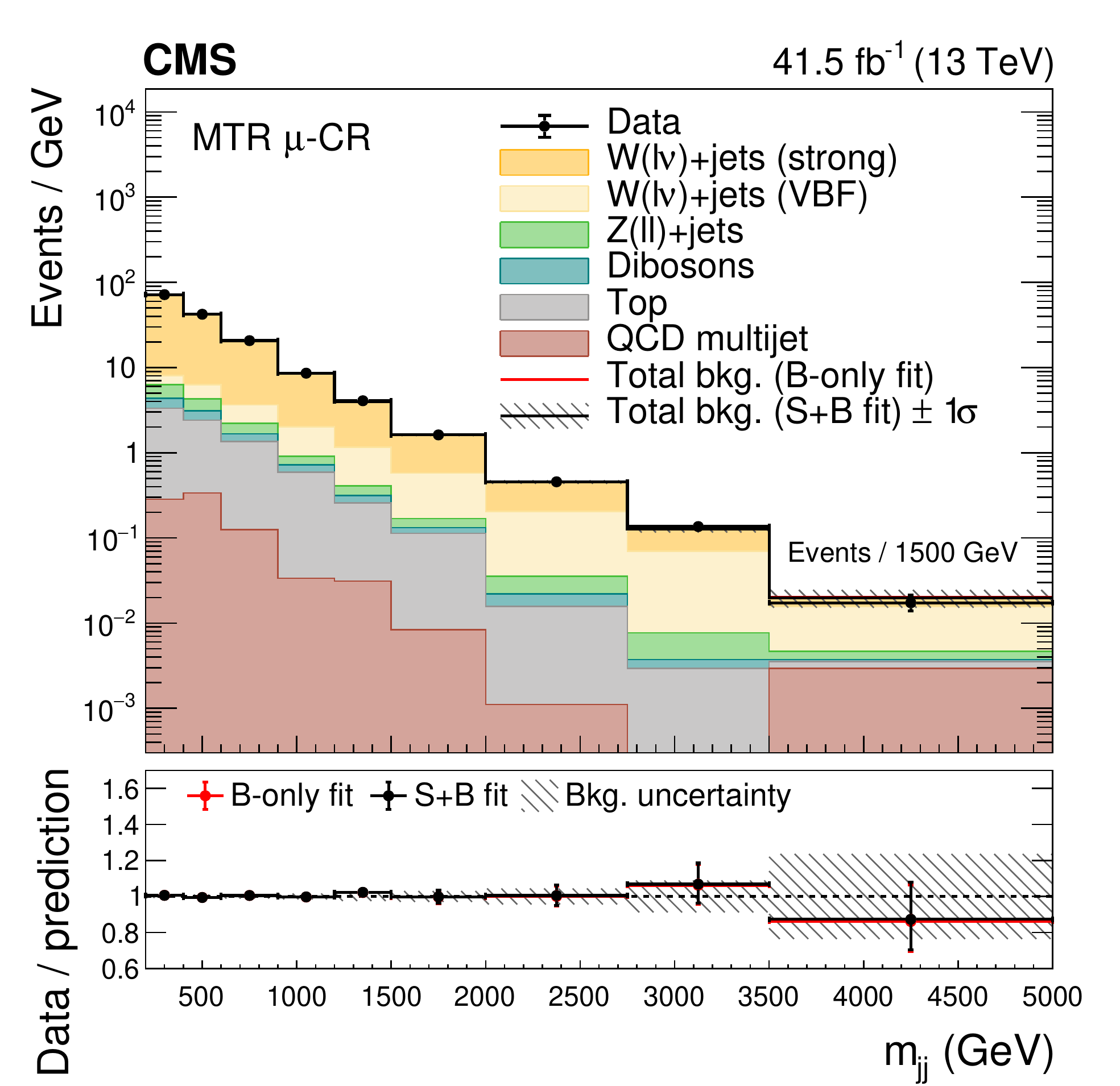}
    \caption{
        The \mjj postfit distributions in the dielectron (upper
        left), dimuon (upper right), single-electron (lower left),
        and single-muon (lower right) CR for the MTR category, with the 2017 sample. The background contributions are estimated from
        a fit to data in the SR and CRs allowing for the signal contribution to vary (S+B fit) and the total background estimated from a fit
        assuming $\brinv=0$ (B-only fit) is also shown.}
    \label{fig:CR_MTR_2017}
\end{figure*}

\begin{figure*}[htb]
    \centering
    \includegraphics[width=0.48\textwidth]{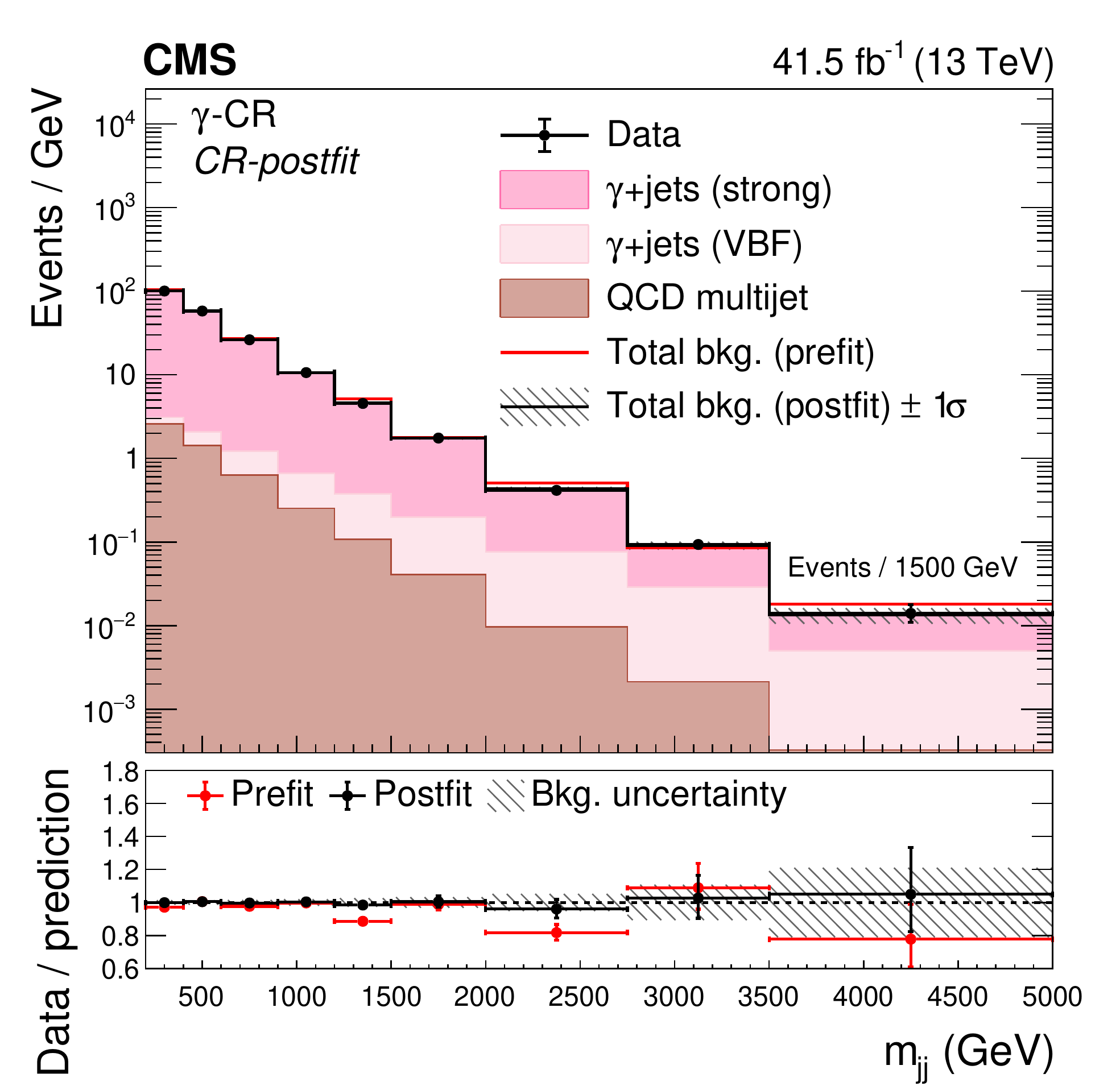}\hfill
    \includegraphics[width=0.48\textwidth]{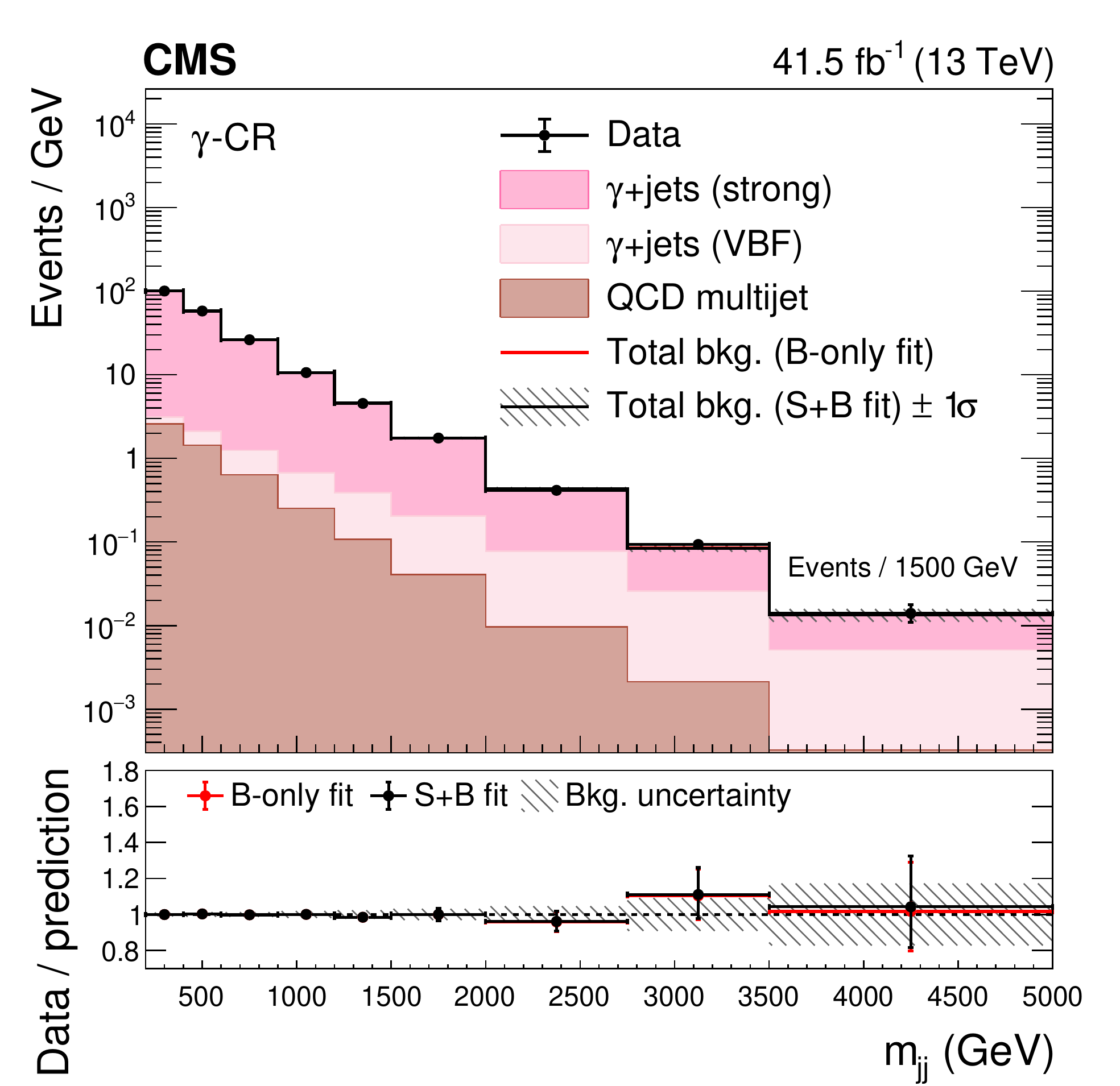}
    \caption{The \mjj CR-postfit (left) and postfit (right) distributions in the photon CR for the MTR category, with the 2017 sample.
        In the right figure, the total background estimated from a fit
        assuming $\brinv=0$ (B-only fit) is also shown.}
    \label{fig:Zgamma_MTR_2017}
\end{figure*}

\begin{figure*}[!htb]
    \centering
    \includegraphics[width=0.48\textwidth]{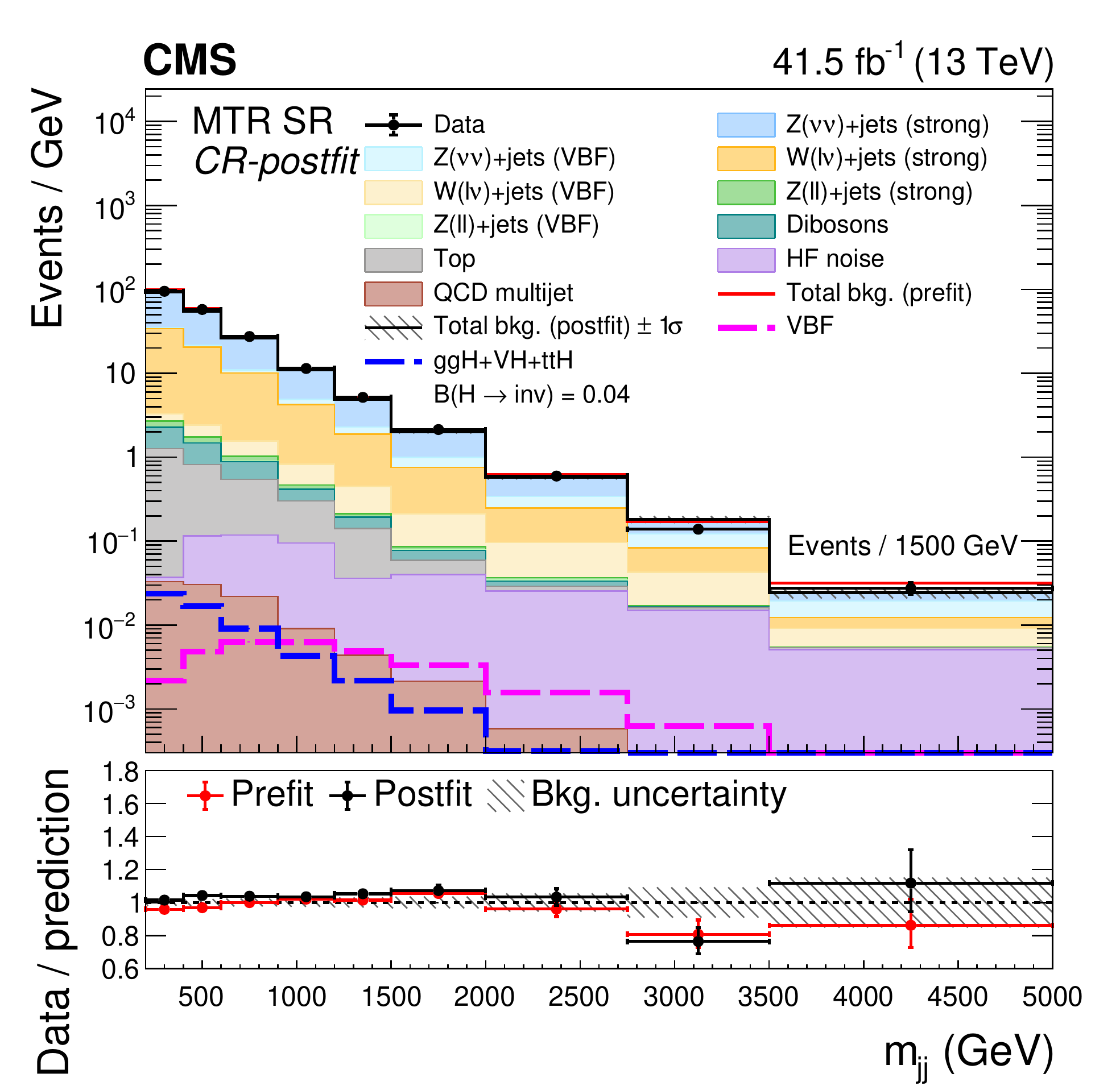}\hfill
    \includegraphics[width=0.48\textwidth]{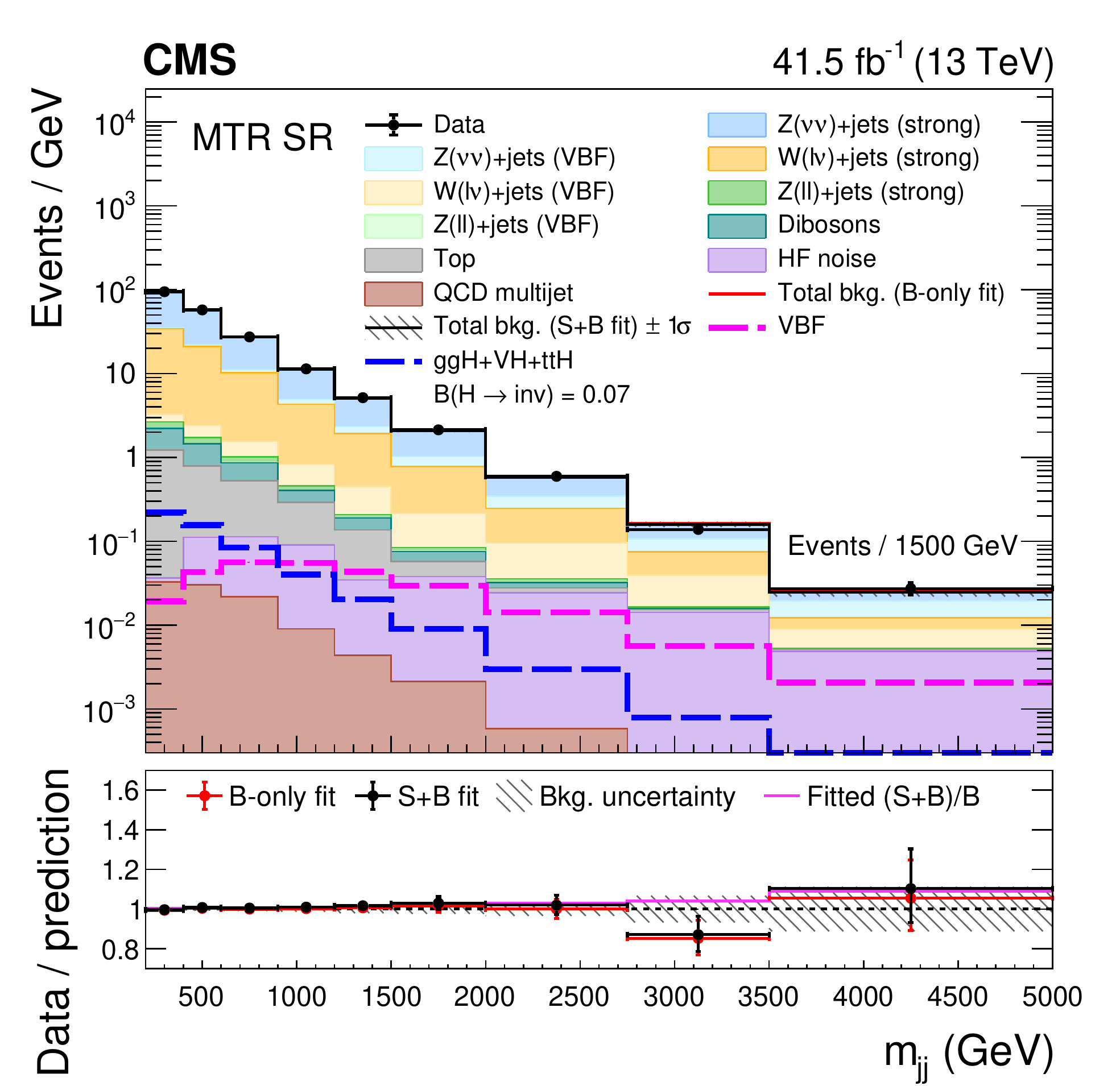}
    \caption{The observed \mjj distribution in the MTR prefit (left) and postfit (right) SR
        compared to the postfit backgrounds, with the 2017 samples.
        The signal processes are scaled by the fitted value of \brinv, shown in the legend.}
    \label{fig:SR_MTR_2017}
\end{figure*}

The results for the MTR 2017 category are shown in Figs.~\ref{fig:CR_MTR_2017_CRonly} to~\ref{fig:SR_MTR_2017} and Table~\ref{tab:yields_MTR_2017}.

\begin{table*}[htb!]
    \centering
    \topcaption[]{
        Expected event yields in each \mjj bin for the different background
        processes in the SR of the MTR category, in the 2017 samples. The
        background yields and the corresponding uncertainties are obtained
        after performing a combined fit across all of the CRs and SR. The
        expected signal contributions for the Higgs boson, produced in the non-\vbf
        and \vbf modes, decaying to invisible particles with a branching
        fraction of $\brinv = 1$, and the observed event yields are also
        reported.  }
    \label{tab:yields_MTR_2017}
    \cmsTable{
        \begin{tabular}{lccccccccc}
            \hline
            \mjj bin range (\GeVns{})                        & 200--400         & 400--600         & 600--900        & 900--1200       & 1200--1500      & 1500--2000      & 2000--2750    & 2750--3500    & $>$3500      \\
            \hline
            $\PZ(\PGn\PGn)+\text{jets}$ (strong)             & $11957.1\pm55.5$ & $7022.4\pm42.2$  & $4855.8\pm34.6$ & $1914.1\pm17.6$ & $826.8\pm11.4$  & $531.3\pm8.5$   & $183.5\pm4.7$ & $39.6\pm4.1$  & $8.3\pm0.9$  \\
            $\PZ(\PGn\PGn)+\text{jets}$ (VBF)                & $202.5\pm4.1$    & $222.2\pm4.1$    & $272.3\pm4.3$   & $197.6\pm3.8$   & $127.2\pm3.2$   & $126.4\pm3.6$   & $74.0\pm2.9$  & $25.3\pm2.9$  & $11.5\pm1.4$ \\
            $\PW(\Pell\PGn)+\text{jets}$ (strong)            & $6247.9\pm57.1$  & $3727.1\pm36.6$  & $2624.7\pm31.6$ & $1052.3\pm15.7$ & $450.0\pm11.7$  & $285.5\pm7.1$   & $116.5\pm4.9$ & $27.1\pm2.7$  & $5.1\pm1.0$  \\
            $\PW(\Pell\PGn)+\text{jets}$ (VBF)               & $122.6\pm7.2$    & $137.9\pm7.5$    & $161.9\pm8.1$   & $109.4\pm5.3$   & $72.3\pm3.5$    & $65.8\pm3.1$    & $45.7\pm2.9$  & $17.5\pm1.8$  & $5.9\pm0.8$  \\
            $\ttbar$ + single \PQt quark                     & $237.6\pm16.0$   & $135.8\pm9.1$    & $124.0\pm8.0$   & $60.1\pm3.7$    & $30.7\pm2.0$    & $9.7\pm0.8$     & $2.7\pm0.3$   & $0.9\pm0.2$   & $0.4\pm0.1$  \\
            Diboson                                          & $201.0\pm24.8$   & $132.9\pm16.0$   & $101.7\pm12.2$  & $34.4\pm4.2$    & $15.8\pm1.9$    & $9.2\pm1.2$     & $3.3\pm0.5$   & $0.3\pm0.1$   & $0.0\pm0.1$  \\
            $\PZ/\Pgg^{*}(\Pell^{+}\Pell^{-})+\mathrm{jets}$ & $86.6\pm3.3$     & $54.9\pm2.1$     & $44.7\pm1.6$    & $15.7\pm0.6$    & $6.0\pm0.4$     & $4.3\pm0.3$     & $2.6\pm0.2$   & $0.5\pm0.1$   & $0.1\pm0.1$  \\
            Multijet                                         & $6.6\pm1.5$      & $6.1\pm1.4$      & $6.6\pm1.5$     & $2.7\pm0.6$     & $1.3\pm0.3$     & $1.1\pm0.2$     & $0.4\pm0.1$   & $0.2\pm0.1$   & $0.1\pm0.1$  \\
            HF noise                                         & $0.8\pm0.1$      & $16.6\pm2.1$     & $28.2\pm3.6$    & $25.1\pm3.2$    & $9.3\pm1.2$     & $18.4\pm2.3$    & $18.2\pm2.3$  & $10.7\pm1.4$  & $7.4\pm0.9$  \\ [\cmsTabSkip]

            $\Pg\Pg\PH(\to \mathrm{inv})$                    & 570.5            & 411.5            & 338.0           & 162.8           & 82.5            & 61.8            & 30.4          & 8.1           & 3.6          \\
            $\PQq\PQq\PH(\to \mathrm{inv})$                  & 56.2             & 125.7            & 245.8           & 244.0           & 191.2           & 217.9           & 156.1         & 62.6          & 45.6         \\
            $\PW\PH(\to \mathrm{inv})$                       & 29.7             & 17.0             & 11.5            & 3.9             & 1.9             & 0.8             & 0.5           & 0.1           & 0.0          \\
            $\PQq\PQq\PZ\PH(\to \mathrm{inv})$               & 14.0             & 6.8              & 4.0             & 1.2             & 0.7             & 0.4             & 0.4           & 0.0           & 0.0          \\
            $\Pg\Pg\PZ\PH(\to \mathrm{inv})$                 & 14.0             & 8.6              & 5.5             & 2.3             & 1.0             & 0.6             & 0.3           & 0.1           & 0.0          \\
            $\PQt\PQt\PH(\to \mathrm{inv})$                  & 3.6              & 2.6              & 1.8             & 0.7             & 0.3             & 0.2             & 0.1           & 0.0           & 0.0          \\ [\cmsTabSkip]

            Total bkg.                                       & $19062.6\pm85.4$ & $11455.8\pm59.5$ & $8220.0\pm50.1$ & $3411.5\pm25.4$ & $1539.4\pm17.3$ & $1051.6\pm12.4$ & $446.9\pm8.3$ & $122.0\pm6.2$ & $38.9\pm2.3$ \\ [\cmsTabSkip]

            Observed                                         & 18945            & 11500            & 8218            & 3419            & 1549            & 1068            & 447           & 104           & 41           \\
            \hline
        \end{tabular}
    }\end{table*}

\cmsClearpage
\subsection*{MTR 2018}

\begin{figure*}[!htb]
    \centering
    \includegraphics[width=0.48\textwidth]{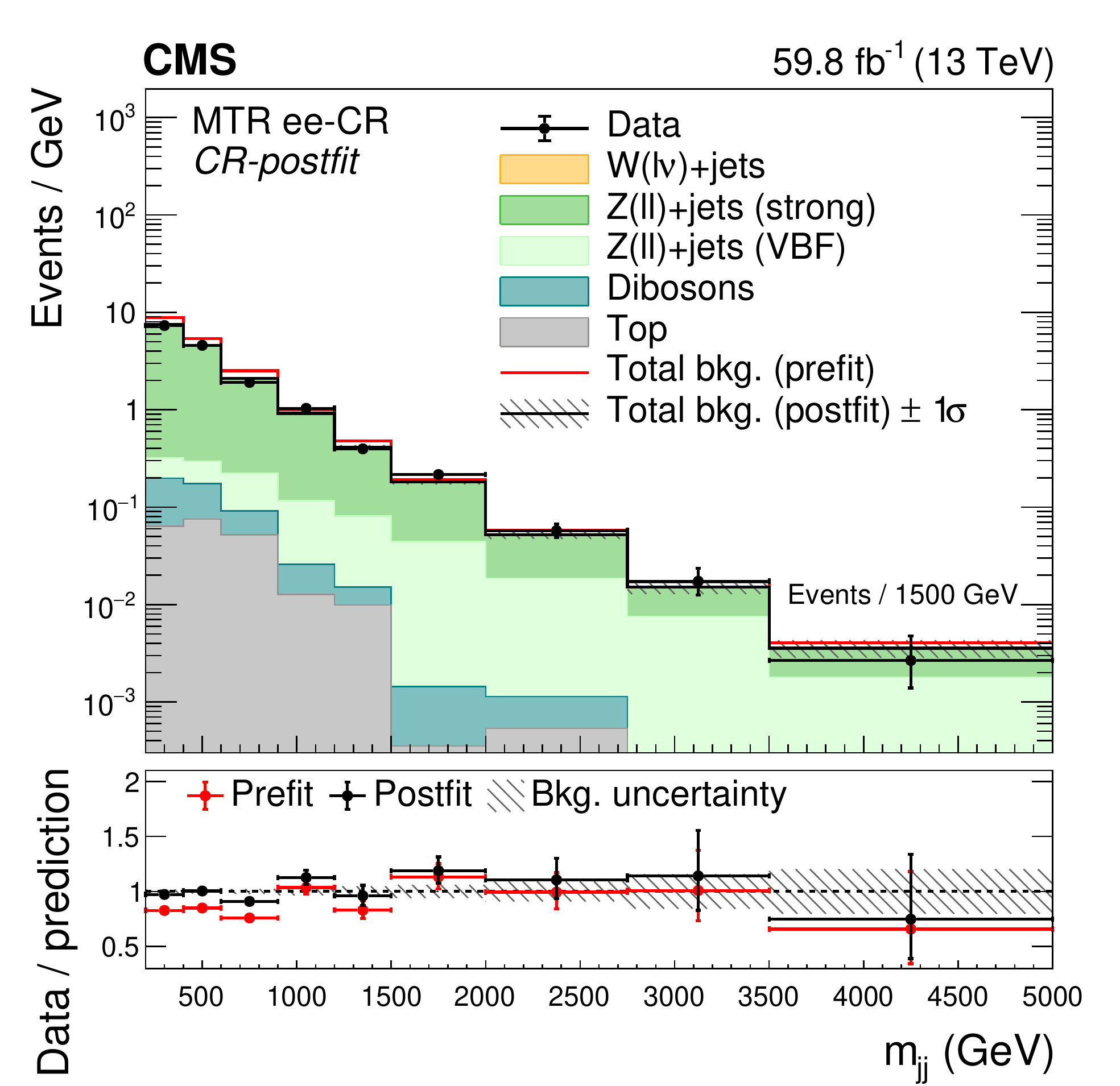}
    \includegraphics[width=0.48\textwidth]{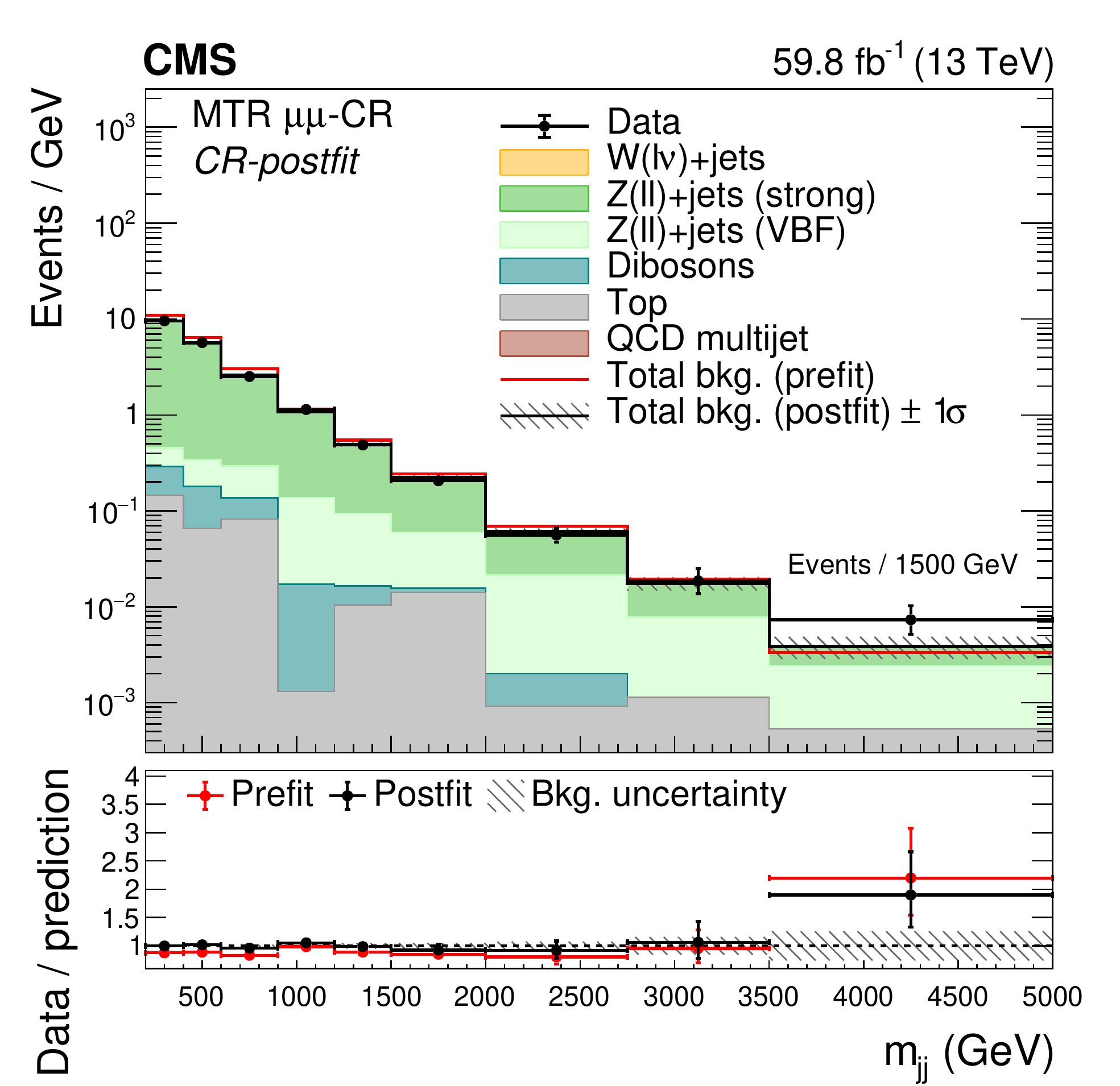}\\
    \includegraphics[width=0.48\textwidth]{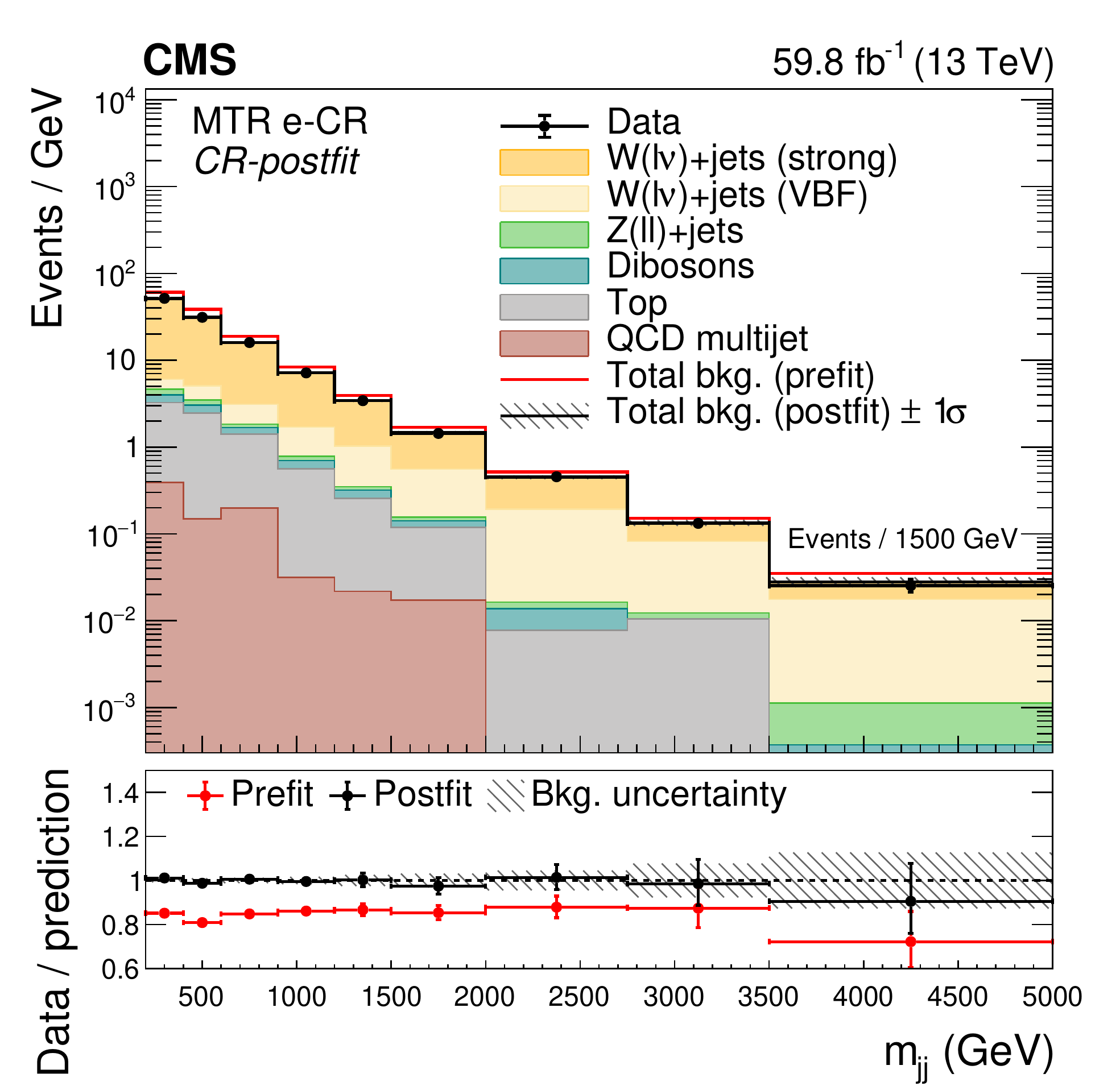}
    \includegraphics[width=0.48\textwidth]{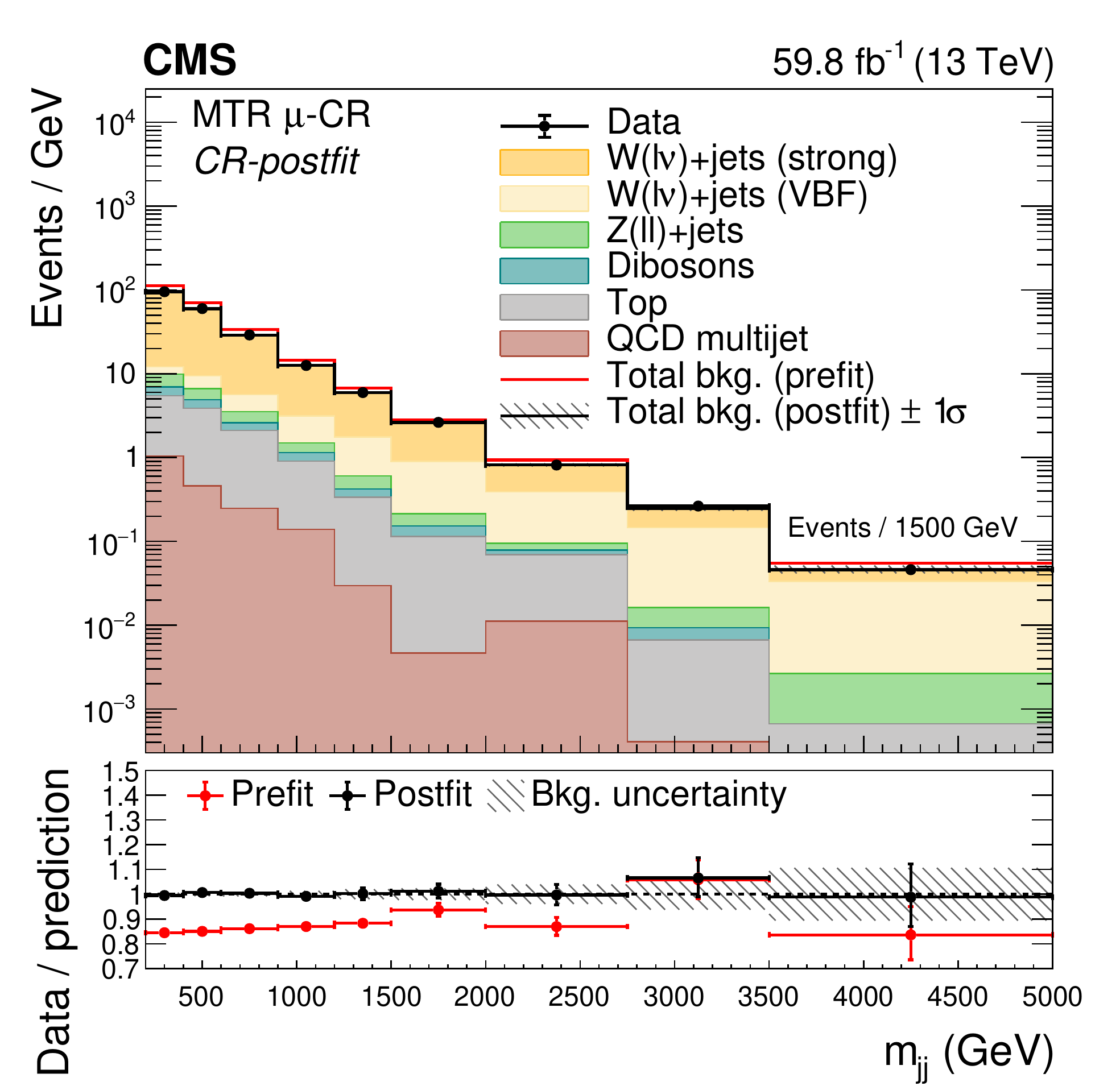}
    \caption{
        The \mjj distributions (prefit and CR-postfit) in the dielectron (upper
        left), dimuon (upper right), single-electron (lower left),
        and single-muon (lower right) CR for the MTR category, with the 2018
        sample.}
    \label{fig:CR_MTR_2018_CRonly}
\end{figure*}

\begin{figure*}[!htb]
    \centering
    \includegraphics[width=0.48\textwidth]{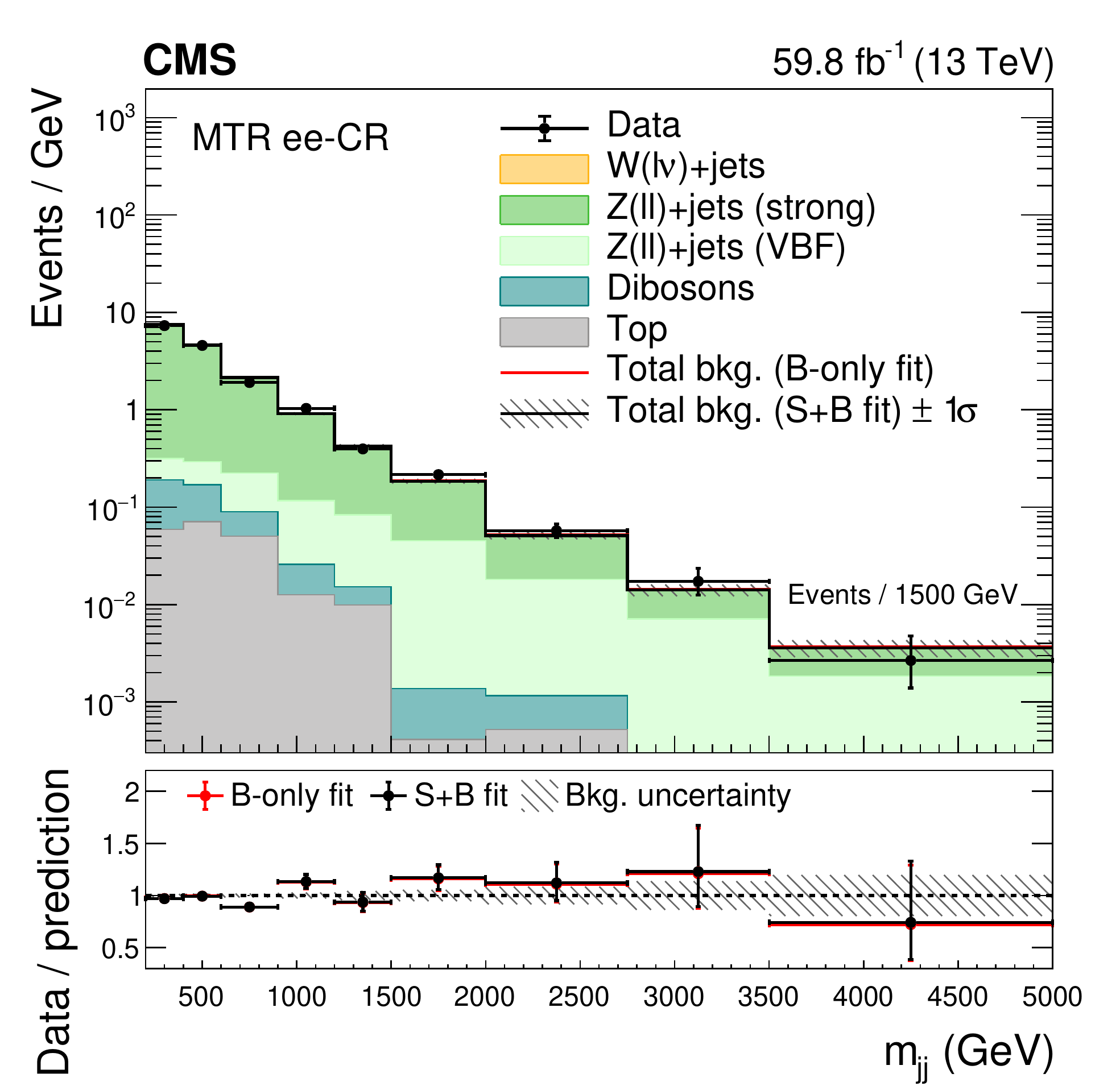}
    \includegraphics[width=0.48\textwidth]{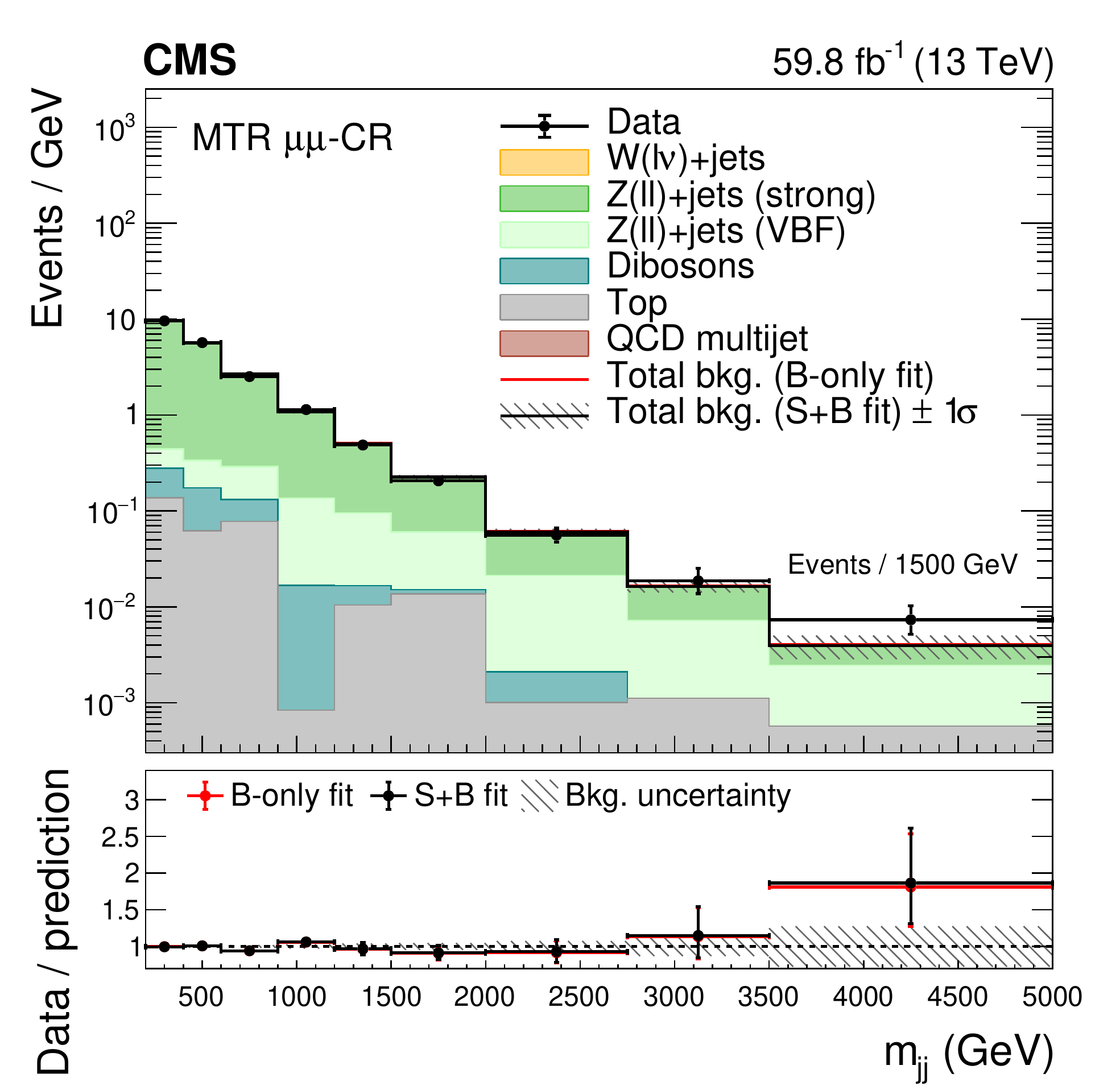}\\
    \includegraphics[width=0.48\textwidth]{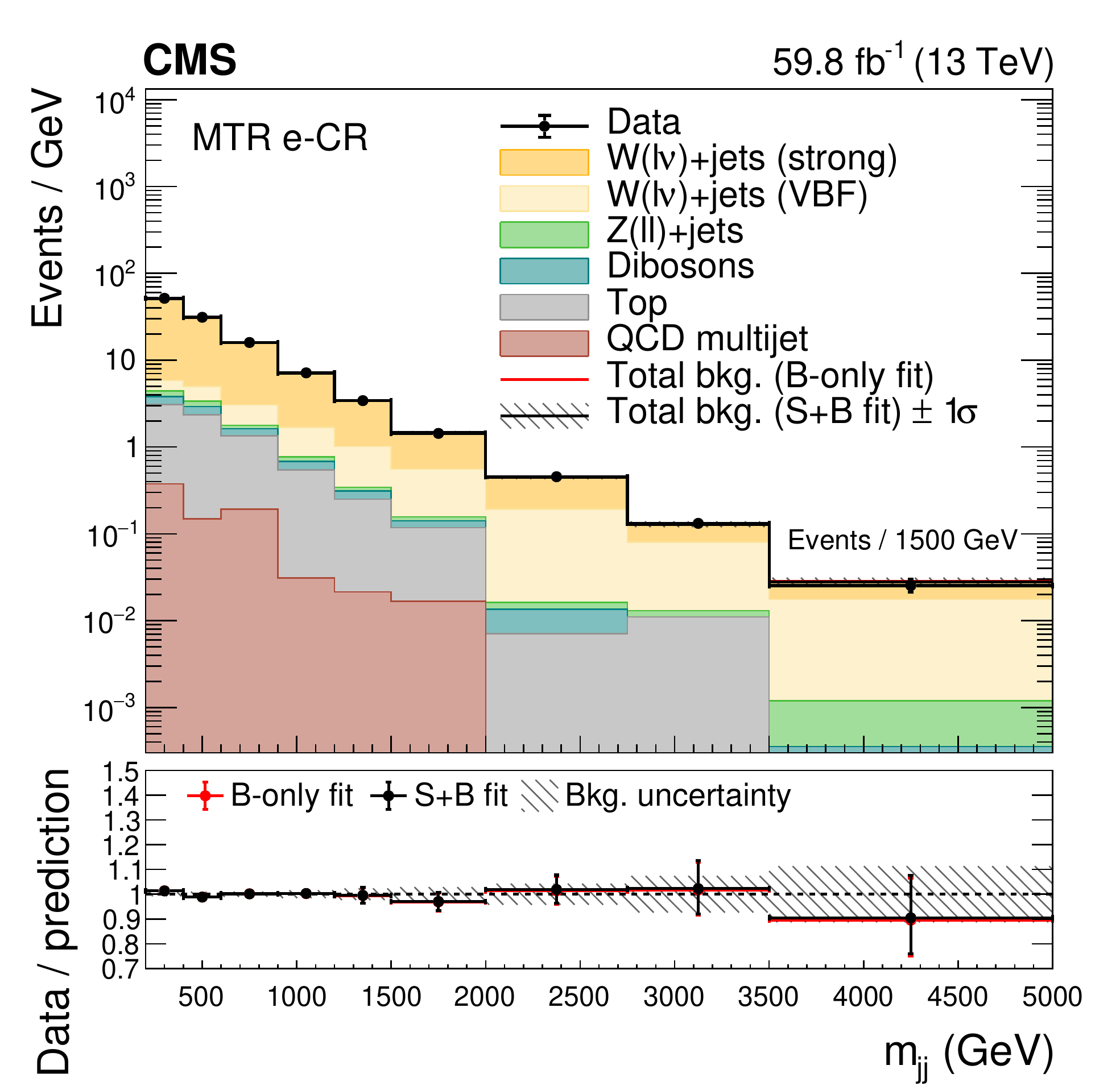}
    \includegraphics[width=0.48\textwidth]{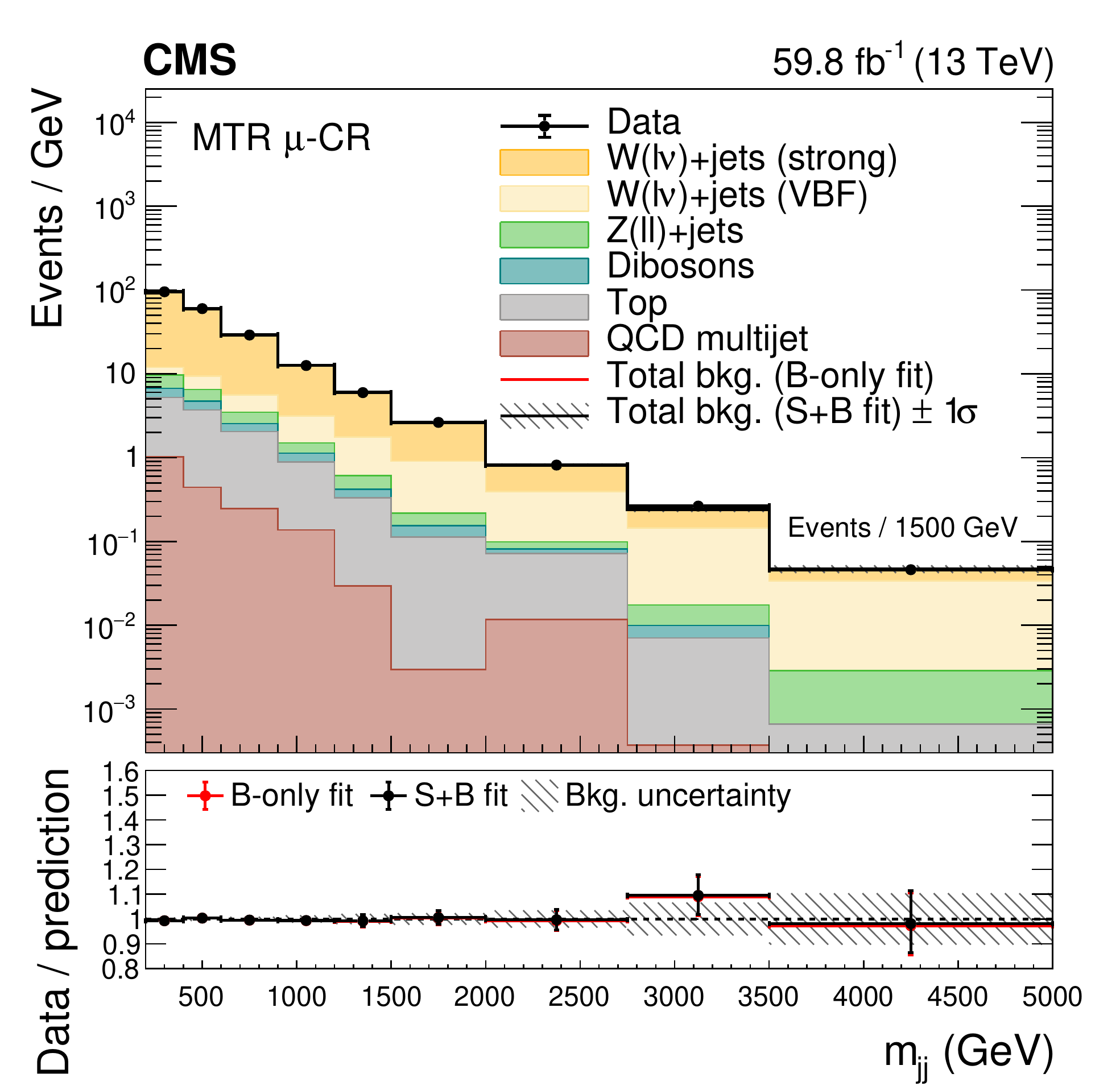}
    \caption{
        The \mjj postfit distributions in the dielectron (upper
        left), dimuon (upper right), single-electron (lower left),
        and single-muon (lower right) CR for the MTR category, with the 2018 sample. The background contributions are estimated from
        a fit to data in the SR and CRs allowing for the signal contribution to vary (S+B fit) and the total background estimated from a fit
        assuming $\brinv=0$ (B-only fit) is also shown.}
    \label{fig:CR_MTR_2018}
\end{figure*}

\begin{figure*}[htb]
    \centering
    \includegraphics[width=0.48\textwidth]{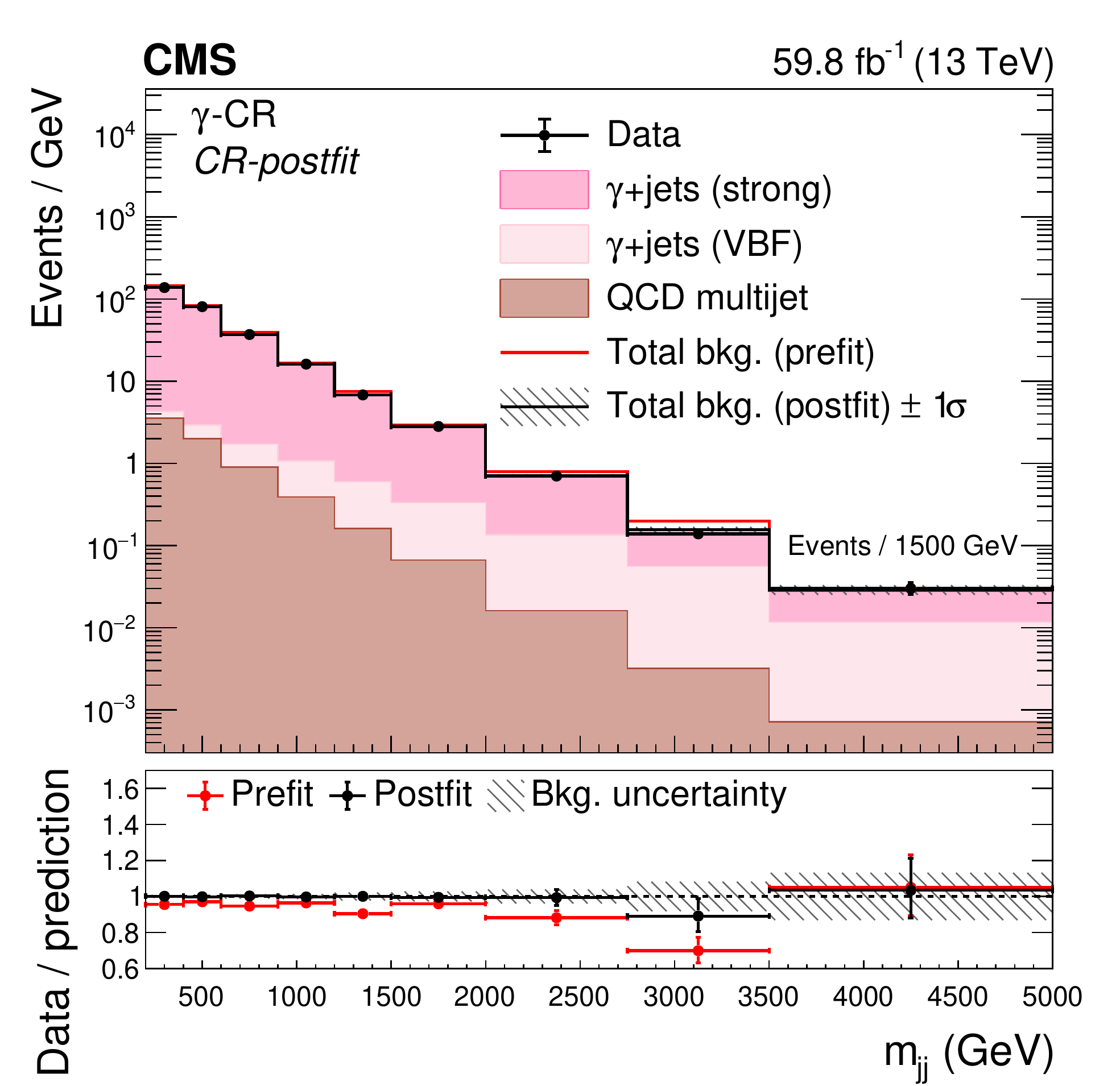}\hfill
    \includegraphics[width=0.48\textwidth]{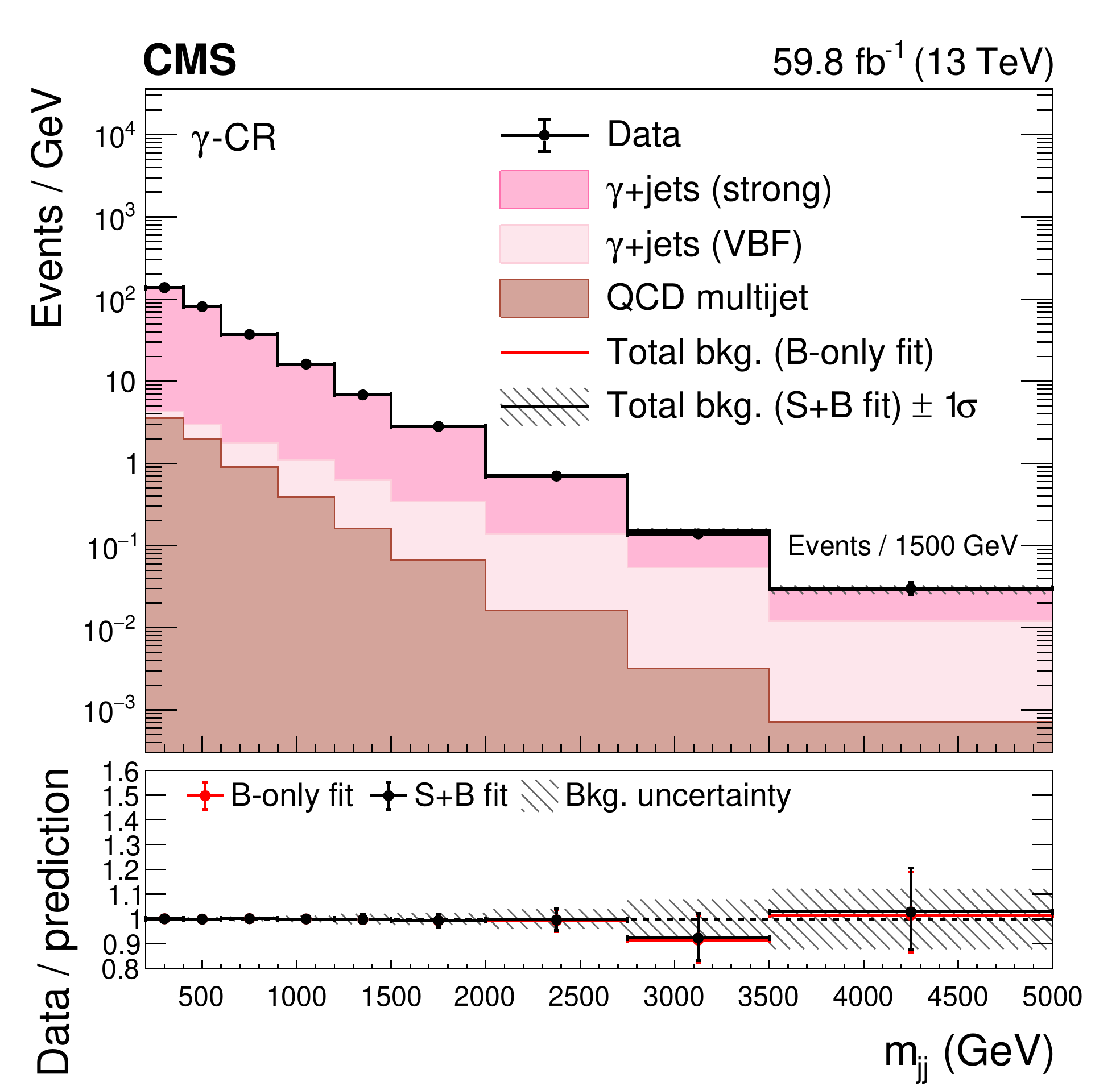}
    \caption{The \mjj CR-postfit (left) and postfit (right) distributions in the photon CR for the MTR category, with the 2018 sample.
        In the right figure, the total background estimated from a fit
        assuming $\brinv=0$ (B-only fit) is also shown.}
    \label{fig:Zgamma_MTR_2018}
\end{figure*}

\begin{figure*}[!htb]
    \centering
    \includegraphics[width=0.48\textwidth]{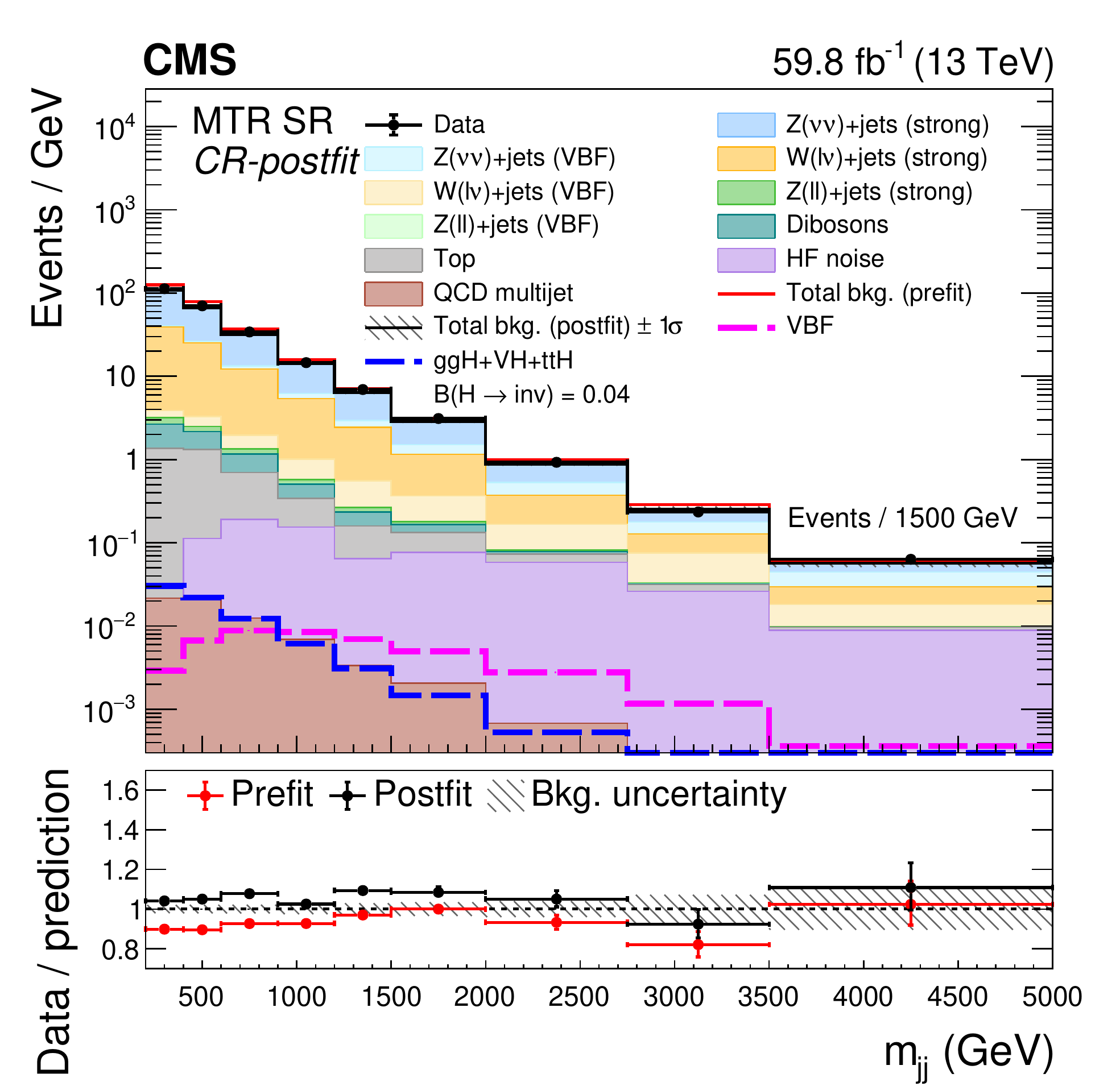}\hfill
    \includegraphics[width=0.48\textwidth]{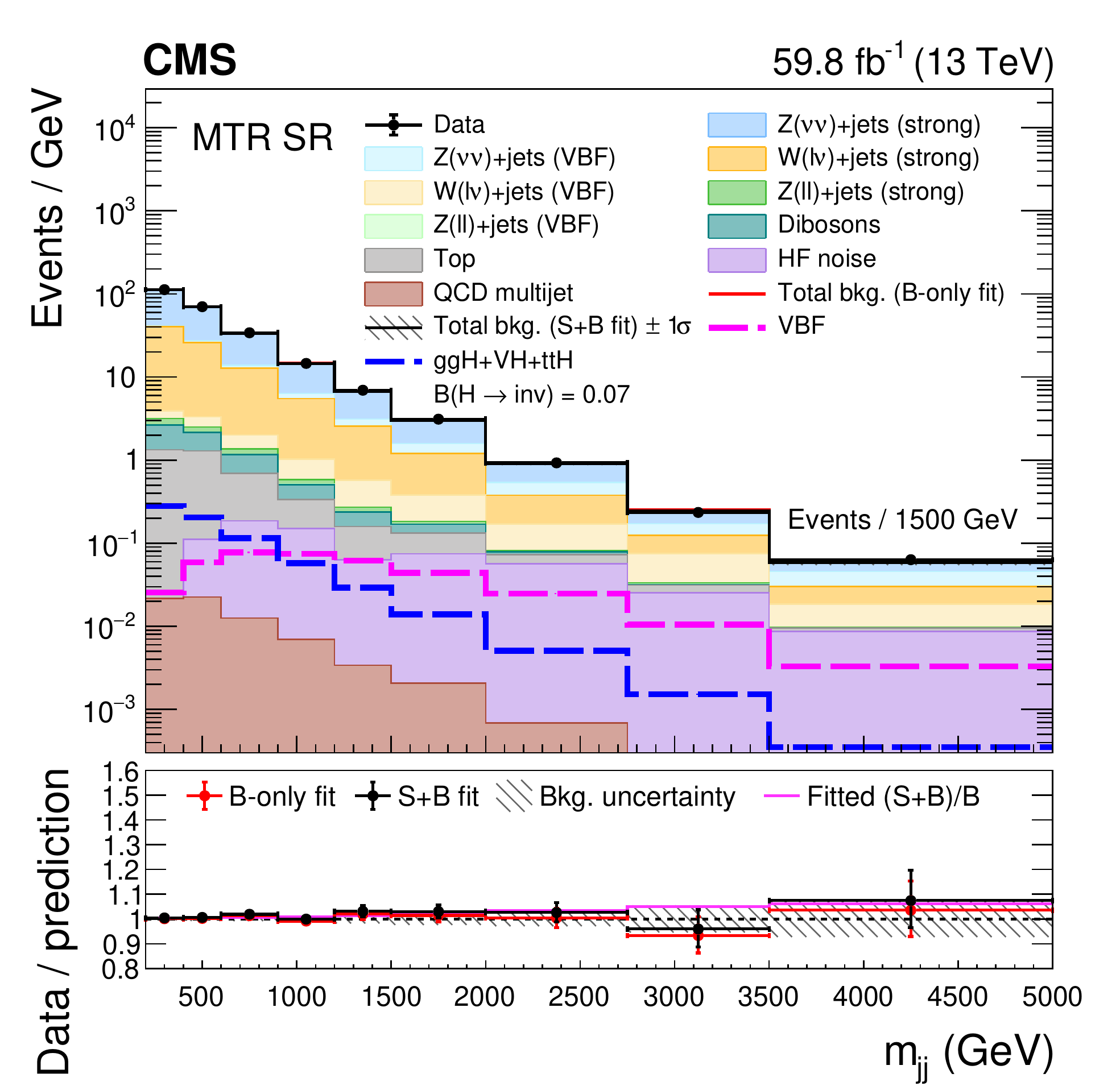}
    \caption{The observed \mjj distribution in the MTR prefit (left) and postfit (right) SR
        compared to the postfit backgrounds, with the 2018 samples. The signal processes are
        scaled by the fitted value of \brinv, shown in the legend.}
    \label{fig:SR_MTR_2018}
\end{figure*}

The results for the MTR 2018 category are shown in Figs.~\ref{fig:CR_MTR_2018_CRonly} to~\ref{fig:SR_MTR_2018} and Table~\ref{tab:yields_MTR_2018}.
 
\begin{table*}[htb!]
    \centering
    \topcaption[]{
        Expected event yields in each \mjj bin for the different background
        processes in the SR of the MTR category, in the 2018 samples. The
        background yields and the corresponding uncertainties are obtained
        after performing a combined fit across all of the CRs and SR. The
        expected signal contributions for the Higgs boson, produced in the non-\vbf
        and \vbf modes, decaying to invisible particles with a branching
        fraction of $\brinv = 1$, and the observed event yields are also
        reported.  }
    \label{tab:yields_MTR_2018}
    \cmsTable{
        \begin{tabular}{lccccccccc}
            \hline
            \mjj bin range (\GeVns{})                        & 200--400         & 400--600         & 600--900         & 900--1200       & 1200--1500      & 1500--2000      & 2000--2750     & 2750--3500    & $>$3500      \\
            \hline
            $\PZ(\PGn\PGn)+\text{jets}$ (strong)             & $14150.5\pm61.3$ & $8498.6\pm45.5$  & $5891.5\pm34.4$  & $2490.3\pm18.1$ & $1096.6\pm12.2$ & $730.4\pm9.4$   & $278.9\pm5.8$  & $55.9\pm2.3$  & $20.5\pm1.2$ \\
            $\PZ(\PGn\PGn)+\text{jets}$ (VBF)                & $228.8\pm4.5$    & $276.0\pm5.1$    & $348.3\pm5.6$    & $254.4\pm4.9$   & $167.7\pm4.3$   & $191.5\pm4.8$   & $123.3\pm4.4$  & $36.8\pm2.1$  & $24.3\pm1.9$ \\
            $\PW(\Pell\PGn)+\text{jets}$ (strong)            & $7323.4\pm51.4$  & $4566.3\pm38.3$  & $3243.7\pm29.5$  & $1357.2\pm17.3$ & $603.5\pm11.1$  & $418.9\pm8.3$   & $160.2\pm5.8$  & $38.3\pm3.0$  & $18.4\pm2.2$ \\
            $\PW(\Pell\PGn)+\text{jets}$ (VBF)               & $145.4\pm7.7$    & $163.6\pm8.4$    & $191.7\pm9.7$    & $133.5\pm6.6$   & $90.7\pm4.7$    & $98.1\pm5.2$    & $66.2\pm3.7$   & $31.9\pm2.6$  & $13.3\pm1.4$ \\
            $\ttbar$ + single \PQt quark                     & $261.3\pm13.9$   & $234.8\pm12.6$   & $151.5\pm8.7$    & $55.2\pm3.5$    & $28.9\pm1.9$    & $29.0\pm2.0$    & $12.2\pm1.2$   & $4.4\pm0.4$   & $1.4\pm0.2$  \\
            Diboson                                          & $264.0\pm31.4$   & $172.2\pm20.7$   & $144.5\pm17.5$   & $51.0\pm6.2$    & $23.6\pm3.0$    & $18.2\pm2.4$    & $4.6\pm0.6$    & $0.4\pm0.1$   & $0.0\pm0.1$  \\
            $\PZ/\Pgg^{*}(\Pell^{+}\Pell^{-})+\mathrm{jets}$ & $105.7\pm2.8$    & $71.4\pm2.0$     & $57.3\pm1.9$     & $22.6\pm0.8$    & $10.1\pm0.4$    & $7.6\pm0.5$     & $2.3\pm0.2$    & $0.9\pm0.1$   & $0.2\pm0.1$  \\
            Multijet                                         & $4.4\pm1.3$      & $4.5\pm1.3$      & $3.8\pm1.1$      & $2.1\pm0.6$     & $1.0\pm0.3$     & $1.0\pm0.3$     & $0.5\pm0.2$    & $0.2\pm0.1$   & $0.1\pm0.1$  \\
            HF noise                                         & $0.0\pm0.1$      & $18.5\pm2.2$     & $54.4\pm6.4$     & $45.2\pm5.3$    & $18.8\pm2.2$    & $38.0\pm4.4$    & $44.0\pm5.1$   & $19.8\pm2.3$  & $13.4\pm1.6$ \\ [\cmsTabSkip]

            $\Pg\Pg\PH(\to \mathrm{inv})$                    & 719.3            & 534.7            & 461.5            & 232.2           & 119.0           & 95.1            & 52.2           & 15.7          & 7.2          \\
            $\PQq\PQq\PH(\to \mathrm{inv})$                  & 74.3             & 171.3            & 340.3            & 327.8           & 269.2           & 321.6           & 271.1          & 115.3         & 72.4         \\
            $\PW\PH(\to \mathrm{inv})$                       & 39.6             & 22.9             & 12.6             & 5.2             & 1.9             & 1.1             & 0.3            & 0.2           & 0.1          \\
            $\PQq\PQq\PZ\PH(\to \mathrm{inv})$               & 18.4             & 8.5              & 3.9              & 1.9             & 0.4             & 0.6             & 0.2            & 0.0           & 0.0          \\
            $\Pg\Pg\PZ\PH(\to \mathrm{inv})$                 & 17.2             & 11.2             & 7.6              & 3.1             & 1.2             & 1.0             & 0.4            & 0.1           & 0.0          \\
            $\PQt\PQt\PH(\to \mathrm{inv})$                  & 4.7              & 3.4              & 2.6              & 1.1             & 0.5             & 0.4             & 0.1            & 0.0           & 0.0          \\ [\cmsTabSkip]

            Total bkg.                                       & $22483.4\pm87.5$ & $14005.9\pm65.1$ & $10086.6\pm51.1$ & $4411.4\pm27.8$ & $2040.8\pm18.1$ & $1532.7\pm15.4$ & $692.3\pm11.4$ & $188.6\pm5.6$ & $91.7\pm3.7$ \\ [\cmsTabSkip]

            Observed                                         & 22505            & 14036            & 10220            & 4374            & 2080            & 1555            & 695            & 176           & 95           \\
            \hline
        \end{tabular}
    }\end{table*}

\cmsClearpage
\subsection*{VTR 2017}
\begin{figure*}[!htb]
    \centering
    \includegraphics[width=0.48\textwidth]{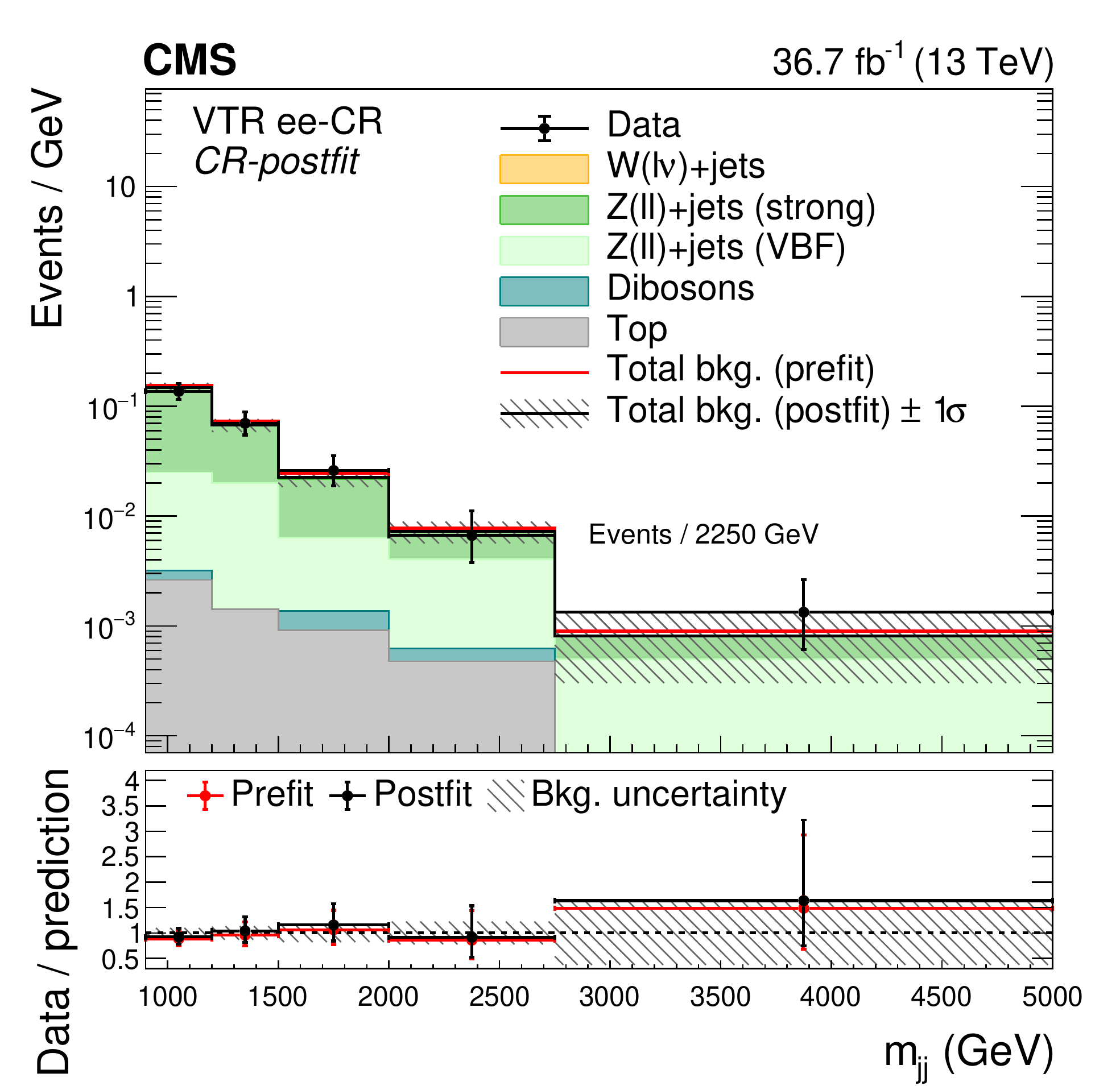}
    \includegraphics[width=0.48\textwidth]{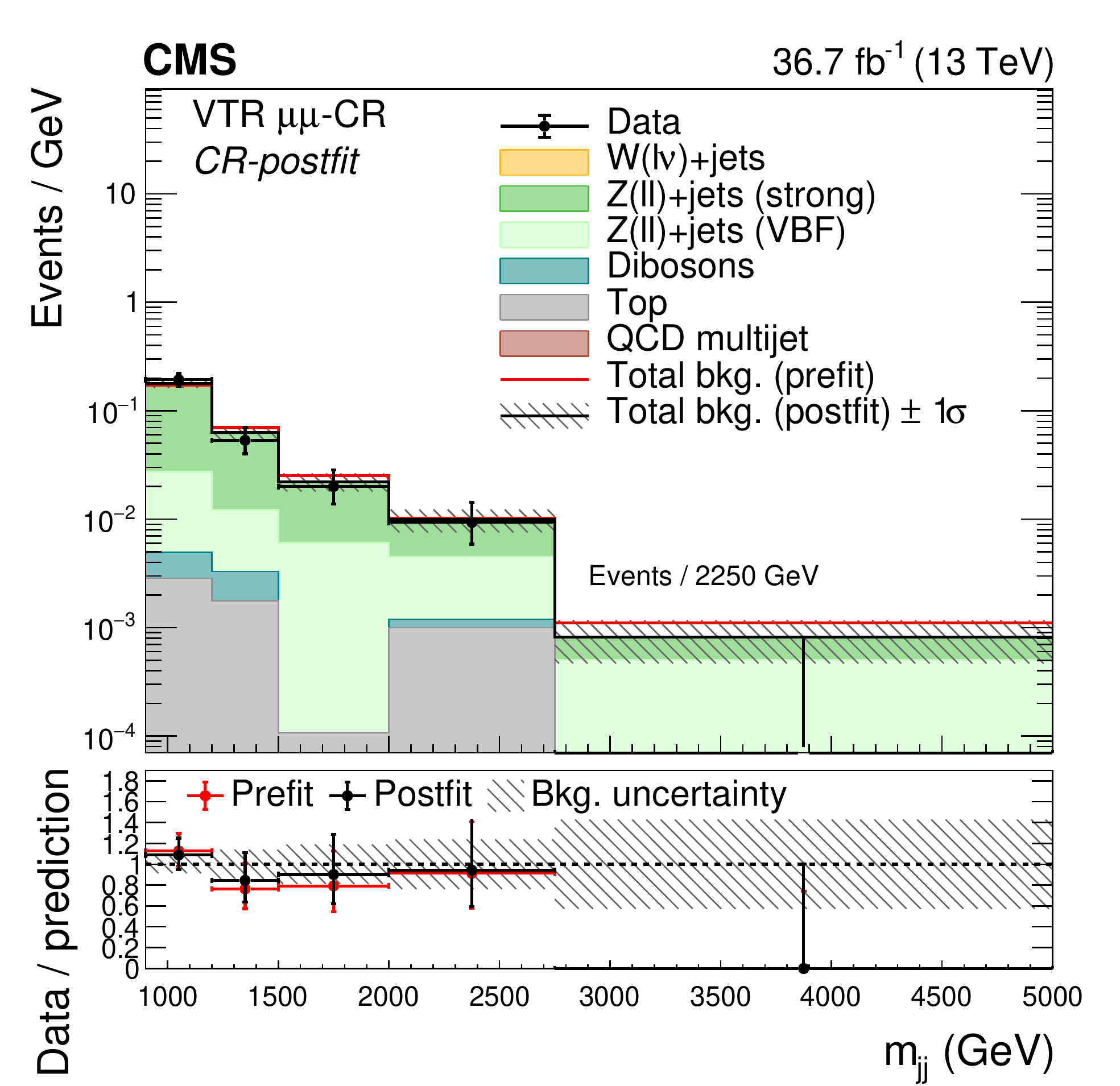}\\
    \includegraphics[width=0.48\textwidth]{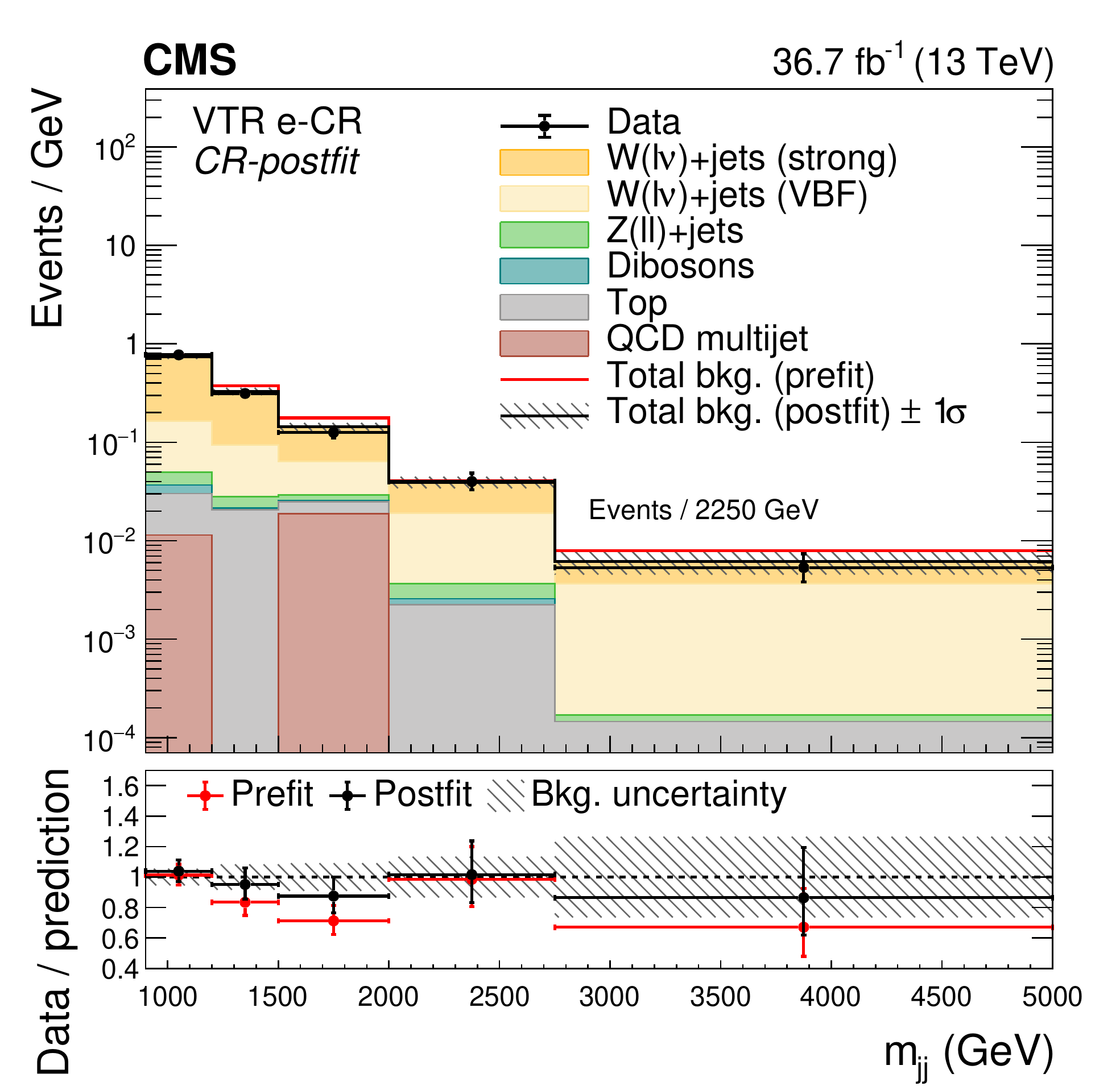}
    \includegraphics[width=0.48\textwidth]{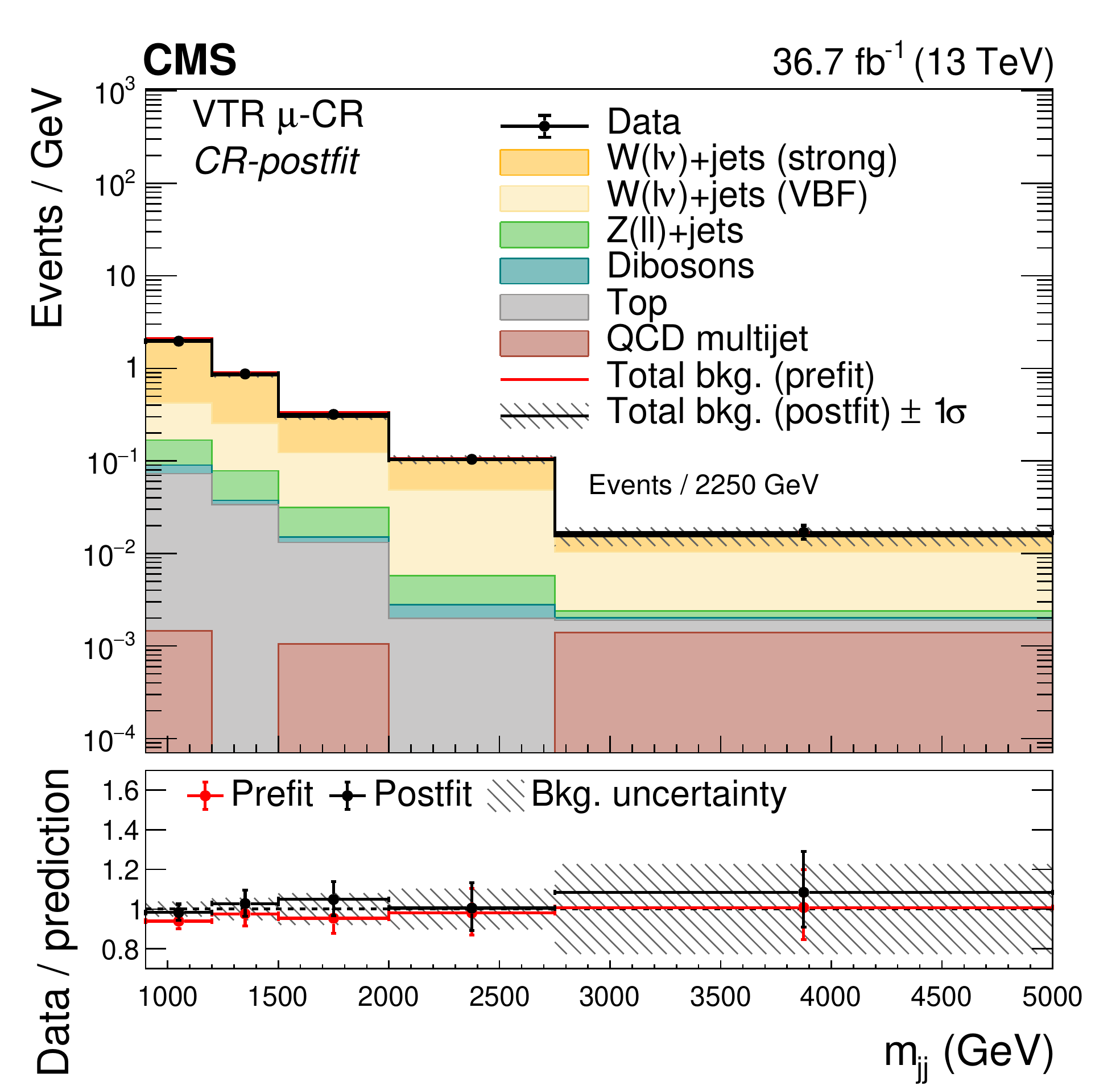}
    \caption{
        The \mjj distributions (prefit and CR-postfit) in the dielectron (upper
        left), dimuon (upper right), single-electron (lower left),
        and single-muon (lower right) CR for the VTR category, with the 2017
        sample.}
    \label{fig:CR_VTR_2017_CRonly}
\end{figure*}

\begin{figure*}[!htb]
    \centering
    \includegraphics[width=0.48\textwidth]{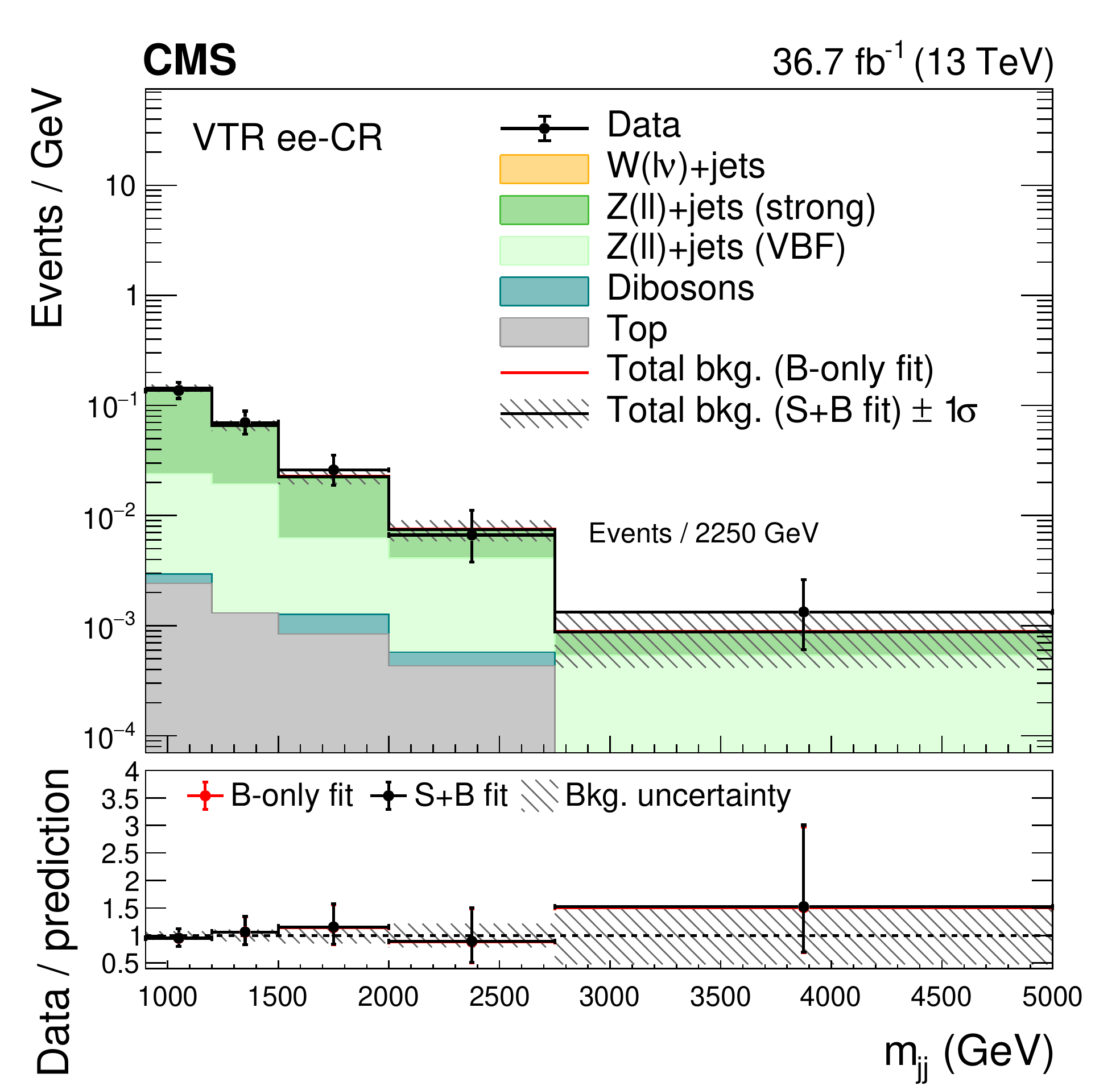}
    \includegraphics[width=0.48\textwidth]{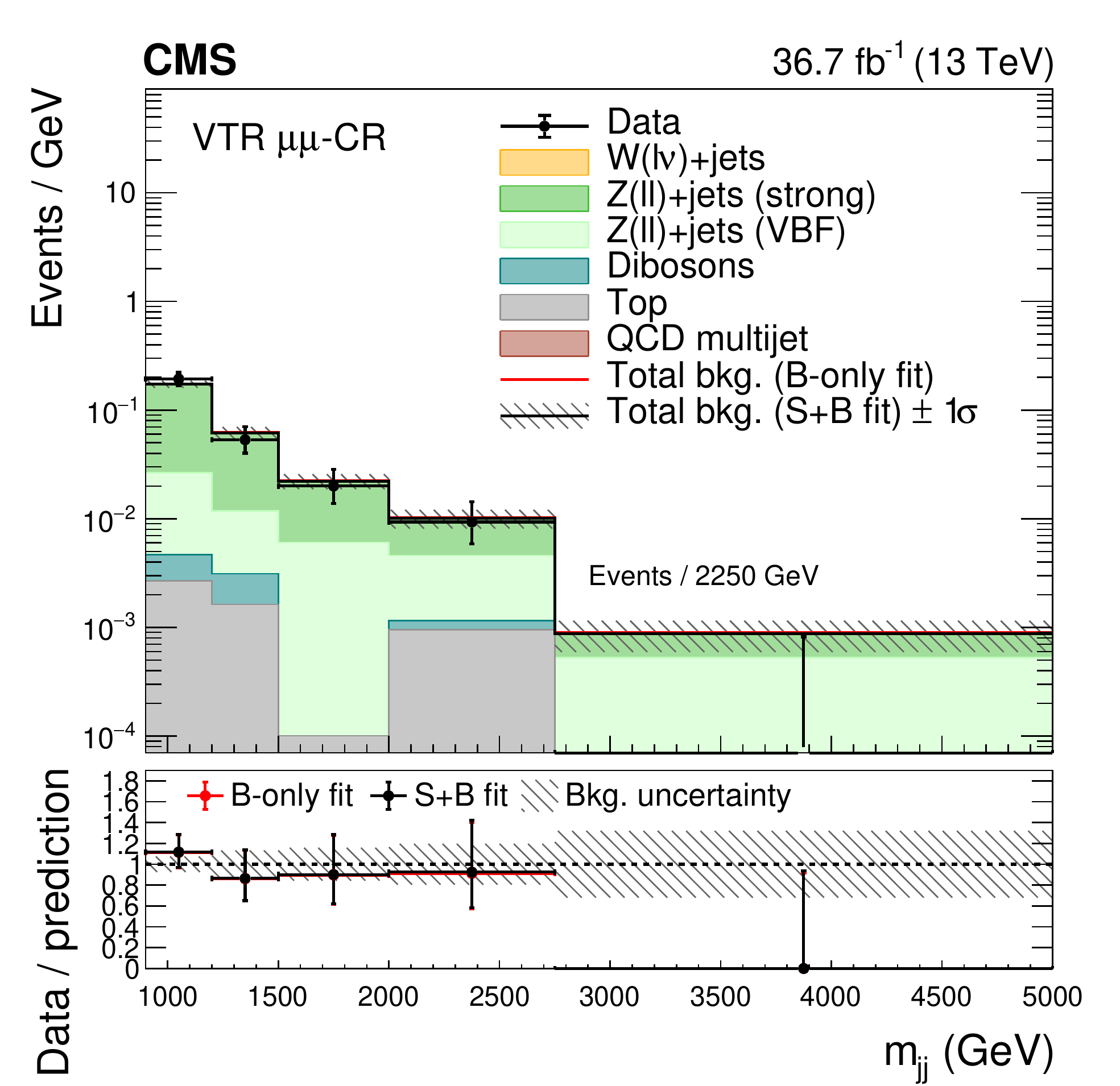}\\
    \includegraphics[width=0.48\textwidth]{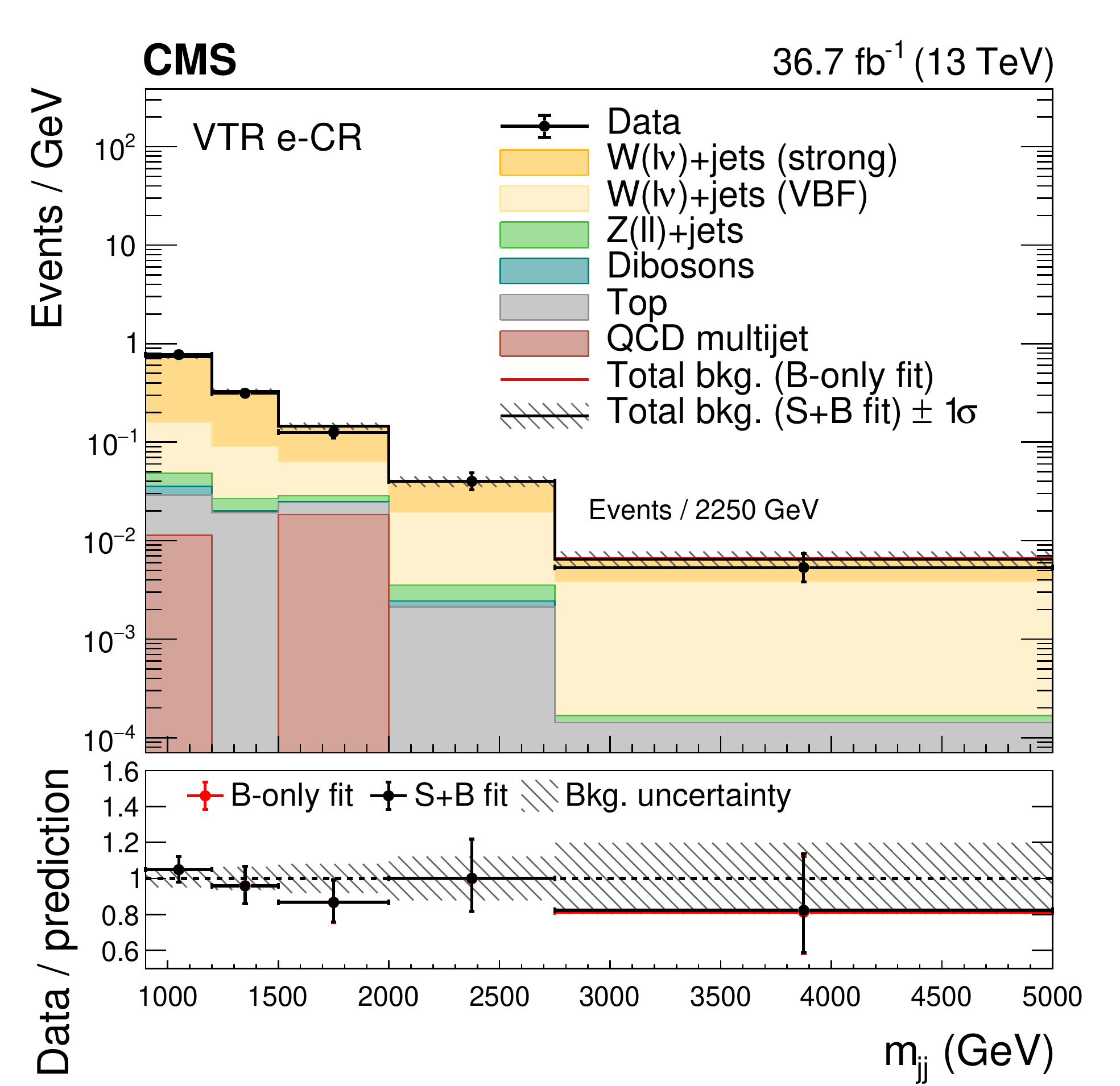}
    \includegraphics[width=0.48\textwidth]{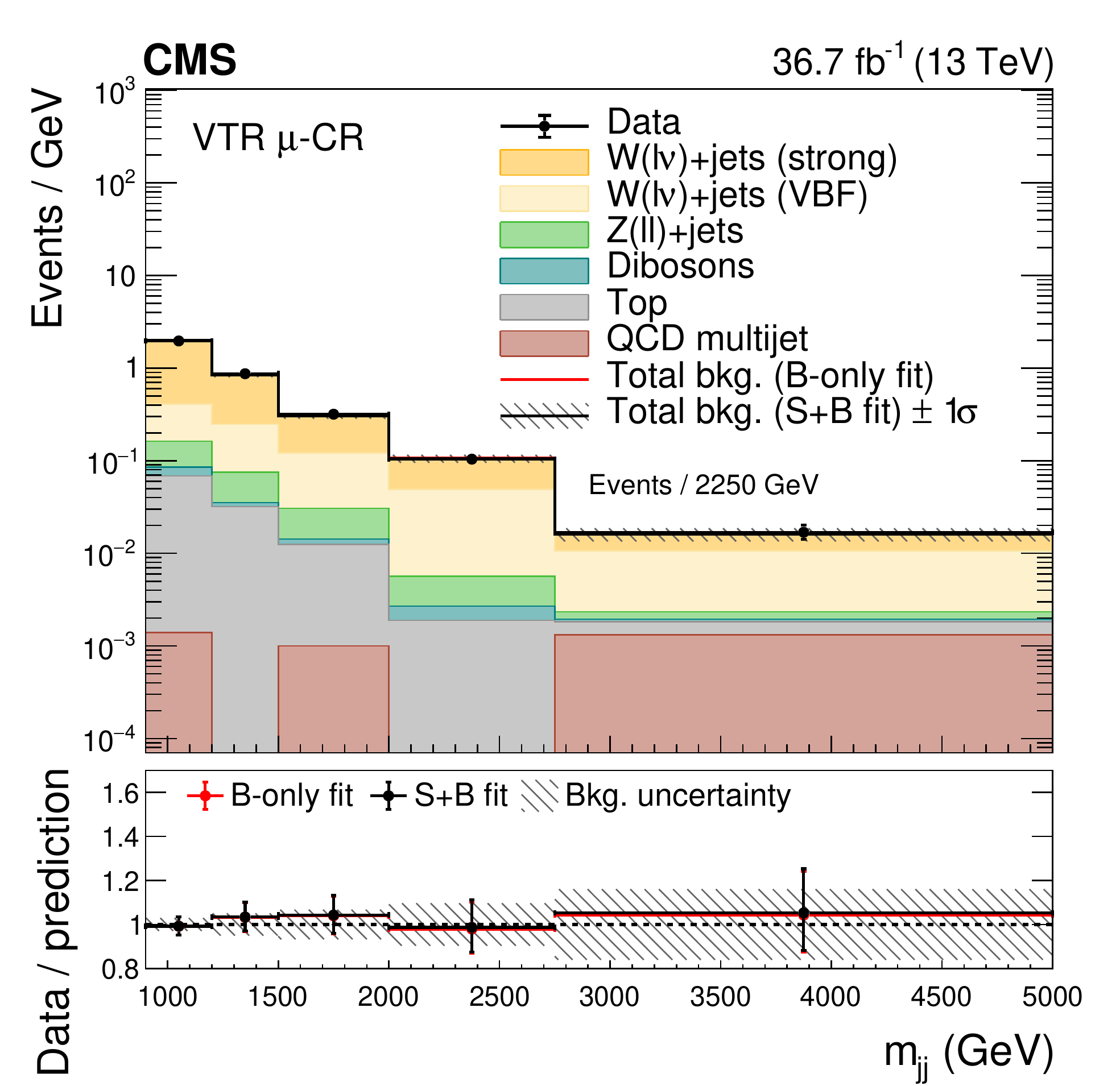}
    \caption{
        The \mjj postfit distributions in the dielectron (upper
        left), dimuon (upper right), single-electron (lower left),
        and single-muon (lower right) CR for the VTR category, with the 2017 sample. The background contributions are estimated from
        a fit to data in the SR and CRs allowing for the signal contribution to vary (S+B fit) and the total background estimated from a fit
        assuming $\brinv=0$ (B-only fit) is also shown.}
    \label{fig:CR_VTR_2017}
\end{figure*}

\begin{figure*}[!htb]
    \centering
    \includegraphics[width=0.48\textwidth]{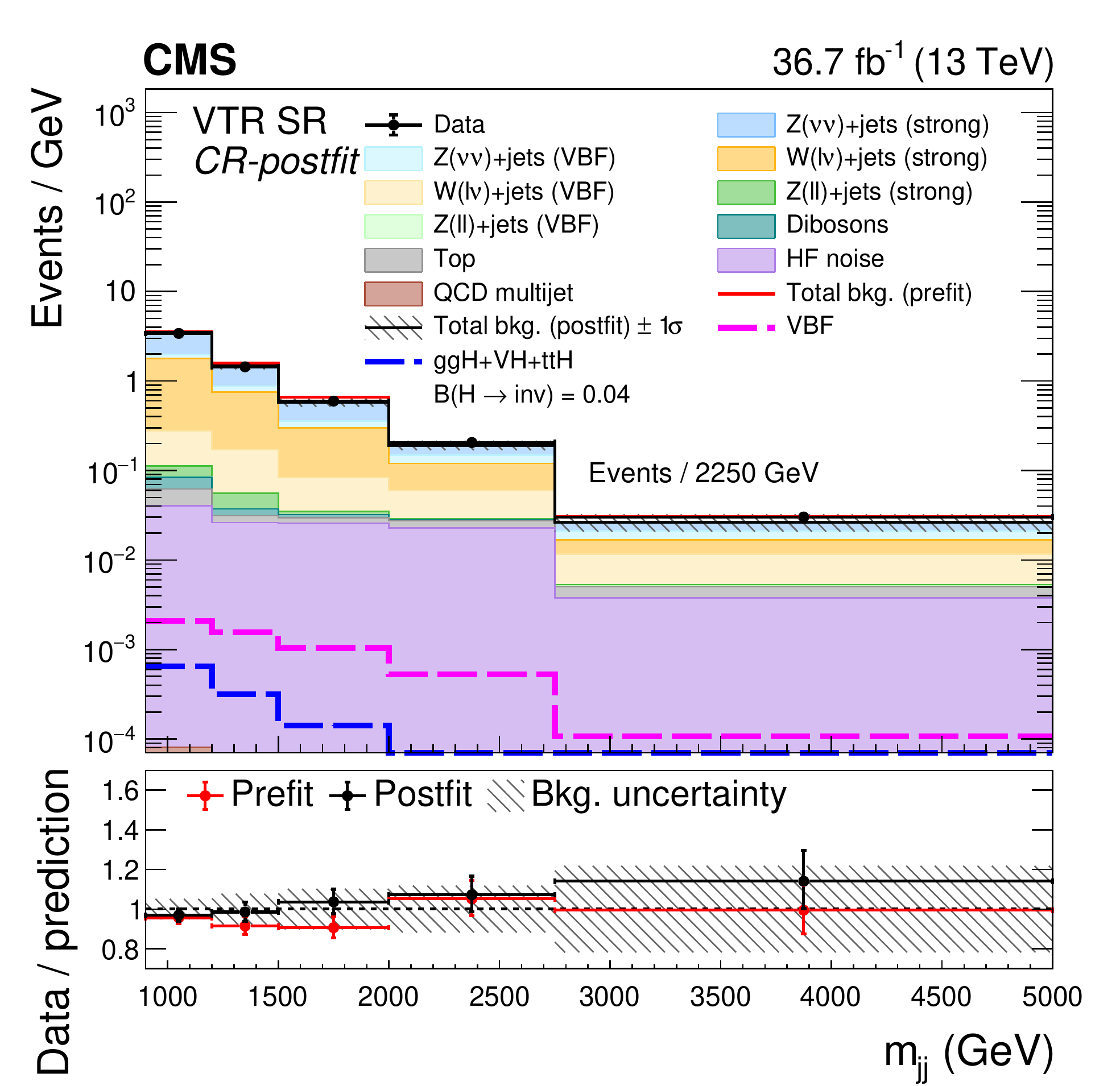}\hfill
    \includegraphics[width=0.48\textwidth]{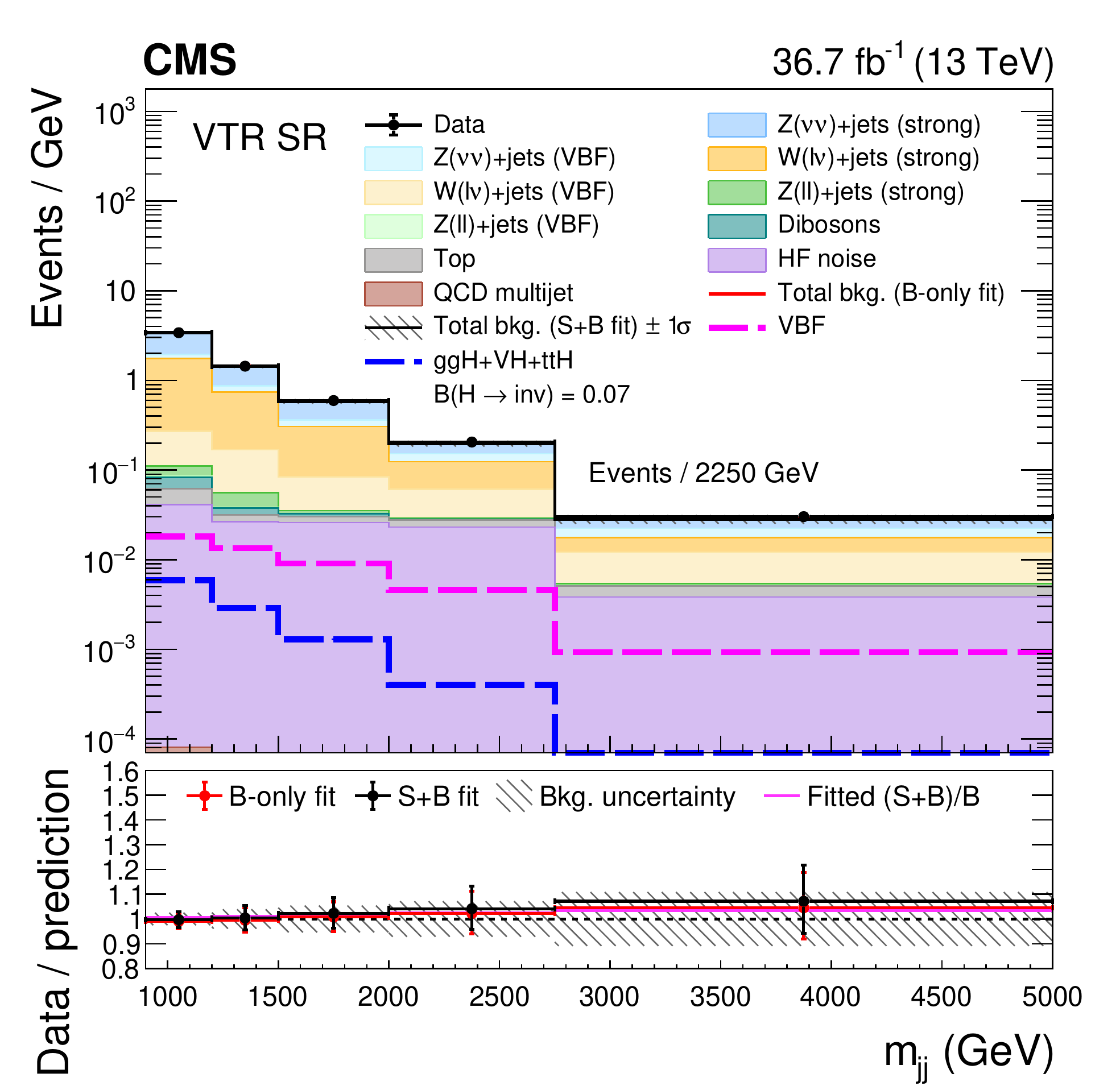}
    \caption{The observed \mjj distribution in the VTR prefit (left) and postfit (right) SR
        compared to the postfit backgrounds, with the 2017 samples. The signal processes are
        scaled by the fitted value of \brinv, shown in the legend.}
    \label{fig:SR_VTR_2017}
\end{figure*}

The results for the VTR 2017 category are shown in Figs.~\ref{fig:CR_VTR_2017_CRonly} to~\ref{fig:SR_VTR_2017} and Table~\ref{tab:yields_VTR_2017}.

\begin{table*}[htb!]
    \centering
    \topcaption[]{
        Expected event yields in each \mjj bin for the different background
        processes in the SR of the VTR category, in the 2017 samples. The
        background yields and the corresponding uncertainties are obtained
        after performing a combined fit across all of the CRs and SR. The
        expected signal contributions for the Higgs boson, produced in the non-\vbf
        and \vbf modes, decaying to invisible particles with a branching
        fraction of $\brinv = 1$, and the observed event yields are also
        reported.  }
    \label{tab:yields_VTR_2017}
        \begin{tabular}{lccccc}
            \hline
            \mjj bin range (\GeVns{})                        & 900--1200       & 1200--1500     & 1500--2000    & 2000--2750    & $>$2750      \\
            \hline
            $\PZ(\PGn\PGn)+\text{jets}$ (strong)             & $442.9\pm9.4$   & $172.1\pm5.3$  & $112.4\pm4.2$ & $34.5\pm2.1$  & $13.5\pm1.2$ \\
            $\PZ(\PGn\PGn)+\text{jets}$ (VBF)                & $58.2\pm2.8$    & $37.3\pm2.2$   & $28.3\pm2.0$  & $22.0\pm2.0$  & $11.2\pm1.5$ \\
            $\PW(\Pell\PGn)+\text{jets}$ (strong)            & $446.9\pm13.1$  & $174.1\pm8.0$  & $112.6\pm6.8$ & $47.9\pm5.0$  & $12.5\pm3.0$ \\
            $\PW(\Pell\PGn)+\text{jets}$ (VBF)               & $48.0\pm3.8$    & $33.9\pm3.0$   & $24.5\pm2.6$  & $24.3\pm2.8$  & $15.7\pm2.2$ \\
            $\ttbar$ + single \PQt quark                     & $6.3\pm0.5$     & $1.5\pm0.2$    & $2.1\pm0.2$   & $3.5\pm0.4$   & $2.9\pm0.8$  \\
            Diboson                                          & $6.3\pm0.8$     & $1.8\pm0.2$    & $1.2\pm0.2$   & $0.2\pm0.1$   & $0.0\pm0.1$  \\
            $\PZ/\Pgg^{*}(\Pell^{+}\Pell^{-})+\mathrm{jets}$ & $8.6\pm0.4$     & $5.5\pm0.3$    & $1.4\pm0.1$   & $0.8\pm0.1$   & $0.6\pm0.1$  \\
            Multijet                                         & $0.0\pm0.1$     & $0.0\pm0.1$    & $0.0\pm0.1$   & $0.0\pm0.1$   & $0.0\pm0.1$  \\
            HF noise                                         & $12.4\pm1.7$    & $8.0\pm1.1$    & $13.1\pm1.8$  & $17.5\pm2.4$  & $8.7\pm1.2$  \\ [\cmsTabSkip]

            $\Pg\Pg\PH(\to \mathrm{inv})$                    & 24.3            & 12.0           & 9.0           & 4.2           & 1.7          \\
            $\PQq\PQq\PH(\to \mathrm{inv})$                  & 79.6            & 59.3           & 66.5          & 50.7          & 30.7         \\
            $\PW\PH(\to \mathrm{inv})$                       & 0.2             & 0.0            & 0.0           & 0.0           & 0.0          \\
            $\PQq\PQq\PZ\PH(\to \mathrm{inv})$               & 0.1             & 0.0            & 0.0           & 0.0           & 0.0          \\
            $\Pg\Pg\PZ\PH(\to \mathrm{inv})$                 & 0.2             & 0.1            & 0.1           & 0.0           & 0.0          \\
            $\PQt\PQt\PH(\to \mathrm{inv})$                  & 0.0             & 0.0            & 0.0           & 0.0           & 0.0          \\ [\cmsTabSkip]

            Total bkg.                                       & $1029.6\pm16.9$ & $434.3\pm10.4$ & $295.7\pm8.8$ & $150.7\pm6.9$ & $65.1\pm4.5$ \\ [\cmsTabSkip]

            Observed                                         & 1020            & 432            & 298           & 154           & 68           \\
            \hline
        \end{tabular}
\end{table*}

\cmsClearpage
\subsection*{VTR 2018}

\begin{figure*}[!htb]
    \centering
    \includegraphics[width=0.48\textwidth]{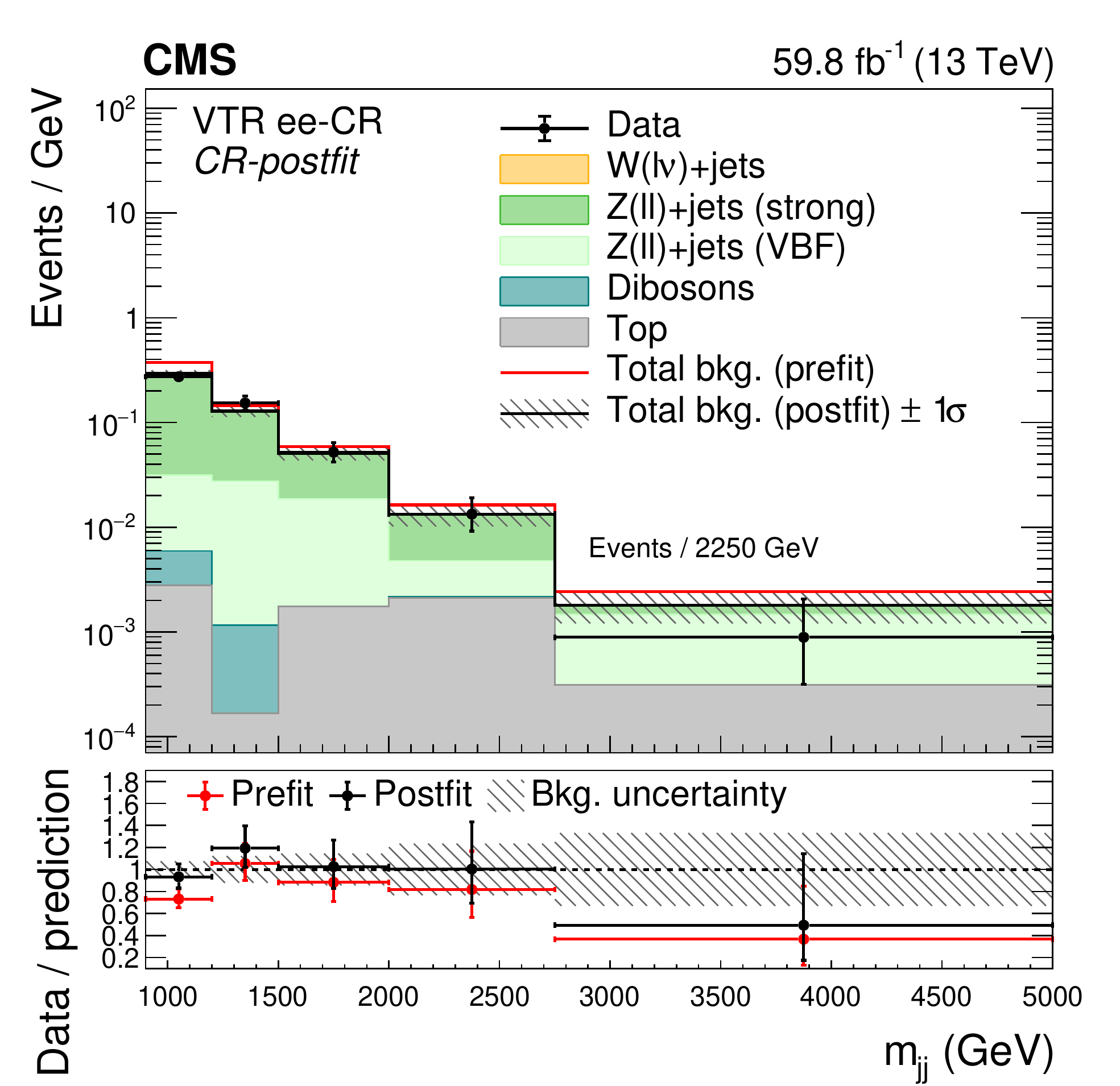}
    \includegraphics[width=0.48\textwidth]{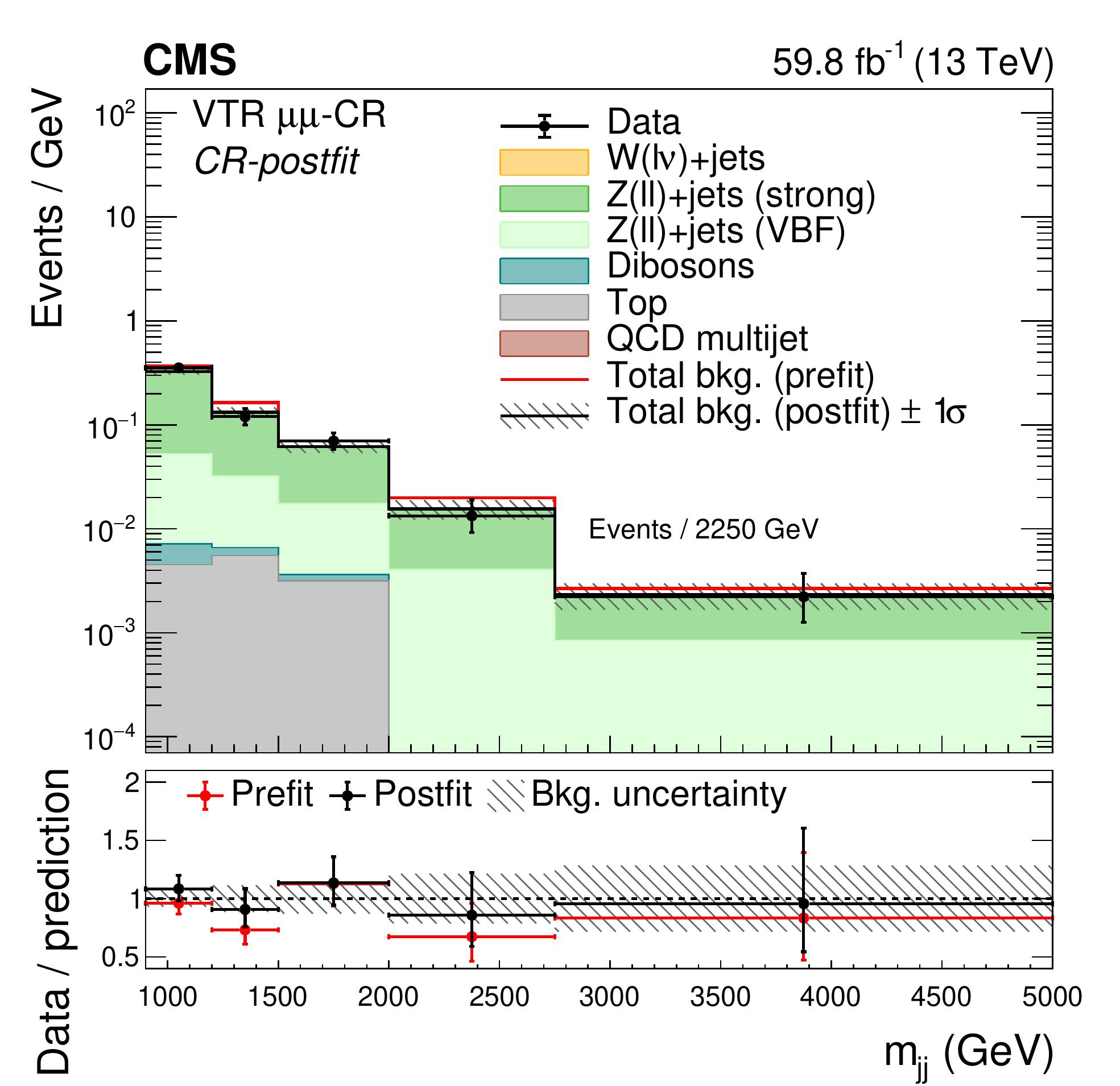}\\
    \includegraphics[width=0.48\textwidth]{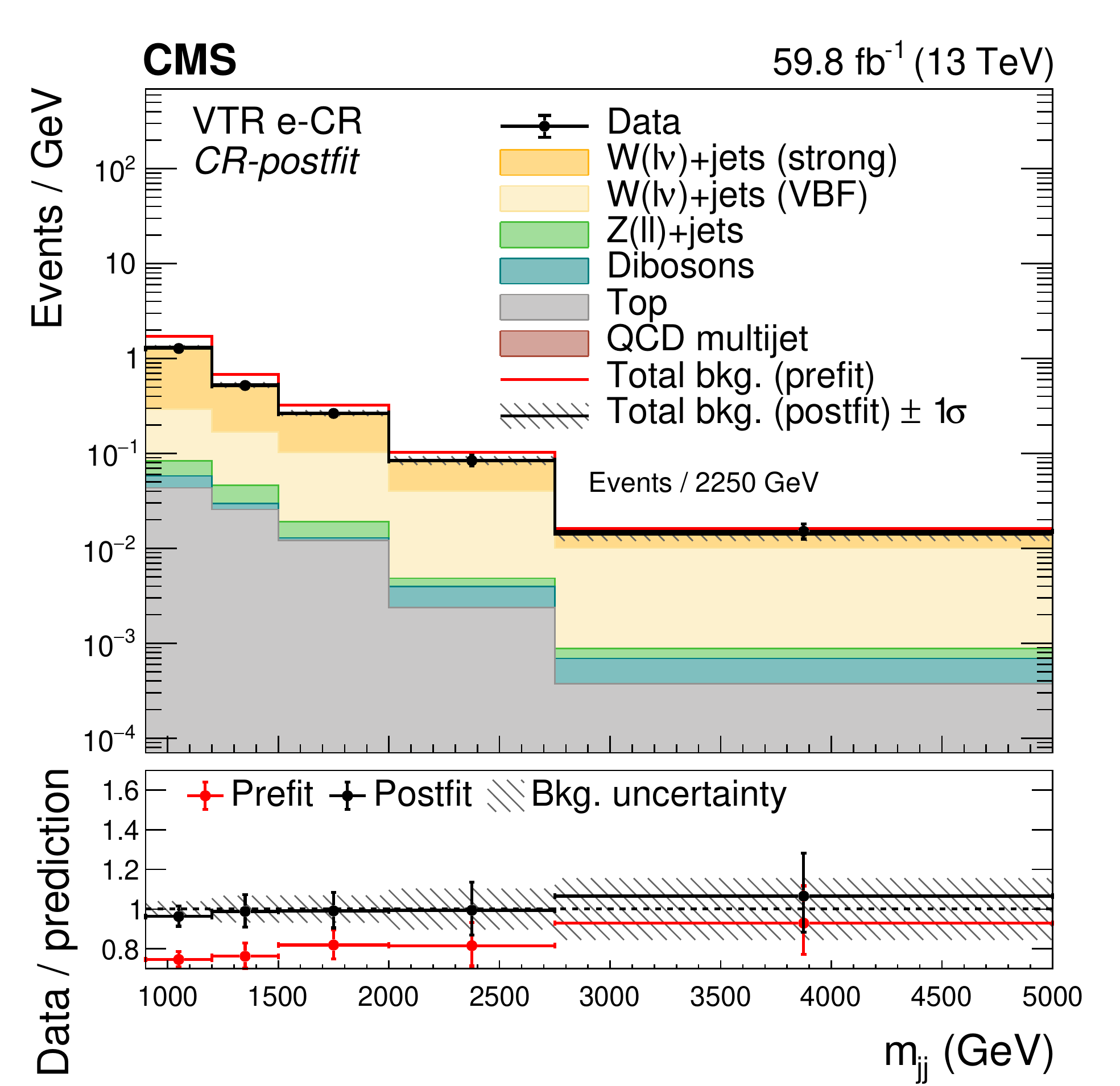}
    \includegraphics[width=0.48\textwidth]{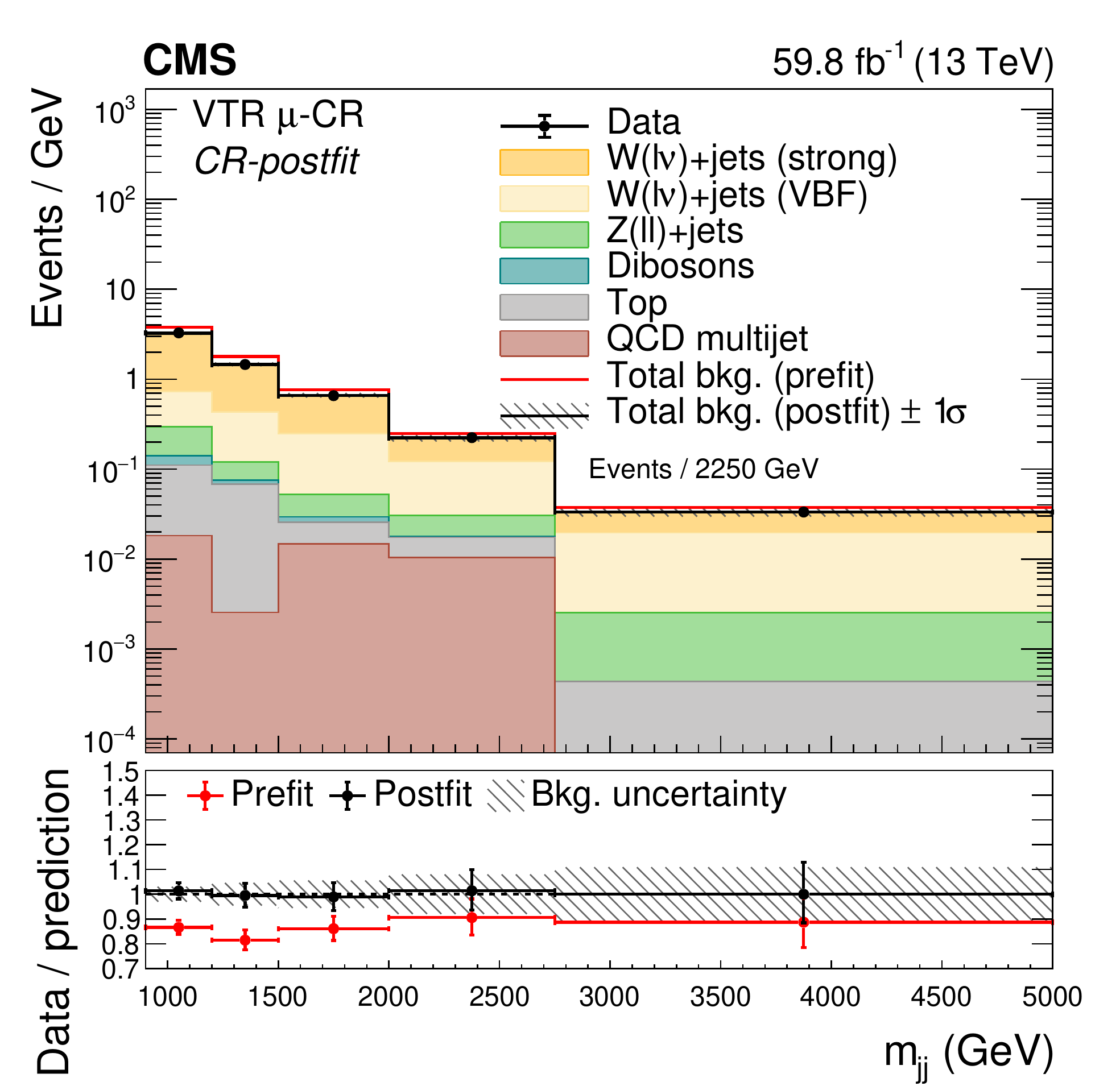}
    \caption{
        The \mjj distributions (prefit and CR-postfit) in the dielectron (upper
        left), dimuon (upper right), single-electron (lower left),
        and single-muon (lower right) CR for the VTR category, with the 2018
        sample.}
    \label{fig:CR_VTR_2018_CRonly}
\end{figure*}

\begin{figure*}[!htb]
    \centering
    \includegraphics[width=0.48\textwidth]{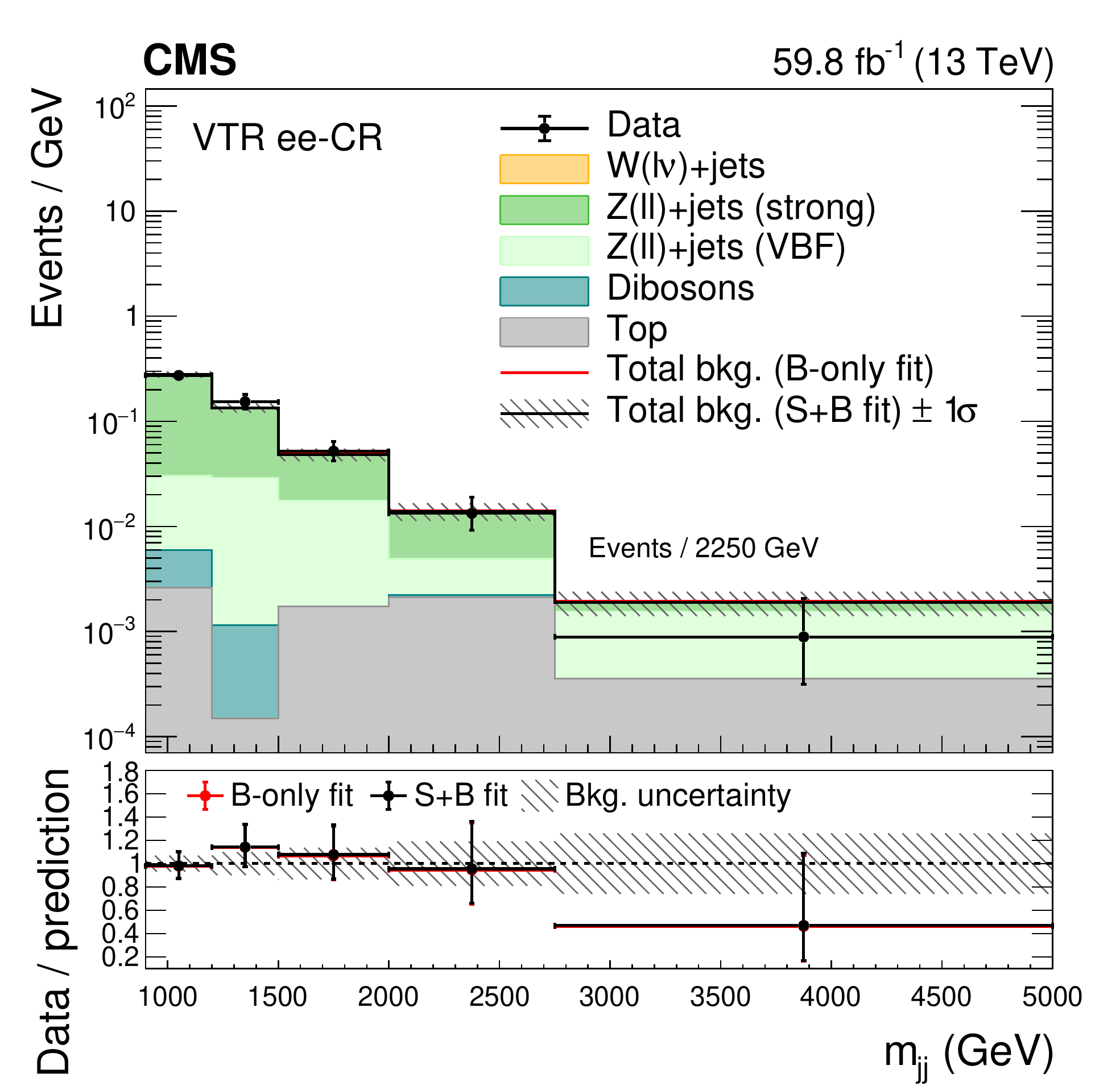}
    \includegraphics[width=0.48\textwidth]{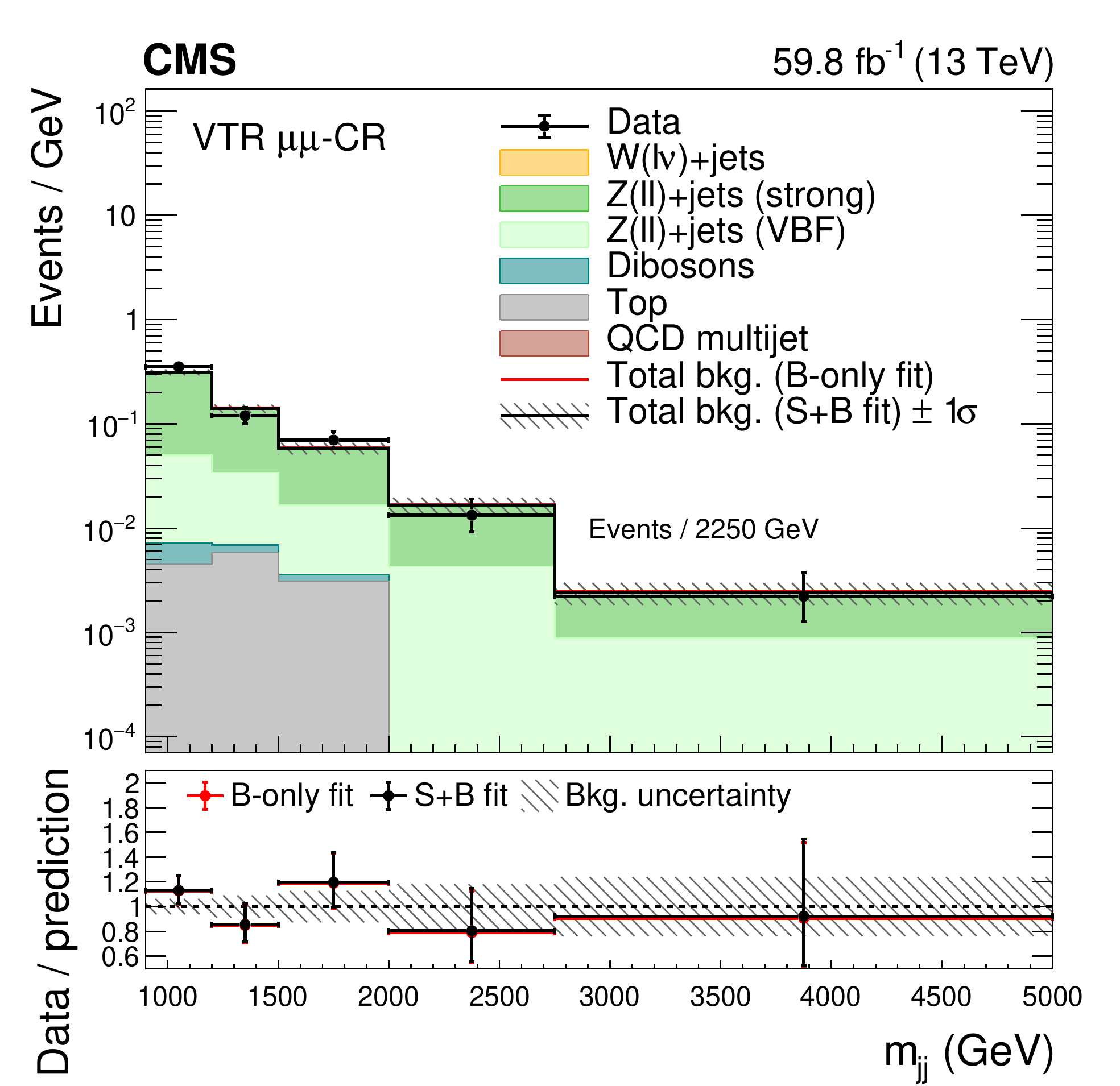}\\
    \includegraphics[width=0.48\textwidth]{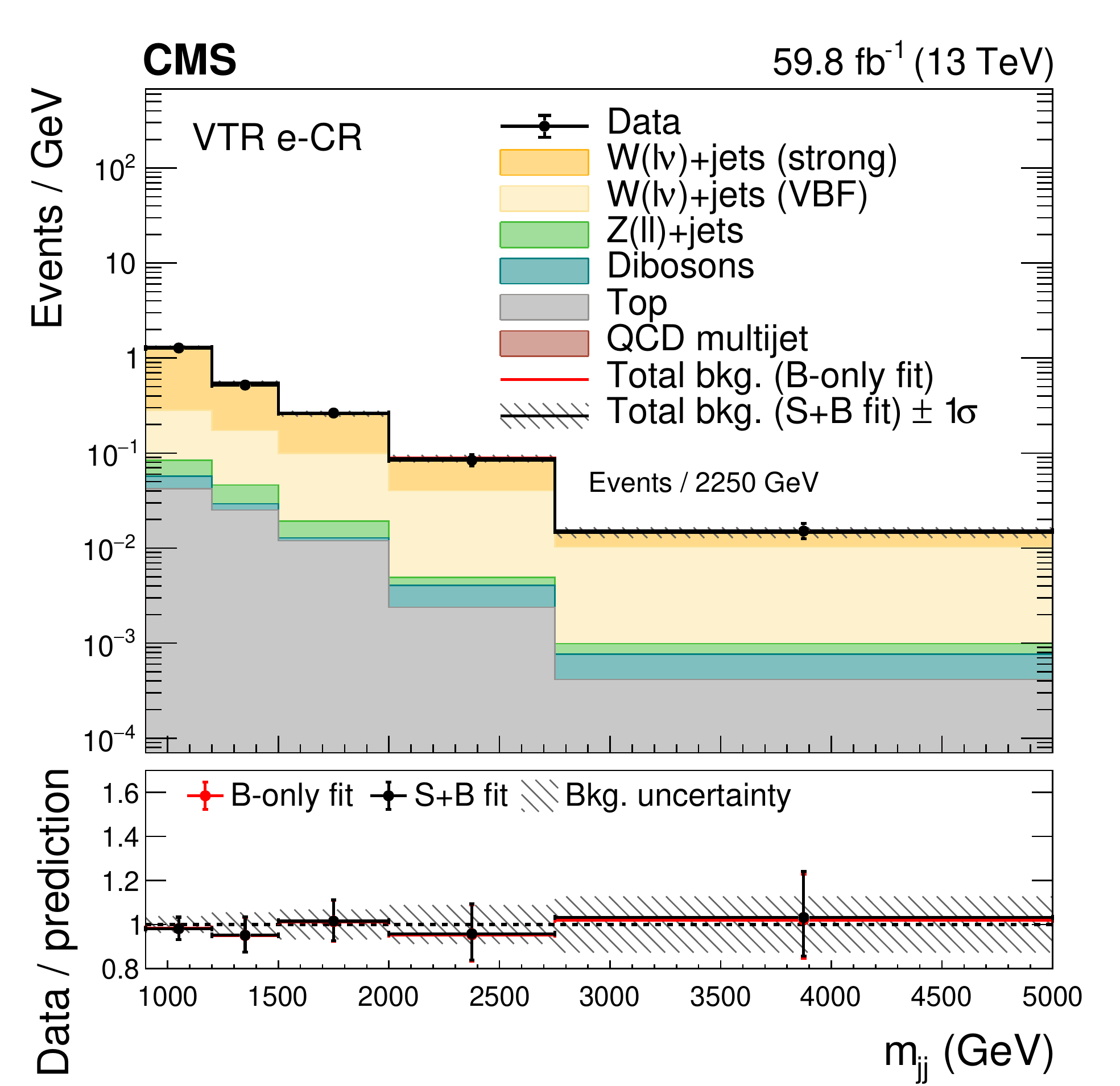}
    \includegraphics[width=0.48\textwidth]{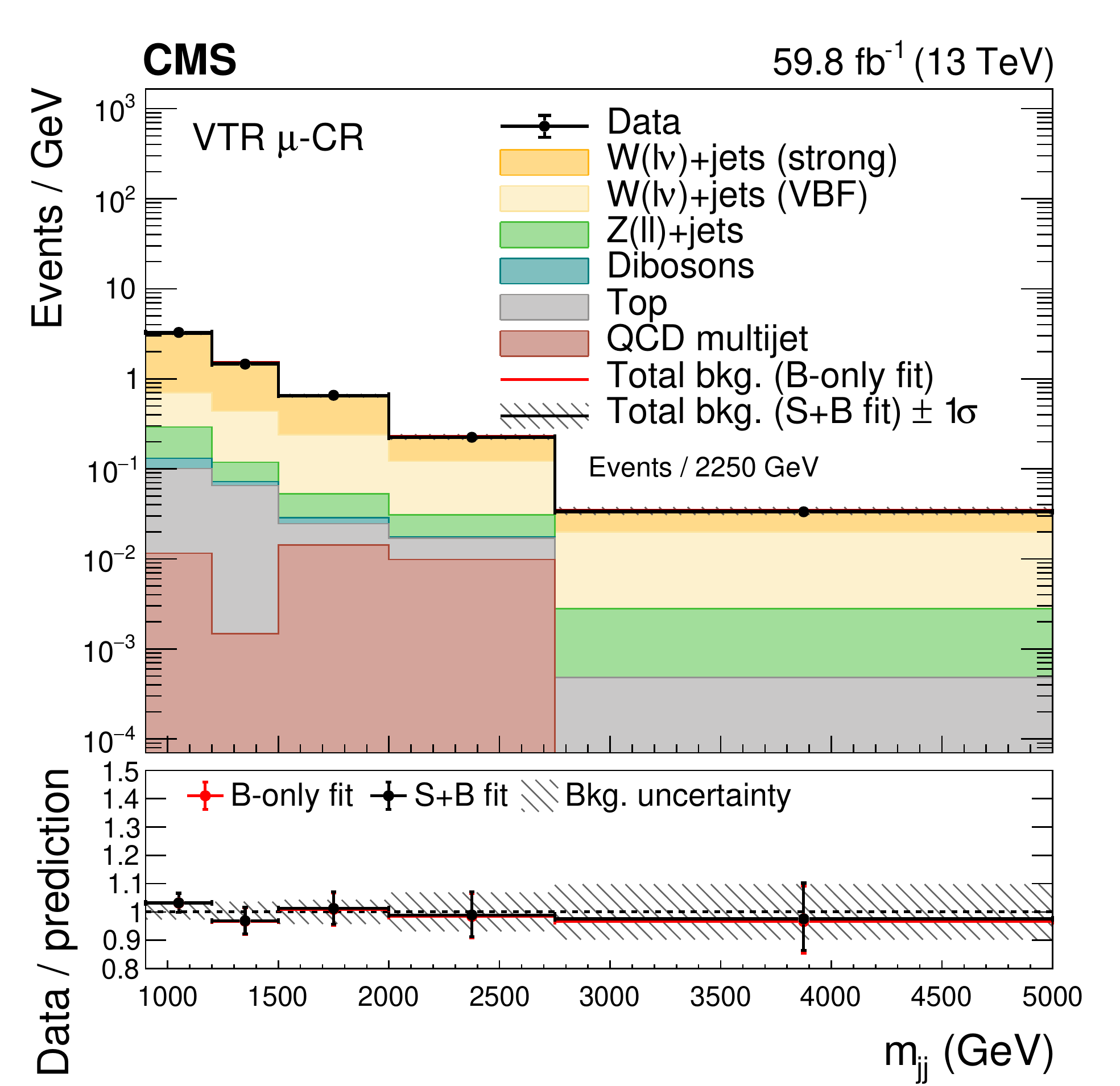}
    \caption{
        The \mjj postfit distributions in the dielectron (upper
        left), dimuon (upper right), single-electron (lower left),
        and single-muon (lower right) CR for the VTR category, with the 2018 sample. The background contributions are estimated from
        a fit to data in the SR and CRs allowing for the signal contribution to vary (S+B fit) and the total background estimated from a fit
        assuming $\brinv=0$ (B-only fit) is also shown.}
    \label{fig:CR_VTR_2018}
\end{figure*}

\begin{figure*}[!htb]
    \centering
    \includegraphics[width=0.48\textwidth]{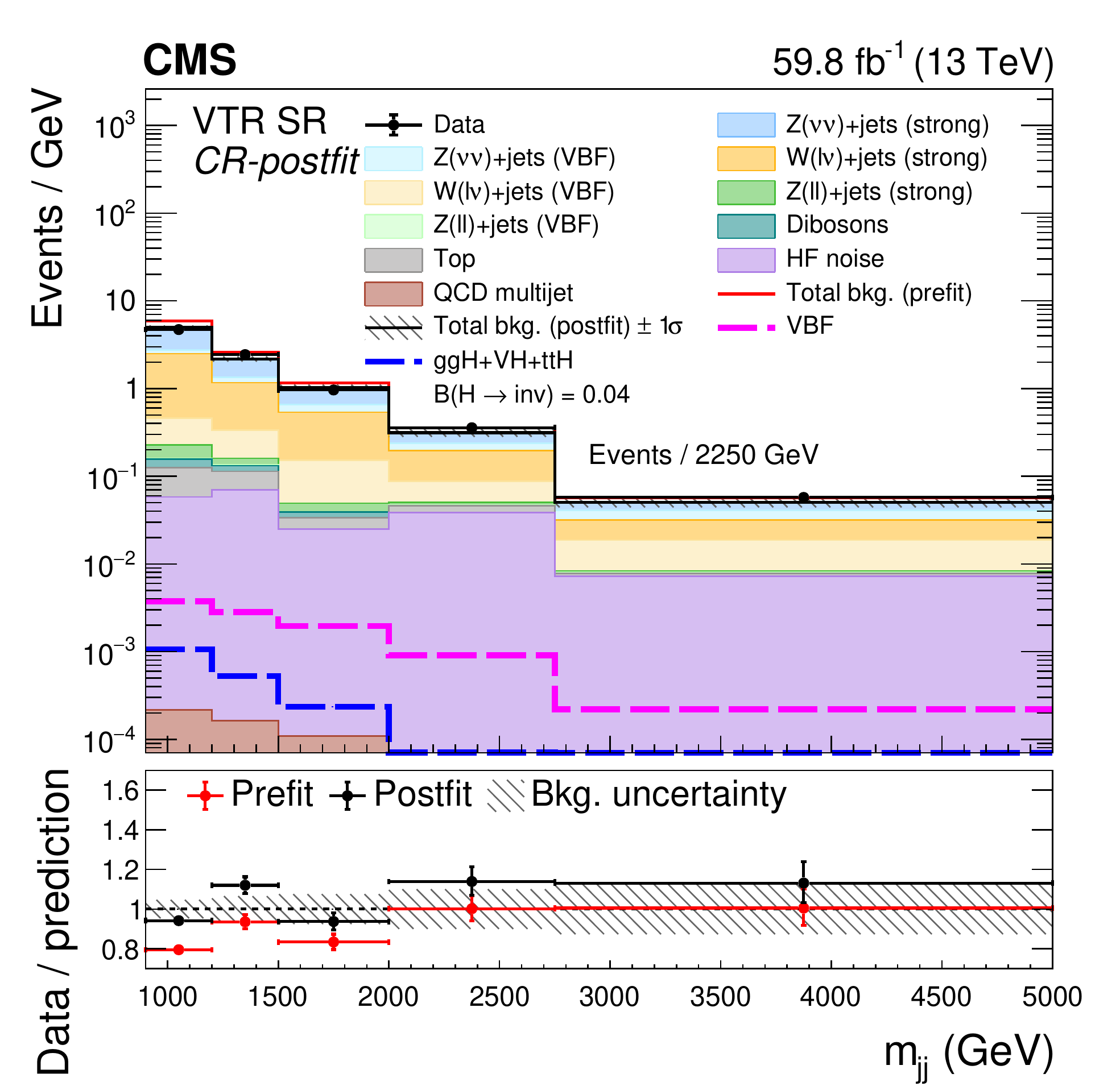}\hfill
    \includegraphics[width=0.48\textwidth]{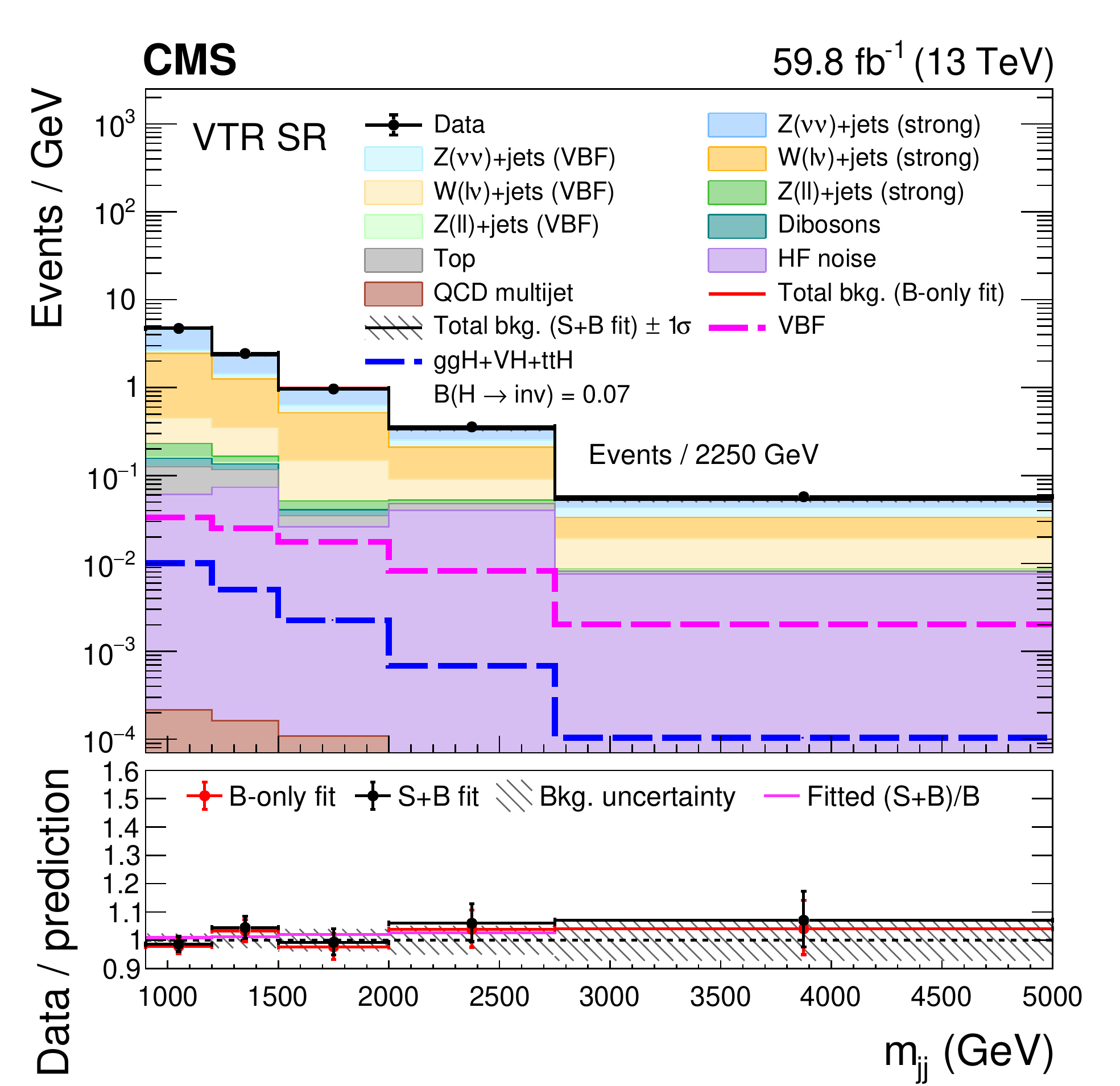}
    \caption{The observed \mjj distribution in the VTR prefit (left) and postfit (right) SR
        compared to the postfit backgrounds, with the 2018 samples. The signal processes are
        scaled by the fitted value of \brinv, shown in the legend.}
    \label{fig:SR_VTR_2018}
\end{figure*}

The results for the VTR 2018 category are shown in Figs.~\ref{fig:CR_VTR_2018_CRonly} to~\ref{fig:SR_VTR_2018} and Table~\ref{tab:yields_VTR_2018}.

\begin{table*}[htb!]
    \centering
    \topcaption[]{
        Expected event yields in each \mjj bin for the different background
        processes in the SR of the VTR category, in the 2018 samples. The
        background yields and the corresponding uncertainties are obtained
        after performing a combined fit across all of the CRs and SR. The
        expected signal contributions for the Higgs boson, produced in the non-\vbf
        and \vbf modes, decaying to invisible particles with a branching
        fraction of $\brinv = 1$, and the observed event yields are also
        reported.  }
    \label{tab:yields_VTR_2018}
    \cmsTable{
        \begin{tabular}{lccccc}
            \hline
            \mjj bin range (\GeVns{})                        & 900--1200       & 1200--1500     & 1500--2000     & 2000--2750     & $>$2750       \\
            \hline
            $\PZ(\PGn\PGn)+\text{jets}$ (strong)             & $632.3\pm11.5$  & $272.2\pm7.7$  & $174.0\pm5.1$  & $62.7\pm2.8$   & $24.5\pm1.5$  \\
            $\PZ(\PGn\PGn)+\text{jets}$ (VBF)                & $74.0\pm3.1$    & $57.9\pm3.6$   & $56.7\pm3.9$   & $34.9\pm3.1$   & $22.4\pm1.9$  \\
            $\PW(\Pell\PGn)+\text{jets}$ (strong)            & $601.8\pm16.1$  & $272.3\pm10.8$ & $187.2\pm8.7$  & $91.2\pm7.6$   & $32.6\pm3.4$  \\
            $\PW(\Pell\PGn)+\text{jets}$ (VBF)               & $66.5\pm5.5$    & $56.0\pm5.2$   & $49.5\pm4.6$   & $29.4\pm3.4$   & $24.9\pm2.6$  \\
            $\ttbar$ + single \PQt quark                     & $19.3\pm1.3$    & $13.0\pm0.8$   & $4.4\pm0.4$    & $5.7\pm0.5$    & $1.4\pm0.2$   \\
            Diboson                                          & $9.3\pm1.1$     & $5.8\pm0.8$    & $2.9\pm0.4$    & $0.2\pm0.1$    & $0.0\pm0.1$   \\
            $\PZ/\Pgg^{*}(\Pell^{+}\Pell^{-})+\mathrm{jets}$ & $22.6\pm0.9$    & $9.0\pm0.4$    & $5.4\pm0.3$    & $3.3\pm0.2$    & $1.0\pm0.1$   \\
            Multijet                                         & $0.1\pm0.1$     & $0.0\pm0.1$    & $0.1\pm0.1$    & $0.0\pm0.1$    & $0.0\pm0.1$   \\
            HF noise                                         & $18.5\pm2.5$    & $22.4\pm3.0$   & $13.3\pm1.8$   & $30.7\pm4.2$   & $17.3\pm2.3$  \\ [\cmsTabSkip]

            $\Pg\Pg\PH(\to \mathrm{inv})$                    & 41.4            & 20.7           & 15.4           & 6.9            & 3.2           \\
            $\PQq\PQq\PH(\to \mathrm{inv})$                  & 147.1           & 110.6          & 128.5          & 90.2           & 66.6          \\
            $\PW\PH(\to \mathrm{inv})$                       & 0.5             & 0.1            & 0.3            & 0.1            & 0.0           \\
            $\PQq\PQq\PZ\PH(\to \mathrm{inv})$               & 0.1             & 0.0            & 0.0            & 0.1            & 0.0           \\
            $\Pg\Pg\PZ\PH(\to \mathrm{inv})$                 & 0.3             & 0.1            & 0.1            & 0.0            & 0.0           \\
            $\PQt\PQt\PH(\to \mathrm{inv})$                  & 0.0             & 0.0            & 0.0            & 0.0            & 0.0           \\ [\cmsTabSkip]

            Total bkg.                                       & $1444.4\pm21.0$ & $708.5\pm15.0$ & $493.5\pm11.9$ & $258.1\pm10.2$ & $124.0\pm5.5$ \\ [\cmsTabSkip]

            Observed                                         & 1413            & 732            & 482            & 268            & 129           \\
            \hline
        \end{tabular}
    }\end{table*}

\cmsClearpage
\subsection*{Separated \PW/\PZ ratios for electron and muon channels}

\begin{figure*}[!htb]
    \centering
    \includegraphics[width=0.48\textwidth]{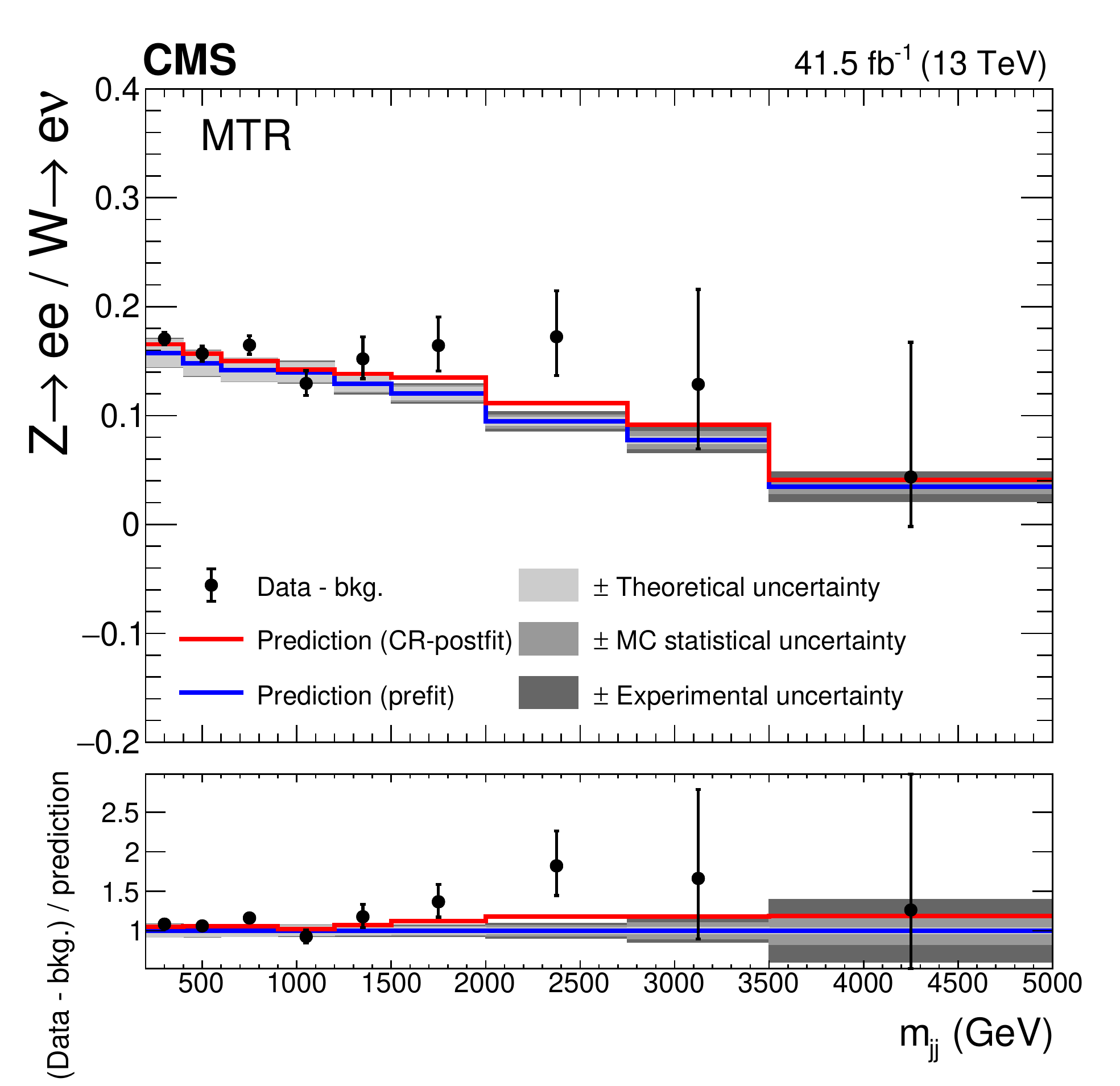}
    \includegraphics[width=0.48\textwidth]{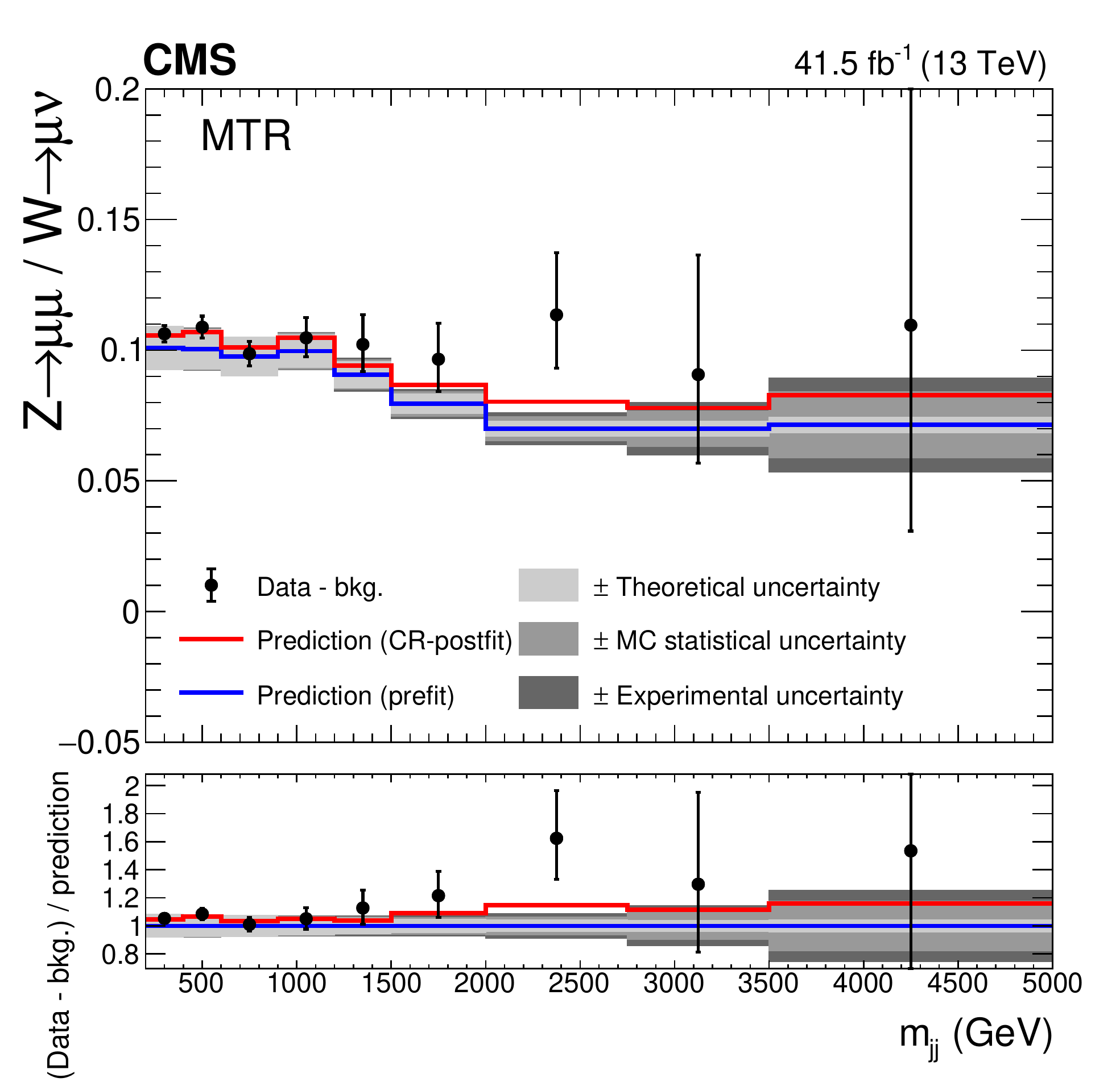}\\
    \centering
    \includegraphics[width=0.48\textwidth]{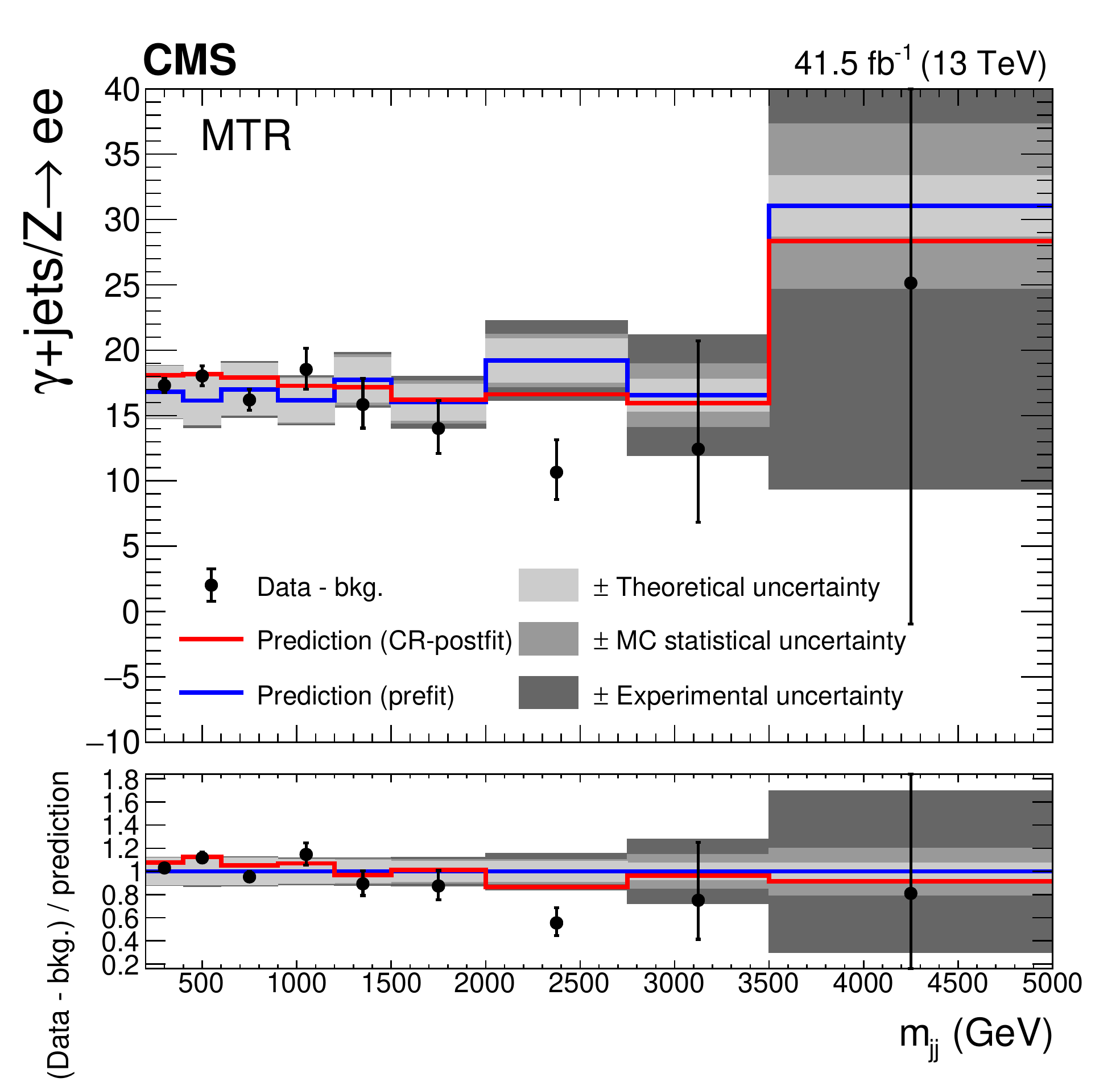}
    \includegraphics[width=0.48\textwidth]{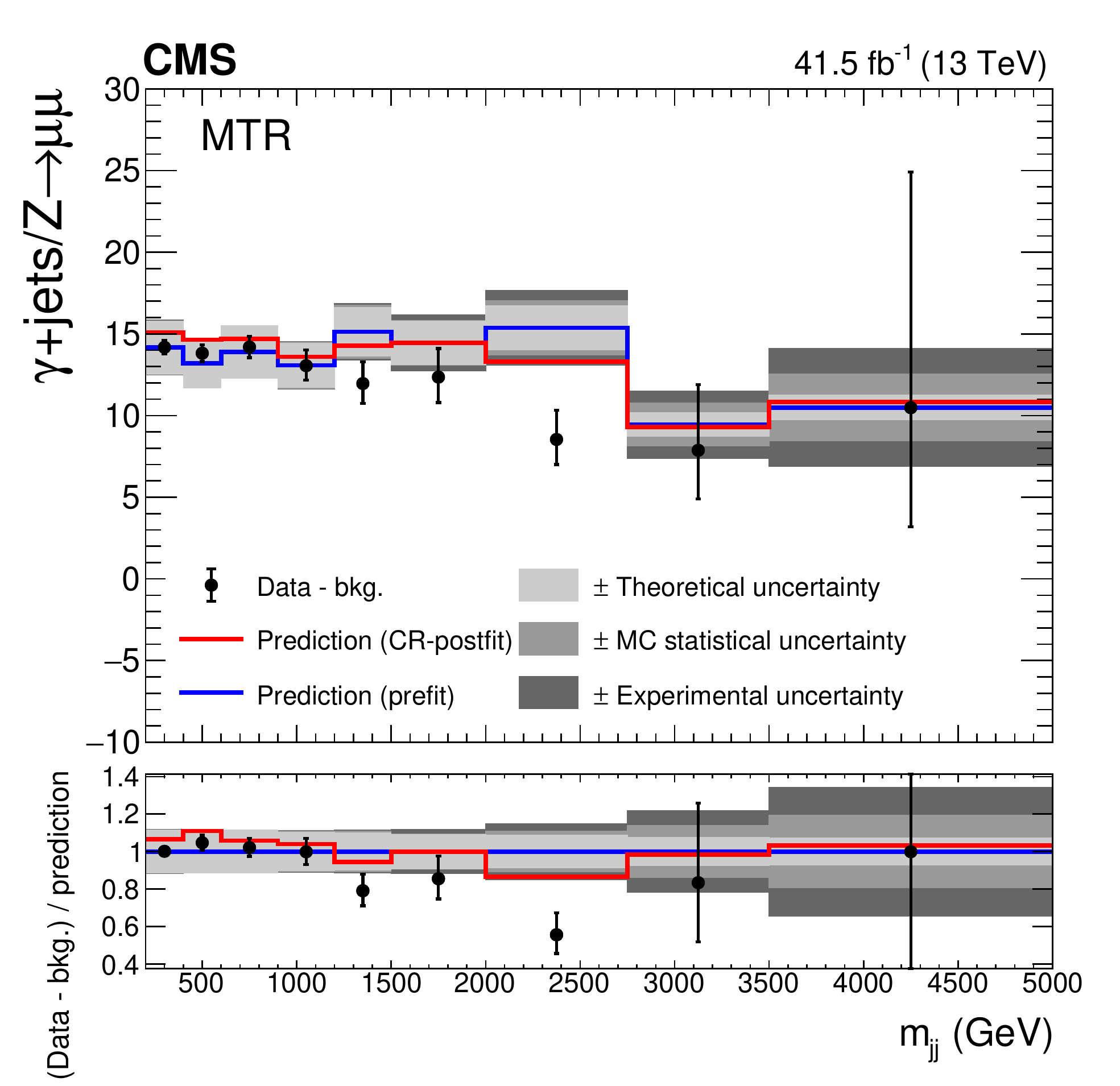}

    \caption{Comparison between data and simulation for the \Zeejets/\Wevjets (upper left), \Zmmjets/\Wmvjets (upper right), \phojets/\Zeejets  (lower left), and  \phojets/\Zmmjets (lower right)  prefit and CR-postfit ratios, as functions of \mjj, for the MTR category 2017 samples. The minor backgrounds in each CR are subtracted from the data using estimates from simulation. The grey bands include the theoretical and experimental systematic uncertainties listed in Table~\tabsysts, as well as the statistical uncertainty in the simulation.}
    \label{fig:WZTF_MTR_2017_eandmu}
\end{figure*}

\begin{figure*}[!htb]
    \centering
    \includegraphics[width=0.48\textwidth]{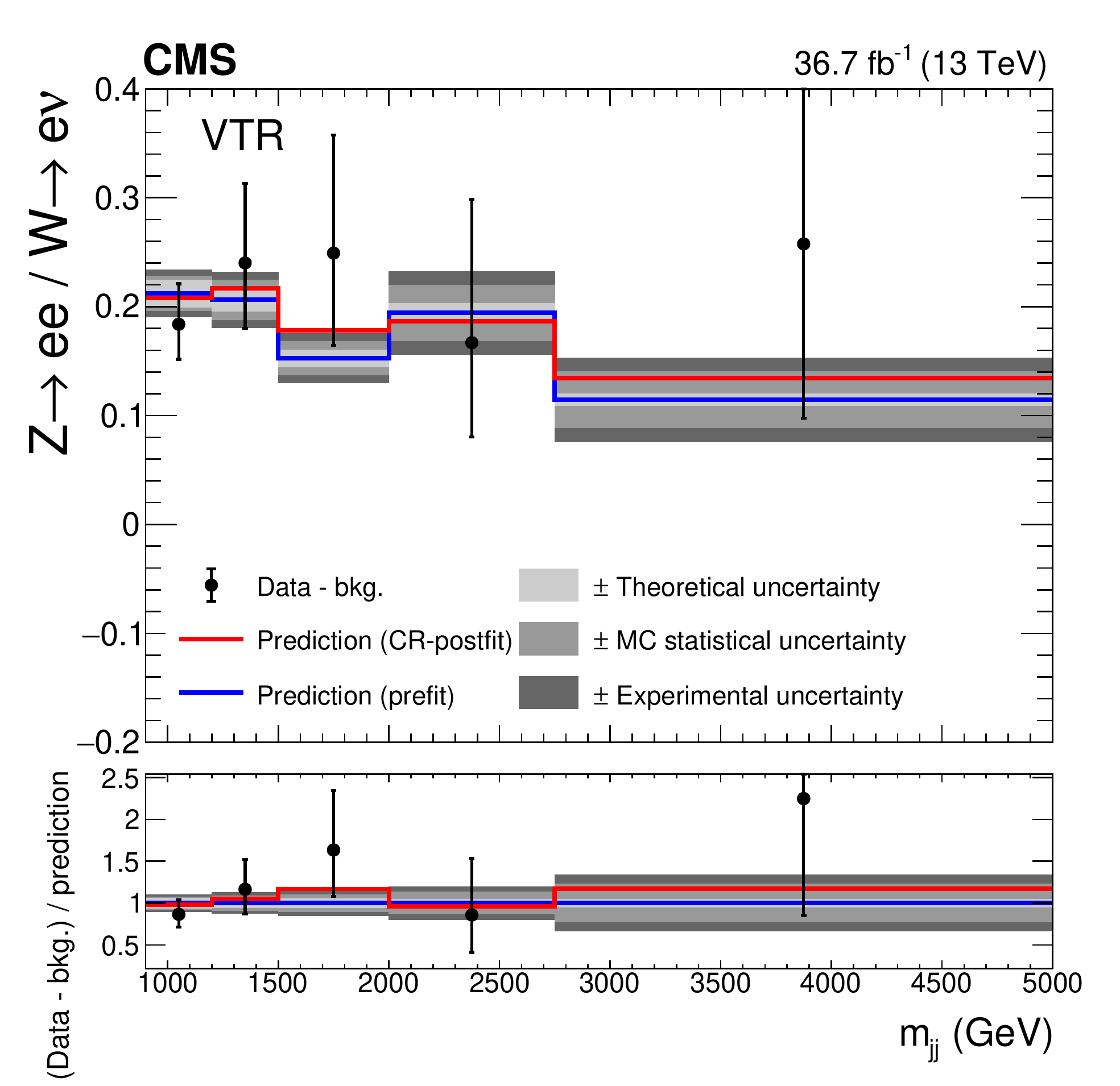}\hfill
    \includegraphics[width=0.48\textwidth]{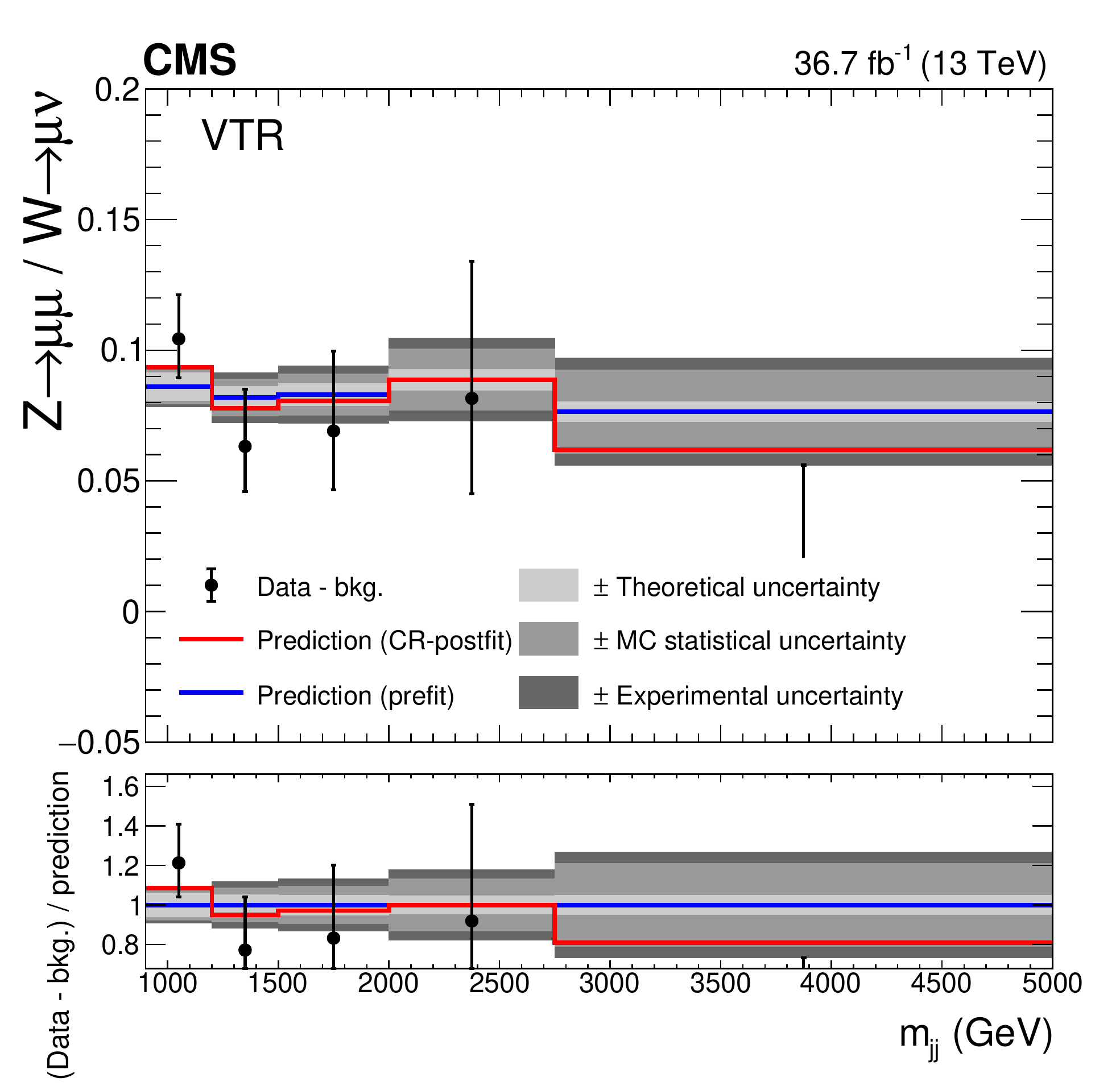}
    \caption{Comparison between data and simulation for the \Zeejets/\Wevjets  (left) and \Zmmjets/\Wmvjets (right) prefit and CR-postfit ratios, as functions of \mjj, for the VTR category 2017 samples. The minor backgrounds in each CR are subtracted from the data using estimates from simulation. The grey bands include the theoretical and experimental systematic uncertainties listed in Table~\tabsysts, as well as the statistical uncertainty in the simulation.}
    \label{fig:WZTF_VTR_2017_eandmu}
\end{figure*}

\begin{figure*}[!htb]
    \centering
    \includegraphics[width=0.48\textwidth]{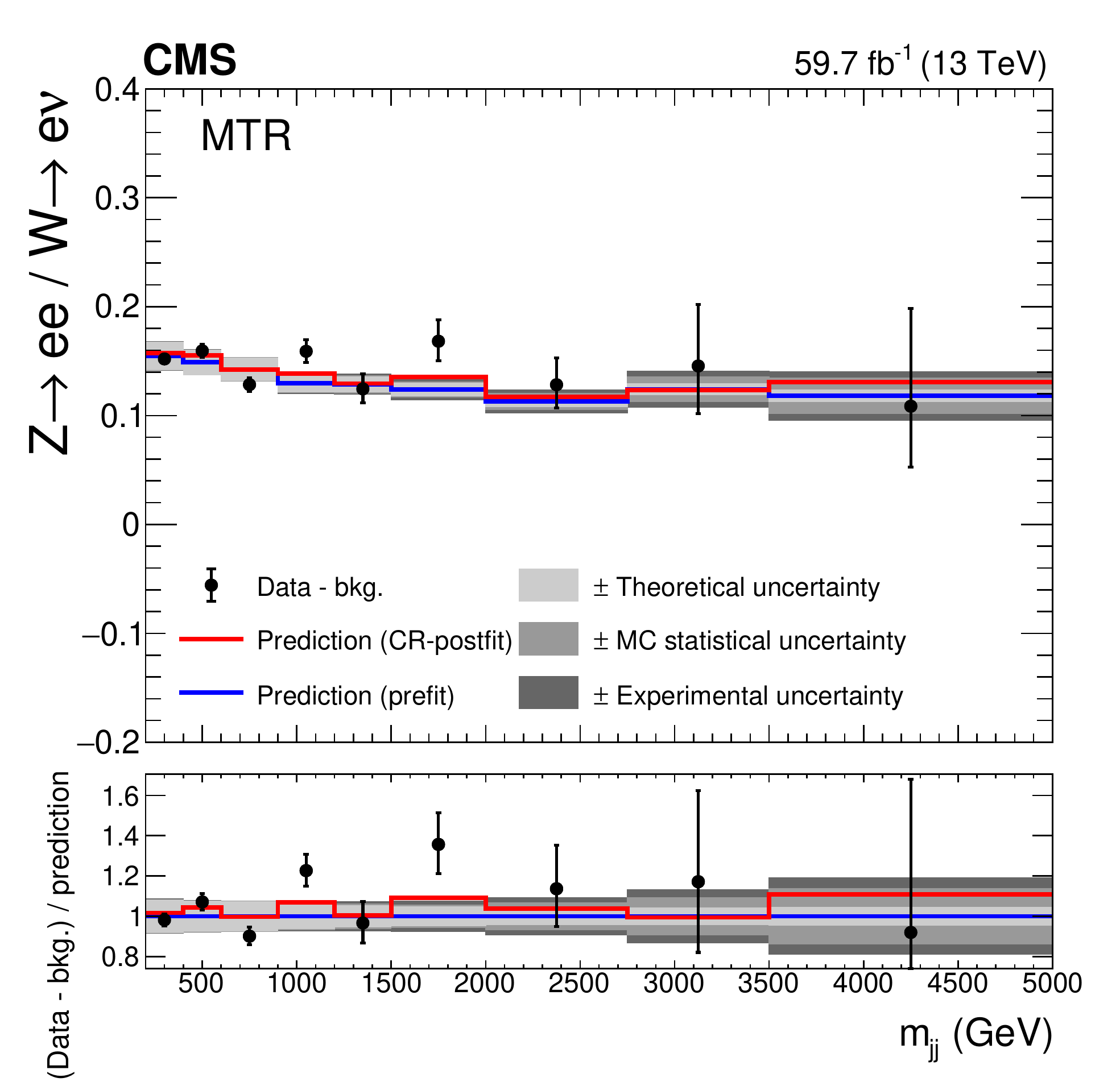}
    \includegraphics[width=0.48\textwidth]{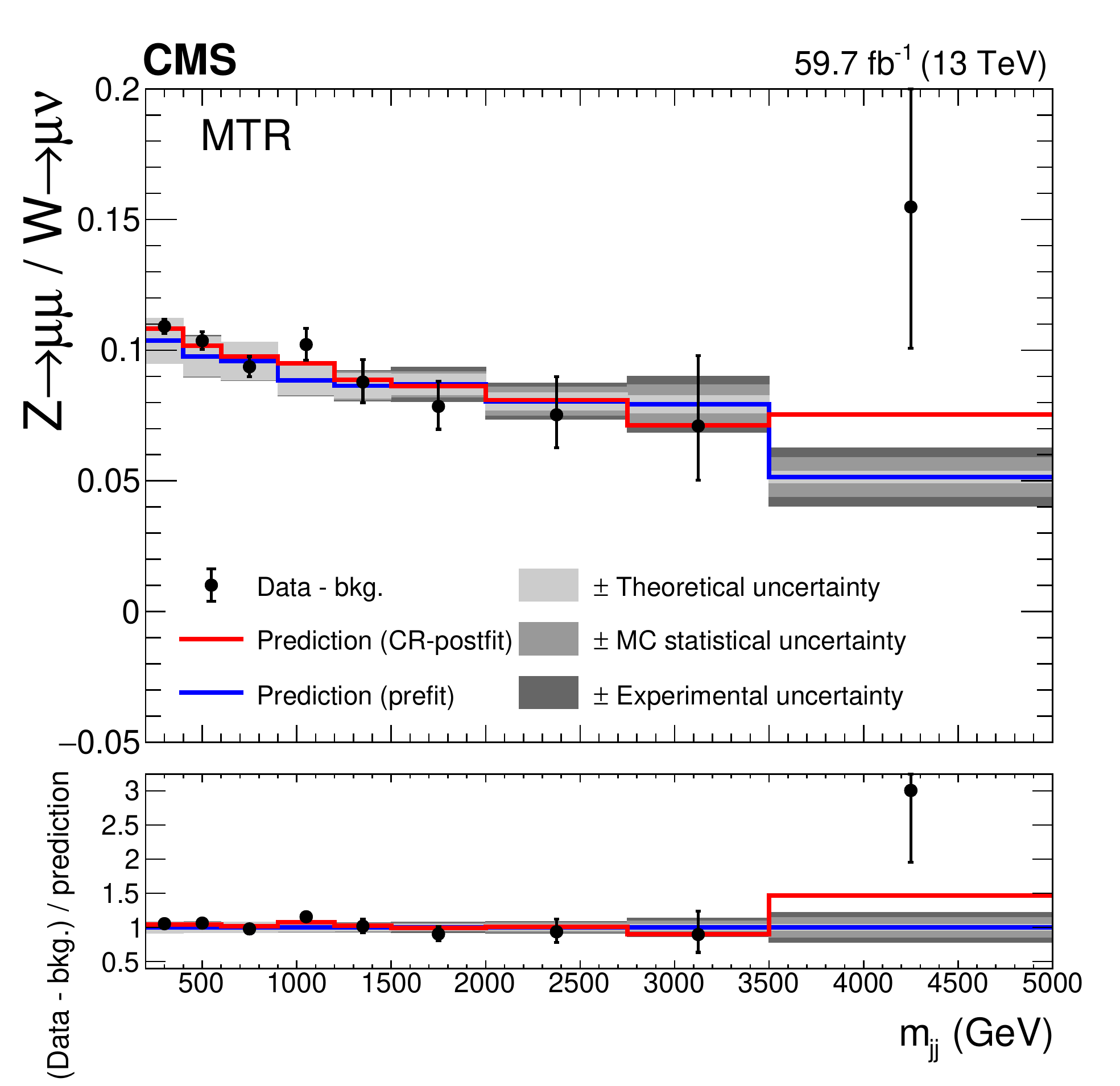}\\
    \centering
    \includegraphics[width=0.48\textwidth]{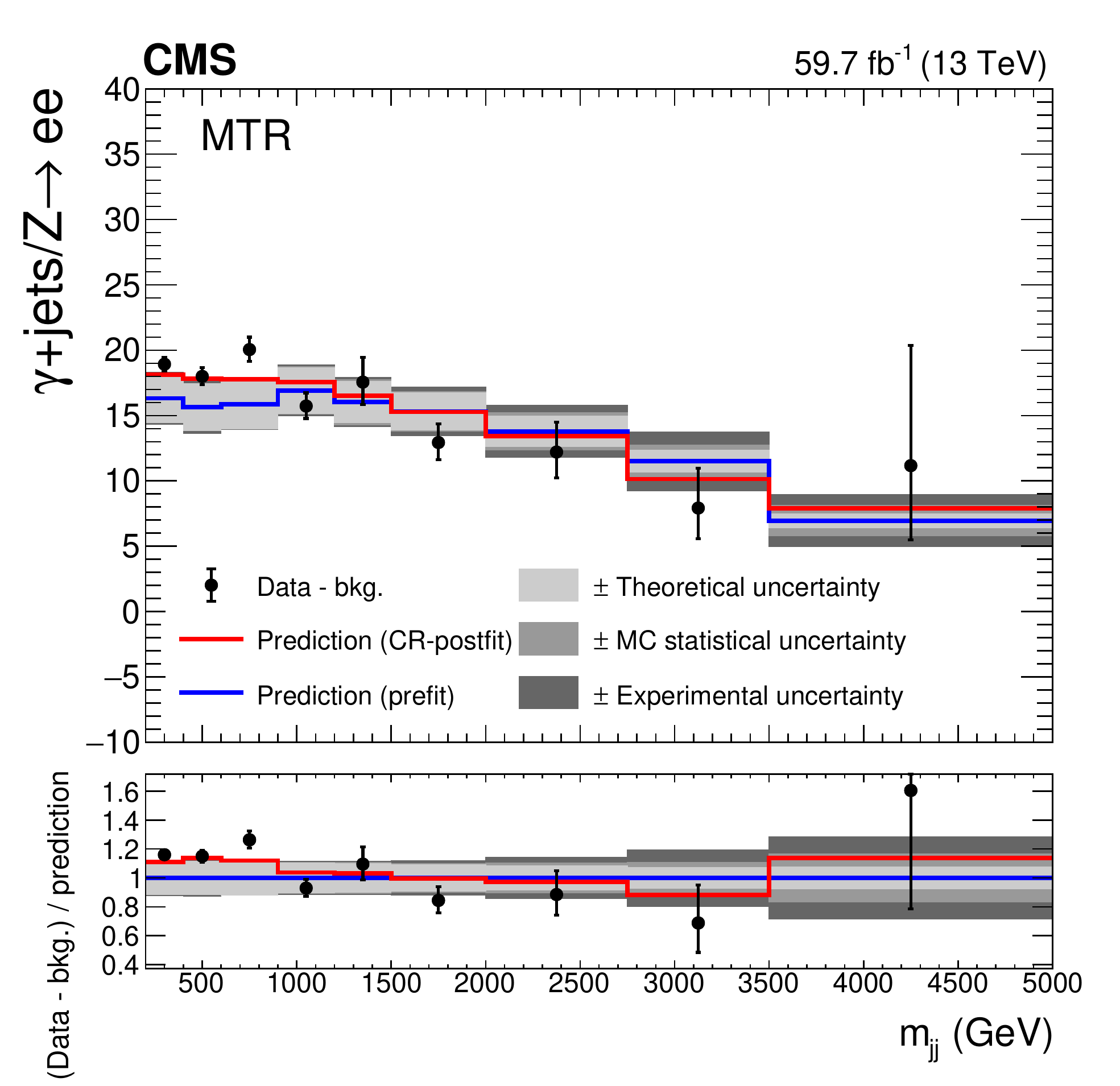}
    \includegraphics[width=0.48\textwidth]{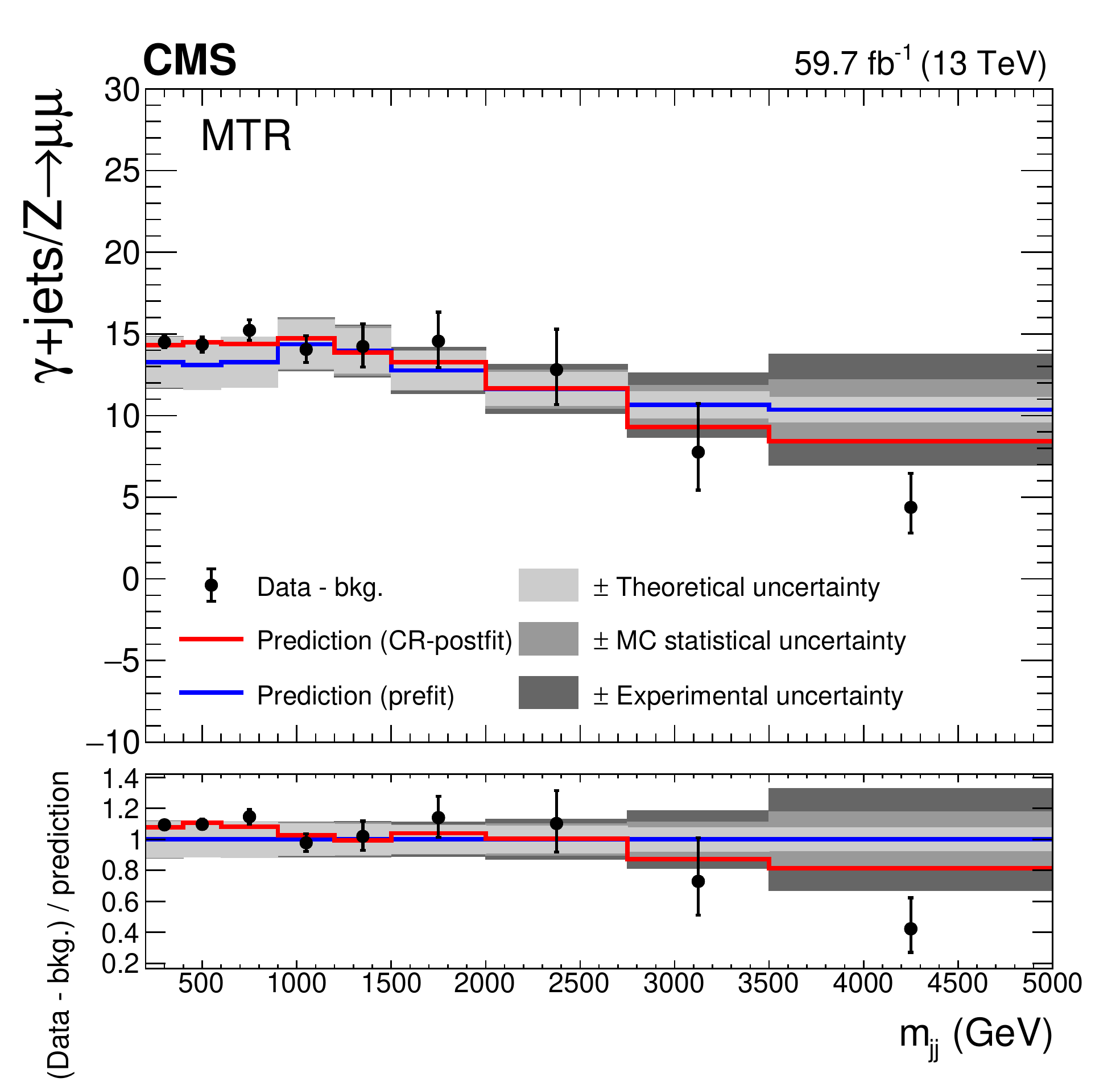}
    \caption{Comparison between data and simulation for the \Zeejets/\Wevjets (upper left), \Zmmjets/\Wmvjets (upper right), \phojets/\Zeejets (lower left), and \phojets/\Zmmjets  (lower right) prefit and CR-postfit ratios, as functions of \mjj, for the MTR category 2018 samples. The minor backgrounds in each CR are subtracted from the data using estimates from simulation. The grey bands include the theoretical and experimental systematic uncertainties listed in Table~\tabsysts, as well as the statistical uncertainty in the simulation.}
    \label{fig:WZTF_MTR_2018_eandmu}
\end{figure*}

\begin{figure*}[!htb]
    \centering
    \includegraphics[width=0.48\textwidth]{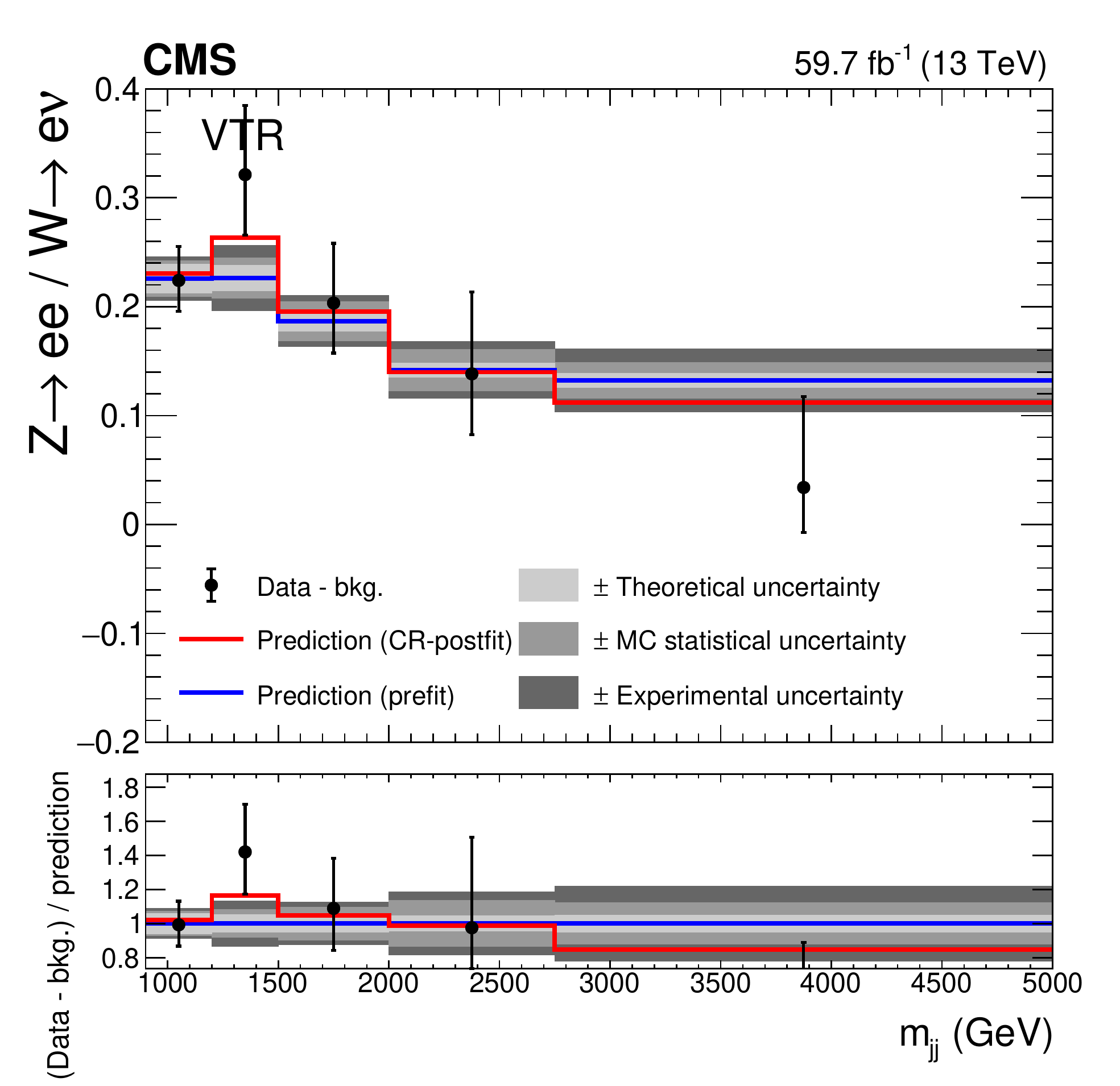}\hfill
    \includegraphics[width=0.48\textwidth]{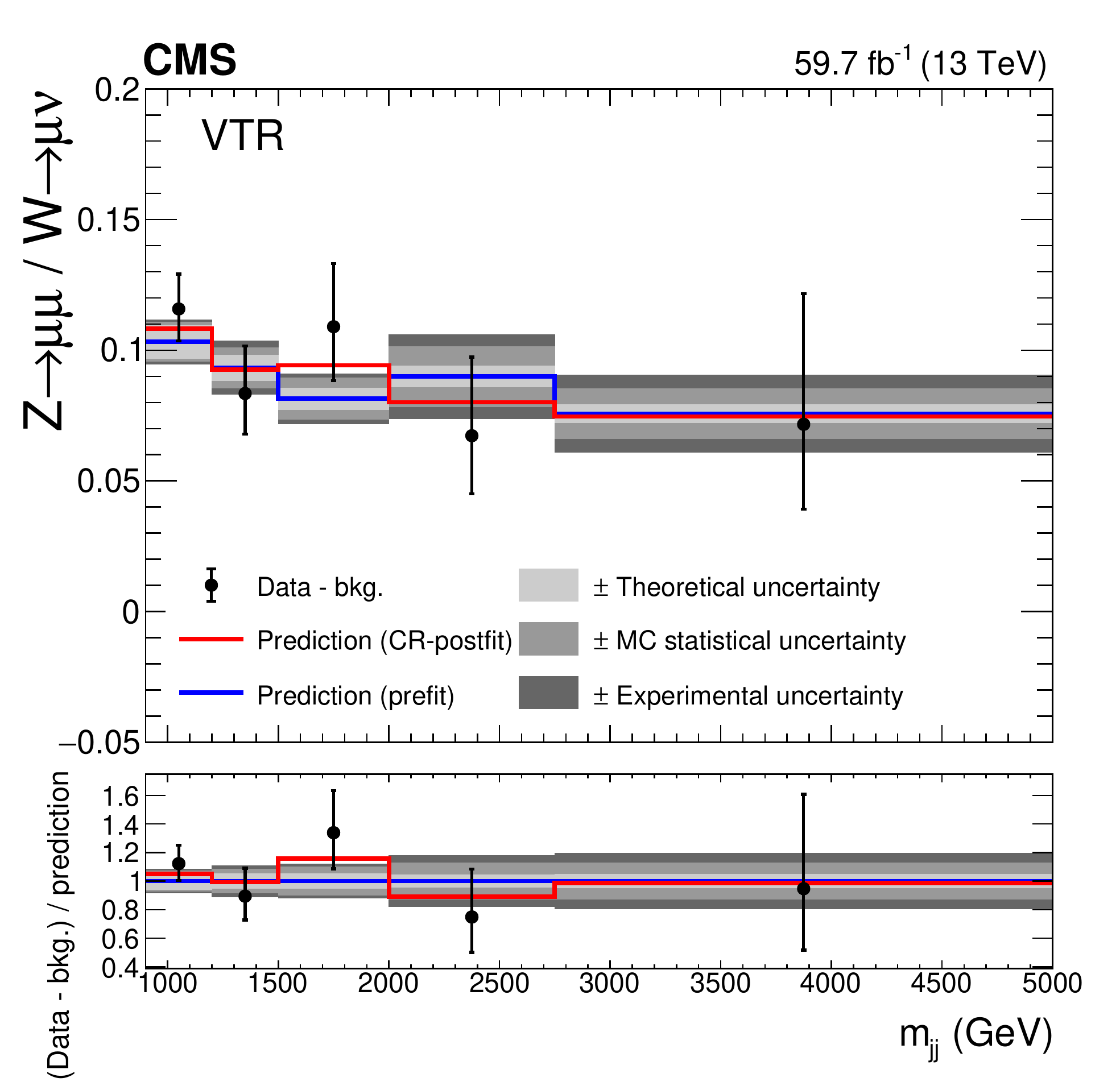}
    \caption{Comparison between data and simulation for the \Zeejets/\Wevjets  (left) and  \Zmmjets/\Wmvjets (right) prefit and CR-postfit ratios, as functions of \mjj, for the VTR category 2018 samples. The minor backgrounds in each CR are subtracted from the data using estimates from simulation. The grey bands include the theoretical and experimental systematic uncertainties listed in Table~\tabsysts, as well as the statistical uncertainty in the simulation.}
    \label{fig:WZTF_VTR_2018_eandmu}
\end{figure*}

Finally, separated \PW/\PZ ratios for electron and muon channels are shown in Figs.~\ref{fig:WZTF_MTR_2017_eandmu} and~\ref{fig:WZTF_VTR_2017_eandmu} for the 2017 samples, and in Figs.~\ref{fig:WZTF_MTR_2018_eandmu} and~\ref{fig:WZTF_VTR_2018_eandmu} for the 2018 samples.

%% file: HIG-20-003-authorlist.tex
\cmsinstitute{Yerevan~Physics~Institute, Yerevan, Armenia}
A.~Tumasyan
\cmsinstitute{Institut~f\"{u}r~Hochenergiephysik, Vienna, Austria}
W.~Adam\cmsorcid{0000-0001-9099-4341}, J.W.~Andrejkovic, T.~Bergauer\cmsorcid{0000-0002-5786-0293}, S.~Chatterjee\cmsorcid{0000-0003-2660-0349}, K.~Damanakis, M.~Dragicevic\cmsorcid{0000-0003-1967-6783}, A.~Escalante~Del~Valle\cmsorcid{0000-0002-9702-6359}, R.~Fr\"{u}hwirth\cmsAuthorMark{1}, M.~Jeitler\cmsAuthorMark{1}\cmsorcid{0000-0002-5141-9560}, N.~Krammer, L.~Lechner\cmsorcid{0000-0002-3065-1141}, D.~Liko, I.~Mikulec, P.~Paulitsch, F.M.~Pitters, J.~Schieck\cmsAuthorMark{1}\cmsorcid{0000-0002-1058-8093}, R.~Sch\"{o}fbeck\cmsorcid{0000-0002-2332-8784}, D.~Schwarz, S.~Templ\cmsorcid{0000-0003-3137-5692}, W.~Waltenberger\cmsorcid{0000-0002-6215-7228}, C.-E.~Wulz\cmsAuthorMark{1}\cmsorcid{0000-0001-9226-5812}
% ...
\vskip\cmsinstskip
% ...
1:~Also at TU Wien, Wien, Austria\\
% ...
